\documentclass[12pt,final,twoside]{book}
\usepackage[a4paper,		%
  left		= 2.5cm,		
  right		= 2.5cm,		
  bindingoffset	= 1.0cm,	
  top		= 3.0cm,		
  bottom	= 3.0cm,		
  includehead	= true,		
  includefoot	= false,	
  showframe	= false			%
]{geometry}					%
\usepackage{amsmath,amsfonts,amsthm}
\usepackage[OT4]{fontenc}		
\usepackage[utf8]{inputenc}		
\usepackage[english]{babel}		%
\usepackage{graphicx}        	
\usepackage{subfigure}			
\usepackage{hyperref}
\usepackage{placeins}

\begin{document}
\begin{titlepage}
  \begin{center}
    \vspace{5.0cm}
    {\Huge Collective phenomena in the early stages of relativistic heavy-ion collisions}\\
    \vspace{3.0cm}
    {\Large \bf Rados\l aw Ryblewski}\\
    \vspace{2.0cm}
    {\large
      The Henryk Niewodnicza\'{n}ski\\
      Institute of Nuclear Physics\\
      Polish Academy of Sciences\\
      Krak\'{o}w, Poland\\
    }
    \vspace{1.0cm}
    \includegraphics[width=0.4 \textwidth]{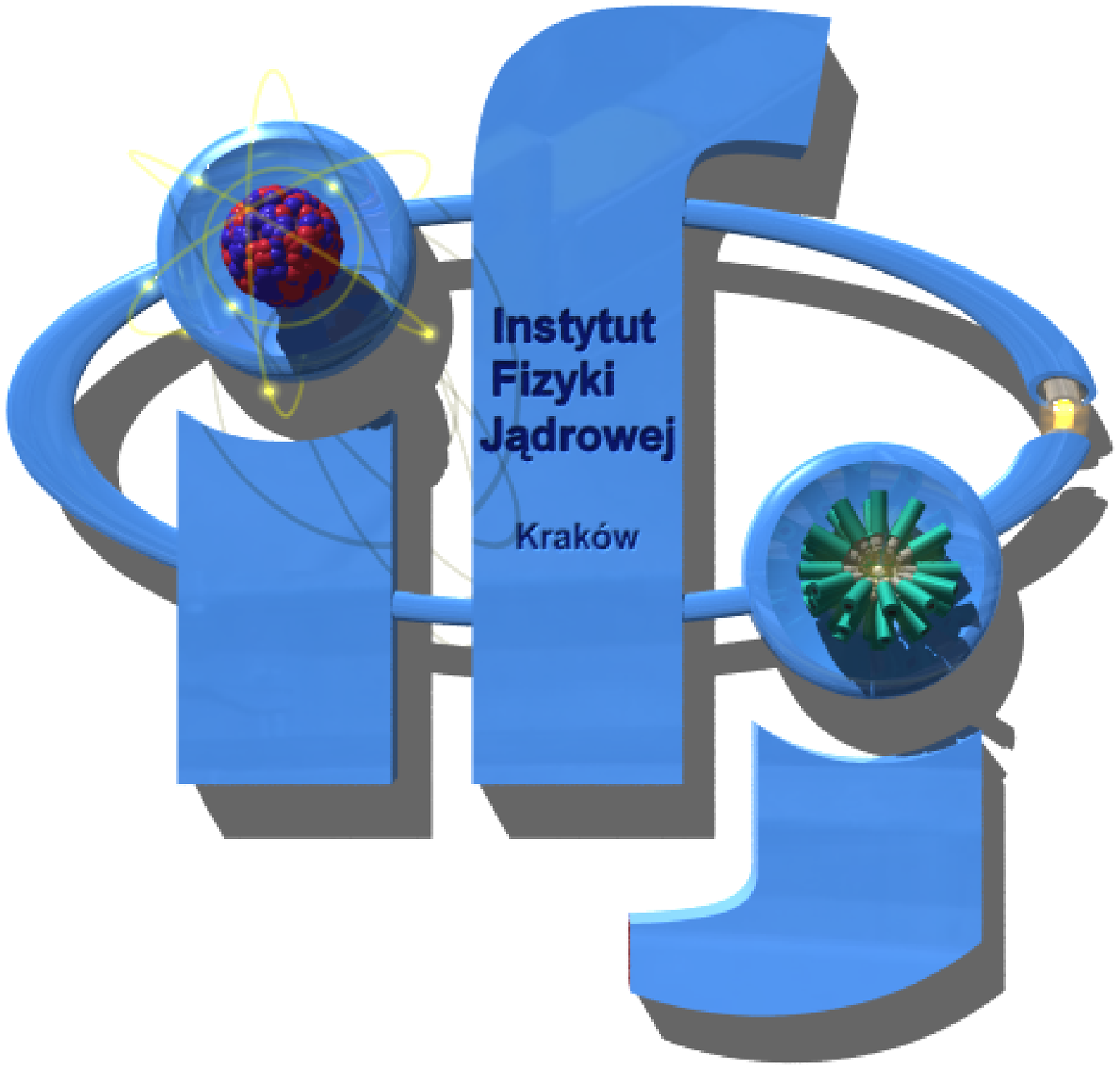}\\
    \vspace{2.0cm}
    {\it Thesis submitted for the Degree of Doctor of Philosophy in Physics}\\
    {\it Prepared under the supervision of Prof. Wojciech Florkowski}\\
    \vspace{2.0cm}
    {Krak\'{o}w, February 2012}
  \end{center}
\end{titlepage}
\pagebreak
\cleardoublepage
\begin{center}
  {\bf \large STRESZCZENIE}
  \end{center}
\par W niniejszej pracy przedstawiono nowy model hydrodynamiczny, \texttt{ADHYDRO} (skrót od \textit{highly-Anisotropic and strongly-Dissipative HYDROdynamics}), stanowiący uogólnienie standardowej hydrodynamiki płynu doskonałego. Umożliwia on efektywny opis zachowania początkowo silnie anizotropowych układów cząstek produkowanych w skrajnie relatywistycznych zderzeniach ciężkich jonów. Model ten oparty jest na przesłankach pochodzących z mikroskopowych modeli początkowych stadiów ewolucji materii, które sugerują, że produkowana materia wykazuje silną anizotropię w przestrzeni pędów. 
\par Bazując na ogólnej postaci funkcji rozkładu przeprowadzono analizę własności termodynamicznych lokalnie anizotropowych układów cząstek. W wyniku tej anali- zy zaproponowano postać uogólnionego równania stanu umożliwiającego integralny opis wielu stadiów ewolucji materii, w którym izotropizacja układu jest opisywana w sposób ciągły. Następnie wyprowadzono ogólną postać równań opisujących ewolucję silnie anizotropowego płynu. Zaproponowano również postać źródła entropii związa- nego z mechanizmem termalizacji układu. W najprostszym przypadku ewolucji jednowymiarowej w kierunku osi wiązki oraz przy założeniu symetrii układu względem pchnięć Lorentza przedyskutowano szereg własności modelu. W szczególności, w granicy małych anizotropii, wykazano zgodność modelu z teorią Israela-Stewarta. 
\par Model \texttt{ADHYDRO} po zintegrowaniu z istniejącym modelem statystycznej hadronizacji \texttt{THERMINATOR} został użyty do systematycznej analizy danych eksperymentalnych pochodzących z eksperymentów prowadzonych na akceleratorze RHIC (Relativistic Heavy Ion Collider) w Brookhaven National Laboratory przy energii \mbox{$\sqrt{s_{\rm NN}}=200$} GeV. W ramach modelu uzyskano opis jedno- i dwucząstkowych obserwabli w zakresie miękkiej fizyki (wartości pędu poprzecznego cząstek nieprzekraczające $3$ GeV). W szczególności przedyskutowano widma cząstek w pędzie poprzecznym, współczynniki przepływu eliptycznego $v_2$ i przepływu ukierunkowanego $v_1$ oraz promienie korelacyjne HBT identycznych pionów. 
\par Przy założeniu istnienia symetrii pchnięć Lorentza w kierunku osi wiązki w ramach \texttt{ADHYDRO} wykazano niewielką czułość obserwabli w obszarze centralnych rapidity na wartości parametrów charakteryzujących początkową fazę anizotropową. Tym samym potwierdzono hipotezę dotyczącą występowania uniwersalnego przepływu poprzecznego w zderzeniach ciężkich jonów bez względu na anizotropię ciśnień tensora energii-pędu układu.
\par Ostatecznie, rezygnując z jakichkolwiek symetrii, zastosowano model \texttt{ADHYDRO} w najogólniejszej (3+1) wymiarowej postaci do opisu danych poza obszarem centralnych rapidity. Ponownie zaobserwowano niewielką czułość obserwabli na szczegóły fazy anizotropowej. Bazując na analizie rezultatów modelu uzyskano oszacowanie na długość fazy anizotropowej na około $1$ fm. Wynik ten dopuszcza zatem istnienie silnych anizotropii w początkowych fazach ewolucji materii produkowanej w skrajnie relatywistycznych zderzeniach ciężkich jonów. Jednocześnie rezultat ten jest zgodny z przewidywaniami wielu obecnie stosowanych modeli mikroskopowych.

%
\cleardoublepage
\begin{center}
  \bf \Large Acknowledgments
\end{center}
\par I would like to thank my supervisor Prof.~Wojciech Florkowski for his patient guidance, encouragement and support. I have been very fortunate to work under his supervision, and I thank him sincerely for his invaluable help and advice that he has provided during the course of this Thesis.
\par I would also like to express my appreciation to all the excellent people of the Department of Theory of Structure of Matter (NZ41), in particular Wojciech Broniowski, Piotr Bożek and Miko\l aj Chojnacki for their help, support and guidance. 
\par Last but not least, I wish to thank all of my family and friends, without whose support, patience and
encouragement I could not have achieved so much.
\vspace{1cm}
\par Research was supported by the Polish Ministry of Science and Higher Education
grant N N202 288638 (2010-2012).

\tableofcontents
\chapter{Introduction}
\label{chapter:intro}
\section[Searching for new states of matter]{Searching for new states \\ of strongly interacting matter}
\label{chapter:intro_search}

Relativistic heavy-ion physics is an interdisciplinary research field connecting \textit{nuclear physics} and \textit{elementary particle physics}. It emerged in the middle of 1980s, when the relativistic beams of heavy ions became available at the Brookhaven National Laboratory and at CERN. The fast development of this field was triggered by a broad interest in the experimental and theoretical studies of properties of strongly interacting matter in the wide range of thermodynamic parameters such as temperature, $T$, and baryon chemical potential, $\mu_B$.

In the framework of \textit{Quantum Chromodynamics} (QCD), which is the fundamental non-abelian gauge field theory of strong interactions, one expects that the strongly interacting matter undergoes {\it deconfinement} and {\it chiral} phase transitions. The experimental verification of those phase transitions is the main aim of the experimental heavy-ion programs. 

In the deconfinement phase transition, the hadronic matter changes into a system of interacting quarks and gluons. The initial theoretical concepts,  based on the phenomenon of the \textit{asymptotic freedom} (discovered by Gross, Politzer and Wilczek) and \textit{color screening}, indicated that this system was weakly interacting --- the asymptotic freedom predicts that the interaction strength between color charged partons (\textit{quarks} and \textit{gluons}) vanishes at short distances. 

Starting from those ideas, Collins and Perry formulated a hypothesis of existence of a new quark state of matter in compressed super-dense star interiors. Also at that time, the limiting temperature introduced by Hagedorn was reinterpreted by Cabibbo and Parisi as the temperature defining the \textit{phase transition} between hadronic and quark matter. In 1975 they drew the first \textit{phase diagram} of strongly interacting matter. It indicated that at high temperature and/or large baryon density the hadronic matter should pass a phase transition to a new state of matter. This new phase was termed by Shuryak a \textit{quark-gluon plasma} (QGP). 

Besides the deconfinement phase transition, at high temperature and/or large baryon density one deals also with the restoration of chiral symmetry, i.e., with the {\it chiral phase transition}. The order parameter of the chiral phase transition is the quark condensate. The standard QCD vacuum is characterized by a non-vanishing value of the quark condensate, which is a consequence of spontaneous breaking of chiral symmetry at $T=\mu_B=0$. If the system passes the chiral phase transition, the quark condensate drops practically to zero.

It was quickly realized that the verification of theoretical predictions concerning properties of strongly interacting matter may be done in laboratory conditions by performing collisions of heavy atomic nuclei at ultra-relativistic energies. In 1980s, various heavy-ion experiments started, collecting data with the required relativistic beam energies. First experiments took place in 1986 at the Alternating Gradient Synchrotron (AGS) at the Brookhaven National Laboratory (BNL) and simultaneously at the Super Proton Synchrotron (SPS) at the European Organization for Nuclear Research (CERN). In 1992 AGS performed $^{197}$Au+$^{197}$Au collisions at the center-of-mass energy per nucleon pair $\sqrt{s_{\rm NN}} = 5$ GeV. Soon, in 1995 at CERN, $^{208}$Pb+$^{208}$Pb events at $\sqrt{s_{\rm NN}} = 17$ GeV were collected. The new XXI century started with the construction of the Relativistic Heavy-Ion Collider (RHIC) at BNL, which was expected to reach the threshold for the phase transition to the QGP phase. Eventually, it has achieved the energy of $\sqrt{s_{\rm NN}} = 200$ GeV, colliding $^{197}$Au beams. In 2010, the largest experimental facility ever built, the Large Hadron Collider (LHC) at CERN, designed for colliding $^{208}$Pb beams, started to collect data at the energy of $\sqrt{s_{\rm NN}} = 2.76$ TeV. 

The experimental results have significantly broadened our general knowledge of the properties of strongly interacting matter. First of all, the multi-particle systems produced in high-energy collisions of heavy ions turn out to be very much different from the systems obtained in more elementary hadron-hadron or electron-positron interactions. It is also generally accepted that due to extremely high densities and sufficiently large sizes of the systems produced in heavy-ion collisions it is possible to ensure the proper conditions for the formation of a quite long-living new state of strongly interacting matter.

Several interesting physical phenomena have been discovered. In particular, at RHIC and at the LHC one observes the strong collective behavior of matter (quantified by splitting of the slopes of the transverse-momentum spectra of various hadron species and by the large elliptic flow coefficient $v_2$), strangeness enhancement, and strong in-medium jet quenching. These facts indicate that quark and gluon degrees of freedom play the main role in the hot and dense medium. However, the data suggests also that the produced system is not weakly interacting. The great success of relativistic perfect-fluid hydrodynamics in the correct description of the data suggests short equilibration times, and consequently large cross sections for quark-gluon interactions. This gives hints that the quark-gluon plasma should be considered as a strongly interacting fluid rather than as a weakly interacting gas.

\begin{figure}[!t]
  \begin{center}
  \subfigure{\includegraphics[angle=0,width=0.5\textwidth]{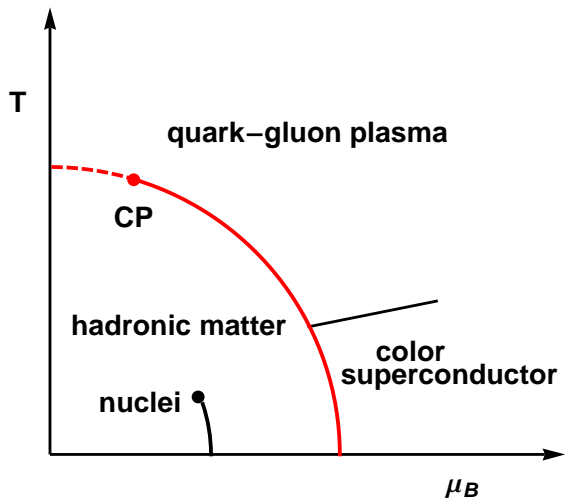}} 
  \end{center}
  \caption{\small Phase diagram of the QCD matter.}
  \label{fig:QCDpd}
\end{figure}

The RHIC experiments suggest the existence of a strongly interacting quark-gluon plasma in the high-temperature region of the QCD phase diagram (above the critical temperature and for negligible baryon chemical potential, see Fig.~\ref{fig:QCDpd}). In this region, the lattice QCD simulations predict a crossover rather than a genuine phase transition. In the crossover phase transition, the thermodynamic variables change suddenly in a narrow range of temperature, however, all changes are smooth and show no discontinuities characteristic for the real phase transitions.
The aim of the other current (STAR, NA61) and planned (FAIR, NICA) experiments is to explore different parts of the QCD phase diagram. The low beam energy scan program at STAR and the NA61 experiment search for signatures of the QCD phase transition at finite values of the baryon chemical potential. They try to identify the position of the critical point that marks the end of a line describing the first order deconfinement  phase transition at finite values of $\mu_B$. The existence of such a line is suggested by many QCD inspired theoretical models. 

\section[Theoretical tools]{Theoretical tools for different stages of \newline relativistic heavy-ion collisions}
\label{chapter:intro_tools}

In order to interpret the multitude of experimental data and formulate conclusions about physical properties of matter, we have to compare the experimental results with theoretical model predictions. Since, in practice, it is impossible to obtain a complete description of such complicated processes within QCD only, we have to develop different \textit{effective models} that are adequate for different stages of the evolution. 

Just before the collision, the coherent nuclear wave functions may be viewed (due to very high energies) as the gluon sheets described in the theory of \textit{color glass condensate} (CGC). The passage of the nuclei through each other at the collision proper time identified with $\tau = 0$ is followed by a subsequent pre-equilibrium stage which includes variety of processes leading to the formation of incoherent distributions of partons. 

One argues that the complex configuration of color electric and magnetic fields named \textit{glasma} is produced just after the collision. Its further dynamics is described by the classical \textit{Yang-Mills theory}. Finally, the glasma fields decay by radiating gluons. Similar picture is established in the framework of \textit{string models} where one considers collisions of hadrons, followed by the spanning of color strings or tubes. In this case, the strings decay producing partons or hadrons. 

The initial dynamics of the collision may be also viewed as reciprocal penetration of partons described by their distribution functions, which leads to multiple parton scatterings. The subsequent non-equilibrium evolution of the system is then described in the framework of \textit{kinetic theory} with collision terms treated in the perturbative limit. This approach is known as the \textit{parton cascade model} (PCM). However, if the system is dense, the particles are off-shell and the use of the \textit{quantum transport theory} becomes necessary (although this is not easy in practice).

The pre-equilibrium dynamics leads eventually to isotropization of momentum distributions of partons, and subsequently to local thermal equilibration of the system. The processes leading to thermalization are not understood well enough at the moment, thus the very fast thermalization, expected at RHIC and the LHC, remains still an unresolved puzzle. In this Thesis we address this problem and propose how to circumvent it. A more detailed discussion of the \textit{early thermalization puzzle} is presented in Chapter \ref{chapter:thermal}.

Multiplicities of particles produced in relativistic heavy-ion collisions are very large. Thus, the use of considerations based on \textit{statistical physics} seems to be well justified. If the system reaches the state of local thermal equilibrium, its further evolution may be described by \textit{relativistic hydrodynamics}. During hydrodynamic evolution the local values of thermodynamic variables change gradually. These changes are connected with the changes of the underlying phase-space distribution functions. If the equilibration rates in the system are not sufficiently fast,  the local thermal equilibrium is not maintained and the hydrodynamic picture fails. In this case, the system's evolution should be described in the framework of the \textit{kinetic theory}.

In a series of publications, it has been shown  that the space-time evolution of matter produced at RHIC may be well described by  relativistic hydrodynamics of the perfect fluid \cite{Kolb:2003dz,Kolb:2002ve,Huovinen:2003fa,Huovinen:2001cy,Teaney:2001av,Hama:2005dz,Hirano:2002ds,Hirano:2007xd,Bass:2000ib,Nonaka:2006yn,Bozek:2009ty,Broniowski:2008vp,Pratt:2008qv} or (if the deviations from equilibrium are noticeable but small) by dissipative hydrodynamics with a small viscosity to entropy ratio \cite{Teaney:2009qa,Heinz:2009xj}.  Relativistic hydrodynamics requires the information about the form of {\it equation of state} (EOS). The latter can be obtained from the {\it lattice QCD simulations} (LQCD). The present results obtained at $\mu_B=0$ predict a smooth crossover phase transition from hadronic matter to the quark-gluon plasma at the critical temperature $T_{c}\approx 170$ MeV. 

The hydrodynamic expansion of the fireball leads to the fast cooling of matter. The initially hot and dense quark-gluon plasma passes the phase transition back to hadronic matter. The hadronic interactions are less and less effective. At a certain stage, the interactions cease and particles freely stream to detectors. This process is called the \textit{kinetic freeze-out}. It is expected that the kinetic freeze-out is preceded by the \textit{chemical freeze-out} --- the stage where hadronic abundances are fixed. Nevertheless, in many approaches the identification of the two freeze-outs (the \textit{single freeze-out} model) turns out to be a reasonable approximation. In this case, all hadronic observables may be obtained with the help of the \textit{Cooper-Frye formula} \cite{Cooper:1974ak}. In particular, this approach allows for the successful description of the HBT \textit{radii} consistently with other soft hadronic observables such as the transverse-momentum spectra and the elliptic flow.

If different hadron species undergo (kinetic) freeze-out at different temperatures,  the modeling of freeze-out becomes a much more complex problem. In this context, there is a growing interest in developing hybrid transport approaches, i.e., the models that link hydrodynamics with  Monte-Carlo simulations which describe the scattering between hadrons in the final state, for example, this is achieved with \textit{ultra-relativistic quantum molecular dynamics} (UrQMD) \cite{Bass:1998ca,Petersen:2008dd}. 

\section{Objectives of the Thesis}
\label{chapter:intro_quest}

\par In this Thesis we analyze the space-time evolution of matter produced in relativistic heavy-ion collisions. {\it We propose to describe this evolution in the framework of a new, hydrodynamics-like model that inherently includes the possibility of having high anisotropy of pressure at the early stage.} We call this new framework the \texttt{ADHYDRO} model (highly-Anisotropic and strongly-Dissipative HYDROdynamics). Using \texttt{ADHYDRO} we try to reproduce the soft hadronic observables measured at RHIC. The presence of the anisotropic phase in our description, for about 1 fm/c at the beginning of the collision, helps us to circumvent the problem of early thermalization --- in order to describe the data we do not have to assume that matter reaches local thermal equilibrium within a fraction of a fermi. On the other hand, a possible connection of \texttt{ADHYDRO} to the underlying kinetic theory suggests that the microscopic collision times are of the order of a few hundredths of a fermi, a result suggesting that our picture may be consistent with the notion of a strongly interacting quark-gluon plasma. 

%
%
%

\par The Thesis is organized as follows. In Chapter \ref{chapter:thermal} we address the problem of fast thermalization in the very early stages of relativistic heavy-ion collisions. We also motivate the development of the \texttt{ADHYDRO} model presented in the Thesis. In Chapter \ref{chapter:aniso} we discuss thermodynamic properties of the locally anisotropic systems of particles and formulate generalized equation of state. In Chapter \ref{chapter:hydro} we derive the hydrodynamic equations determining the evolution of matter in \texttt{ADHYDRO}. Subsequently, in Chapter \ref{chapter:lbim} we discuss different aspects of the model in the simple case of purely longitudinal and boost-invariant evolution. The Chapter \ref{chapter:initial} contains the information about the initial conditions for realistic hydrodynamic simulations. In Chapter \ref{chapter:freeze} we include the freeze-out prescription. The results of the model for the boost-invariant systems are presented in Chapter \ref{chapter:boostinv} and the results for the most general non-boost-invariant case are shown in Chapter \ref{chapter:nonboostinv}. We close the Thesis with the Summary and Conclusions.

\par Throughout the Thesis we use the natural units where $c=\hbar=k_{\rm B}=1$. We also use the notation where the metric tensor has the following signature $g^{\mu\nu}=\mathrm{diag}(+1,-1,-1,-1)$.

\vspace{1cm}

The results presented in this Thesis were published in the following papers:
{\small
\begin{itemize}
  \item[1.] W. Florkowski and R. Ryblewski,\\
  {\it Dynamics of anisotropic plasma at the early stages of relativistic heavy-ion collisions},\\
  Acta Phys. Pol. {\bf B40} (2009) 2843, arXiv:0901.4653 [nucl-th].
  \item[2.] W. Florkowski and R. Ryblewski,\\
  {\it Highly-anisotropic and strongly-dissipative hydrodynamics for early stages of relativistic heavy-ion collisions},\\
  Phys. Rev. {\bf C83} (2011) 034907, arXiv:1007.0130 [nucl-th].
  \item[3.] R. Ryblewski and W. Florkowski,\\
  {\it Non-boost-invariant motion of dissipative and highly anisotropic fluid},\\  
  J. Phys. {\bf G38} (2011) 015104, arXiv:1007.4662 [nucl-th].
  \item[4.] R. Ryblewski and W. Florkowski,\\
  {\it Highly anisotropic hydrodynamics – discussion of the model assumptions and forms of the initial conditions},\\
  Acta Phys. Pol. {\bf B42} (2011) 115, arXiv:1011.6213 [nucl-th].
  \item[5.] R. Ryblewski and W. Florkowski,\\
  {\it Highly-anisotropic and strongly-dissipative hydrodynamics with transverse expansion},\\
  Eur. Phys. J. {\bf C71} (2011) 1761, arXiv:1103.1260 [nucl-th]. 
  \item[6.] R. Ryblewski,\\
  {\it Flow characteristics and strangeness production in the framework of highly-anisotropic and strongly-dissipative hydrodynamics},\\
  to be published in Acta Phys. Pol. {\bf B}, arXiv:1111.1909 [nucl-th]. 
\end{itemize}
}
%

\chapter{Early thermalization puzzle}
\label{chapter:thermal}
In this Chapter, the {\it early thermalization puzzle} in relativistic heavy-ion collisions is discussed. Different theoretical approaches describing early stages of the collisions are reviewed briefly and different results for the thermalization time-scales are presented. Finally, several recent attempts to circumvent the problem of early thermalization are introduced. In particular, we provide motivation for the development of our framework of highly-anisotropic and strongly-dissipative hydrodynamics (\texttt{ADHYDRO}) which will be presented in greater detail in the next Chapters.

\section{Evidence for early thermalization}
\label{chapter:evid}

The results collected in the heavy-ion experiments at RHIC are interpreted as the evidence that matter produced in relativistic heavy-ion collisions reaches {\it local thermal equilibrium} in a very short time\footnote{The physical processes leading to local equilibrium are named by us shortly as ``thermalization'' or ``equilibration'' processes.}. The main argument for such early equilibration comes from the observation that the space-time evolution of matter is very well explained by the relativistic hydrodynamics of perfect fluid  \cite{Kolb:2003dz,Kolb:2002ve,Huovinen:2003fa,Huovinen:2001cy,Teaney:2001av,Hama:2005dz,Hirano:2002ds,Hirano:2007xd,Bass:2000ib,Nonaka:2006yn,Bozek:2009ty,Broniowski:2008vp,Pratt:2008qv} or by viscous hydrodynamics with a small viscosity to entropy ratio \cite{Teaney:2009qa,Heinz:2009xj}. In particular, a very good description of the soft region of the transverse-momentum spectra of hadrons has been achieved in those frameworks. The shape of the spectra as well as their azimuthal dependence are interpreted as the evidence for the {\it radial} and {\it elliptic} flow of matter (the latter is described  by the elliptic flow parameter $v_2$). The key ingredient of the successful hydrodynamic calculations is {\it an early starting time assumed for the hydrodynamic evolution}, $\tau_{0} < 1$ fm/c. Since the initial starting time $\tau_{0}$ is identified typically  with the thermalization time $\tau_{\rm th}$,  the question arises whether such incredibly fast thermalization of matter can be explained by  microscopic models of the very early stages of heavy-ion collisions.

Fast equilibration and perfect-fluidity are naturally explained by the concept that the produced matter is a {\it strongly coupled quark-gluon plasma} (sQGP) \cite{Shuryak:2004cy,Gyulassy:2004zy}. Strong interactions between quarks and gluons might be responsible for the fast thermalization rate and small viscosity. Unfortunately, the large value of the coupling constant prohibits the use of the standard perturbative methods in this case. That is why the methods based on the AdS/CFT correspondence (Anti de Sitter/Conformal Field Theory correspondence) have become now a very attractive tool to analyze the early dynamics of heavy-ion collisions. Of course, we have also at our disposal LQCD which allows us to study thermodynamic properties of the QCD matter \cite{Karsch:2003jg,Borsanyi:2010cj}. In fact, deviations of the lattice results from the Stefan-Boltzmann limit in the temperature range $T \sim (1-3) T_{\rm c}$ (where $T_c \sim$ 170 MeV is the critical temperature) may be interpreted as the evidence for a strong coupling. 

On the other hand, we should keep in mind that the physics of relativistic heavy-ion collisions has been triggered by the idea that at extremely high temperatures, $T \gg T_{\rm c}$, the asymptotic freedom property of QCD leads to color screening and deconfinement  \cite{Collins:1974ky}. In this case, QGP should behave as a  {\it weakly interacting gas of quasiparticles} (wQGP) treated with perturbative QCD (pQCD) techniques. It seems hard to find reasonable explanations for sufficiently fast equilibration of such a weakly interacting system. However, several approaches have been proposed in order to understand the early time behavior of wQGP.

The thermalization of wQGP was broadly studied in the framework of Monte-Carlo parton cascade models (PCM) where the equilibration of the system was understood typically as the effect of multiple hard and semi-hard rescatterings of deconfined partons. Originally, PCM included only lowest-order binary ($2\rightarrow 2$) collisions  \cite{Boal:1986nr}. They turned out to be insufficient to provide fast equilibration. Moreover, the local equilibration of the system was observed only in the central region in coordinate space. Further improvements of PCM models led to inclusion of more subtle effects such as: dilated emission from excited partons, the balance between radiative emissions and reverse absorption processes, and the interference of soft gluons \cite{Geiger:1991nj,Geiger:1992si,Geiger:1992ac,Geiger:1994he}. They have been proved to significantly affect the thermalization rate. Nevertheless, the estimates for equilibration time-scales have delivered quite long times, for example, about $1.8$ fm/c for Au+Au collisions at the RHIC top energy. Recently, the importance of multi-particle processes have been emphasized. The particle production and absorption processes ($2\rightarrow 3$) \cite{Xu:2004mz,Xu:2007qn} together with the three-particle collisions ($3\rightarrow 3$) \cite{Xu:2004gw} have turned out to be significant in speeding up the equilibration. It has been demonstrated that the overall kinetic equilibration may be achieved much earlier but still later than \mbox{$1$ fm/c}.


At the very early stages of heavy-ion collisions, the momentum distribution of produced partons is expected to be highly anisotropic. Using theoretical methods known from the electron-ion plasma physics in a framework of the quark-gluon transport theory, one can show that the anisotropies of QGP lead to kinetic instabilities with respect to the formation of unstable plasma modes \cite{Mrowczynski:1994xv,Arnold:2004ti}. For QGP, only magnetic instabilities known as filamentation or chromo-Weibel instabilities \cite{Weibel:1959zz} are relevant. These strong instabilities drive the system towards the locally isotropic state much faster \cite{Rebhan:2004ur,Dumitru:2005gp,Strickland:2007fm} than the ordinary perturbative scattering processes in the weak coupling regime. This, in turn, indirectly helps to achieve the genuine  equilibration~\footnote{We note that {\it locally isotropic distributions} are not necessarily  equilibrium distributions. On the other hand, the local equilibrium requires that the distributions have the form of Boltzmann distributions (or Fermi-Dirac/Bose-Einstein distributions for quantum statistics) which are isotropic in the local rest frame.}.

A quite new and original mechanism of thermalization of QGP has been formulated during studies of statistical mechanics of the Yang-Mills classical mechanics. It has been postulated that equilibration may be understood as the effect of chaotic dynamics of the non-Abelian highly excited classical color fields coupled or not to the classical colored particles \cite{Biro:1993qc,Sengupta:1999jy,Bannur:2005wz}. In this case, the thermalization time estimated from the Lyapunov exponent is expected to be in the range $0.4-1.0$ fm/c.

\medskip

The microscopic models of the early stages of relativistic heavy-ion collisions use quite often the concepts of color strings or color-flux tubes. The idea of fast thermalization is particularly difficult to reconcile with the results of such models. For example, in the color glass condensate (CGC) approach that is often used to describe initial stages of relativistic collisions of heavy nuclei \cite{El:2007vg}, the momentum distribution functions are far away from the equilibrium ones \cite{Kovner:1995ja,Bjoraker:2000cf}.

On the basis of the QCD saturation mechanism incorporated in the CGC model, the \textit{bottom-up thermalization scenario} \cite{Baier:2000sb} was proposed. Since for very large nuclei and/or sufficiently high collision energies  the saturation scale $Q_s$ is much larger than the QCD scale $\Lambda_{\rm QCD}$ (at RHIC one expects $Q_s\sim 1$ GeV), the weak coupling techniques can be used in the calculations \cite{Kovchegov:2009he}. At the very early times, $ \tau\sim Q_s^{-1}$, small $x$ gluons with  the typical transverse momentum of the saturation scale $Q_s$ are freed from the incoming nuclei.  Subsequently, in the time interval $Q_s^{-1}< \tau < \alpha^{-5/2}Q_s^{-1}$, the soft gluons are emitted. After the time $\tau\sim \alpha^{-5/2}Q_s^{-1}$ the soft gluons form a thermal bath and draw the energy from hard gluons. The stop of temperature growth identified with full thermalization happens at $\tau\sim \alpha^{-13/5}Q_s^{-1}$. At that time the primary hard gluons loose their all energy. Straightforward calculations done in this approach for the most central collisions at RHIC give the equilibration time of $2.6$ fm/c or even larger \cite{El:2007vg,Baier:2002bt}.

The CGC model itself gives quite interesting picture of the temporal behavior of the momentum anisotropies of the matter produced in heavy-ion collisions. For the very early proper times, $\tau \ll Q_s^{-1}$, the dominant gluon fields produced in the collisions are classical and can be described by the classical Yang-Mills equations. In this case, one predicts the energy-momentum tensor with negative longitudinal pressure \cite{Lappi:2006hq,Fukushima:2007ja}. The pressure asymmetry persists also at later times, $\tau \gg Q_s^{-1}$, where the analytic perturbative approaches \cite{Kovchegov:2005ss} and the full numerical simulations \cite{Krasnitz:2002mn} yield the energy-momentum tensor with the longitudinal pressure significantly lower than the transverse pressure. This results indicates that the hydrodynamic description of the gluon system as a perfect fluid may start only at later times . 

\medskip
In the color-flux-tube models, the production of partons is described by the Schwinger tunneling mechanism in strong color fields \cite{Bialas:1987en,Florkowski:2003mm}. The Schwinger mechanism leads to transverse-momentum distributions which have Gaussian shapes. However, if the string tension fluctuates, the Gaussian distributions are changed effectively into the exponential distributions \cite{Bialas:1999zg}. The resulting forms of the spectra  resemble the Boltzmann distributions in the transverse-momentum space. Thus, the produced system may be viewed as a system where only transverse degrees of freedom are thermalized. This phenomenon may be considered as {\it apparent transverse thermalization}, since thermal-like spectra are caused by the specific mechanism of particle production and not by the parton rescattering processes. 


An intriguing thermalization scenario explaining the equilibration of both the transverse and longitudinal degrees of freedom is based on the Hawking-Unruh effect \cite{Kharzeev:2005iz,Castorina:2007eb} that states that an observer moving with an acceleration $a$ experiences the influence of a thermal bath with the effective temperature $a/2\pi$.  In the Color Glass Condensate theory the typical acceleration in the strong color fields is of the order of $Q_{s}$, which leads to the thermalization time $2\pi/Q_{s} \sim 1$ fm/c.


We conclude our discussion of the early thermalization problem with the statement that microscopic models of heavy-ion collisions cannot explain naturally the thermalization time-scales shorter than $1$ fm/c. Moreover, several models suggest that thermalization of the transverse degrees of freedom takes place before the complete three-dimensional thermalization is achieved\footnote{Speaking about the complete or full thermalization we have in mind always the approach towards {\it local} equilibrium.}. As a consequence, these models predict significant momentum anisotropies during early stages of the collisions. It means that the standard hydrodynamic picture may be inadequate during very early times of heavy-ion collisions. 

These observations have initiated development of effective, hydrodynamics-like models which describe the whole evolution of matter and relax the assumption of the early thermalization. Typically,  this is done by imposing existence of a short initial pre-equilibrium phase followed by more standard hydrodynamic evolution. These models are discussed below in Section \ref{chapter:recent}.

\section{Relaxing the early thermalization assumption}
\label{chapter:recent}

Significant differences between the results of various theoretical models of the early stages of heavy-ion collisions reflect our lack of precise knowledge concerning the processes of particle production. The early stages are, however, of great importance as they may have an important impact on further evolution of matter. Based on the qualitative hints coming from the {\it microscopic models} discussed above in Section \ref{chapter:evid}, we may study different {\it effective scenarios} for the early stages. By comparing the experimental consequences of such scenarios, we may try to check which hypothesis is the most likely to be realized in Nature. 

Of the special interest are the physical scenarios where the assumption of the early thermalization of matter is relaxed. In practice, this means that one does not start the hydrodynamic description at an early time which is a fraction of a fermi but the early stage is described by an effective non-equilibrium model which is matched with the hydrodynamic description at a later time.  

In particular, it has been proposed that the early-stage dynamics consists of the partonic free-streaming (FS) phase initiated at $\tau_{\rm 0}\ll 1$ fm/c and followed by a sudden equilibration (SE) process. Typically, the sudden thermalization is realized by applying the Landau matching conditions at a certain transition time $\tau_{\rm tr}>\tau_{\rm 0}$. The remaining part of the evolution is determined by standard relativistic hydrodynamics. The first FS+SE approximation was proposed by Kolb, Sollfrank, and Heinz \cite{Kolb:2000sd}. Recently, it has been reconsidered by several authors \cite{Sinyukov:2006dw,Jas:2007rw,Gyulassy:2007zz}. It has been demonstrated that the final observables  are insensitive to the presence of the early
free-streaming phase if this phase lasts no longer than 1 fm/c \cite{Broniowski:2008qk}.

According to the results coming from the string and color-flux-tube models, the pre-equilibrium stage may be viewed in a different way, namely, as the {\it transversally thermalized parton system}. It is conceivable that the initial quark-gluon matter is produced already in a state close to the thermodynamic equilibrium in the transverse direction only, while its longitudinal dynamics is described as free streaming of two dimensional clusters \cite{Bialas:2007gn}. This observation resulted in the development of the model that is quite similar to the FS+SE framework. In this approach, the initial dynamics of the parton system is described by {\it transverse hydrodynamics} (TH)\cite{Ryblewski:2010tn}. Similarly to the FS+SE approach, the system that is initially not completely thermalized undergoes sudden  equilibration and its subsequent evolution is described by standard hydrodynamics. Interestingly, also in this framework the insensitivity of final results with respect to details of the initial transverse-hydrodynamics stage has been observed.

The two approaches discussed above, which have turned out to be successful in describing the experimental data, show that the models incorporating high initial anisotropy should also account for the fact that matter is eventually fully thermalized. Therefore, there is a need for models which can, in a comprehensive way, describe the gradual transition between early highly-anisotropic stages and later fully thermalized stages governed by the perfect-fluid hydrodynamics. Of course, in this context the use of the Landau matching conditions seems to be a very crude approximation.

Certainly, for small deviations from equilibrium the most appropriate description should be based on the framework of the second order viscous hydrodynamics \cite{Israel:1979wp,Muronga:2001zk,Muronga:2003ta}. However, at the earliest stages of the collision the anisotropies are very large. In this regime neither perfect-fluid hydrodynamics nor viscous hydrodynamics are formally applicable. In particular, viscous corrections become compatible with the leading terms, which may lead to negative pressure. Of course, these stages can be described by models based on the kinetic theory, but the latter has its own well known limitations.

The idea of developing the hydrodynamics-like approach that is suitable for uniform description of several stages of evolution of matter has motivated our development of the \texttt{ADHYDRO} model. This is an effective model which is able to describe strongly-dissipative dynamics of initially highly-anisotropic systems. \texttt{ADHYDRO} works in three different regimes. At very early stages, it describes the system with extremely high momentum anisotropy. Close to equilibrium, where the anisotropies are small, \texttt{ADHYDRO} agrees with the viscous hydrodynamics~\footnote{This is strictly valid in the case of purely longitudinal expansion. If the transverse expansion is compatible with the longitudinal one, it is possible to switch from the \texttt{ADHYDRO} description to the viscous hydrodynamics in the consistent way, as shown in Ref. \cite{Florkowski:2011jg}.}. For very large evolution time the system is (locally) isotropic in momentum space and our model gives standard perfect-fluid hydrodynamic description.

\chapter[Locally anisotropic systems of particles]{Locally anisotropic\\systems of particles}
\label{chapter:aniso}
%
\section[Locally anisotropic phase-space distribution functions]{Locally anisotropic phase-space distribution functions}
\label{sect:aniso-distr}
%
In this Thesis we consider {\it locally anisotropic systems of particles}. We assume that such systems are characterized by the anisotropy of their momentum distributions analyzed in the local rest frame. The local rest frame at the space-time point $x$ is the reference frame in which a small element of the considered system, located at this space-time point, is at rest.  Locally anisotropic systems of particles may be described by the phase-space distribution functions whose dependence on different momentum components in the local rest frame is different. 

We restrict our considerations to the case where the anisotropy of the system is connected with the difference of pressures in the longitudinal and transverse directions. These directions are always defined with respect to the collision axis~\footnote{ It is practical to introduce the standard Cartesian coordinate system where the longitudinal \textit{z}-axis is identified with the direction of the beam, while the $x$ and $y$ axes are chosen in such a way that the $x$ and $z$ axes span the reaction plane.}. The asymmetry in momentum may be realized by introducing two different space-time dependent scales, $\lambda_\perp$ and $\lambda_\parallel$, that may be interpreted as the {\it transverse and longitudinal temperatures}, respectively. The explicitly covariant form of such a non-equilibrium distribution function is  
\begin{equation}
f  = f\left( \frac{\sqrt{(p \cdot U)^2 - (p \cdot V)^2 }}{\lambda _\perp },
\frac{|p \cdot V|}{\lambda _\parallel  }\right),
\label{Fcov}
\end{equation}
where $p^{\mu}$ denotes the four-momentum. The four-vector $U^\mu$ describes the fluid four-velocity
\begin{equation}
U^\mu = \gamma (1, v_x, v_y, v_z), \quad \gamma = (1-v^2)^{-1/2}.
\label{U}
\end{equation}
In comparison with standard equilibrium distributions, we introduce here a new four-vector $V^\mu$. Its appearance is related to the special role played by the beam axis ($z$-axis) and it is defined in the following way
\begin{equation}
V^\mu = \gamma_z (v_z, 0, 0, 1), \quad \gamma_z = (1-v_z^2)^{-1/2}.
\label{V}
\end{equation}
The four-vectors $U^\mu$ and $V^\mu$ satisfy simple normalization conditions
\begin{eqnarray}
U^2 = 1, \quad V^2 = -1, \quad U \cdot V = 0.
\label{UVnorm}
\end{eqnarray}
In the local rest frame (LRF) of the fluid element the four-vectors $U^\mu$ and $V^\mu$ have simple forms
\begin{eqnarray}
 U^\mu = (1,0,0,0), \quad V^\mu = (0,0,0,1), 
 \label{UVLRF}
\end{eqnarray}
and the formula (\ref{Fcov}) reduces to
\begin{equation}
f = f\left( \frac{m_\perp}{\lambda_\perp},\frac{|p_\parallel|}{\lambda_\parallel}\right).
\label{FLRFmassive}
\end{equation}
For massless particles $m_\perp=p_\perp$, and we may also write
\begin{equation}
f = f\left( \xi_\perp,|\xi_\parallel| \right).
\label{FLRFmassless}
\end{equation}
Here, for our convenience, we have introduced two dimensionless parameters
\begin{equation}
\xi_\perp=p_\perp/\lambda_\perp, \quad     \xi_\parallel=p_\parallel/\lambda_\parallel.
\label{parametersxi}
\end{equation}

For example, we may consider the Boltzmann distribution that is stretched or squeezed in momentum space,
\begin{eqnarray}
f_0 
&=& \exp\left[ -\sqrt{\frac{(p \cdot U)^2 - (p \cdot V)^2}{\lambda_\perp^2} + \frac{(p \cdot V)^2}{\lambda_\parallel^2}}\right] \nonumber \\
&=& \exp\left[ -\frac{1}{\lambda_\perp}\sqrt{(p \cdot U)^2 + \xi (p \cdot V)^2}\right],
\label{f0}
\end{eqnarray}
where $\xi$ is expressed by the {\it pressure anisotropy parameter} $x$,
\begin{equation}
x = \xi+1 = \left( \frac{\lambda_\perp}{\lambda_\parallel} \right)^2. 
\label{Xaniso}
\end{equation}
For isotropic systems $x=1$ ($\xi=0$) and
\begin{eqnarray}
f_0 = f_{\rm cl} = \exp\left[ -\frac{p \cdot U}{\lambda_\perp}\right].
\label{f0}
\end{eqnarray}
%
\section{Moments of anisotropic distributions}
\label{sect:aniso-moments}
%
The knowledge of the distribution function (\ref{Fcov}) allows us to calculate one-particle characteristics of the system under consideration. In particular, the first and second moments of the distribution function yield the particle density, energy density, and the two pressures.  Since at the very beginning of the evolution the considered system is expected to consist mainly of gluons, in this Section we restrict our analysis to massless particles by setting $m = 0$. 

Using the standard definitions of the particle number current $N^\mu$ and the energy-momentum tensor $T^{\mu \nu}$,
\begin{eqnarray}
N^\mu &=& \int \frac{d^3p}{(2\pi)^3 \, p^0} \, p^{\mu} f
\label{Ndef}
\end{eqnarray}
and
\begin{eqnarray}
T^{\mu \nu} &=& \int \frac{d^3p}{(2\pi)^3 \, p^0} \, p^{\mu} p^\nu f,
\label{Tdef} 
\end{eqnarray}
and the definition of the entropy current
\begin{equation}
S^{\mu} = - g_0\int \frac{d^3p}{(2 \pi)^3} \frac{p^\mu}{ p^0} \Phi\left[ f \right],
\label{Sdef}
\end{equation}
where
\begin{equation}
\Phi\left[ f \right] = \frac{f}{g_0} \ln \frac{f}{g_0}  - \frac{1}{\epsilon} \left(1 +  \frac{\epsilon f}{g_0} \right)\ln\left(1 +  \frac{\epsilon f}{g_0} \right),
\label{Phidef}
\end{equation}
we obtain the following decompositions (the method leading to Eqs.~(\ref{Naniso})--(\ref{Saniso}) is explained in Ref.~\cite{Florkowski:2009sw}):
\begin{eqnarray}
N^\mu &=&  n \, U^\mu ,
\label{Naniso}
\end{eqnarray}
\begin{eqnarray}
T^{\mu \nu} = \left( \varepsilon  + P_{\perp}\right) U^{\mu}U^{\nu} - P_{\perp} \, g^{\mu\nu} - (P_{\perp} - P_{\parallel}) V^{\mu}V^{\nu}, 
\label{Taniso}
\end{eqnarray}
and
\begin{equation}
S^{\mu} = \sigma \, U^\mu .
\label{Saniso}
\end{equation}
Here $g_0$ is the degeneracy factor related to internal quantum numbers of particles that form the fluid and $p^0=E_p=\sqrt{m^2 + p_\perp^2 + p_\parallel^2}$. For simplicity, in Eq.~(\ref{Phidef}) we have introduced $\epsilon = + 1 $ for bosons and $\epsilon = - 1$ for fermions. In the limit $\epsilon \to 0$ we obtain the classical Boltzmann definition.

Equations (\ref{Naniso})--(\ref{Saniso}) contain {\it non-equilibrium} parameters: particle density $n$, entropy density $\sigma$, energy density $\varepsilon$, transverse pressure $P_\perp$, and longitudinal pressure $P_\parallel$ \footnote{Below, the parameters such as $n$, $\sigma$, $\varepsilon$, $P_\perp$ and $P_\parallel$ will be called and treated as thermodynamics-like or thermodynamic parameters, although, strictly speaking, these quantities do not describe the equilibrium state.}.

In the isotropic case, the longitudinal and transverse pressures are equal, $P_\perp=P_\parallel=P_{\rm eq}$, and we recover the standard form of the energy-momentum tensor valid for the perfect fluid. In general, when the system is out of equilibrium, the last term in (\ref{Taniso}) is different from zero. Then, in the local rest frame of the fluid element we find~\footnote{The Lorentz transformation leading to LRF is defined explicitly in \cite{Ryblewski:2010ch}.} 
\begin{equation}
T^{\mu \nu} =  \left(
\begin{array}{cccc}
\varepsilon & 0 & 0 & 0 \\
0 & P_\perp & 0 & 0 \\
0 & 0 & P_\perp & 0 \\
0 & 0 & 0 & P_\parallel
\end{array} \right).
\label{TARR}
\end{equation}
Hence, as expected, the structure (\ref{Taniso}) allows for different pressures in the longitudinal and transverse directions. For massless particles the condition that the energy-momentum tensor is traceless leads us to the following expression 
\begin{equation}
\varepsilon = 2 P_\perp + P_\parallel.
\label{Ttrace}
\end{equation}
From Eq.~(\ref{Ttrace}) we can infer that thermodynamics-like parameters describing the anisotropic system are functions of two independent parameters. For example, the energy density is a function of two pressures, $\varepsilon=\varepsilon (P_\perp,P_\parallel)$, or equivalently of two temperatures, $\varepsilon=\varepsilon (\lambda_\perp,\lambda_\parallel)$. This feature distinguishes anisotropic systems from equilibrated matter where the energy density for baryon-free matter is a function of isotropic pressure only, $\varepsilon=\varepsilon(P_{\rm eq})$. For similar out-of-equilibrium systems with non-vanishing baryon number, three parameters are needed to specify the local state of the system.
%
\section{Pressure anisotropy}
\label{sect:aniso-relax}
%
The internal consistency of the approach based on the  distribution function (\ref{Fcov}) allows us to derive the following concise structure of the thermodynamic parameters describing the anisotropic system of particles (for more details see \cite{Florkowski:2009sw})
\begin{eqnarray}
\varepsilon (x,n) &=& \left( \frac{n}{g} \right)^{4/3} R(x), 
\label{eaniso} \\ 
P_{\perp} (x,n)   &=& \left( \frac{n}{g} \right)^{4/3} \left[ \frac{R(x)}{3} + x R'(x) \right], 
\label{PTaniso}\\
P_{\parallel}(x,n)&=& \left( \frac{n}{g} \right)^{4/3} \left[ \frac{R(x)}{3} - 2 x R'(x) \right], 
\label{PLaniso}
\end{eqnarray}
where the function $R(x)$ is defined by the integral
\begin{equation}
R(x) = x^{-1/3} \int \frac{d\xi_\perp\,\xi_\perp\,d\xi_\parallel}{2\pi^2}
\sqrt{\xi_\parallel^2 + x \xi_\perp^2} f(\xi_\perp,\xi_\parallel),
\label{Rinteg}
\end{equation}
and $g$ is a constant defined by the expression
\begin{equation}
g =  \int \frac{d\xi_\perp\,\xi_\perp\,d\xi_\parallel}{2\pi^2}
 f(\xi_\perp,\xi_\parallel).
\label{ginteg}
\end{equation}
The range of the integration over $\xi_\perp$ and $\xi_\parallel$ is always between 0 and infinity. The function $R^\prime(x)$ in Eqs.~(\ref{PTaniso}) and (\ref{PLaniso}) is the derivative of $R(x)$ with respect to $x$. 

The {\it pressure relaxation function} $R(x)$ describes bulk properties of the system characterized by the anisotropy $x$.  The particular form of $R(x)$ follows directly from the microscopic calculation if the distribution function (\ref{Fcov}) is known. {\it At the more phenomenological level of description, a special form of the function $R(x)$ can be assumed and treated as an ansatz describing the anisotropic system} \cite{Florkowski:2008ag}. The anisotropy parameter $x$ and the non-equilibrium entropy density $\sigma$ are the most convenient independent variables used in our further analysis. 

\medskip

Equations  (\ref{eaniso})--(\ref{PLaniso}) impose general physical constraints on the form of $R(x)$. For $x \ll 1$ \mbox{($\lambda_\perp \ll \lambda_\parallel$)} the function $R(x)$ behaves like $x^{-1/3}$. In this limit $P_{\perp} = 0$ and $\varepsilon = P_\parallel$. Similarly, for $x \gg 1$ \mbox{($\lambda_\perp \gg \lambda_\parallel$)} the function $R(x)$ behaves like $x^{1/6}$, implying that $P_\parallel = 0$ and $\varepsilon = 2 P_{\perp}$. This behavior is expected if we interpret the parameters $\lambda_\perp$ and $\lambda_\parallel$ as the transverse and longitudinal temperatures, respectively. Since color fields are neglected in this approach both pressures are expected to be positive, $P_\perp \geq 0$ and $P_\parallel \geq 0$. Furthermore, close to equilibrium the longitudinal and transverse pressures are expected to equilibrate, $P_{\perp} = P_{\parallel}$. Thus, according to Eqs.~(\ref{PTaniso}) and (\ref{PLaniso}) we obtain another key restriction for $R(x)$, namely,
\begin{equation}
 R^\prime(1)=0.
\label{cond1}
\end{equation}
Based on the RHIC experimental data, we assume that the system eventually achieves the perfect fluid limit, hence we may write that $\lim \limits_{\tau \to \infty} x = 1$. 

The particle and entropy densities introduced in Eqs.~(\ref{Naniso}) and (\ref{Saniso}) are defined by the following integrals
\begin{eqnarray}
n &=& \frac{\lambda_\perp^2 \,\lambda_\parallel}{2 \pi^2} \int\limits_0^\infty  
d\xi_\perp\,\xi_\perp\, \int\limits_{0}^\infty d\xi_\parallel \, f\left(\xi_\perp,\xi_\parallel \right),\label{partdens} \\
\sigma &=& \frac{g_0 \lambda_\perp^2 \,\lambda_\parallel}{2 \pi^2} \int\limits_0^\infty  
d\xi_\perp\,\xi_\perp\, \int\limits_{0}^\infty d\xi_\parallel \, \Phi\left[ f\left(\xi_\perp,\xi_\parallel \right) \right]. \label{entrdens}
\end{eqnarray}
Comparison of Eqs.~(\ref{partdens}) and (\ref{entrdens}) indicates that the particle density and the entropy density are proportional to each other, with the proportionality constant depending on the specific choice of the parton distribution function $f$.
\section{Special forms of anisotropic distribution functions}
\label{sect:aniso-ansatz}
%
So far we have not considered any explicit forms of the distribution functions. In this Section, we turn to the discussion of the anisotropic distribution functions satisfying the ansatz (\ref{Fcov}) and having the following form 
\begin{equation}
f = g_0 \left[ \exp \left(\sqrt{\xi_\perp^2 + \xi_\parallel^2 }\right)- \epsilon \right]^{-1}.
\label{F}
\end{equation}
Equation (\ref{F}) may be regarded as a simple generalization of the equilibrium distributions in the case of vanishing baryon number. In a more general case, we should take into account also the net baryon number density $n_B$ and introduce the baryon chemical potential $\mu_B$ in (\ref{F}). However, if we focus on the central rapidity region in the ultra-relativistic limit (which is the case when analyzing the RHIC and LHC experimental data) we may neglect the baryon chemical potential $\mu_B$. Hence, the thermodynamic parameters remain the functions of entropy density and anisotropy.

The function $f(\epsilon=0)$ may be considered as the Boltzmann equilibrium distribution that has been stretched (or squeezed) in the longitudinal direction. For completeness, it is also interesting to consider the stretched (or squeezed) Bose-Einstein distribution $f(\epsilon=+1) \equiv f_{BE}$ or  Fermi-Dirac distribution, $f(\epsilon=-1) \equiv f_{FD}$. We note that in the local equilibrium limit, Eq.~(\ref{F}) with $\epsilon=0$ is reduced to the standard \textit{J\"uttner} formula where $\lambda_\perp = \lambda_\parallel = T$.

With the definition (\ref{F}), we recover the structure (\ref{eaniso})--(\ref{PLaniso}) \cite{Ryblewski:2010ch}. In the case of the classical Boltzmann statistics, the relaxation function (\ref{Rinteg}) is given by the form \footnote{Note that for $x < 1$ the function $(\arctan\sqrt{x-1})/\sqrt{x-1}$ should be replaced by $(\hbox{arctanh}\sqrt{1-x})/\sqrt{1-x}$}
\begin{equation}
R(x) = \frac{3\, g_0\, x^{-\frac{1}{3}}}{2 \pi^2} \left[ 1 + \frac{x \arctan\sqrt{x-1}}{\sqrt{x-1}}\right]
\label{RB}
\end{equation}
and the constant (\ref{ginteg}) is simply 
\begin{equation}
g = \frac{g_0}{\pi^2}.
\label{gB}
\end{equation}
Equations (\ref{partdens}) and (\ref{entrdens}) relate the entropy and particle densities in the following way 
\begin{equation}
\sigma = 4 n.
\label{sB}
\end{equation}
In the similar way we may consider the Bose-Einstein and Fermi-Dirac distributions, see summary in Table~\ref{table:aniso} . 
\begin{table}[t]
  \begin{center}
    \begin{small}
      \begin{tabular}{cccc}
       \hline
        $\displaystyle \vphantom{\frac{1}{2}}$
         &  classical particles & bosons &  fermions \\  \hline  \\
        $\displaystyle \vphantom{\frac{1}{2}}$ $R(x)$&  $R_{\rm cl}(x)=\frac{3\, g_0\, x^{-\frac{1}{3}}}{2 \pi^2} \left( 1 + \frac{x \arctan\sqrt{x-1}}{\sqrt{x-1}}\right)$       &  $\zeta(4) R_{\rm cl}(x)$  & $\frac{7 \zeta(4)}{8} R_{\rm cl}(x)$ \\  
        [2ex]
        $\displaystyle \vphantom{\frac{1}{2}}$ $g$   &  $\frac{g_0}{\pi^2}$       & $\zeta(3) \frac{g^{BE}_0}{\pi^2}$ & $\zeta(3) \left(1-2^{-2} \right) \frac{g^{FD}_0}{\pi^2} $  \\  
        [2ex]
        $\displaystyle \vphantom{\frac{1}{2}}$  $\sigma(n)$     & $\sigma = 4 n$ & $\sigma =  \frac{2\pi^4}{45 \zeta(3)} n $ & $\sigma = \frac{7\pi^4}{135 \zeta(3)} n$ \\ \\ \hline
      \end{tabular}
    \end{small}
  \end{center}
  \caption{\small Thermodynamic properties of the anisotropic systems for different statistics. The rows contain the following information: $R(x)$ --- pressure relaxation function (\ref{Rinteg}), $g$ --- constant defined by Eq.~(\ref{ginteg}), $\sigma(n)$ --- relation between entropy density and particle density calculated from Eqs.~(\ref{partdens}) and (\ref{entrdens}). 
}
  \label{table:aniso}
\end{table}
\begin{figure}[t]
\begin{center}
\hspace{0.508cm}\subfigure{\includegraphics[angle=0,width=0.61\textwidth]{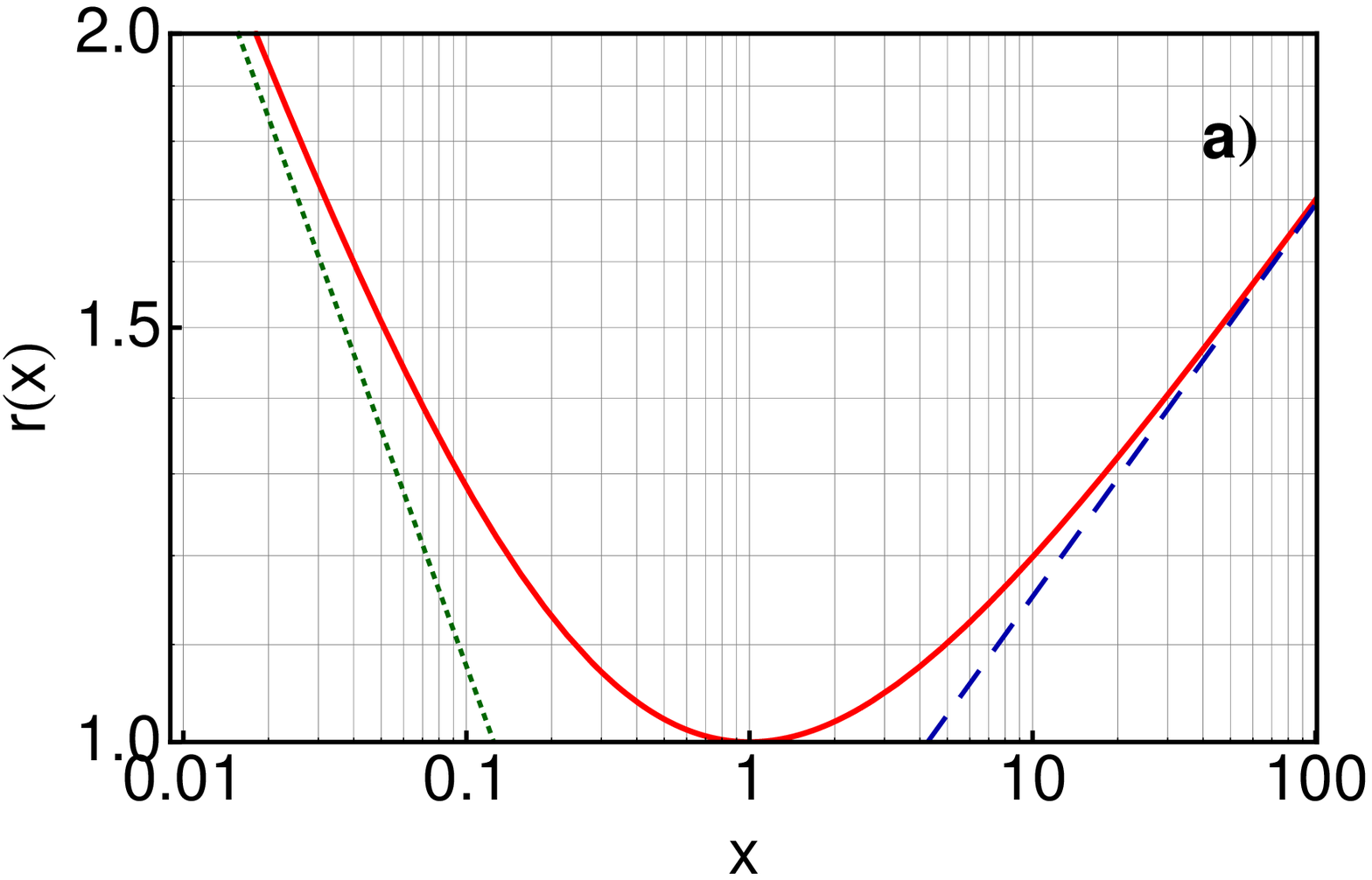}} \\
\subfigure{\includegraphics[angle=0,width=0.65\textwidth]{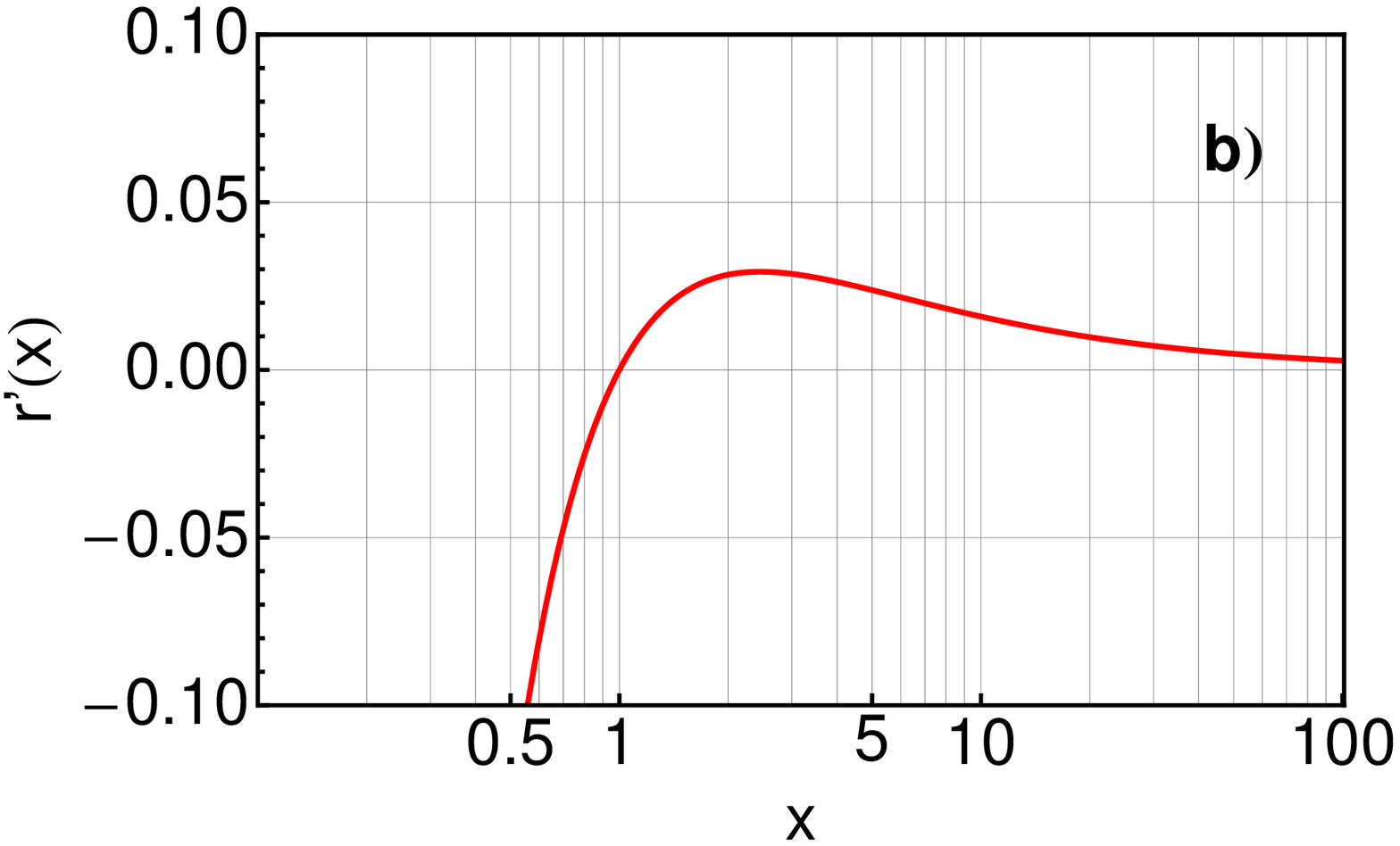}} 
\end{center}
\caption{{\bf a)} The function $r(x)$ (solid red line) shown together with the asymptotics $r(x) \sim (\pi / 4) \, x^{1/6}$ (blue dashed line) and $r(x) \sim (1/2)\, x^{-1/3}$ (green dotted line). {\bf b)} The derivative $r^{\prime}(x)$ of the pressure relaxation function (solid red line).}
\label{fig:randrp}
\end{figure}

The results presented in Table~\ref{table:aniso} may be used in Eqs.~(\ref{eaniso})--(\ref{PLaniso}). In this way we find that the numerical factor $\varepsilon/\varepsilon_{BE}$  (giving the ratio of the classical energy density to the Bose-Einstein energy density for the fixed $\sigma$ and $x$) is very close to unity, $\varepsilon/\varepsilon_{BE} \approx 1.027$ . It means that the equations defining the energy density and pressures in terms of $\sigma$ and $x$ for the classical and quantum systems are practically the same. Therefore, in our further studies we use the classical statistics given by Eq.~(\ref{F}). Moreover, in order to account for the internal degrees of freedom of gluons we set $g_0 = 16$ (spin $\times$ color). 

We stress that physical restrictions concerning the pressure relaxation function, discussed in Section \ref{sect:aniso-relax}, exclude from the consideration certain forms of distribution functions that formally satisfy the ansatz (\ref{Fcov}). For example, the anisotropic distribution functions of the factorized form $f_2 = g_0 \exp\left( -\xi_\perp \right) \exp\left( - |\xi_\parallel | \right)$ do not satisfy the condition (\ref{cond1}). 

In the upper part of Fig.~\ref{fig:randrp} we present the rescaled pressure relaxation function $r(x) = \pi^2 R(x)/ (3 g_0)$ while in the lower part of Fig.~\ref{fig:randrp} we have plotted its derivative $r^\prime(x)$. We can see that according to (\ref{cond1}) $r^\prime(x)$ vanishes when $P_{\perp}=P_{\parallel}$. 
\begin{figure}[t!]
\begin{center}
\subfigure{\includegraphics[angle=0,width=0.65\textwidth]{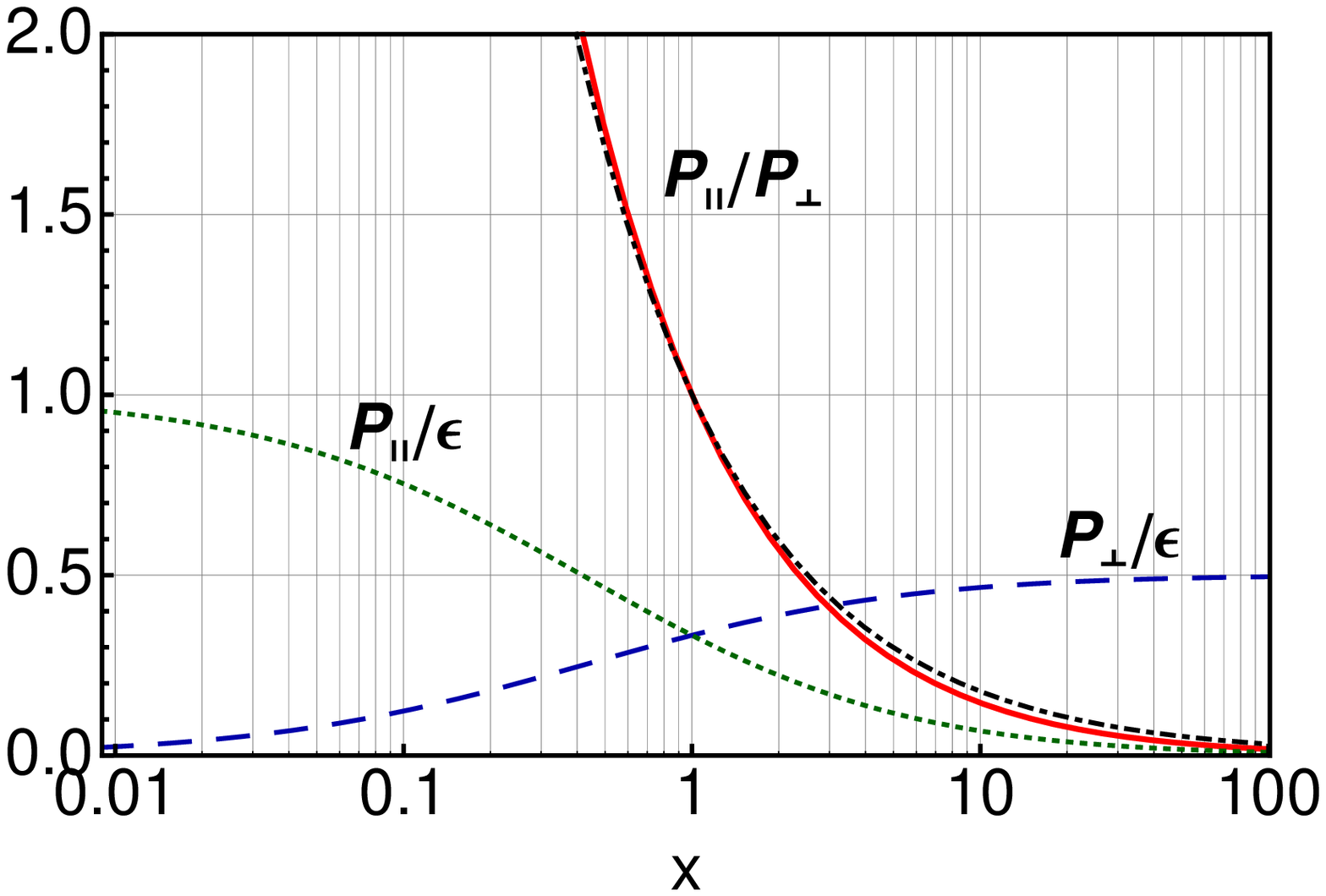}} 
\end{center}
\caption{The ratios $P_{\parallel}/P_{\perp}$ (solid red line), $P_{\parallel}/\varepsilon$ (green dotted line), and  $P_{\perp}/\varepsilon$ (blue dashed line) shown as functions of the variable $x$, the results for the pressure relaxation function (\ref{RB}). We also show approximation of $P_{\parallel}/P_{\perp}$ by the function $x^{-3/4}$ (black dot-dashed line).}
\label{fig:ratios}
\end{figure}
In Fig.~\ref{fig:ratios} we show the ratios: $P_{\parallel}/P_{\perp}$ (solid red line), $P_{\parallel}/\varepsilon$ (green dotted line), and $P_{\perp}/\varepsilon$ (blue dashed line) for the case of $R(x)$ given by Eq.~(\ref{RB}). The considered ratios are functions of $x$ only. In agreement with previously given remarks we see that $\varepsilon=P_{\parallel}$ for $x=0$ and $\varepsilon=2 P_{\perp}$ for $x\to\infty$. The equilibration of pressures is reached for $x=1$. To a good approximation $P_{\parallel}/P_{\perp} \approx x^{-3/4}$, thus, the anisotropy parameter $x$ may be treated as the direct measure of the pressure anisotropy.

\section{Generalized equation of state}
\label{sect:aniso-eos}
%
The equations of perfect-fluid hydrodynamics form a closed system of equations only if they are supplemented with the equation of state that specifies thermodynamic quantities as functions of one parameter~\footnote{If the baryon-free matter is considered.}. Similarly, our framework requires that all thermodynamic quantities should be expressed by two arbitrarily chosen parameters, e.g., by the non-equilibrium entropy density $\sigma$ and the anisotropy parameter $x$. Such relations play a role of the {\it generalized equation of state} and allow us to close the system of dynamic equations (the latter will be formulated in Chapter \ref{chapter:hydro}). 
  
If we consider a system of partons described by the anisotropic phase-space distribution function given by Eq.~(\ref{F}) and obtained by squeezing (or stretching) the classical Boltzmann distribution, we have to use Eqs.~(\ref{eaniso})--(\ref{PLaniso}) together with Eqs.~(\ref{RB})--(\ref{sB}). Equations (\ref{eaniso})--(\ref{PLaniso}) describe the non-equilibrium state defined by the values of the non-equilibrium entropy density $\sigma$ and the anisotropy parameter $x$. Equations  (\ref{eaniso})--(\ref{PLaniso}) approach naturally the equilibrium limit if $x \to 1$, since $r(1)=1$ and $r^\prime(1)=0$. In the equilibrium the following relations are valid: $\varepsilon_{\rm id} = 3 g_0 T^4/\pi^2$, $P_{\rm id} = g_0 T^4/\pi^2$, and $\sigma_{\rm id} = 4 g_0 T^3/\pi^2$ \cite{Florkowski:2010zz}. Hence, Eqs.~(\ref{eaniso})--(\ref{PLaniso}) may be written equivalently in the form,
\begin{eqnarray}
\varepsilon (x,\sigma)&=&  \varepsilon_{\rm id}(\sigma) r(x), \label{epsilon2a}  \\ \nonumber 
P_\perp (x,\sigma)&=&  P_{\rm id}(\sigma) \left[r(x) + 3 x r^\prime(x) \right],  \label{PT2a}  \\ \nonumber 
P_\parallel (x,\sigma)&=&  P_{\rm id}(\sigma) \left[r(x) - 6 x r^\prime(x) \right], \label{PL2a} 
\end{eqnarray}
where $\varepsilon_{\rm id}(\sigma)$ and $P_{\rm id}(\sigma)$ are {\it equilibrium expressions} for the energy density and pressure. 

\begin{figure}[!t]
  \begin{center}
  \subfigure{\includegraphics[angle=0,width=0.65\textwidth]{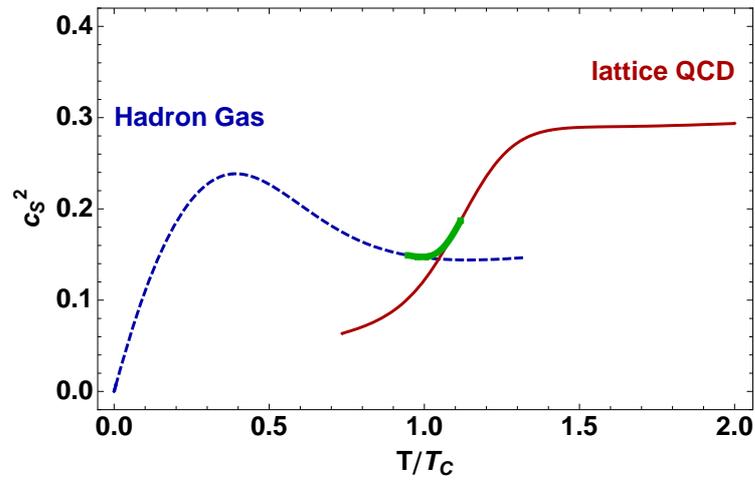}} 
  \end{center}
  \caption{\small Temperature dependence of the square of the sound velocity for strongly interacting baryon-free matter obtained in Ref. \cite{Chojnacki:2007jc}. The results of the lattice simulations of QCD are represented by the solid line, and the results obtained with the hadron-gas model are represented by the dashed line. The thick solid line describes a simple interpolation between these two calculations in the temperature region close to $T_{\rm c}=170$ MeV.}
  \label{fig:eos}
\end{figure}

Equation of state (\ref{epsilon2a}) is very likely to be realized at the very early stages of the collisions, when the produced system is highly anisotropic and described by the distribution functions discussed in this Chapter. On the other hand, as the system expands, the interactions between its constituents thermalize it, so the system's thermodynamic variables approach the realistic EOS and $x$ tends to unity. The realistic EOS for the strongly interacting matter, encoded in the temperature dependent sound velocity, is shown in Fig.~\ref{fig:eos}. At low temperatures it is consistent with the hadron-gas model with all experimentally identified resonances. On the other hand, at high temperatures it coincides with the LQCD results.

The two limiting cases discussed above suggest that one may use the generalized equation of state of the form
\begin{eqnarray}
\varepsilon (x,\sigma)&=&  \varepsilon_{\rm qgp}(\sigma) r(x), \label{epsilon2b}  \\ \nonumber 
P_\perp (x,\sigma)&=&  P_{\rm qgp}(\sigma) \left[r(x) + 3 x r^\prime(x) \right], \label{PT2b}   \\ \nonumber 
P_\parallel (x,\sigma)&=&  P_{\rm qgp}(\sigma) \left[r(x) - 6 x r^\prime(x) \right]. \label{PL2b} 
\end{eqnarray}
Here, the equation of state of the ideal relativistic gas, defined by the functions $\varepsilon_{\rm id}(\sigma)$ and $P_{\rm id}(\sigma)$ has been replaced by the formulas characterizing the realistic equation of state for vanishing baryon chemical potential, i.e., by the functions $\varepsilon_{\rm qgp}(\sigma)$ and $P_{\rm qgp}(\sigma)$ corresponding to the sound velocity function depicted in
Fig.~\ref{fig:eos}.
\par We have to stress that there is no microscopic explanation for the form of Eqs.~(\ref{epsilon2b}).
The main reason for this is that we cannot provide simple quasi-particle picture for the realistic QCD equation of state. Despite this Eqs.~(\ref{epsilon2b}) have many attractive features. Firstly, as discussed above, they reproduce the two attractive limits and they interpolate between them in the continuous way. Moreover there is no additional assumptions needed (such as, e.g., the Landau matching conditions) in order to describe the change from one thermodynamic regime to another one. 
Additionally, Eqs.~(\ref{epsilon2b}) naturally include the phase transition from the QGP to the hadron gas. In this way Eqs.~(\ref{epsilon2b}) are able to describe different stages of evolution of matter in heavy-ion collisions in a uniform way.

\vfill

\chapter[ADHYDRO model]{Highly-anisotropic hydrodynamics with strong dissipation}
\label{chapter:hydro}

In this Chapter we introduce the dynamic equations that determine the evolution of a highly-anisotropic fluid. Our approach is based on the energy-momentum conservation law and on the postulate that specifies the entropy production in the system. The resulting equations define the \texttt{ADHYDRO} model. 

\section[Space-time dynamics of highly-anisotropic systems]{Space-time dynamics \\ of highly-anisotropic systems}
\label{sect:hydro-general}
%
In Chapter \ref{chapter:aniso} we argued that the system formed at the very early stages of relativistic heavy-ion collision may be described by the anisotropic phase-space distribution function (\ref{Fcov}). We assume now that the space-time dynamics of such a system is governed by the following equations,
\begin{eqnarray}
\partial_\mu T^{\mu \nu} &=& 0, \label{enmomcon} \\
\partial_\mu S^{\mu} &=& \Sigma. \label{engrow}
\end{eqnarray}
Equation (\ref{enmomcon}) expresses the energy and momentum conservation law. It has an analogous form to the equation that defines dynamics of the perfect fluid. On the other hand, Eq.~(\ref{engrow}) describes the entropy production determined by the entropy source $\Sigma$. In the standard perfect-fluid approach, the right-hand-side of Eq.~(\ref{engrow}) vanishes and Eq.~(\ref{engrow}) becomes the condition of adiabaticity of the flow.  We note that in the perfect-fluid hydrodynamics Eq.~(\ref{engrow}) follows directly from Eq.~(\ref{enmomcon}) and the appropriate form of the energy-momentum tensor. 

In our approach, Eq.~(\ref{engrow}) should be treated as an ansatz. In Chapter \ref{chapter:aniso} we showed that the energy-momentum tensor $T^{\mu \nu}$ and the entropy flux $S^{\mu}$ appearing in Eqs.~(\ref{enmomcon})--(\ref{engrow}) have the structure defined by Eqs.~(\ref{Taniso}) and (\ref{Saniso}). In this case, the specific form of the entropy source $\Sigma$ allows us to close the system of equations and determines the dynamics of the anisotropic fluid \footnote{This issue will be discussed in more detail later in Section \ref{sect:hydro-entropy}.}.

The projections of Eq.~(\ref{enmomcon}) on the four-vectors $U_\nu$ and $V_\nu$ yield 
\begin{eqnarray}
U^\mu \partial_\mu \varepsilon &=& - \left( \varepsilon+P_\perp \right) \partial_\mu U^\mu 
+ \left( P_\perp-P_\parallel \right) U_\nu  V^\mu \partial_\mu V^\nu, \label{enmomconU} \\
V^\mu \partial_\mu P_\parallel &=& - \left( P_\parallel-P_\perp \right) \partial_\mu V^\mu 
+ \left( \varepsilon+ P_\perp \right) V_\nu  U^\mu \partial_\mu U^\nu. \label{enmomconV}
\end{eqnarray}
Equation (\ref{enmomconU}) expresses the energy conservation, whereas Eq.~(\ref{enmomconV}) describes the conservation of the longitudinal momentum.

\section{General three-dimensional expansion}
\label{sect:hydro-full}

In the general case, where matter expands in the longitudinal and transverse directions without any symmetry constraints, we may use the following parametrization of the four-velocity of the fluid $U^\mu$ and the four-vector $V^{\mu}$ 
\begin{eqnarray}
U^\mu &=& (u_0 \cosh \vartheta, u_x, u_y, u_0 \sinh \vartheta), \label{U3+1} \\
V^\mu &=& (	 \sinh \vartheta, 0, 0,  \cosh \vartheta). \label{V3+1}
\end{eqnarray}
In Eqs.~(\ref{U3+1}) and (\ref{V3+1}) we have introduced the two components of the transverse four-velocity of the fluid, $u_x$ and $u_y$, and the longitudinal fluid rapidity $\vartheta$. Using the normalization condition (\ref{UVnorm}) we also find  
\begin{eqnarray}
u_0 = \sqrt{1+u_\perp^2}, \quad \quad u_\perp = \sqrt{u_x^2 + u_y^2}. \label{u0}
\end{eqnarray}
If the parametrizations (\ref{U3+1}) and (\ref{V3+1}) are used, the differential operators appearing in (\ref{enmomconU}) and (\ref{enmomconV}) take the following form
\begin{eqnarray}
U^\mu \partial_\mu &=& {\bf u}_\perp \cdot {\bf \nabla}_\perp + u_0 L_1, \\
\partial_\mu U^\mu &=& {\bf \nabla}_\perp \cdot {\bf u}_\perp + L_1 u_0 + u_0 L_2 \vartheta, \\
V^\mu \partial_\mu &=&  L_2, \\
\partial_\mu V^\mu &=&  L_1 \vartheta, \\
U_\nu V^\mu \partial_\mu V^\nu &=& u_0 L_2 \vartheta, \\
V_\nu U^\mu \partial_\mu U^\nu &=& - u_0 \left( {\bf u}_\perp \cdot 
{\bf \nabla}_\perp + u_0 L_1  \right) \vartheta,
\label{op1}
\end{eqnarray}
where the two linear differential operators $L_1$ and $L_2$ are defined by the expressions
\begin{eqnarray}
L_1 &=& \cosh Y \partial_\tau - \sinh Y \frac{\partial_\eta}{\tau}, \label{op21}\\
L_2 &=& - \sinh Y \partial_\tau + \cosh Y \frac{\partial_\eta}{\tau}.\label{op22}
\end{eqnarray}
In Eqs.~(\ref{op21})--(\ref{op22}) we have introduced the space-time rapidity $\eta$ and the proper time $\tau$,
\begin{eqnarray}
\eta &=& \frac{1}{2} \ln \frac{t+z}{t-z}, \label{eta}\\
\tau &=& \sqrt{t^2 - z^2}. \label{tau}
\end{eqnarray}
The difference between the longitudinal fluid rapidity $\vartheta$ and the space-time rapidity $\eta$ is defined as 
\begin{equation}
Y = \eta - \vartheta.
\label{Yetatheta}
\end{equation}
We use the boldface notation for two-dimensional vectors in the transverse plane: ${\bf u}_\perp = (u_x,u_y)$ and 
${\bf  \nabla}_{\perp}=(\partial_x,\partial_y )$. 

Besides Eqs.~(\ref{enmomconU}) and (\ref{enmomconV}), the two additional equations describing transverse dynamics of the system are needed. They can be chosen, for example, as the two linear combinations: $U_{1} \partial_{\mu} T^{\mu 1} +   U_{2} \partial_{\mu} T^{\mu 2} = 0$ and $U_{2} \partial_{\mu} T^{\mu 1} -  U_{1} \partial_{\mu} T^{\mu 2} = 0$. In this case we find
\begin{eqnarray}
	& & \mathcal{D} u_{\perp} = - \frac{u_{\perp}}{\varepsilon + P_{\perp}} 
\left[ \frac{ {\bf u}_\perp \cdot {\bf \nabla}_\perp P_{\perp}}{u_\perp^2}  
+ \mathcal{D} P_{\perp} +  (P_{\perp} - P_{\parallel}) U_\nu V^\mu \partial_\mu V^\nu \right],
	\label{HydEqEuler1}
	 \\
	 & & \mathcal{D} \left( \frac{u_x}{u_y} \right) = \frac{1}{u_y^2 (\varepsilon + P_{\perp})} \left( u_x \partial_y - u_y \partial_x \right)P_{\perp}.
\label{HydEqEuler2}
\end{eqnarray}
Here $\mathcal{D} = U^\mu \partial_\mu$ is the total time derivative  and $\theta = \partial_\mu U^\mu$ is the volume expansion rate. Using this notation, the entropy production equation, see Eq.~(\ref{engrow}), can be put in the compact form
\begin{equation}
\mathcal{D} \sigma + \sigma \theta = \Sigma.
\label{entrprod}
\end{equation}

Equation (\ref{entrprod}) together with Eqs.~(\ref{enmomconU}), (\ref{enmomconV}), (\ref{HydEqEuler1}) and (\ref{HydEqEuler2}) form a set of hydrodynamic equations describing dynamics of the anisotropic fluid. {\it We refer shortly to this system of equations as to the \texttt{ADHYDRO} equations or as to the \texttt{ADHYDRO} model or framework}. 

If the generalized equation of state $\varepsilon(\sigma,x)$, see Sect. \ref{sect:aniso-eos}, and the entropy source $\Sigma(\sigma,x)$, see Sect. \ref{sect:hydro-entropy}, are specified, the \texttt{ADHYDRO} equations form a closed system of five equations for five unknown functions: two components of the fluid velocity $u_x$ and $u_y$, the longitudinal rapidity of the fluid $\vartheta$, the non-equilibrium entropy density $\sigma$ and the anisotropy parameter $x$. These functions depend on  transverse coordinates $x,y$, the space-time rapidity $\eta$, and the proper time $\tau$. One has to solve the \texttt{ADHYDRO} equations numerically for a given initial condition specified at a certain initial time $\tau_0$ (to be discussed in Chapter \ref{chapter:initial}).

\section{Entropy source}
\label{sect:hydro-entropy}
%
The entropy source $\Sigma$ describes the entropy growth due to equilibration of pressures in the system. As we have already mentioned above, Eqs.~(\ref{enmomcon}) and  (\ref{engrow}) may be solved only if the function $\Sigma(\sigma,x)$ is specified. The functional form $\Sigma(\sigma, x)$ must be delivered as the external input for the anisotropic hydrodynamics. 

The simplest ansatz for $\Sigma(\sigma,x)$ satisfying general physical assumptions may be proposed in the following form \cite{Ryblewski:2010ch,Ryblewski:2010bs,Ryblewski:2011aq}
\begin{equation}
\Sigma = \Sigma_0(\sigma,x) = \frac{(\lambda_{\perp}-\lambda_{\parallel})^{2}}{\lambda_{\perp} \lambda_{\parallel}}\frac{\sigma}{\tau_{\rm eq}} = \frac{(1-\sqrt{x})^{2}}{\sqrt{x}}\frac{\sigma}{\tau_{\rm eq}},
\label{en1}
\end{equation}
where $\tau_{\rm eq}$ is a time-scale parameter introduced to control the rate of the processes leading to equilibration of the system. The expression on the right-hand-side of Eq.~(\ref{en1}) has several appealing features. First of all it is positive, as expected on the grounds of the second law of thermodynamics. Moreover, it has a correct dimension, since $\Sigma$ is proportional to the entropy density $\sigma$. In this way Eq.~(\ref{en1}) does not destroy the scale invariance of the perfect fluid hydrodynamics, which allows for multiplication of $\sigma$ in the evolution equations by an arbitrary constant. In the natural way, the entropy source $\Sigma$ defined by (\ref{en1}) vanishes in equilibrium, where $x=1$, thus Eqs.~(\ref{enmomcon})--(\ref{engrow}) can be reduced to the equations of perfect fluid hydrodynamics in the limit $x \to 1$. The expression (\ref{en1}) is also symmetric with respect to the interchange of $\lambda_\perp$ with $\lambda_\parallel$ (consequently $\Sigma$ does not change if $x \to 1/x$). Finally, for small deviations from equilibrium, where $|x-1| \ll 1$, we find
\begin{equation}
\Sigma_0 (x) \approx \frac{(x-1)^{2}}{4 \tau_{\rm eq}} \sigma.
\label{en1exp}
\end{equation}
The quadratic dependence of the entropy source displayed in (\ref{en1exp}) is characteristic for the 2nd order viscous hydrodynamics (to be discussed in Chapter \ref{sect:hydro-is}). 
%
\section{Anisotropy evolution}
\label{sect:hydro-anisotropy}
%

If both the generalized equation of state and the entropy production term are defined, one can derive a compact expression for the time evolution of the anisotropy parameter $x$. Of course, this expression should be considered together with other dynamic equations. Nevertheless, its form turns out to be useful in the analysis of the time evolution of the system.

In the case described by Eqs.~(\ref{epsilon2b}) we use Eq.~(\ref{enmomconU}) and find
\begin{eqnarray}
   \mathcal{D} x &=&  \frac{3 x P_{\rm qgp}}{\varepsilon_{\rm qgp}} \left( 3 U_\nu V^\mu \partial_\mu V^\nu -\theta \right) - \left(1 + \frac{P_{\rm qgp}}{\varepsilon_{\rm qgp}}\right) \frac{H(x)}{\tau_{\rm eq}},
   \label{x1}
\end{eqnarray}
where 
\begin{equation}
H(x) =  \frac{r(x)}{r'(x)} \frac{\Sigma}{\sigma} \tau_{\rm eq}.
\label{H}
\end{equation}
Similarly, in the case described by Eqs.~(\ref{epsilon2a}) we find
\begin{eqnarray}
\mathcal{D} x &=& x \left( 3 U_\nu V^\mu \partial_\mu V^\nu -\theta \right) -\frac{4}{3 \tau_{\rm eq}} H(x).
\label{x2}
\end{eqnarray}
Equation (\ref{x2}) may be treated as the special case of (\ref{x1}) where we set
\mbox{$\varepsilon_{\rm qgp} = 3 P_{\rm qgp}$.} 

\chapter{Purely longitudinal boost-invariant motion}
\label{chapter:lbim}
In this Chapter we study a simple case of purely longitudinal, boost-invariant expansion. This situation is easy to investigate and we can obtain useful information about the behavior of anisotropic systems by applying simple analytic methods. This will help us to understand the behavior of more complicated systems which either expand transversally or do not exhibit boost invariance.   

\section{Implementation of boost invariance}
\label{sect:hydro-bi}

The measurements done by the BRAHMS Collaboration \cite{Bearden:2004yx} at RHIC showed that the rapidity distributions of charged particles are almost constant in the midrapidity region ($|\mathrm{y}| \leq 1$). This observation suggests that all measured quantities, to a good approximation, are boost-invariant {\it in this region}. The boost-invariance implies that all thermodynamic variables are functions of the proper time $\tau$ and transverse coordinates only, while the longitudinal flow has the form
\begin{equation}
v_z \equiv \tanh \vartheta = \frac{z}{t}.
\label{bjorkenflow}
\end{equation}
Since $t=\tau \cosh \eta $ and $z=\tau \sinh \eta $,  Eq.~(\ref{bjorkenflow}) is equivalent to the simple condition
\begin{equation}
\vartheta = \eta,
\label{thetaeqeta}
\end{equation}
i.e., $Y = 0$. With this assumption at hand, it is sufficient to find the solutions of our model for $z=0$ ($\tau=t$), since solutions for other values of $z$ may be obtained by simple Lorentz transformations. 

To illustrate the main features of our model with simple examples, we neglect now the transverse expansion of the fluid. This is a reasonable approximation for the early stages of the evolution, where the longitudinal expansion dominates,  \mbox{$u_x=u_y=0$} ($u_0=1$). We also assume now that the system may be treated as homogeneous in the transverse plane, hence, we neglect dependence of thermodynamic quantities on the transverse variables $x$ and $y$. These assumptions are characteristic for the Bjorken model \cite{Bjorken:1982qr}, where the behavior of the perfect fluid has been considered. In our approach the perfect-fluid dynamics is replaced by the hydrodynamics of the highly-anisotropic fluid.   

In the considered one-dimensional and boost-invariant case Eq.~(\ref{enmomconV}) is automatically fulfilled, since $\mathcal{D} = \partial_\tau$, $\theta=1/\tau$, and $\tau V^\mu \partial_\mu V^\nu = U^\nu$. After neglecting transverse expansion, also Eqs. (\ref{HydEqEuler1}) and (\ref{HydEqEuler2}) are satisfied automatically. Therefore, we are left with Eqs. (\ref{enmomconU}) and (\ref{entrprod}) which are reduced to the expressions
\begin{eqnarray}
 & &\frac{d\varepsilon}{d\tau} = -\frac{\varepsilon + P_\parallel}{\tau},\label{enmomconU_LBI} \\
 & &\frac{d\sigma}{\sigma d\tau} + \frac{1}{\tau} = \frac{\Sigma}{\sigma}.\label{entrprod_LBI}
\end{eqnarray}
Equations (\ref{enmomconU_LBI}) and (\ref{entrprod_LBI}) are two equations for two unknown functions, $\sigma (\tau)$ and $x(\tau)$, which may be solved numerically.
\section{Asymptotic limits}
\label{sect:hydro-al}

As we have discussed above in Sect.~\ref{sect:hydro-anisotropy}, Eqs.~(\ref{enmomconU_LBI}) and (\ref{entrprod_LBI}) may be used to derive an equation describing the evolution of pressure anisotropy $x$. For the purely longitudinal boost-invariant motion,  Eq.~(\ref{x2}) simplifies to an ordinary differential equation for $x$, 
\begin{equation}
\frac{dx}{d\tau}  =  \frac{2 x}{\tau} - \frac{4 H(x)}{3\tau_{\rm eq}},
\label{x3}
\end{equation}
which with the help of Eq.~(\ref{H}) may be also rewritten as
\begin{equation}
r'(x) \left(\frac{dx}{d\tau}  -  \frac{2 x}{\tau}\right) = - r(x) \frac{4 \Sigma}{3 \sigma}.
\label{x4}
\end{equation}

First of all, it is useful to consider the isentropic flow where $\Sigma(\sigma,x)=0$. In this case, from Eq.~(\ref{entrprod_LBI}) we obtain the Bjorken solution 
\begin{equation}
\sigma = \frac{\sigma_0 \tau_0}{\tau},
\label{sigmaBj}
\end{equation}
where the parameters $\sigma_0$ and $\tau_0$ are arbitrary constants. For $\Sigma(\sigma,x)=0$ the right-hand-side of Eq.~(\ref{x4}) vanishes. This implies that either $x=1$ or $x = x_0 \tau^2/\tau_0^2$ (where $x_0$ is another integration constant). 

The case $x=1$ corresponds to the standard, perfect-fluid hydrodynamics. Indeed, for an ideal relativistic gas in equilibrium we have $P_\parallel = P_\perp = P_{\rm eq}$ and  Eq.~(\ref{enmomconU_LBI}) implies $\varepsilon(\tau) \sim \tau^{-4/3}$. In the second case we obtain a simple longitudinal free-streaming solution (see our detailed discussion in Section \ref{sect:hydro-fs}). This situation is equivalent to the case where the equilibration time is set to infinity,  $\tau_{\rm eq} = \infty$. In addition we can consider also the case where $P_\parallel = 0$. In this situation Eq.~(\ref{enmomconU_LBI}) describes the evolution of the energy density valid for transverse hydrodynamics \cite{Bialas:2007gn,Ryblewski:2010tn}, $\varepsilon(\tau) \sim \tau^{-1}$.

\section{Consistency with Israel-Stewart theory}
\label{sect:hydro-is}

For a near-equilibrium system, the distribution function can be decomposed into an equilibrium part plus a small correction part. The methodology of including such corrections in the consistent way leads to the framework of relativistic viscous hydrodynamics of the first order (non-causal Navier-Stokes theory) or the second order (causal Israel-Stewart formalism \cite{Israel:1979wp,Muronga:2001zk,Muronga:2003ta}). When deviations from equilibrium are small, which in our approach may be quantified by the condition  $|\xi| \ll 1$, and the motion is purely longitudinal, one can show that our model is consistent with the Israel-Stewart theory.

As before, we consider again the baryon-free systems. The 2nd order dissipative fluid is characterized by the energy-momentum tensor 
\begin{equation}
T^{\mu \nu} = \varepsilon U^\mu U^\nu + P^{\mu \nu},
\label{vTmunu}
\end{equation}
where the pressure tensor $P^{\mu \nu}$ is defined by the formula
\begin{equation}
P^{\mu \nu} = - (P_{\rm eq} + \Pi) \Delta^{\mu \nu} + \pi^{\mu \nu}.
\label{vPmunu}
\end{equation}
Here $\Delta^{\mu \nu} = g^{\mu \nu} - U^\mu U^\nu$ is the tensor projecting on the space perpendicular to the four-velocity $U^\mu$, $P_{\rm eq}$ is the isotropic pressure, $\Pi$ is the \textit{viscous bulk pressure}, and $\pi^{\mu \nu}$ is the \textit{viscous shear tensor}. 

The energy-momentum tensor (\ref{Taniso}), which forms the basis of our approach, may be rewritten in the form (\ref{vTmunu}) if we make the following identification  
\begin{equation}
P^{\mu \nu} = - P_\perp \Delta^{\mu \nu} - (P_\perp - P_\parallel) V^\mu V^\nu.
\label{aPmunu}
\end{equation}
Let us ignore now the bulk viscosity, since for massless particles or ultra-relativistic particles the bulk viscosity vanishes, $\Pi =0$. If there is no transverse expansion, we may compare Eqs.~(\ref{vPmunu}) and (\ref{aPmunu}) in the Landau local rest frame, and make the identification of the longitudinal and transverse pressures: $P_\perp = P_{\rm eq} + \overline{\pi}/2$, $P_\parallel = P_{\rm eq} -\overline{\pi}$, where $P_{\rm eq}=\varepsilon_{\rm eq}/3 = \varepsilon/3$ and $\overline{\pi}\equiv \pi^{33}$ is the $33$ component of the fluid shear tensor. Then, the expansion for small $\xi$ gives \cite{Martinez:2009ry,Martinez:2010sc}
\begin{equation}
\frac{\overline{\pi}}{\varepsilon_{\rm eq}} = 2 x \frac{R^\prime(x)}{R(x)} \approx \frac{8}{45} \xi.
\label{IS1}
\end{equation}
By differentiating Eq.~(\ref{IS1}) with respect to the proper time we obtain
\begin{equation}
\frac{45}{8} \left( \frac{1}{\varepsilon}\frac{d\overline{\pi}}{d\tau}-\frac{\overline{\pi}}{\varepsilon^2}\frac{d\varepsilon}{d\tau}  \right) = \frac{d\xi}{d\tau}.
\label{IS3}
\end{equation}
Incorporating the leading order terms in $\xi$ of Eqs.~(\ref{enmomconU_LBI}) and (\ref{x3}) in Eq.~(\ref{IS3}) one finds
\begin{equation}
\frac{d\overline{\pi}}{d\tau} + \frac{4 \overline{\pi}}{3 \tau}- \frac{16}{45} \frac{\varepsilon}{\tau} = - \frac{15 \overline{\pi}}{4 \tau_{\rm eq}}.
\label{IS4}
\end{equation}
We can match our model with Israel-Stewart theory by making the following identifications
\begin{equation}
\frac{1}{\tau_{\rm eq}} = \frac{4}{15 \tau_\pi}, \quad   \tau_\pi = \frac{5 \eta_\pi}{T \sigma_{\rm eq}},
\label{IS4a}
\end{equation} 
where $\tau_{\pi}$ is the \textit{shear relaxation time}, $\eta_\pi$ is \textit{the shear viscosity}, $T$ is the temperature in the isotropic system, and $\sigma_{\rm eq}$ is the equilibrium entropy density. In this way, Eq.~(\ref{IS4}) takes the form of the 2nd order viscous hydrodynamic equation for $\overline{\pi}$
\begin{equation}
\frac{d\overline{\pi}}{d\tau} = - \frac{4 \overline{\pi}}{3 \tau} + \frac{4\eta_\pi}{3 \tau_\pi \tau} - \frac{\overline{\pi}}{\tau_\pi}.
\label{IS5}
\end{equation}
The entropy production in the Israel-Stewart theory is given by the formula
\begin{equation}
\partial_\mu S^\mu = \frac{\pi_{\alpha \beta} \pi^{\alpha \beta}}{2 \eta_\pi T},
\label{IS5}
\end{equation} 
and for the longitudinal expansion considered in this Section, $\pi_{\alpha \beta} \pi^{\alpha \beta} = 3 \overline{\pi}^2/2$. Substituting Eqs.~(\ref{IS1}) and (\ref{IS4a}) into Eq.~(\ref{IS5}) we obtain
\begin{equation}
\partial_\mu S^\mu = \sigma_{\rm eq} \frac{\xi^2}{4 \tau_{\rm eq}},
\label{IS5a}
\end{equation} 
which is consistent with our main ansatz for the entropy production if deviations from equilibrium are small.

One should stress that deviations from equilibrium in the Israel-Stewart theory have to be small, so that the dissipative fluxes are small compared to equilibrium variables, $\sqrt{\pi_{\mu \nu} \pi^{\mu \nu}} \ll P_{\rm eq}$. If this condition is not fulfilled, the use of the Israel-Stewart framework cannot be formally justified. 

\section[Deriving entropy production from kinetic theory]{Deriving entropy production\newline from kinetic theory}
\label{sect:hydro-ms}

Our formulation of the highly-anisotropic hydrodynamics in one-dimensional case was followed independently by Martinez and Strickland \cite{Martinez:2010sc,Martinez:2010sd}, who derived equations analogous to Eqs.~(\ref{enmomconU_LBI})--(\ref{entrprod_LBI}) starting from the relativistic Boltzmann equation 
\begin{equation}
p^\mu \partial_\mu f(x,p) = {\cal C}[f(x,p)].
\label{kineq2}
\end{equation}
Here ${\cal C}[f(x,p)]$ is the collision kernel which accounts for all microscopic interaction processes among particles. In the Martinez-Strickland model \cite{Martinez:2010sc,Martinez:2010sd}, the evolution of matter is described by two coupled ordinary differential equations: the first one for the momentum anisotropy $\xi$, and the second one for the typical hard momentum scale $p_{\rm hard}$ (average momentum in the parton distribution function). These equations are obtained by taking the $0$th and $1$st moments of the Boltzmann equation (\ref{kineq2}).

In general, the entropy production may be calculated directly from the definition (\ref{Sdef}) with the help of Eq.~(\ref{kineq2}). For classical particles, it is given by the expression
\begin{equation}
\partial_\mu S^\mu = - \int \frac{d^3p}{(2 \pi)^3 p^0}  {\cal C}[f(x,p)](\ln f(x,p) + 1).
\label{entrkin}
\end{equation}
Martinez and Strickland used the relaxation time approximation for the collision kernel.  This allows for a direct calculation of the right-hand-side of Eq.~(\ref{entrkin}) and gives the entropy source $\Sigma_{MS}$ (analogous to our $\Sigma$),
\begin{eqnarray}
\Sigma_{{\rm MS}} &=& \sigma \Gamma 
\left[\frac{R^\prime_{{\rm MS}}(\xi)}{R_{{\rm MS}}(\xi)} +\frac{2}{3(1+\xi)}\right] 
H_{{\rm MS}}(\xi) \nonumber \\ 
&=& \sigma \Gamma \, \frac{R'(x)}{R(x)} \, H_{{\rm MS}}(\xi(x)).
\label{en3}
\end{eqnarray}
In Eq.~(\ref{en3}), the parameter $\Gamma$ denotes the inverse of the relaxation time, $\Gamma = 1/\tau^{\rm MS}_{\rm eq}$. The function $H_{{\rm MS}}(\xi(x))$ appearing in Eq.~(\ref{en3}) is an analog of our function $H(x)$,
\begin{eqnarray}
H_{{\rm MS}}(\xi) = \frac{3 (\xi +1) \left(\sqrt{\xi +1} R_{{\rm MS}}^{3/4}(\xi )-1\right) R_{{\rm MS}}(\xi )}{2 R_{{\rm MS}}(\xi )+ 3 (\xi +1) R_{{\rm MS}}'(\xi )},
\label{Hms}
\end{eqnarray}
and $R_{{\rm MS}}(\xi(x))$ is an analog of our $R(x)$,
\begin{eqnarray}
R_{{\rm MS}}(\xi) = \frac{1}{2} \left[ \frac{1}{1+\xi} + \frac{\arctan\sqrt{\xi}}{\sqrt{\xi}}\right]
= \frac{\pi^2}{3 g_0} (1+\xi)^{-2/3} R(1+\xi).
\end{eqnarray}
For small anisotropies, one can check that the Martinez-Strickland approach coincides with our model if
\begin{equation}
\Gamma = \frac{1}{\tau^{\rm MS}_{\rm eq}} = \frac{15}{2 \tau_{\rm eq} } .
\label{Gammataueq}
\end{equation}

%
\section{Longitudinal free-streaming}
\label{sect:hydro-fs}

In the case of anisotropic distribution functions (see Section \ref{sect:aniso-distr}) it is interesting to consider the pure longitudinal free transport (alias {\it free-streaming}) of particles, which is the regime in which particles do not interact completely with each other. In this case, the distribution functions should satisfy the collisionless Boltzmann kinetic equation
\begin{equation}
p^\mu \partial_\mu f(x,p) = 0.
\label{kineq}
\end{equation}
We assume that the effects of free-streaming do not change the shape of the anisotropic distribution function (\ref{Fcov}) and the time evolution of the parameters $\lambda_\perp$, $\lambda_\parallel$, and $U^\mu$ is determined by the conservation laws. 

We parametrize the four-momentum of a particle in the standard way
\begin{eqnarray}
p^\mu &=& \left(E_p, {\bf p}_\perp, p_\parallel \right) =
\left(m_\perp \cosh \mathrm{y}, {\bf p}_\perp, m_\perp \sinh \mathrm{y} \right), \label{parmom}
\end{eqnarray} 
where we have introduced the transverse mass $m_\perp = \sqrt{m^2 + p_\perp^2}$  and the rapidity of a particle
\begin{eqnarray}
\mathrm{y} = \frac{1}{2} \ln \frac{E_p+p_\parallel}{E_p-p_\parallel} \label{particlerap} .
\end{eqnarray}
For simplicity, in the following analysis we assume that particles are massless, $m=0$. For the pure longitudinal boost-invariant expansion we use Eqs.~(\ref{Fcov}), (\ref{U3+1}), (\ref{V3+1}), and (\ref{parmom}), and we rewrite Eq.~(\ref{kineq}) in the form 
\begin{equation}
\left[ \cosh(\mathrm{y}-\eta) \frac{\partial}{\partial\tau}
+ \frac{\sinh(\mathrm{y}-\eta)}{\tau} \frac{\partial}{\partial \eta} \right] f\left(w,v\right) = 0,
\label{kineq1}
\end{equation}
where we have introduced the variable $w = p_\perp/\lambda_\perp(\tau)$ and $v = p_\perp \sinh(y-\eta)/\lambda_\parallel(\tau)$. After simple algebra we obtain
\begin{equation}
\frac{\partial f}{\partial w} \frac{d\lambda_\perp}{\lambda_\perp^2 d\tau}
+ \frac{\sinh(\mathrm{y}-\eta)}{\lambda_\parallel^2} \frac{\partial f}{\partial v}
\left[\frac{d\lambda_\parallel}{d\tau} + \frac{\lambda_\parallel}{\tau}  \right] =0.
\end{equation}
This equation is satisfied by any form of the function $f$ provided $\lambda_\perp = \lambda_\perp^0$ and \mbox{$\lambda_\parallel = \lambda_\parallel^0 \tau_0/\tau$}, where $\tau_0$, $\lambda_\perp^0$ and $\lambda_\parallel^0$ are constants. In the considered case the anisotropy parameter $x$ has the form
\begin{equation}
x = \left( \frac{\lambda_\perp^0 }{\lambda_\parallel^0 \tau_0} \right)^2 \, \tau^2,
\label{xquadra}
\end{equation}
where $x_0=\left( \lambda_\perp^0 / \lambda_\parallel^0 \right)^2$ specifies the initial anisotropy of the system. Thus, we see that our approach with $\tau_{\rm eq} \to \infty$ includes the boost-invariant free-streaming as a special case.

\begin{figure}[t!]
\begin{center}
\subfigure{\includegraphics[angle=0,width=0.59\textwidth]{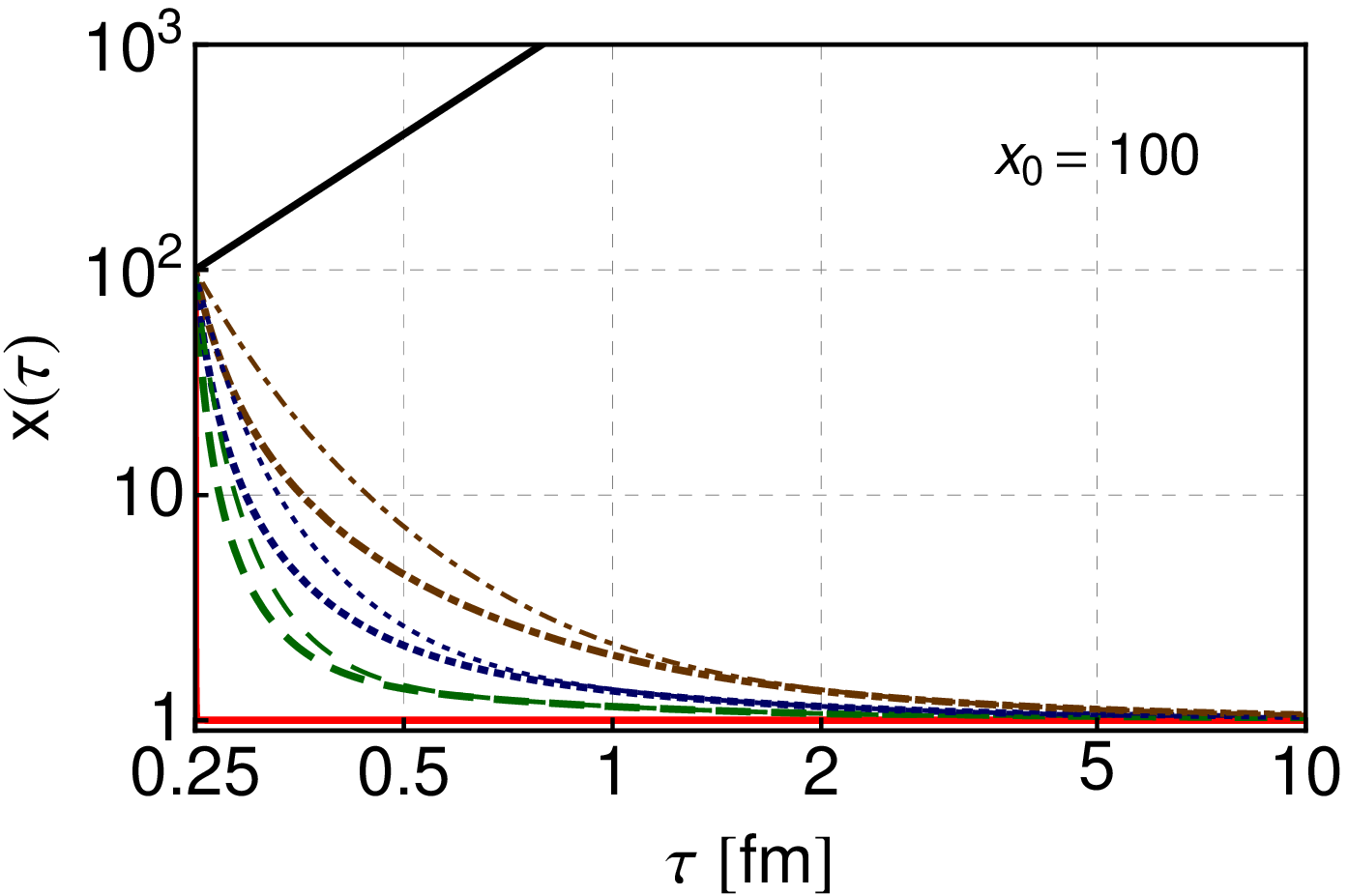}} \\
\subfigure{\includegraphics[angle=0,width=0.58\textwidth]{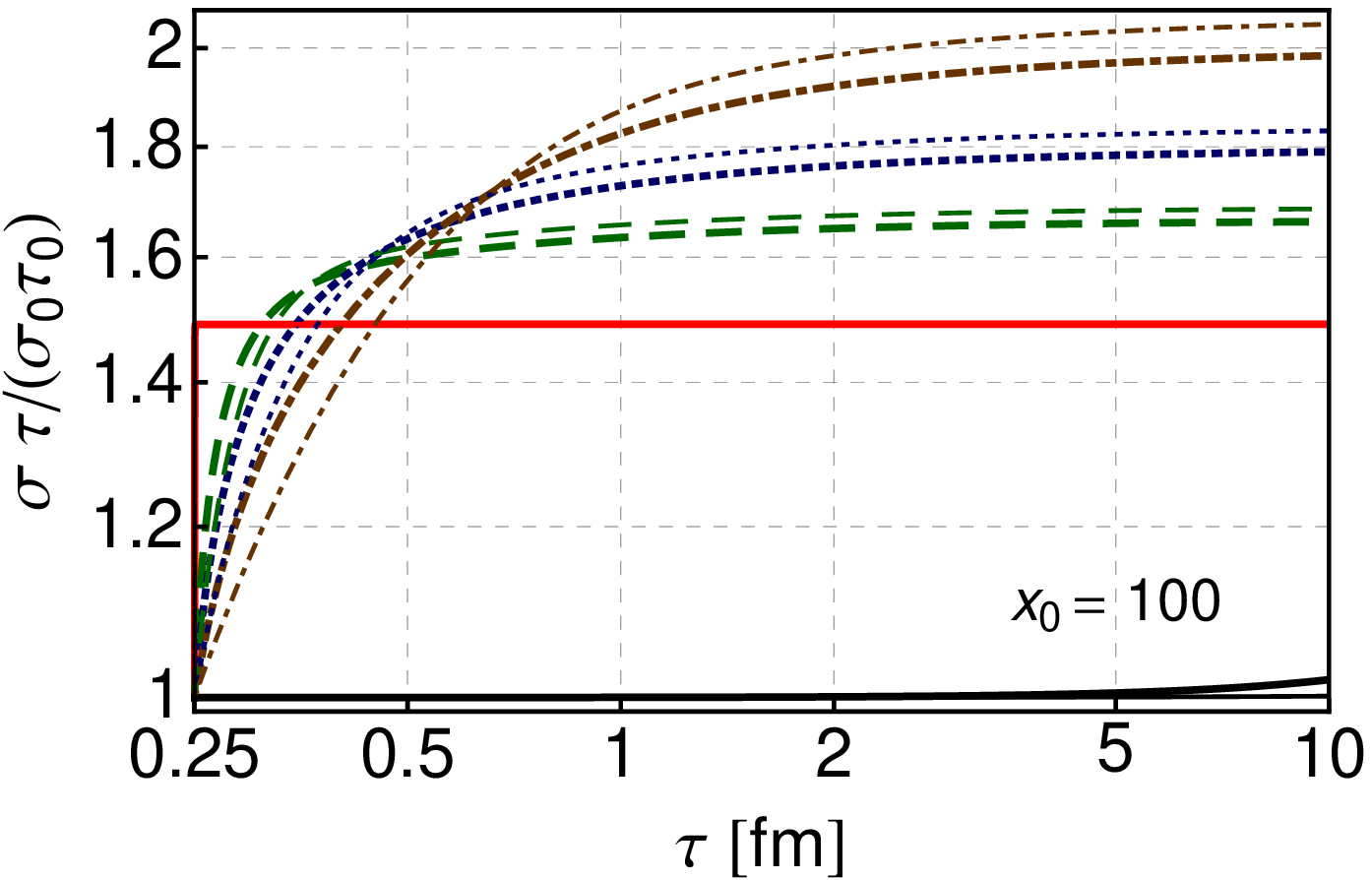}} 
\end{center}
\caption{Time dependence of the asymmetry parameter $x$ (upper part) and the entropy density normalized to the values obtained from the Bjorken model \mbox{$\sigma = \sigma_0 \tau_0/\tau$} (lower part) for the large initial anisotropy parameter, $x_0=100$. Thick lines correspond to the entropy source (\ref{en1}) and thin lines correspond to the source (\ref{en3}). The results are shown for different values of the time-scale parameter $\tau_{\rm eq}$: $0$ (solid red lines), $0.25$ fm (dashed green lines), $0.5$ fm (dotted blue lines), $1$~fm (dashed-dotted brown lines), $\infty$ (solid black lines). The initial time is fixed to \mbox{$\tau_0=0.25$ fm.}}
\label{fig:transvdom}
\end{figure}

%
\section[Results for longitudinal boost-invariant motion]{Results for purely longitudinal boost-invariant motion}
\label{sect:hydro-r}

We solve Eqs.~(\ref{enmomconU_LBI}) and (\ref{entrprod_LBI}) numerically with the entropy source defined by Eq.~(\ref{en1}) with the initial conditions $x=x_0$ and $\sigma=\sigma_0$ specified at the initial proper time $\tau=\tau_0$. For comparison, we also use the entropy source defined by Eq.~(\ref{en3}). The time-scale parameter $\tau_{\rm eq}$ is set equal to $0$, $0.25$ fm, $0.5$ fm, $1$~fm, and $\infty$\footnote{The value $\tau_{\rm eq}=0$ means that in the numerical calculations we take  $\tau_{\rm eq}= 10^{-3}$ fm, similarly, the value $\tau_{\rm eq}=\infty$ corresponds in the numerical calculations to  $\tau_{\rm eq}=10^5$ fm.}. The values of the relaxation time $\tau_{\rm eq}$ and the parameter $\Gamma$ are connected by Eq.~(\ref{Gammataueq}). In all the numerical calculations, the initial conditions are specified at the initial proper time $\tau_0=0.25$ fm. We consider three cases: 

\begin{itemize}
\item[{\bf i)}] $x_0=100$, where the transverse pressure dominates over the longitudinal pressure during the very early stages of the evolution (as suggested by the microscopic models), in this case the momentum shape is oblate, i.e., it is stretched in the transverse direction and squeezed in the longitudinal direction, 

\item[{\bf ii)}] $x_0=0.032$ \footnote{This value has been chosen because $r(100)\approx r(0.032)$ and the case with $x_0=0.032$ may be treated as a counterpart of the case  $x_0=100$, with the exchanged roles of the transverse and longitudinal pressures.}, where the longitudinal pressure is much larger than the transverse pressure, in this case the momentum shape is prolate, and finally,

\item[{\bf iii)}] the initial system is isotropic, $x_0=1$. This case is the closest to the perfect-fluid hydrodynamics. In all the cases the system's evolution is studied in the time interval $0.25 \le \tau \le 10$ fm.

\end{itemize}

\noindent  In Fig.~\ref{fig:transvdom} we show our results for the case \textbf{i)}. In the upper part of Fig.~\ref{fig:transvdom} we show the functions $x(\tau)$ for the entropy source (\ref{en1}) and (\ref{en3}), represented by thick and thin lines, respectively. The results are shown for different values of the time-scale parameter: $\tau_{\rm eq} =0$ (solid red lines), $\tau_{\rm eq} =0.25$ fm (dashed green lines), $\tau_{\rm eq} =0.5$ fm (dotted blue lines), $\tau_{\rm eq} =1.0$ fm (dashed-dotted brown lines), $\tau_{\rm eq} =\infty$ (solid black lines). In the cases $\tau_{\rm eq} = 0.25, 0.5, 1.0$ fm we observe fast changes of $x(\tau)$ caused by the fact that $H(x)$ and $H_{\rm MS}(x)$ are very large for $x \gg 1$ ($H(x) \sim 6 x^{3/2}$). For $\tau \gg \tau_{\rm eq}$ the function $x(\tau)$ saturates and approaches unity. In addition, we show the result for $\tau_{\rm eq} = 0$, which corresponds to \textit{sudden isotropization} of the system. The initial highly-anisotropic system becomes isotropic almost instantaneously. Another interesting situation takes place for $\tau_{\rm eq} = \infty$, where the quadratic behavior (\ref{xquadra}) typical for free streaming is observed. In the latter case, the anisotropy grows with time and the system never reaches equilibrium.

\begin{figure}[t!]
\begin{center}
\hspace{0.55cm}\subfigure{\includegraphics[angle=0,width=0.58\textwidth]{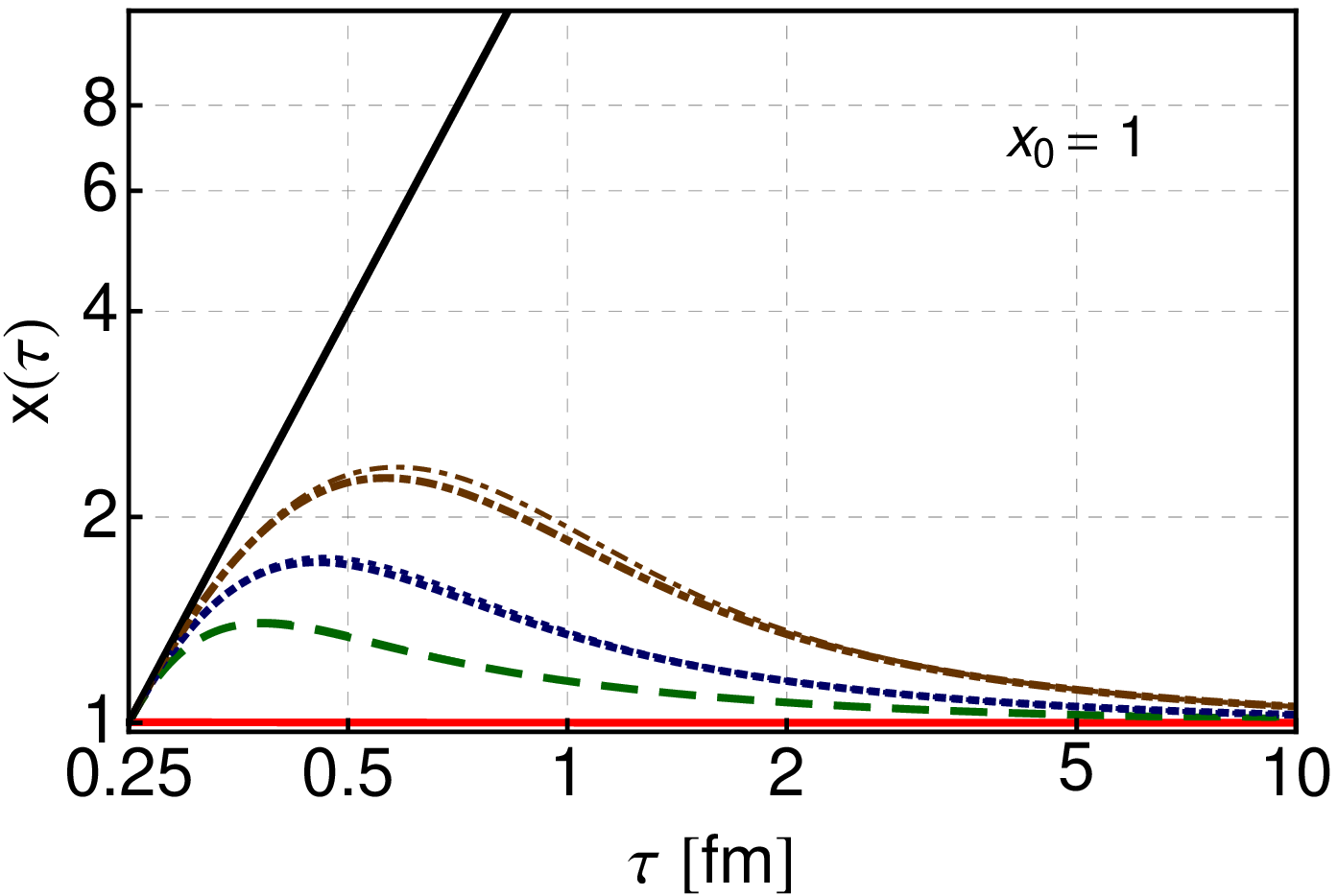}} \\
\subfigure{\includegraphics[angle=0,width=0.62\textwidth]{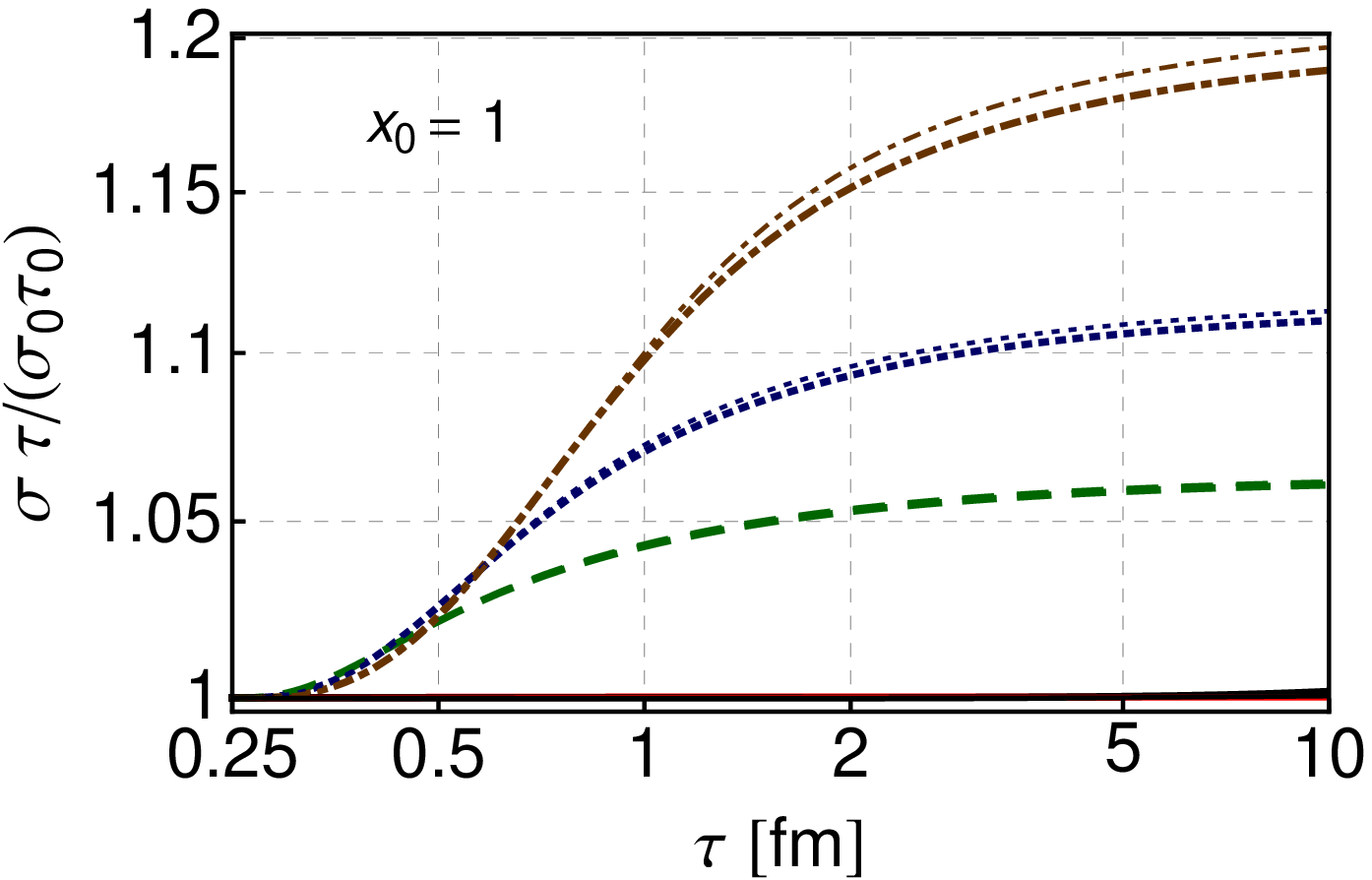}} 
\end{center}
\caption{Same as Fig.~\ref{fig:transvdom}, but the calculations done for the initial anisotropy $x_0=1$.}
\label{fig:isotropic}
\end{figure}

In the lower part of Fig.~\ref{fig:transvdom} we show the corresponding plots of the entropy density normalized by the Bjorken solution (\ref{sigmaBj}). During the equilibration process, large amount of entropy is produced. For $\tau_{\rm eq} =0.25, 0.5, 1$ fm, the entropy increases by about $65\%$, $80\%$ and $100\%$, respectively. For $\tau \gg \tau_{\rm eq}$ the normalized  entropy saturates, indicating that the flow attains the form of the Bjorken flow. In the case where $\tau_{\rm eq} = \infty$, we also obtain the Bjorken-like dependence (\ref{sigmaBj}) where $\sigma \sim \lambda_{\perp}^2 /\tau$ and the transverse temperature becomes independent of time. This is an expected result, since without the longitudinal pressure no longitudinal work is done. If, in addition, there is no transverse expansion, the system cannot cool down and its temperature remains constant. In the case $\tau_{\rm eq} = 0$ the sudden jump of entropy by about $50 \%$ is observed.

In Fig.~\ref{fig:isotropic} we show our results for the initially isotropic case \textbf{iii)}. The notation is the same as in Fig.~\ref{fig:transvdom}. In the upper part of Fig.~\ref{fig:isotropic}, for $\tau_{\rm eq} =0.25, 0.5, 1$ fm we observe the non-monotonic behaviour of the functions $x(\tau)$. This is caused by the fact that $H(x=0)=0$ and free streaming dominates the early evolution. In this case, the anisotropy grows with time ($dx/d\tau > 0$). This is an expected feature, since even if the starting distribution is isotropic, the system becomes anisotropic if there are no interactions to sustain the equilibrium distribution. With increasing $\tau$, the second therm on the right-hand-side of Eq.~(\ref{x3}) starts to dominate over the first therm and the anisotropy decreases  ($dx/d\tau < 0$). For $\tau \gg \tau_{\rm eq}$ the system becomes isotropic and the term $2 x/\tau$ becomes negligible. In the lower part of Fig.~\ref{fig:isotropic} we observe the corresponding entropy production for $\tau_{\rm eq} =0.25, 0.5, 1$ fm. Again, the longer is the anisotropic stage, the more entropy is produced.

\begin{figure}[t!]
\begin{center}
\subfigure{\includegraphics[angle=0,width=0.58\textwidth]{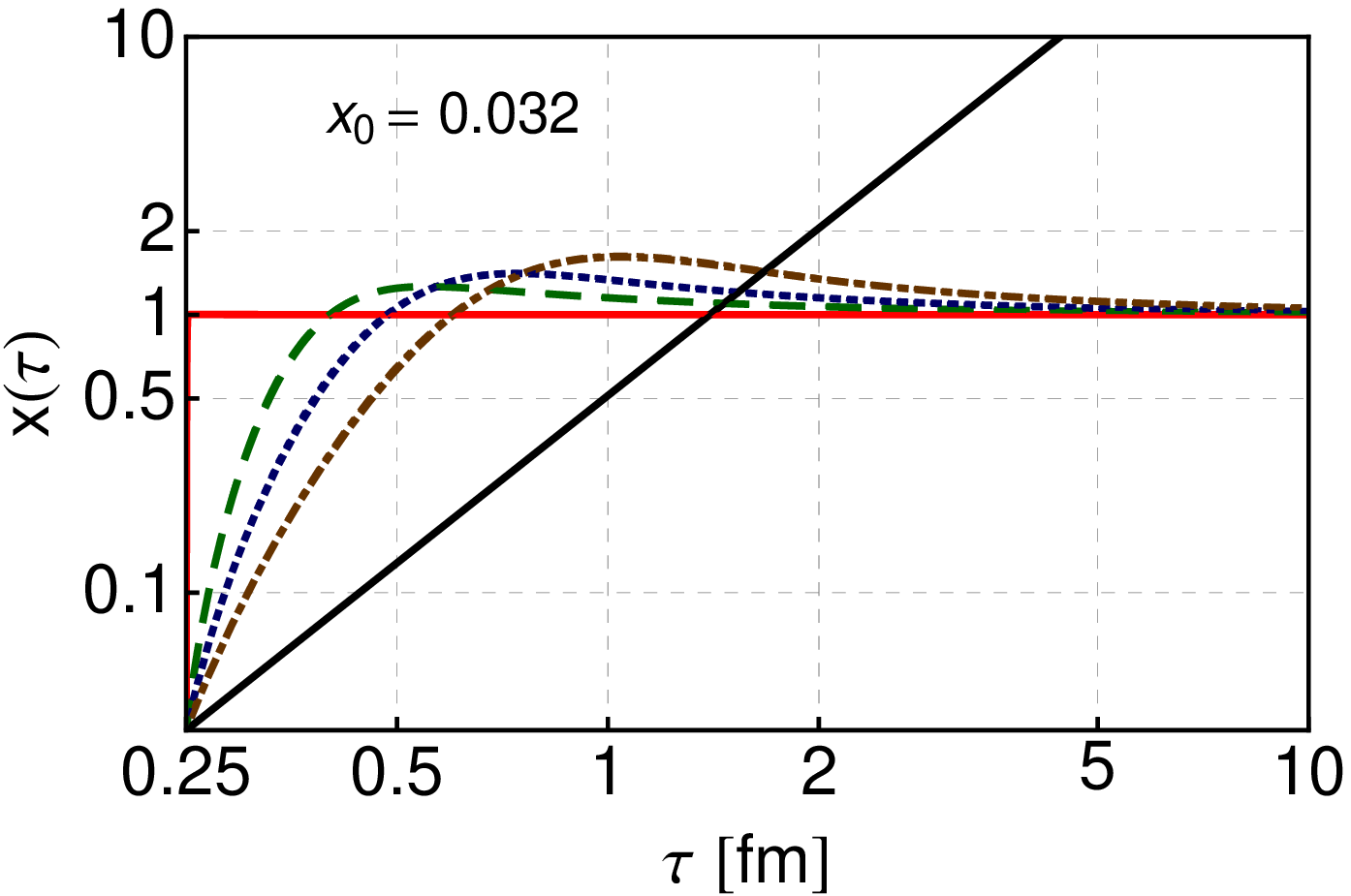}} \\ 
\subfigure{\includegraphics[angle=0,width=0.58\textwidth]{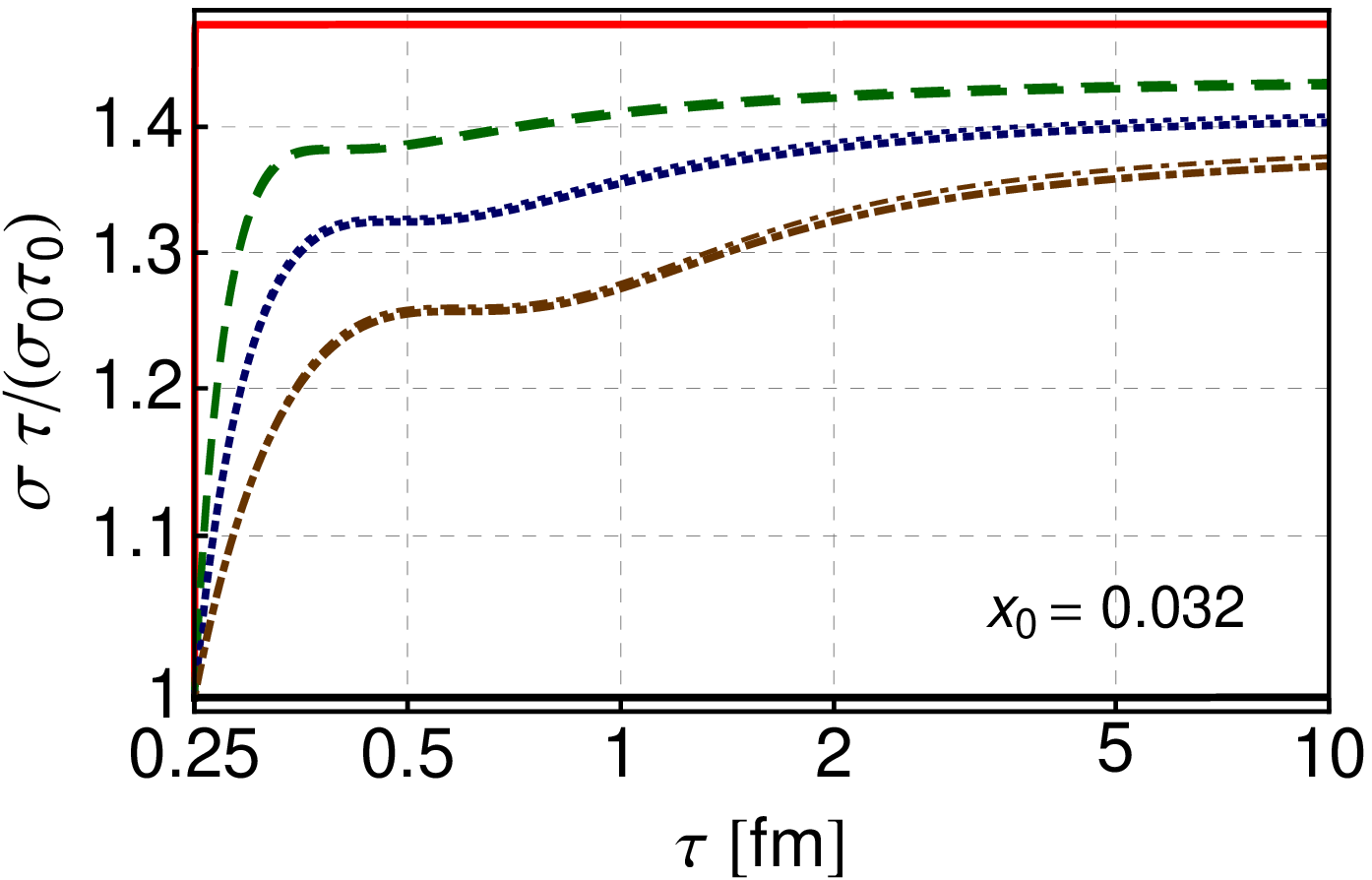}} 
\end{center}
\caption{Same as Fig.~\ref{fig:transvdom} and \ref{fig:isotropic}, but the calculations done for the initial anisotropy $x_0=0.032$.}
\label{fig:longitdom}
\end{figure}

Finally, for completeness, in Fig.~\ref{fig:longitdom} we show our results for the case \textbf{ii)}.  For $\tau_{\rm eq} = 0.25, 0.5, 1.0$ fm we again observe the non-monotonic behaviour of $x(\tau)$. Initially, the dissipation tends to isotropize the system. However, as the equilibrium is reached the dissipative processes stop, $H(0) = 0$ and the free-streaming stage starts to dominate the dynamics. The subsequent evolution proceeds very much similarly to the case \textbf{iii)}. In the lower part of Fig.~\ref{fig:longitdom}, for $\tau_{\rm eq} =0.25, 0.5, 1$ fm, we observe the corresponding entropy production. We can see that when the system is near equilibrium the entropy production stops. Moreover, contrary to the case \textbf{i)}, the higher is the value of $\tau_{\rm eq}$, the less entropy is produced.

Comparing our results presented in  Figs.~\ref{fig:transvdom}, \ref{fig:isotropic}, and \ref{fig:longitdom}, we observe that the time dependence of $x$ is very much similar in the cases (\ref{en1}) and (\ref{en3}). It suggests that the particular form of the entropy source has small effect on the time evolution. Far from equilibrium, the cases (\ref{en1}) and (\ref{en3}) differ significantly but the large entropy production leads to fast decrease of $x$. As soon as $x$ becomes small these two schemes are equivalent, since for small $x$ Eqs.~(\ref{en1}) and (\ref{en3}) have the same $x$ dependence.

\par We also note that $\tau_{\rm eq} = 0.25, 0.5, 1.0$ fm correspond to $\tau^{\rm MS}_{\rm eq} \approx 0.03, 0.067, 0.13$ fm. Hence, our choice of the parameters corresponds to very short relaxation times used in the kinetic theory (in the way done by Martinez and Strickland). This may sound reasonable if we consider the plasma as a strongly interacting system.
\chapter{Initial conditions}
\label{chapter:initial}
In order to make realistic comparisons with the data, we should include the transverse expansion of matter and, possibly, relax the assumption of boost-invariance. In this Chapter we introduce the initial conditions used in our model for such situations. 

\section[Glauber initial conditions for boost-invariant systems]{Glauber initial conditions \\ for boost-invariant systems}
\label{section:gini}
%
In order to solve the \texttt{ADHYDRO} equations discussed in Chapter \ref{chapter:hydro} in the boost-invariant 2+1 dimensional [(2+1)D] case, it is necessary to specify initial conditions at a given transverse space point ${\bf x}_\perp$ for the initial proper time $\tau=\tau_0$.  The proper time $\tau=0$ corresponds to the moment when the centers of the colliding nuclei pass through the $z=0$ plane. In the numerical calculations we choose \mbox{$\tau_0 =$ 0.25 fm}. Before this time, the parton interactions are expected to produce at least a {\it partially thermalized} system whose later behavior may be described by \texttt{ADHYDRO}.  For boost-invariant simulations the initial conditions are defined by four functions: $\sigma(\tau_0,{\bf x}_\perp)$, $x(\tau_0,{\bf x}_\perp)$, $u_x(\tau_0,{\bf x}_\perp)$, and $u_y(\tau_0,{\bf x}_\perp)$.

We assume that the initial energy density in the transverse plane, $\varepsilon_0(\tau_0,{\bf x}_\perp)$, is proportional to the normalized density of sources $\tilde{\rho}(b,{\bf x}_\perp)$, where $b$ is the impact parameter corresponding to a given centrality class, 
\begin{equation}
\varepsilon_0({\bf x}_\perp) =\varepsilon(\tau_0,{\bf x}_\perp) = \varepsilon_{\rm i} \, \tilde{\rho}(b,{\bf x}_\perp).
\label{ei1}
\end{equation}
The parameter $\varepsilon_{\rm i}$ is the {\it initial energy density at the center of the system in most central collisions}. Its value may be estimated from the standard hydrodynamic calculations. We use the value \mbox{$\varepsilon_{\rm i} = 86.76$ GeV/fm$^3$}. If the considered matter were in equilibrium, its temperature would be equal to $T_{\rm}=485$ MeV.

The normalized density of sources, $\tilde{\rho}(b,{\bf x}_\perp)$, is constructed as a linear combination of the {\it wounded-nucleon density} $\rho_{\rm W}(b,{\bf x}_\perp)$ and the {\it density of binary collisions} $\rho_{\rm B}(b,{\bf x}_\perp)$,
\begin{equation}
\rho(b,{\bf x}_\perp) =  \frac{1-\kappa}{2}\rho_{\rm W} \left(b,{\bf x}_\perp \right) 
+ \kappa \rho_{\rm B} \left(b,{\bf x}_\perp \right),
\label{dsources}
\end{equation}
\begin{equation}
 \tilde{\rho}(b,{\bf x}_\perp) = \frac{\rho(b,{\bf x}_\perp)}{\rho(0,0)}.
 \label{dsourcest}
\end{equation}
\begin{figure}[t!]
\begin{center}
\vspace{-1cm} \subfigure{\includegraphics[angle=0,width=0.65\textwidth]{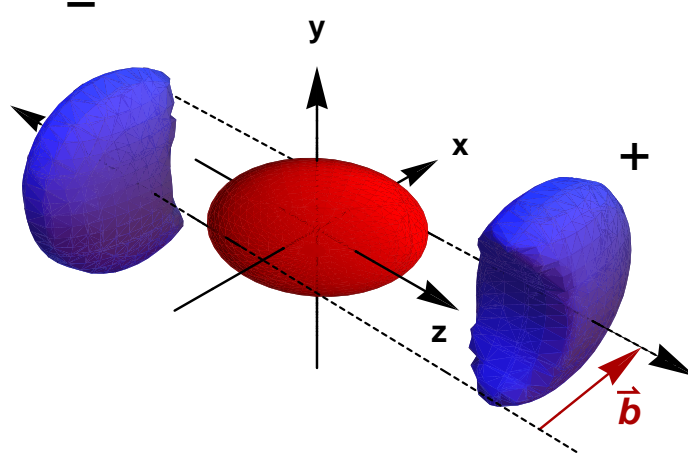}} \vspace{-2cm}
\end{center}
\caption{Geometry of the collision of two symmetric nuclei. The collision centrality is specified by the impact vector ${\bf b}$.}
\label{fig:col}
\end{figure}
The choice of the parameter $\kappa$ ($0 \leq \kappa \leq 1$) controls the mixing between $\rho_{\rm W}$ and $\rho_{\rm B}$. In the case $\kappa = 0$ we obtain the standard wounded-nucleon model \cite{Bialas:1976ed}. The parameter $\kappa$ should be chosen in such a way as to reproduce the experimentally observed dependence of the charged particle multiplicity on the collision centrality. Typically one uses $\kappa=0.14$ for RHIC \cite{Back:2004dy} and $\kappa=0.15$ for the LHC \cite{Bozek:2010er}. 

The wounded-nucleon and binary-collision densities may be obtained from the optical limit of the Glauber model. In this approach, for a collision with the centrality specified by the impact vector ${\bf b}$ (see Fig.~\ref{fig:col}), we use the formula for the transverse density of wounded nucleons being a sum of the contributions from the nuclei with the mass numbers $A$ (projectile) and $B$ (target). Thus, assuming only symmetric collisions where $A=B$ and large nuclei ($A,B \gg 1$) we may write
\begin{eqnarray}
  \rho_{\rm W} \left(b, {\bf x}_\perp\right) &=&
  \rho_{\rm W}^{+} \left(b, {\bf x}_\perp \right)           
+ \rho_{\rm W}^{-} \left(b, {\bf x}_\perp \right),  \nonumber \\
  \rho_{\rm W}^{+} \left(b, {\bf x}_\perp \right) &=&  
\mathcal{T} \left( {\bf x}_\perp - {\bf b}/2 \right) 
\left[1 - \exp \left(- \sigma_{\rm in}\, \mathcal{T} \left({\bf x}_\perp + {\bf b}/2 \right) \right) \right],
 \label{eqn:initial_rhoWN} \\   
   \rho_{\rm W}^{-} \left(b, {\bf x}_\perp \right) &=& 
\mathcal{T} \left(  {\bf x}_\perp + {\bf b}/2 \right) 
\left[1 - \exp \left(- \sigma_{\rm in}\, \mathcal{T}\left( {\bf x}_\perp - {\bf b}/2 \right) \right) \right], 
 \nonumber
\end{eqnarray}
and the density of binary collisions is 
\begin{equation}
  \rho_{\rm B} \left(b, {\bf x}_\perp  \right) = \sigma_{\rm in}\,
    \mathcal{T}\left({\bf x}_\perp +{\bf b}/2 \right)
    \mathcal{T}\left({\bf x}_\perp -{\bf b}/2 \right).
  \label{eqn:initial_rhoBC}
\end{equation}
In Eqs.~(\ref{eqn:initial_rhoWN}) and (\ref{eqn:initial_rhoBC}) $\sigma_{\rm in}$ is the total inelastic nucleon-nucleon cross section which is a function of the collision energy. For the RHIC Au+Au ($A=B=197$) collisions at $\sqrt{s_{\rm NN}}=200$ GeV one chooses $\sigma_{\rm in}=42$ mb \cite{Chojnacki:2007rq} and for the LHC Pb+Pb ($A=B=208$) collisions at $\sqrt{s_{\rm NN}}=2760$ GeV one can take $\sigma_{\rm in}=62$ mb \cite{Bozek:2010er}.  

\begin{figure}[t!]
\begin{center}
\subfigure{\includegraphics[angle=0,width=0.5\textwidth]{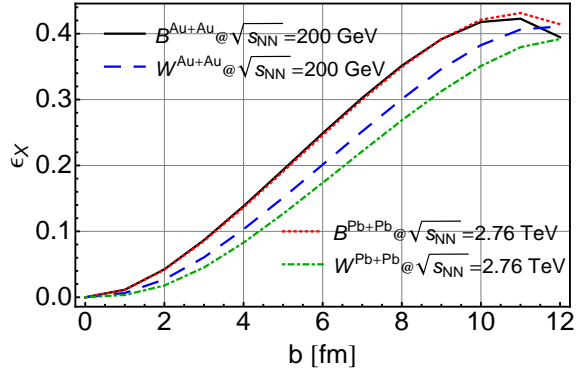}} 
\end{center}
\caption{Spatial eccentricity $\varepsilon_{x}(\tau) = (\langle y^2 -x^2 \rangle)/(\langle y^2 + x^2 \rangle)$ of the wounded-nucleon density $\rho_{\rm W}(b)$ and the binary-collisions density $\rho_{\rm B}(b)$ as functions of the impact parameter $b=|{\bf b}|$ calculated for Au+Au collisions at RHIC and for Pb+Pb collisions at the LHC.}
\label{fig:exc}
\end{figure}

The role of $\sigma_{\rm in}$ is more important in the definition (\ref{eqn:initial_rhoWN}), since it influences the shape of profiles. In (\ref{eqn:initial_rhoBC}) it enters just as a multiplication factor. Thus, the eccentricity of $\rho_{\rm W}(b,{\bf x}_\perp)$ for RHIC is larger than that for the LHC, whereas for not too peripheral collisions the spatial eccentricity of $\rho_{\rm B}(b,{\bf x}_\perp)$ is in general the same, see Fig.~\ref{fig:exc}.
The density of binary collisions yields a much steeper profile (hard component) than the density of the wounded nucleons (soft component).

The function $\mathcal{T}({\bf x}_\perp)$ is the nuclear thickness function 
\begin{equation}
  \mathcal{T}({\bf x}_\perp) = \int dz\, \rho_{\rm WS}\left({\bf x}_\perp,z\right).
  \label{eqn:initial_TA}
\end{equation}
The function $\rho_{\rm WS}\left({\bf x}_\perp,z \right)$ in (\ref{eqn:initial_TA}) is the nuclear density function given by the Woods-Saxon spherical profile
\begin{equation}
  \rho_{\rm WS} \left({\bf x}_\perp,z \right) = \frac{\rho_0}{1+\exp\left[ \left( \sqrt{{\bf x}_\perp^2 + z^2} - R_A \right)/a \right]},
  \label{eqn:initial_W-S}
\end{equation}
where $R_A$ is the radius of the nucleus with the mass number $A$ calculated from the formula 
\begin{equation}
  R_A=1.12 A^{1/3}-0.86 A^{-1/3},
  \label{eqn:initial_R_A}
\end{equation}
$a$ is the surface diffuseness ($a = 0.54$ fm for both, gold and lead nuclei) and $\rho_0$ is the nuclear saturation density chosen in such a way that $\rho_{\rm WS} \left({\bf x}_\perp,z\right)$ is normalized to the number of nucleons in the nucleus.

For the initial anisotropy parameter $x(\tau_0,{\bf x}_\perp)$, which we take as independent of ${\bf x}_\perp$ and denote simply as $x_0$, we consider again three different options: $x_0=100$, $x_0=1$, and $x_0=0.032$. The case $x_0=1$ is, of course, the closest to that described by the standard perfect-fluid hydrodynamics. The case $x_0 > 1$ corresponds to the initial situation where the transverse pressure is much larger than the longitudinal pressure. In this case the momentum shape is {\it oblate} (the momentum distribution is stretched in the transverse direction and squeezed in the longitudinal direction with respect to the beam axis). This type of the initial conditions is considered in the Color Glass Condensate (CGC) approach where the distribution functions in the longitudinal direction are described by the Dirac delta function, $\delta(p_\parallel)$ at $z=0$ \cite{Kovner:1995ja,Bjoraker:2000cf},  see also Ref. \cite{El:2007vg}. The values $x_0 < 1$  correspond to the {\it prolate} momentum shape. This type of the initial conditions has been analyzed, for example, in Refs. \cite{Jas:2007rw,Randrup:2003cw}. 

By fixing both the initial energy density and the initial anisotropy parameter, we determine the initial entropy density profile from Eq.~(\ref{epsilon2b}), namely
\begin{equation}
 \sigma_0({\bf x}_\perp) = \sigma(\tau_0,{\bf x}_\perp) = 
\varepsilon_{\rm qgp}^{-1} 
\left[ \frac{\varepsilon_{\rm i} \, \tilde{\rho}(b,{\bf x}_\perp)}{r(x_0)} \right].
\label{sig1}
\end{equation}
Here $\varepsilon_{\rm qgp}^{-1}(\varepsilon)$ is the inverse function to $\varepsilon_{\rm qgp}(\sigma)$. The $b$-dependence displayed on the right-hand-side of Eq.~(\ref{sig1}) induces centrality dependence of the initial entropy density profiles. 

We note that the entropy described by Eq.~(\ref{sig1}) corresponds to the entropy produced in the proper time interval $0 \leq \tau \leq \tau_0$, just before the \texttt{ADHYDRO} description is turned on. For $\tau > \tau_0$, the \texttt{ADHYDRO} model accounts for the additional entropy production described by the entropy source term $\Sigma$.
Since the overall entropy produced in a given class of collisions is connected with the final multiplicity, we may consider it as a fixed quantity. Therefore, if the initial anisotropy $x_0$ is much different from unity, the initial entropy density $\sigma_{\rm 0}$ is expected to be small. In this case, the large amount of entropy is produced in the phase described by \texttt{ADHYDRO}. On the other hand, if the initial anisotropy is close to unity, the subsequent entropy production should be negligible.

Our form of the initial conditions is completed by the assumption that there is no transverse flow present at $\tau = \tau_0$, i.e., we set $u_x(\tau_0,{\bf x}_\perp)=0$ and $u_y(\tau_0,{\bf x}_\perp)=0$.
 
\section[Tilted initial conditions]{Tilted initial conditions \\ for non-boost-invariant systems}
\label{section:tini}

The experimental data collected at RHIC show that the boost-invariance is realized only in a relative narrow central range
of rapidities. Thus, the boost-invariant (2+1)D approach discussed in Section \ref{section:gini} is not sufficient to perform realistic predictions for observables outside the midrapidity region ($|\mathrm{y}|>1$). In order to investigate in more detail the influence of partial thermalization on the behavior of matter produced in the whole range of rapidities in relativistic heavy-ion collisions, we have to perform full-geometry non-boost-invariant 3+1 dimensional [(3+1)D] simulations based on the differential equations introduced in Chapter \ref{chapter:hydro}. 

For a general non-boost-invariant case, the initial conditions for \texttt{ADHYDRO} are defined by five functions: $\sigma(\tau_0,\eta,{\bf x}_\perp)$, $x(\tau_0,\eta,{\bf x}_\perp)$, $u_x(\tau_0,\eta,{\bf x}_\perp)$, and $u_y(\tau_0,\eta,{\bf x}_\perp)$, and $\vartheta(\tau_0,\eta,{\bf x}_\perp)$. The initial entropy density profile is defined in the similar way as in the (2+1)D calculations,
\begin{equation}
 \sigma_0(\eta,{\bf x}_\perp) = \sigma(\tau_0,\eta,{\bf x}_\perp) = 
\varepsilon_{\rm qgp}^{-1} 
\left[ \varepsilon_{\rm i} \, \tilde{\rho}(b,\eta,{\bf x}_\perp) \right],
\label{sig2}
\end{equation}
however, in the (3+1)D calculations we choose the parameter $\varepsilon_{\rm i}$  separately for
different physical situations (such as different initial anisotropy profiles and/or different values of the time-scale parameter $\tau_{\rm eq}$) in such a way as to reproduce well the final multiplicity.

The normalized density of sources $\tilde{\rho}(b,\eta,{\bf x}_\perp)$ from (\ref{sig2}), is constructed now as a sum of the two terms:
\begin{eqnarray}
\rho(b,\eta,{\bf x}_\perp) &=&  (1-\kappa)\left[
\rho_{\rm W}^{+} \left(b,{\bf x}_\perp \right) f^{+} \left(\eta\right) +
\rho_{\rm W}^{-} \left(b,{\bf x}_\perp \right) f^{-} \left(\eta\right) 
\right]  \nonumber \\
& & \quad \quad  \quad  \quad +\, \kappa \rho_{\rm B} \left(b,{\bf x}_\perp \right) f \left(\eta\right).
\label{dsources2}
\end{eqnarray}
Here, the function $f \left(\eta\right)$ is the initial longitudinal profile
\begin{equation}
f \left(\eta\right)= \exp 
\left[   
- \frac{(\eta - \Delta\eta)^2}{2 \sigma_\eta^2} \theta(|\eta|-\Delta\eta)
\right].
\label{longitprof}
\end{equation}
The first term in (\ref{dsources2}) introduces contributions to the density of sources from the forward- ($+$) or backward-moving  ($-$) participant nucleons. This formula assumes a preferred emission from the participating nucleons along the direction of their motion. The second term introduces a symmetric contribution from the binary collisions. The half-width of the central plateau, $\Delta\eta$, and the  half-width of Gaussian tails,  $\sigma_\eta$, are fitted to reproduce the RHIC multiplicity of charged particles in pseudorapidity. 

The initial density of sources produced by a single forward- or backward-moving participant nucleon with rapidity $\mathrm{y}_b = \ln (\sqrt{s_{\rm NN}}/m_{\rm N})$, where $m_{\rm N}$ is the nucleon mass, is proportional to
\begin{equation}
f^{+(-)} \left(\eta\right)= f \left(\eta\right) f_{F(B)} \left(\eta\right),
\label{fplusminus}
\end{equation}
where
\begin{equation}
f_F(\eta)=
\begin{cases} 0 & \eta< -\eta_m \\
\frac{\eta+\eta_m}{2\eta_m}   & -\eta_m \le \eta \le \eta_m \\
1 & \eta_m<\eta
\end{cases}, \quad \quad
f_B(\eta)=f_F(-\eta).
\label{fForBack}
\end{equation}
The range of rapidity correlations equals $\eta_m =\mathrm{y}_b-\eta_s \simeq 3.36$ and the shift in the rapidity $\eta_s = 2$ is a phenomenological parameter. With the assumption (\ref{fForBack}), the hydrodynamic description outside the region $(-\eta_m,\eta_m)$ is not reliable. 

Similarly to (2+1)D calculations, for the initial anisotropy profile, $x(\tau_0,\eta,{\bf x}_\perp)$, we choose three different values: $x_0=100$, $x_0=1$, and $x_0=0.032$. Moreover we check the possibility of the spatial dependence of the initial anisotropy parameter, $x_0=x_0( \tilde{\rho}(\eta,{\bf x}_\perp))$. We assume also that there is no transverse flow present initially, therefore, we set $u_x(\tau_0,\eta,{\bf x}_\perp)=0$ and $u_y(\tau_0,\eta,{\bf x}_\perp)=0$, and the initial longitudinal rapidity of the fluid follows the simple Bjorken scaling $\vartheta(\tau_0,\eta,{\bf x}_\perp)=\eta$.
\chapter{Freeze--out}
\label{chapter:freeze}
%
In this Chapter we present first the method used in  \texttt{ADHYDRO} to extract the freeze-out hypersurface from hydrodynamic evolution. Then, the Cooper-Frye prescription in the form used by the Monte-Carlo event generator \texttt{THERMINATOR} \cite{Kisiel:2005hn,Chojnacki:2011hb} to describe particle production on a given freeze-out hypersurface is derived. Finally, calculations of physical observables with the help of \texttt{THERMINATOR} are explained in more detail.
%
\section{Single-freeze-out approximation}
\label{section:stat}
%
\par The rapid expansion of dense and hot matter (described in our approach by \texttt{ADHYDRO}) implies that the density of particles decreases and the mean free path of particles grows. When the time between interactions of particles becomes larger than the expansion rate (described by divergence of the four-velocity field)
\begin{equation}
\tau_{\rm coll} \geq \tau_{\rm exp},
\label{eqn:freeze}
\end{equation}
the collision processes cannot maintain the local equilibrium any longer. The strongly interacting system of particles undergoes a transition to the system consisting of essentially free-streaming particles. This complex dynamic process, called {\it freeze-out}, is quite often treated approximately as if it took place locally on the three-dimensional freeze-out hypersurface. 

Typically, the freeze-out condition (\ref{eqn:freeze}) is realized by assuming that hadrons completely decouple at the space-time points $x$ where the entropy density drops below a certain freeze-out entropy density $\sigma_{\rm f}$. For systems in local thermal equilibrium described by the equation of state $\sigma(T)$ this condition may be formulated in terms of temperature,
\begin{equation}
T(\sigma(x)) = T_{\rm f} = T(\sigma_{\rm f}).
\label{eqn:tempfreeze}
\end{equation}
A set of freeze-out points $x$ extracted from the hydrodynamic evolution by the constraint (\ref{eqn:tempfreeze}) forms the freeze-out hypersurface. One can subsequently use it to generate primordial particles employing the Cooper-Frye formalism.

In general, the inelastic processes in the expanding system cease earlier than the elastic ones. Therefore, one can distinguish \textit{chemical freeze-out} (the stage when inelastic processes stop and hadron abundances are fixed) from subsequent \textit{kinetic freeze-out} (the stage when all types of interactions cease and the momentum distributions of hadrons are frozen). Of course, the latter may be considered as the true freeze-out process. In the case of high energy collisions at RHIC, which are the subject of our studies, we can use quite efficient simplification called the \textit{single-freeze-out} approach \cite{Broniowski:2001we,Broniowski:2001uk,Broniowski:2002nf}. In this approach, one assumes that the kinetic freeze-out coincides with the chemical freeze-out, $T_{\rm kin}= T_{\rm chem} = T_{\rm f}$. In our studies we adopt this scenario.
%
\section{Cooper-Frye formalism}
\label{section:cfform}
%
\par The momentum distribution of primordial hadrons emitted on the freeze-out hypersurface $\Sigma$ may be calculated from the Cooper-Frye formula~\cite{Cooper:1974mv}
\begin{equation}
\frac{dN}{d{\rm y} d^2p_T} = \overline{g} \int d\Sigma_\mu(x) p^\mu f(x, p),
\label{eqn:cf-for}
\end{equation}
where $d\Sigma^\mu$ is the element of the three-dimensional freeze-out hypersurface submerged in the four-dimensional Minkowski space, $f(x, p)$ is the phase-space distribution function of the emitted particles and $\overline{g}=2 s +1$ is the spin degeneracy factor, where $s$ denotes the particle's spin. For systems in local thermal equilibrium described by the perfect fluid hydrodynamics one assumes equilibrium Fermi-Dirac ($\epsilon=-1$) or Bose-Einstein ($\epsilon=+1$) statistics, i.e., one uses
the distribution functions
\begin{equation}
  f(x, p) = \left\{ \exp\left[\frac{p_\mu U^\mu(x) -  \mu}{T(x)}\right] -\epsilon \right\}^{-1}
  \label{eqn_PdotU0}
\end{equation}
depending on the scalar product of particle four-momentum $p^\mu$ and the four-velocity $U^\mu$ of the fluid element. Clearly, if deviations from equilibrium at freeze-out are large, the distributions (\ref{eqn_PdotU0}) should be replaced by the appropriate non-equilibrium distributions. In our calculations, for a typical choice of parameters, only initial parts of freeze-out hypersurfaces exhibit large anisotropies. Thus, as long as they weakly contribute to total final multiplicities they may be simply neglected in the calculation of the observables. 

\par One can determine thermodynamic parameters on the freeze-out hypersurface with the help of \textit{thermal models}  \cite{Rafelski:2001hp,BraunMunzinger:2001ip,Turko:2007ri,Florkowski:2001fp,Baran:2003nm} which analyze relative hadronic abundances. At RHIC, this type of study \cite{Baran:2003nm} indicates the non-zero values of chemical potentials. Thus in the generation of hadrons we introduce the chemical potential $\mu$
\begin{equation}
 \mu= B \mu_B + I_3 \mu_{I_3} + S \mu_S + C \mu_C,
  \label{chem_pot}
\end{equation}
where $B$, $I_3$, $S$, $C$ is the baryon number, third component of isospin, strangeness, and charm of the emitted hadron, respectively. Similarly, $\mu_B$, $\mu_{I_3}$, $\mu_S$, and $\mu_C$ denote the corresponding chemical potentials.

The particle's four-momentum may be parametrized in the standard way
\begin{equation}
  p^{\mu} = \left( m_T \cosh {\rm y}, p_T \cos \phi_p, p_T \sin \phi_p, m_T \sinh {\rm y} \right),
  \label{particle_mom}
\end{equation}
where $m_T=\sqrt{m^2+p_T^2}$ is the transverse mass, $p_T$ is the transverse momentum, ${\rm y}$ is the (longitudinal) rapidity of the emitted particle, and $\phi_p$ is the azimuthal angle of the transverse momentum. Together with the parametrization (\ref{U3+1}), one may write
\begin{equation}
  p_\mu U^\mu = \sqrt{1 + u_x^2 + u_y^2}\, m_T \cosh (\vartheta - {\rm y}) - p_T ( u_x \cos\phi_p + u_y \sin\phi_p ).
  \label{pdotu}
\end{equation}
The form of $d\Sigma^\mu$ may be obtained with the help of the formula known from differential geometry \cite{Misner:1974qy}
\begin{equation}
d\Sigma_\mu = \varepsilon_{\mu \alpha \beta \gamma}
\frac{\partial x^\alpha}{\partial a} \frac{\partial x^\beta}{\partial b} \frac{\partial x^\gamma}{\partial c }
d a d b d c,
\label{d3Sigma}
\end{equation}
where $\varepsilon_{\mu \alpha \beta \gamma}$ is the antisymmetric Levi-Civita tensor with $\varepsilon_{0 1 2 3}=+1$. The quantity $d\Sigma^\mu(x)$ defines a four-vector that is perpendicular to the hypersurface at point $x$. Its norm is equal to the volume of the hypersurface element. The variables $a$, $b$ and $c$ introduce a coordinate system in Minkowski space parameterizing the positions of points on the freeze-out hypersurface. Their ordering is chosen in such a way that $d\Sigma^\mu$ points in the direction of decreasing temperature.
 
\par
\begin{figure}[t]
\begin{center}
\subfigure{\includegraphics[angle=0,width=0.575\textwidth]{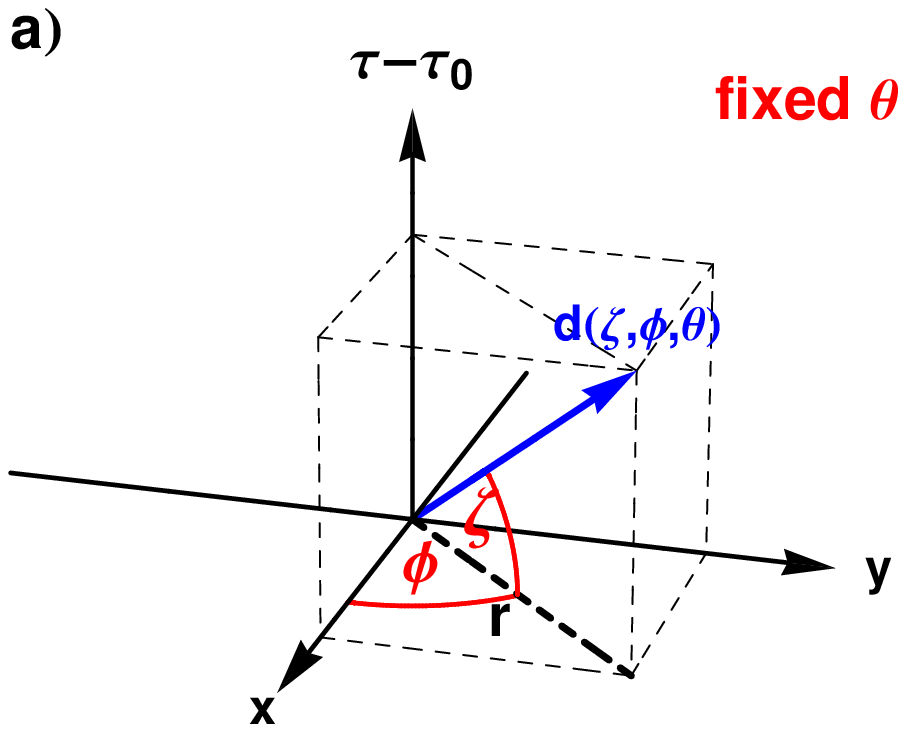}}  \\
\subfigure{\includegraphics[angle=0,width=0.575\textwidth]{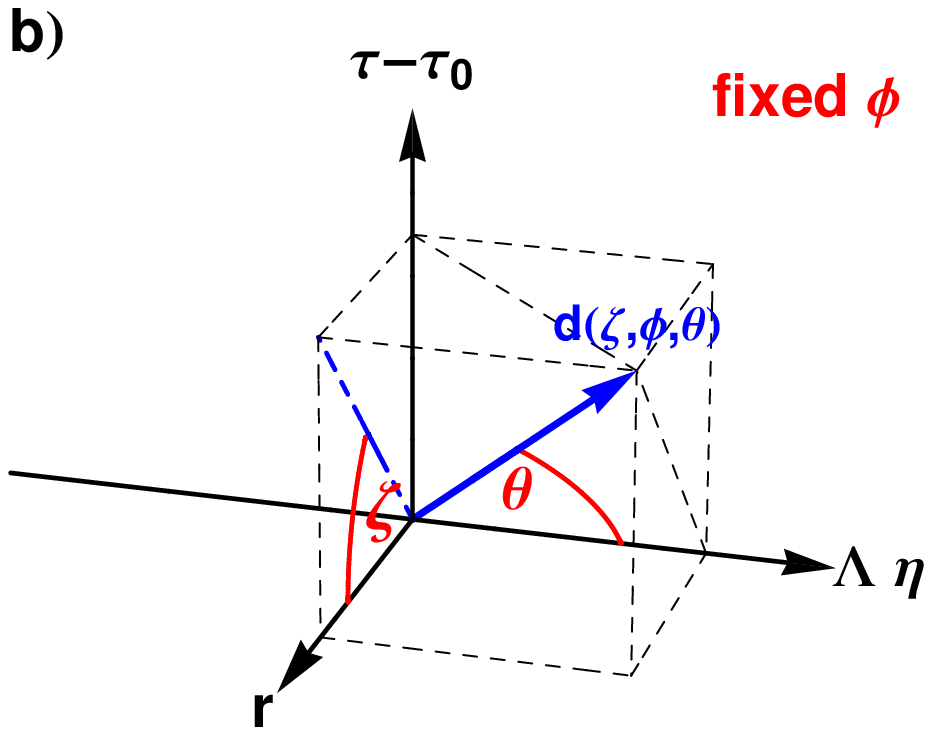}} 
\end{center}
\caption{\small Coordinate system used to parametrize the freeze-out hypersurface extracted from a (3+1)D dimensional hydrodynamic code, see Eqs.~(\ref{3d-par1})--(\ref{3d-angles}). The part (\textbf{a}) shows the view in $x-y-\tau$ space with the fixed value of $\theta$, while the part (\textbf{b}) shows the view in $r-\Lambda \eta-\tau$ space with the fixed value of $\phi$. }
\label{fig:freezeangles}
\end{figure}

 The freeze-out hypersurface obtained from the most general (3+1)D hydrodynamic evolution (without assuming any symmetries) may be parametrized in the following way~\cite{Bozek:2009ty} (see Fig.~\ref{fig:freezeangles})
\begin{eqnarray}
  t &=& \left(\tau_0 + d(\zeta,\phi,\theta) \sin\theta \sin\zeta \right)
         \cosh\frac{d(\zeta,\phi,\theta) \cos\theta}{\Lambda}, \nonumber \\
  x &=& d(\zeta,\phi,\theta) \sin\theta \cos\zeta \cos\phi, \nonumber \\
  y &=& d(\zeta,\phi,\theta) \sin\theta \cos\zeta \sin\phi, \nonumber \\
  z &=& \left(\tau_0 + d(\zeta,\phi,\theta) \sin\theta \sin\zeta \right)
          \sinh\frac{d(\zeta,\phi,\theta) \cos\theta}{\Lambda}.
\label{3d-par1}
\end{eqnarray}
This parametrization leads to simple formulas for the space-time rapidity $\eta$, longitudinal proper time $\tau$, and the transverse distance $r$ \footnote{In the calculations we remove the emission points with $\tau < 1$ fm/c. In this region the system is anisotropic and consists mainly from gluons.},
\begin{eqnarray}
\eta &=& \frac{d(\zeta,\phi,\theta) \cos\theta}{\Lambda},
\nonumber \\
 \tau &=& \tau_0 + d(\zeta,\phi,\theta) \sin\theta \sin\zeta,
 \nonumber \\
 r &=&  d(\zeta,\phi,\theta) \cos\zeta.
\label{3d-par2}
\end{eqnarray}
The parameters $a$, $b$ and $c$ in Eq. (\ref{d3Sigma}) are identified with three angles $\zeta$, $\phi$ and  $\theta$, respectively, which are restricted to the ranges:
\begin{eqnarray}
0 \leq &\zeta & \leq \pi/2, \nonumber \\
0 \leq &\phi & < 2 \pi, \nonumber \\
0 \leq &\theta & \leq \pi.
\label{3d-angles}
\end{eqnarray}
The quantity  $d(\zeta,\phi,\theta)$ describes the distance between a point on the freeze-out hypersurface and the coordinate system's origin $(\tau=\tau_0,x=0,y=0,\eta=0)$.  The parameter $\tau_0$ is the initial proper time for hydrodynamic evolution, at which the initial conditions are specified. The parameter $\Lambda$ introduces a scale which changes dimensionless space-time rapidity $\eta$ into a quantity with dimension of fermi. The parametrization (\ref{3d-par1}) works for all typical freeze-out hypersurfaces where the distance $d$ is a function of $\zeta$, $\phi$, and $\theta$.

\par Substitution of (\ref{3d-par1}) in (\ref{d3Sigma}) leads to the following form of the Cooper-Frye integration measure
\begin{eqnarray}
  d\Sigma_\mu p^\mu 	
&=& \frac{d^2 \tau \sin\theta }{\Lambda} 
\Biggl[ \Biggr.  
\frac{\partial d}{\partial\zeta} \cos\zeta 
\left(
p_T \sin\zeta \cos\left(\phi - \phi_p\right)- m_T \cos\zeta \cosh\left(\vartheta - \eta \right) 
\right) 
\nonumber\\
&+& \!\! \cos\zeta \sin\theta \left(d \sin\theta - \frac{\partial d}{\partial\theta} \cos\theta \right)
\left(
 p_T \cos\zeta \cos\left(\phi - \phi_p\right) + m_T \sin\zeta \cosh\left(\vartheta - \eta \right)
\right) 
\nonumber\\
&+& \!\! \cos\zeta \sin\theta \left(d \cos\theta + \frac{\partial d}{\partial\theta}   \sin\theta \right) \frac{\Lambda}{\tau} m_T  \sinh\left(\vartheta - \eta \right) 
\nonumber\\
&+& \!\! \frac{\partial d}{\partial\phi}\,\, p_T\, \sin(\phi - \phi_p) 
\Biggl. \Biggr]  
d \zeta d\phi d\theta.
\label{3d-CF-intmeas}
\end{eqnarray}
Using formulas (\ref{pdotu}) and (\ref{3d-CF-intmeas}) in (\ref{eqn:cf-for}) we obtain obtain a six-dimensional particle distribution which one can use to generate both stable hadrons and unstable resonances on the freeze-out hypersurface.
%
\section{Extraction of physical observables}
\label{section:observables}
%
\par After performing hydrodynamic calculations with \texttt{ADHYDRO}, the freeze-out hypersurface is extracted. This procedure yields four functions: $d(\zeta,\phi,\theta)$, $u_x(\zeta,\phi,\theta)$, $u_y(\zeta,\phi,\theta)$ and $\vartheta(\zeta,\phi,\theta)$ which are subsequently used by the \texttt{THERMINATOR} Monte-Carlo generator as an input. \texttt{THERMINATOR} generates physical events using the Cooper-Frye formula (\ref{eqn:cf-for}). In the single event all known hadron species (stable and unstable) are generated on the freeze-out hypersurface. Subsequently, the unstable primordial particles decay through strong and electro-weak interactions. The final particle distributions consist of stable primordial particles and contributions from resonance decays. The complete information about all generated particles and their decays is kept in the program. This information allows for the calculation of soft hadronic observables typically measured in heavy-ion experiments.

\par Probably, the most important information about the evolution of matter created in relativistic heavy-ion collisions is contained in the momentum distributions of identified particles. Experimental data on transverse-momentum distributions are often corrected for contributions from weak decays (the contributions from weak decays to the hadronic yields and spectra are experimentally determined and subtracted). The detailed output of \texttt{THERMINATOR} allows for applying such corrections in the model calculations in the easy way, making comparisons to the data more realistic.

\par In non-central collisions of two heavy nuclei, the experimentally observed momentum distribution of particles is azimuthally anisotropic and may be expressed as a Fourier series in the momentum azimuthal angle, $\phi_p = \arctan (p_y/p_x)$, with respect to reaction plane \cite{Voloshin:1994mz}
\begin{equation}
  \frac{d^3N}{ d^2 p_Td\mathrm{y}} = \frac{d^2N}{2 \pi p_T \, dp_T d\mathrm{y} } \left( 1 + \sum\limits_{n=1}^{\infty} 2 v_n(p_T,\mathrm{y}) \cos(n \phi_p)\right).
  \label{eqn:FourierExp}
\end{equation}
In the hydrodynamic calculations where the initial conditions are defined by a smooth density profile taken from the optical Glauber formula (\ref{dsources}) the sine terms disappear due to symmetry with respect to the reaction plane. In event-by-event calculations where the fluctuations of initial conditions are present this is not the case. The coefficient $v_{n}(p_T,\mathrm{y})$ is the $n$-th harmonic differential flow which quantitatively characterizes the momentum azimuthal anisotropy. The $v_1(p_T,\mathrm{y})$ coefficient is called the \textit{directed flow}, and the $v_2(p_T,\mathrm{y})$ coefficient is called the \textit{elliptic flow}. Due to symmetry reasons the $v_1(p_T,\mathrm{y})$ vanishes for $\mathrm{y}\approx 0$. The $v_3$ coefficient is called the \textit{triangular flow}.  The higher harmonics are typically negligible. For instance, the $v_4$ coefficient is predicted by hydrodynamic calculations to be at the level of $\sim 0.1\%$.

\par Having complete information about the particles generated by the \texttt{THERMINATOR} code one can perform the femtoscopic analysis. The calculation of the correlation function for the identical pions is performed using the two-particle Monte-Carlo method described in Ref.~\cite{Kisiel:2006is,Kisiel:2006yv} and implemented in \texttt{THERMINATOR 2} \cite{Chojnacki:2011hb} code. In the numerical procedure the ideal Bose-Einstein correlation function is evaluated using the formula
\begin{eqnarray}
  C({\bf q}, {\bf k}) = \frac{\sum\limits_{i} \sum\limits_{j \neq i}
    \delta_\Delta \left( {\bf q} -                  {\bf p}_i + {\bf p}_j        \right) \,\,
    \delta_\Delta \left( {\bf k} - \frac{1}{2}\left[{\bf p}_i + {\bf p}_j \right]\right) \,
    \left|\Psi({\bf q}^{*}, {\bf r}^{*}) \right|^2
  }{\sum\limits_i \sum\limits_{j \neq i}
    \delta_\Delta \left( {\bf q} -                  {\bf p}_i + {\bf p}_j        \right) \,\,
    \delta_\Delta \left( {\bf k} - \frac{1}{2}\left[{\bf p}_i + {\bf p}_j \right]\right)
  }.
  \label{eqn:freeze_cfbysum}
\end{eqnarray}
The calculation of the correlation function (\ref{eqn:freeze_cfbysum}) is performed with the help of the box function $\delta_{\Delta}$ defined in the following way
\begin{equation}
  \delta_{\Delta}({\bf p}) =
  \begin{cases}
    1, & \mbox{if}\,\, |p_x| \leq \frac{\Delta}{2}, |p_y| \leq \frac{\Delta}{2}, |p_z| \leq \frac{\Delta}{2} \\
    0, & \mbox{otherwise},
  \end{cases}
  \label{eqn:freeze_deltadelta}
\end{equation}
where one uses the bin resolution $\Delta = 5~{\rm MeV}$.
In Eq.~(\ref{eqn:freeze_cfbysum}) we use the momentum difference of the pions,
\begin{equation}
q = (q_0, {\bf q}) = \left(E_{p_1} - E_{p_2}, {\bf p}_1 - {\bf p}_2 \right), 
\label{vq}
\end{equation}
and the average momentum of the pion pair, 
\begin{equation}
k = \left( k_0, {\bf k} \right) = {{1} \over {2}} 
\left( E_{p_1} + E_{p_2}, {\bf p}_1 + {\bf p}_2 \right).
\label{vK}
\end{equation}
We also consider the space-time separation between pions
\begin{equation}
\Delta x = \left( t, {\bf r} \right) = \left( t_{1} - t_{2}, {\bf r}_1 - {\bf r}_2 \right).
\label{vr}
\end{equation}
In the HBT analysis we use the standard Bertsch-Pratt system of coordinates \cite{Bertsch:1988db,Pratt:1986cc}, where the \textit{long} direction is parallel to beam axis, the \textit{out} axis coincides with the transverse momentum of the pion pair $k_{T}$ and the \textit{side} axis is perpendicular to \textit{out} and \textit{long} axes. We perform the analysis in the longitudinally co-moving system (LCMS) in which $k_{\rm long}=0$. For each pair of pions the wave function in Eq.~(\ref{eqn:freeze_cfbysum}),  
\begin{equation}
\Psi({\bf q}^{*}, {\bf r}^{*}) = \frac{1} {\sqrt{2}} (e^{i {\bf q}^* {\bf r}^*} 
+ e^{-i {\bf q}^* {\bf r}^*}),
\label{psiq}
\end{equation}
is calculated in the pair rest frame (PRF). The ${\bf q}^*$ and ${\bf r}^*$ denote the relative momentum and position calculated in PRF. In the calculation of expression (\ref{eqn:freeze_cfbysum}) we do not include Coulomb effects. Using the Bertsch-Pratt coordinates  $k_T, q_{\rm out}, q_{\rm side}, q_{\rm long}$ the correlation function (\ref{eqn:freeze_cfbysum}) may be approximated by the formula 
\begin{eqnarray}
& & C\left(k_\perp,q_{\rm out},q_{\rm side},q_{\rm long} \right) = 1 + 
\lambda \exp\left[
-R^2_{\rm out}(k_\perp) q^2_{\rm out} \right.
\nonumber \\
& & \hspace{4cm} \left.
-R^2_{\rm side}(k_\perp) q^2_{\rm side}
-R^2_{\rm long}(k_\perp) q^2_{\rm long}
\right],
\label{cfgaus}
\end{eqnarray}
where the HBT correlation radii $R_{\rm out}$, $R_{\rm side}$ and $R_{\rm long}$ are the widths of the Gaussian source.
From the fit we obtain parameters $R_{\rm out}$, $R_{\rm side}$ and $R_{\rm long}$ as a function of $k_{T}$. In order to make a comparison of our results with the available data from STAR we consider four $k_{T}$ bins $(0.15-0.25)$, $(0.25-0.35)$, $(0.35-0.45)$, and $(0.45-0.60)$~GeV.
\chapter[Boost-invariant systems]{Boost-invariant systems with transverse expansion}
\label{chapter:boostinv}

\par The data from RHIC and the LHC shows that the observables measured in the central rapidity region are approximately invariant with respect to the Lorentz boost transformations along the beam axis. Thus, in this Chapter, we use \texttt{ADHYDRO} assuming the boost-invariance along the beam direction. In this way, we restrict our analysis to the midrapidity region ($|\mathrm{y}| < 1 $) of ultra-relativistic heavy-ion collisions. We perform calculations for central, \mbox{$c=0-5$\%} ($b=2.26$ fm), and mid-central collisions,  \mbox{$c=20-30$\%} ($b=7.16$ fm) and \mbox{$c=20-40$\%} ($b=7.84$ fm), and compare the model results with the available data from Au+Au collisions at the energy of \mbox{$\sqrt{s_{\rm NN}}=200$ GeV} at RHIC. We check whether the existence of high anisotropies of pressure at the early stages of evolution may be probed with the midrapidity observables. 

\section{Central collisions}
\label{sect:resbi_cc}

\par First we consider central Au+Au collisions at the energy of $\sqrt{s_{\rm NN}}=200$ GeV at RHIC. We choose the centrality class $c=0-5$\%, which corresponds to the impact parameter $b=2.26$ fm.  We check three scenarios of the initial momentum anisotropy: 

\begin{itemize}

\item[\textbf{i)}] $x_{\rm 0}=100$ (dashed blue lines), which means that the system is initially {\it oblate} in the momentum space, the fact predicted by the microscopic models of heavy-ion collisions, 

\item[\textbf{ii)}] $x_{\rm 0}=1$ (solid black lines), which corresponds to initially {\it isotropic} systems in the momentum space, this case gives the closest description to perfect fluid hydrodynamics, and 

\item[\textbf{iii)}] $x_{\rm 0}=0.032$ (dotted green lines), which means that the system is initially {\it prolate} in momentum space, which is less-likely to be realized in nature than the case \textbf{i)}, however, we consider this case for completeness. 

\end{itemize}

\noindent The results of the \texttt{ADHYDRO} model are compared to the reference perfect-fluid calculations performed with the \texttt{LHYQUID} code \cite{Chojnacki:2006tv,Chojnacki:2007rq}. In our calculations we use the initial conditions described in Chapter \ref{section:gini}. The value of the initial energy density \mbox{$\varepsilon_{\rm i} = 86.76$ GeV/fm$^3$} in Eq.~(\ref{ei1}) is the same for all analyzed cases. For the system in local thermal equilibrium, it corresponds to the initial central temperature \mbox{$T_{\rm i} = 485$ MeV}. Following Eq.~(\ref{sig1}), the initial central entropy density is \mbox{$\sigma_{\rm i} = \sigma(\tau_0,{\bf x}_\perp=0)= 227$ fm$^{-3}$} for the case \mbox{$x_0=1$}, and  \mbox{$\sigma_{\rm i} = 151$ fm$^{-3}$} for the cases $x_0=100$ and $x_0=0.032$. The value of the time-scale parameter $\tau_{\rm eq} = 0.25$ fm used in our calculations guarantees that the system reaches local thermal equilibrium in the whole fireball volume approximately after 1 fm (this value for thermalization time is easily acceptable by most of the microscopic model calculations).  

\begin{figure}[t]
\begin{center}
\subfigure{\includegraphics[angle=0,width=0.45\textwidth]{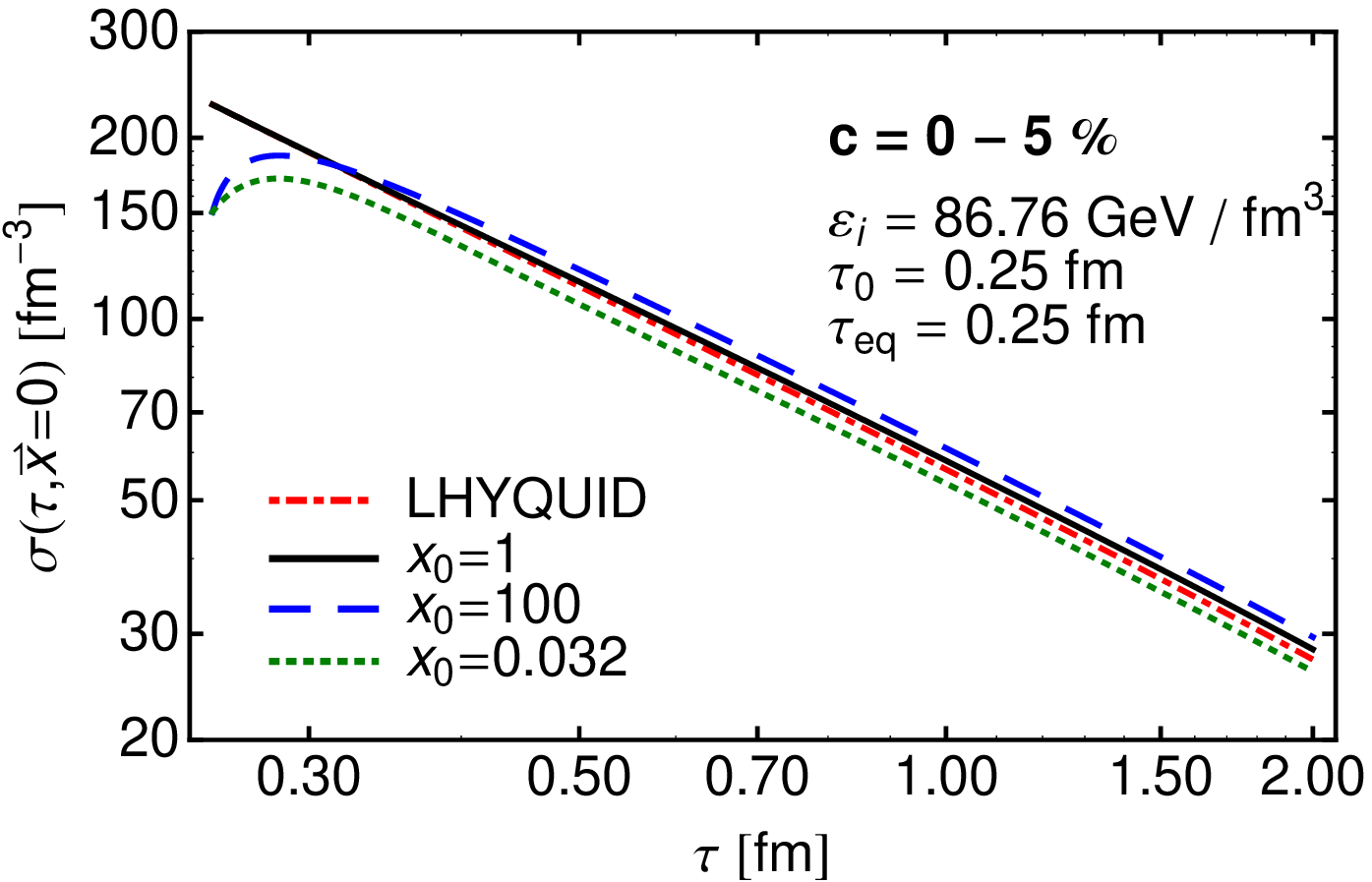}} 
\subfigure{\includegraphics[angle=0,width=0.45\textwidth]{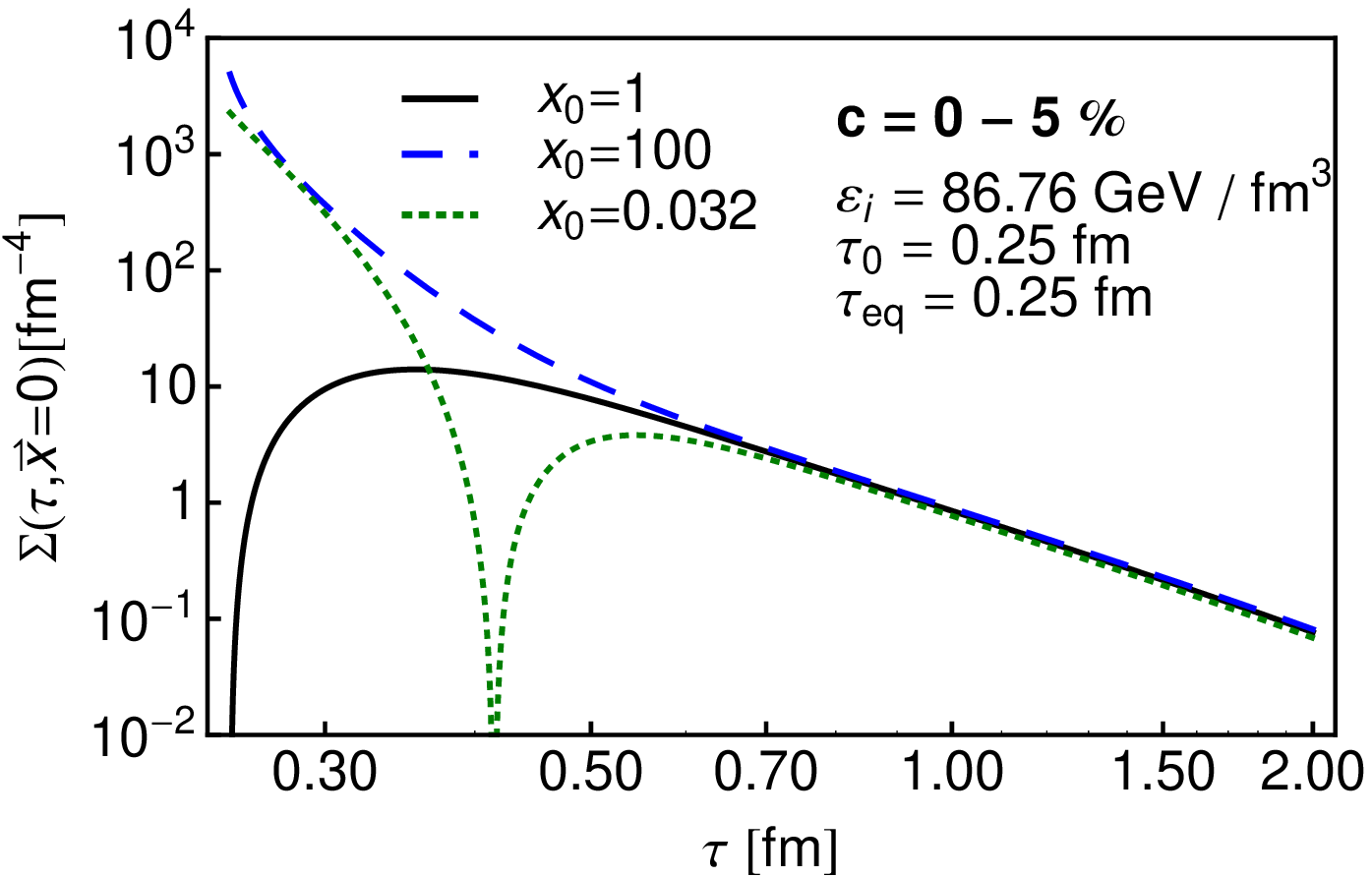}} 
\end{center}
\caption{\small Left panel: Time dependence of the central entropy density for the centrality class $c=0-5$\%. Results obtained for three values of the initial anisotropy parameter: $x_0=100$ (dashed blue line), $x_0=1$ (solid black line), and $x_0=0.032$ (dotted green line) as discussed in the text. The results are compared to the reference perfect-fluid calculations (dashed-dotted red line). Right panel: Time dependence of the entropy production in the center of the system. The prefect-fluid hydrodynamics is entropy conserving, hence, it is not included in the plot.}
\label{fig:s}
\end{figure}

\par In the left panel of Fig.~\ref{fig:s} we present the time dependence of the central entropy density $\sigma(\tau,{\bf x}_\perp=0)$ for three considered cases of the initial anisotropy parameter: $x_0=100$ (dashed blue line), $x_0=1$ (solid black line), and $x_0=0.032$ (dotted green line). We compare our results to the perfect-fluid case (dashed-dotted red line). We observe that in the cases $x_0=100$ and $x_0=0.032$ the entropy density  grows initially. This behavior is different from the entropy-conserving perfect-fluid case where $\sigma(\tau,{\bf x}_\perp=0)$ decreases like $1/\tau$ (dashed-dotted red line). For $\tau \approx $ 0.3~fm $\sigma(\tau,{\bf x}_\perp=0)$ reaches maximum and for \mbox{$\tau > $ 0.5 fm} it starts to behave similarly as in the perfect-fluid case. The case $x_0=1$ gives results similar to the perfect-fluid case. The entropy densities at the end of evolution are similar, however, small differences are noticeable. It means that the rescaling of the initial entropy density $\sigma_{\rm i}$ according to Eq.~(\ref{sig1}) assures only approximately the same final multiplicity and entropy density in all considered cases. 

\begin{figure}[h]
\begin{center}
\subfigure{\includegraphics[angle=0,width=0.45\textwidth]{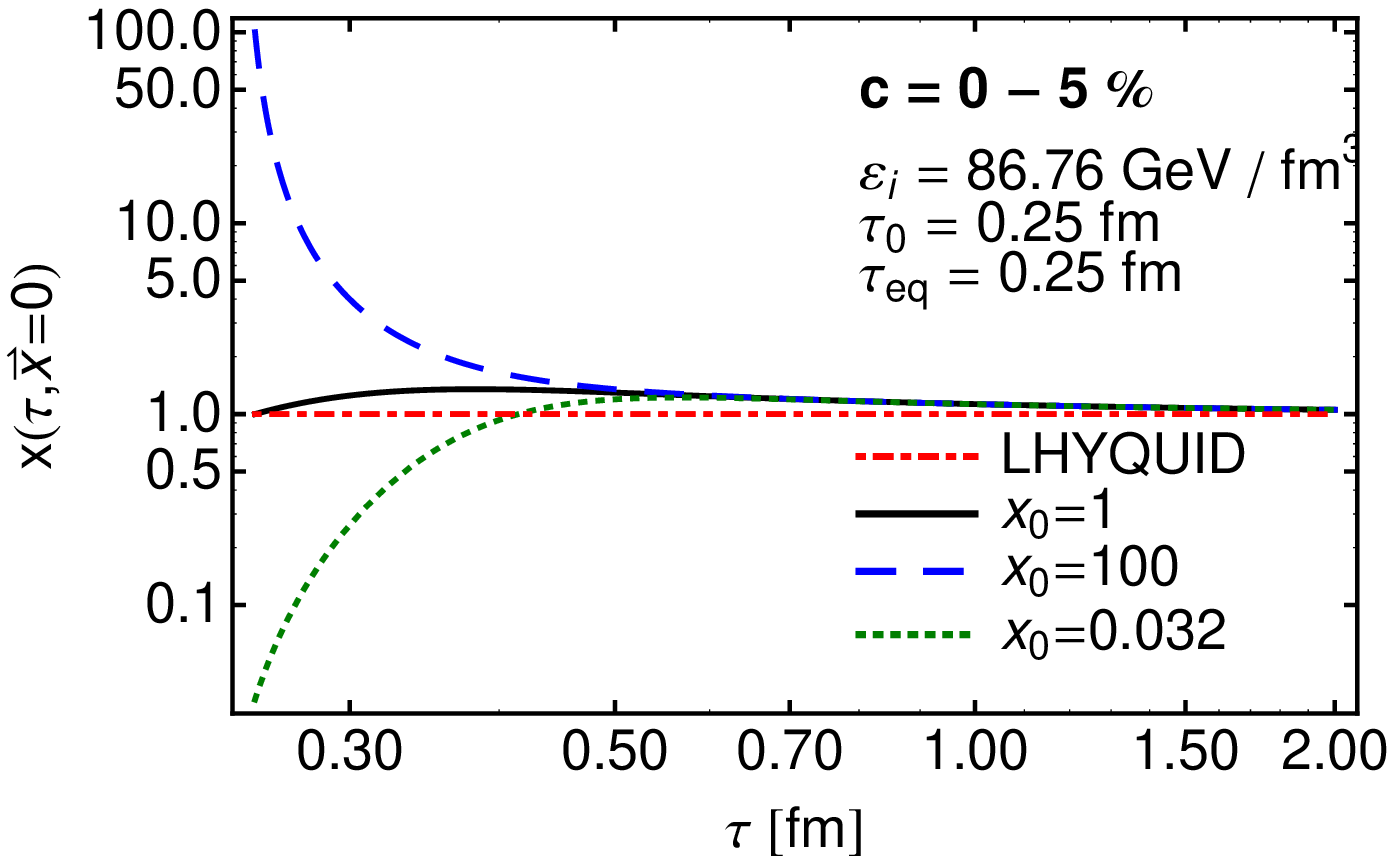}} 
\subfigure{\includegraphics[angle=0,width=0.45\textwidth]{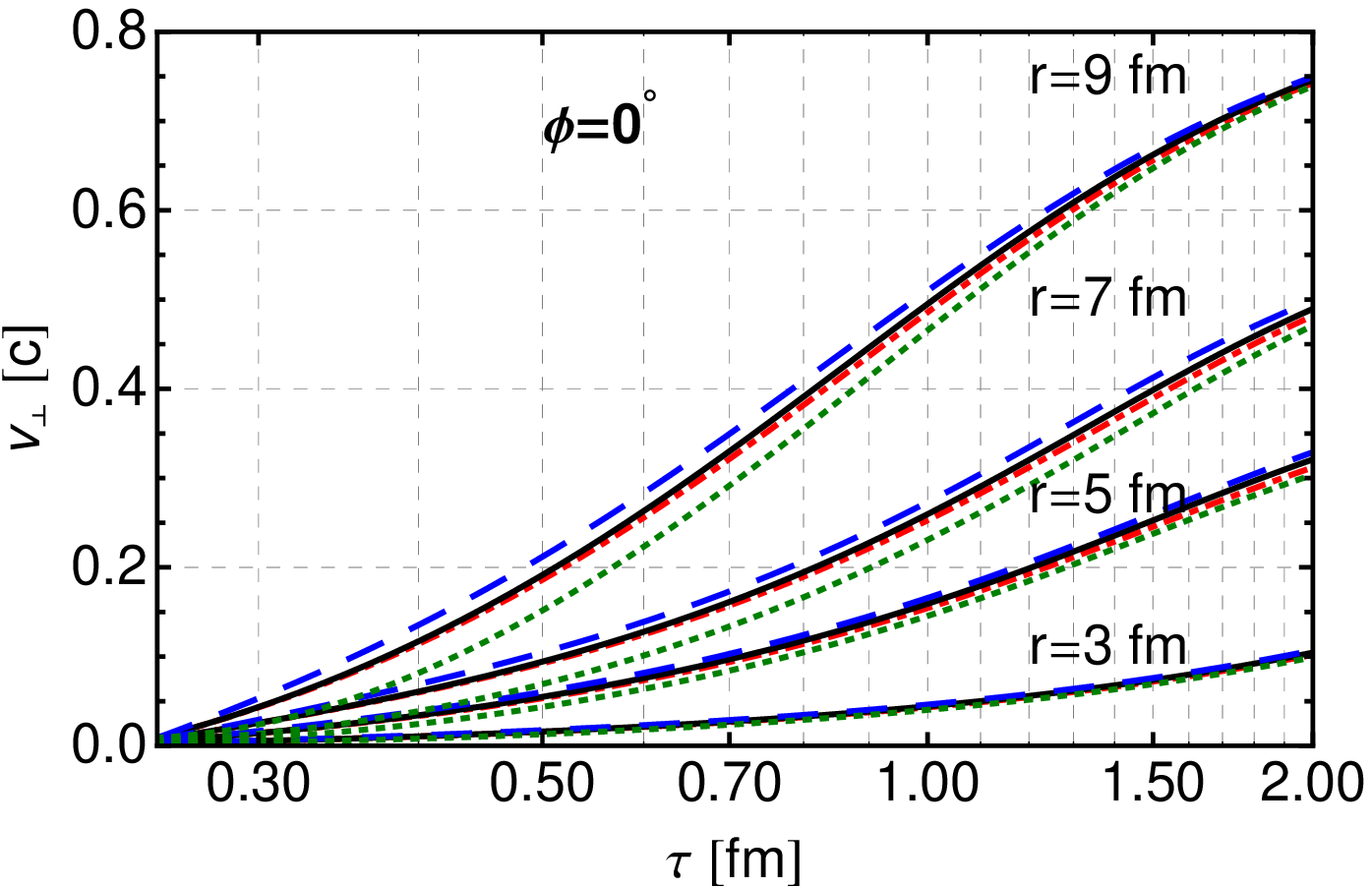}} 
\end{center}
\caption{\small Left panel: Time dependence of the anisotropy parameter at the center of the system for the centrality class $c=0-5$\% for three values of $x_{0}$. Notation is the same as in Fig.~\ref{fig:s}. Right panel: In-plane ($\phi = 0$) time dependence of the radial velocity $v_{\perp}$ for different values of radial distance.}
\label{fig:x}
\end{figure}

\par In the right panel of Fig.~\ref{fig:s} we show the time dependence of the entropy source $\Sigma(\tau,{\bf x}_\perp=0)$ defined by Eq.~(\ref{en1}). The notation is the same as in the left panel. The function $\Sigma(\tau,{\bf x}_\perp=0)$ contains information complementary to that encoded in $\sigma(\tau,{\bf x}_\perp=0)$. Due to strong initial anisotropy, in the case $x_0=100$ the initial entropy production is very large. As the pressure anisotropy decreases (see the corresponding time dependence of the anisotropy parameter $x(\tau,{\bf x}_\perp=0)$ in the left panel of Fig.~\ref{fig:x}) the production of entropy becomes smaller. In the case $x_0=0.032$, the entropy production is also quite large initially. However, it quickly drops to zero when $x$ passes unity (see also Chapter \ref{sect:hydro-r}). Subsequently, it increases again and becomes similar to the values found for the case $x_0=100$~\footnote{Different entropy productions in the cases $x_0=100$ and $x_0=0.032$ are responsible for slightly different final entropy densities.}.  When $x_0=1$, the anisotropy remains always close to unity. Hence, the entropy production is very small in this case. We note that in all the cases the anisotropy parameter  tends asymptotically to unity, which means that for large times the entropy source vanishes and the system approaches the perfect-fluid regime.

\begin{figure}[t]
\begin{center}
\subfigure{\includegraphics[angle=0,width=0.4\textwidth]{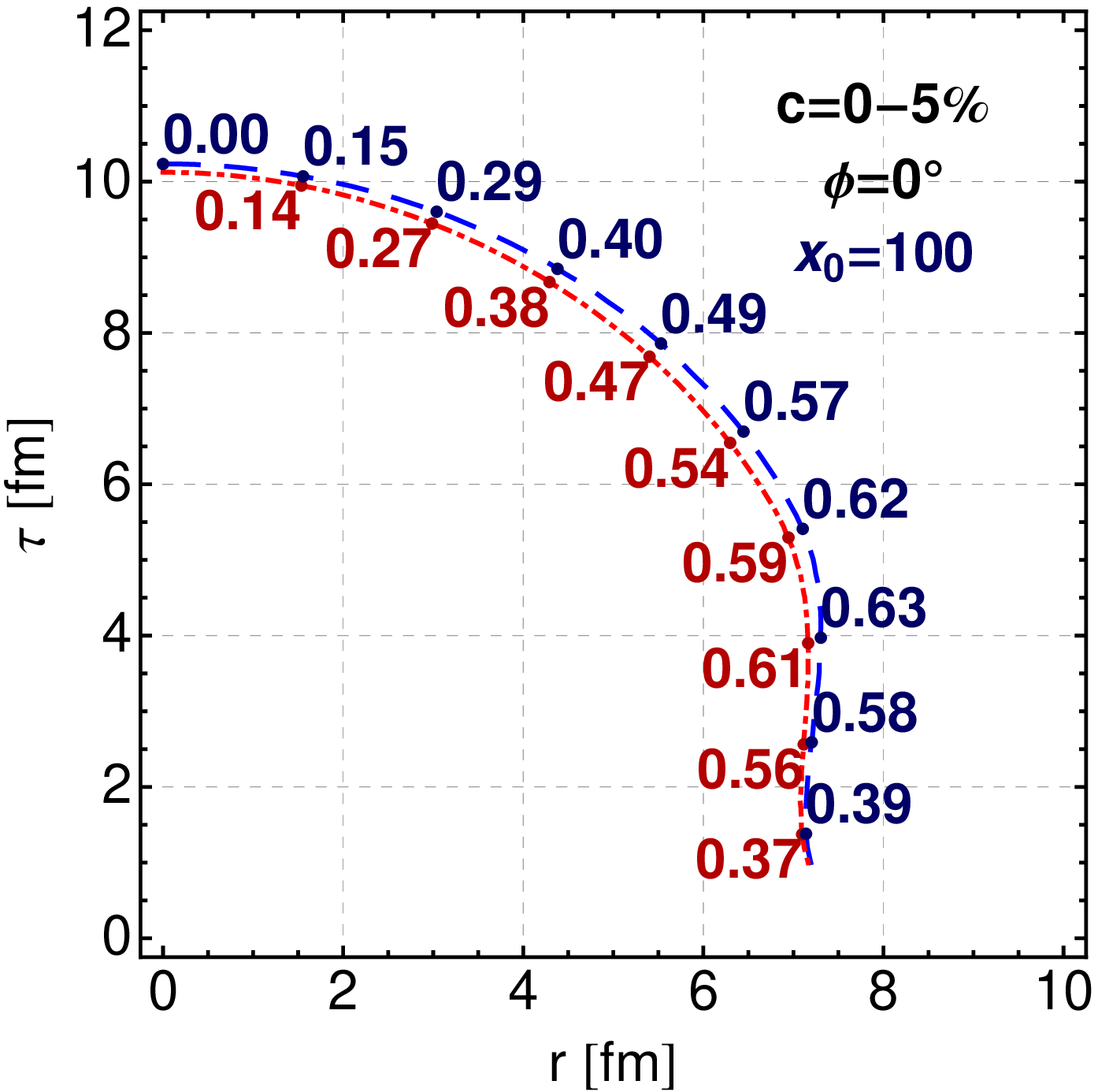}} 
\subfigure{\includegraphics[angle=0,width=0.4\textwidth]{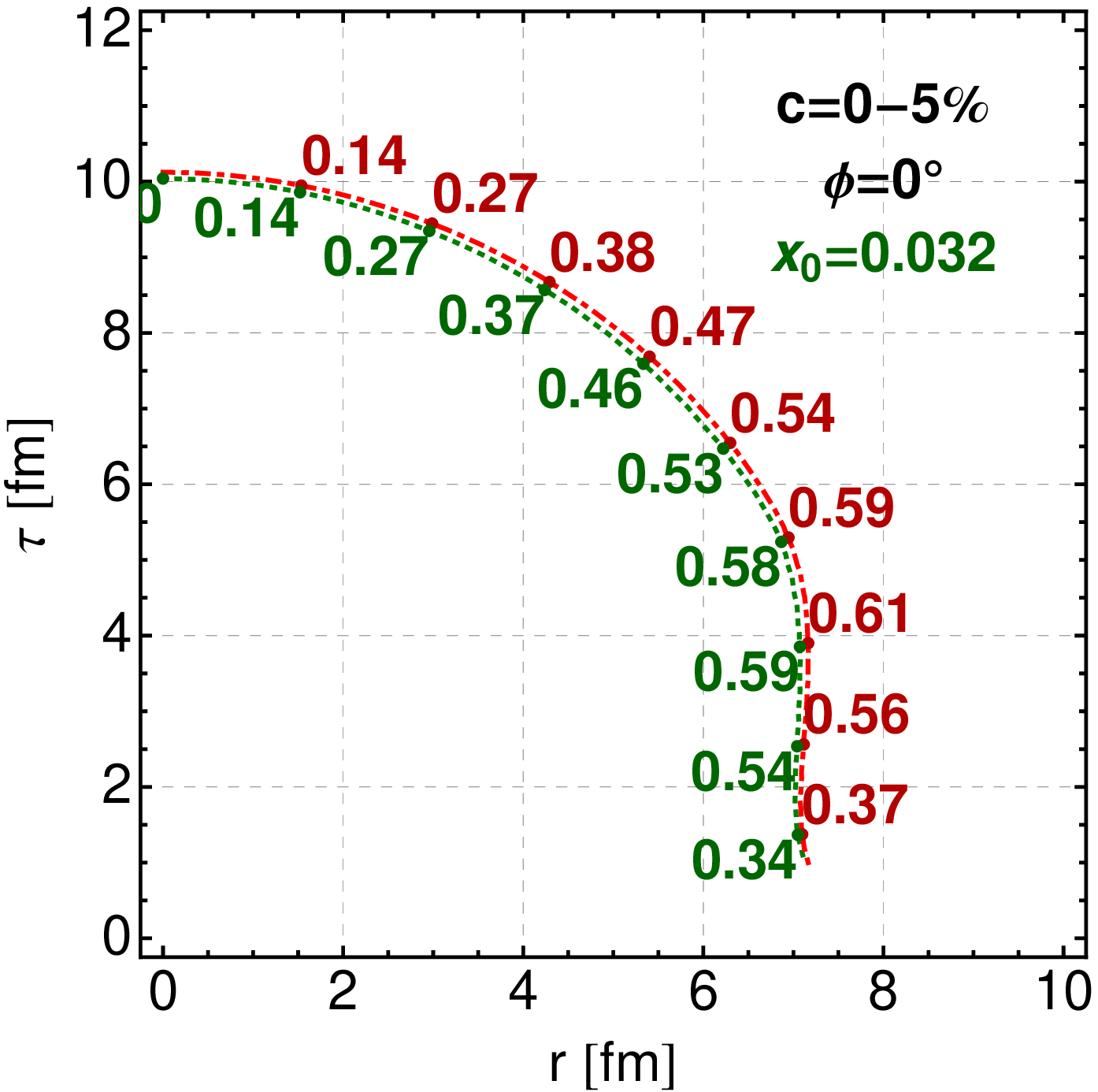}}
\subfigure{\includegraphics[angle=0,width=0.4\textwidth]{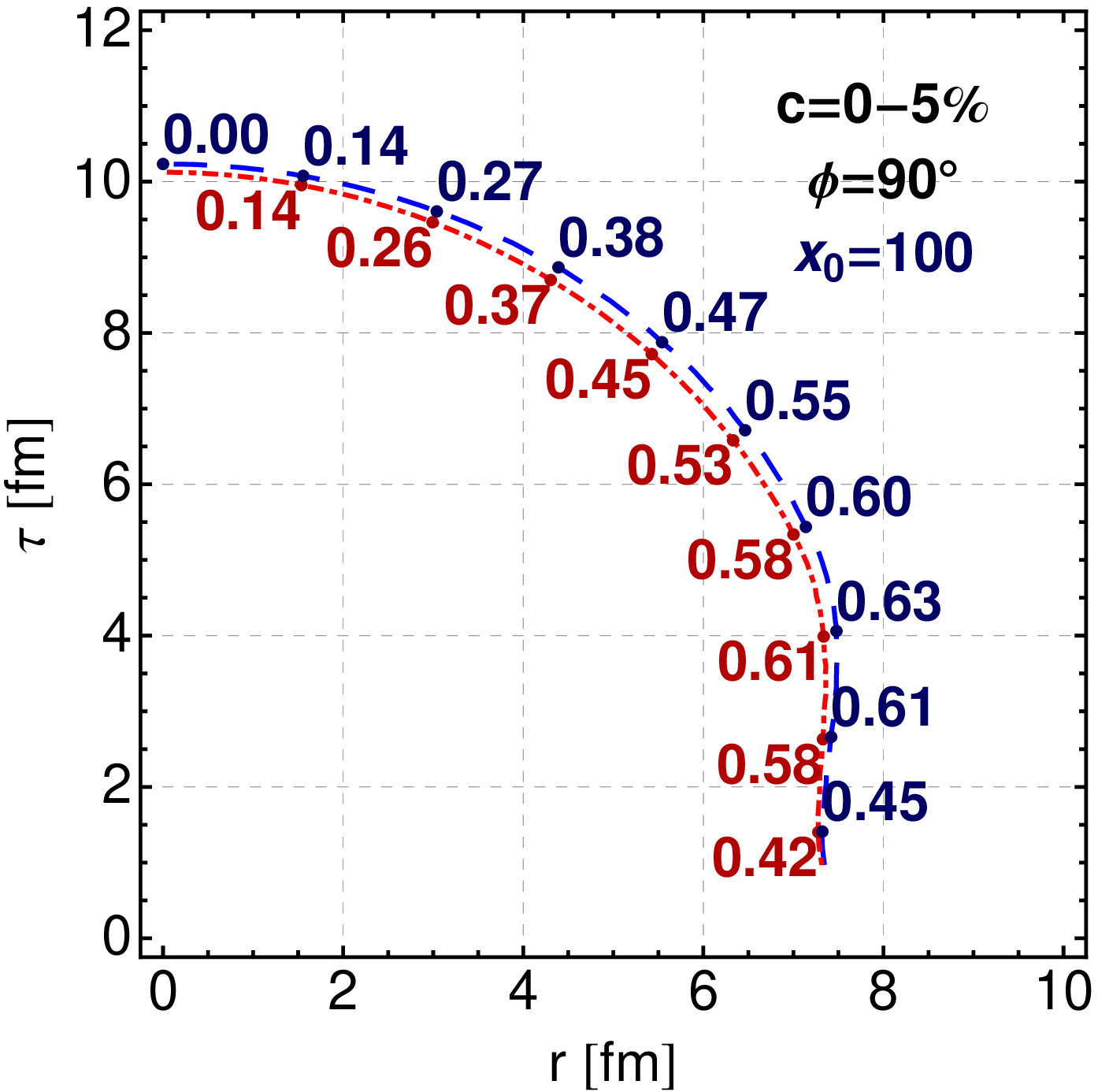}} 
\subfigure{\includegraphics[angle=0,width=0.4\textwidth]{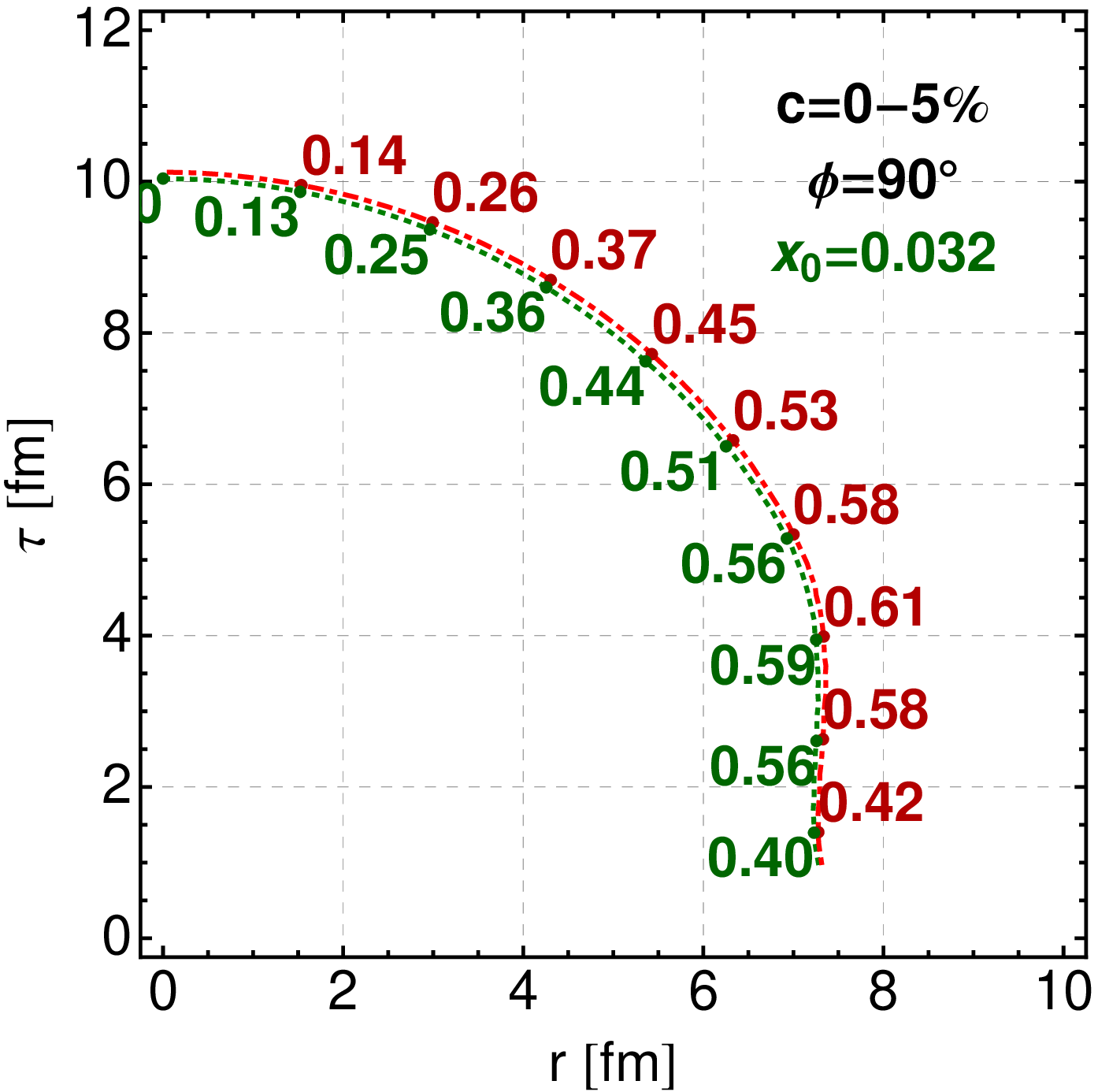}}
\end{center}
\caption{\small In-plane ($\phi = 0$) and out-of-plane ($\phi = \pi/2$) freeze-out hypersurfaces for the cases $x_0=100$ (dashed blue lines) and $x_0=0.032$ (dotted green lines) shown together with the values of the velocity $v_{\perp}(\zeta)$ for the centrality $c=0-5$\%. \texttt{ADHYDRO} results are compared to the perfect-fluid case (dashed-dotted red lines). The freeze-out temperature equals $T_{\rm f}=150$ MeV.}
\label{fig:hiper0005}
\end{figure}

\par The initial large anisotropy in the case $x_0=100$ is equivalent to the large $P_{\perp}/P_{\parallel}$ ratio which implies strong hydrodynamic flow in the transverse direction at the early stages of the evolution (small longitudinal pressure means that no work is done in the longitudinal direction and, consequently, thermal energy is used to build the transverse flow only). This effect is, however, compensated by the smaller value of the initial entropy density obtained from Eq.~(\ref{sig1}). Thus, our results obtained with \texttt{ADHYDRO} are very much similar to those  obtained in the standard hydrodynamics. 
This property is seen in the right panel of Fig.~\ref{fig:x}, where we plot in-plane  ($\phi = 0$)  radial velocity $v_{\perp}$ as a function of proper time $\tau$ for different values of the distance $r$, for the cases \textbf{i)}, \textbf{ii)}, and \textbf{iii)}~\footnote{Since for $c=0-5$\% the evolution is almost azimuthally symmetric in the transverse plane, the out-of-plane ($\phi = \pi/2$) behavior is very similar to the one shown in the right panel of Fig.~\ref{fig:x}.}. 

\begin{figure}[t]
\centering
\begin{minipage}{5.75cm}
\subfigure{\includegraphics[angle=0,width=\textwidth]{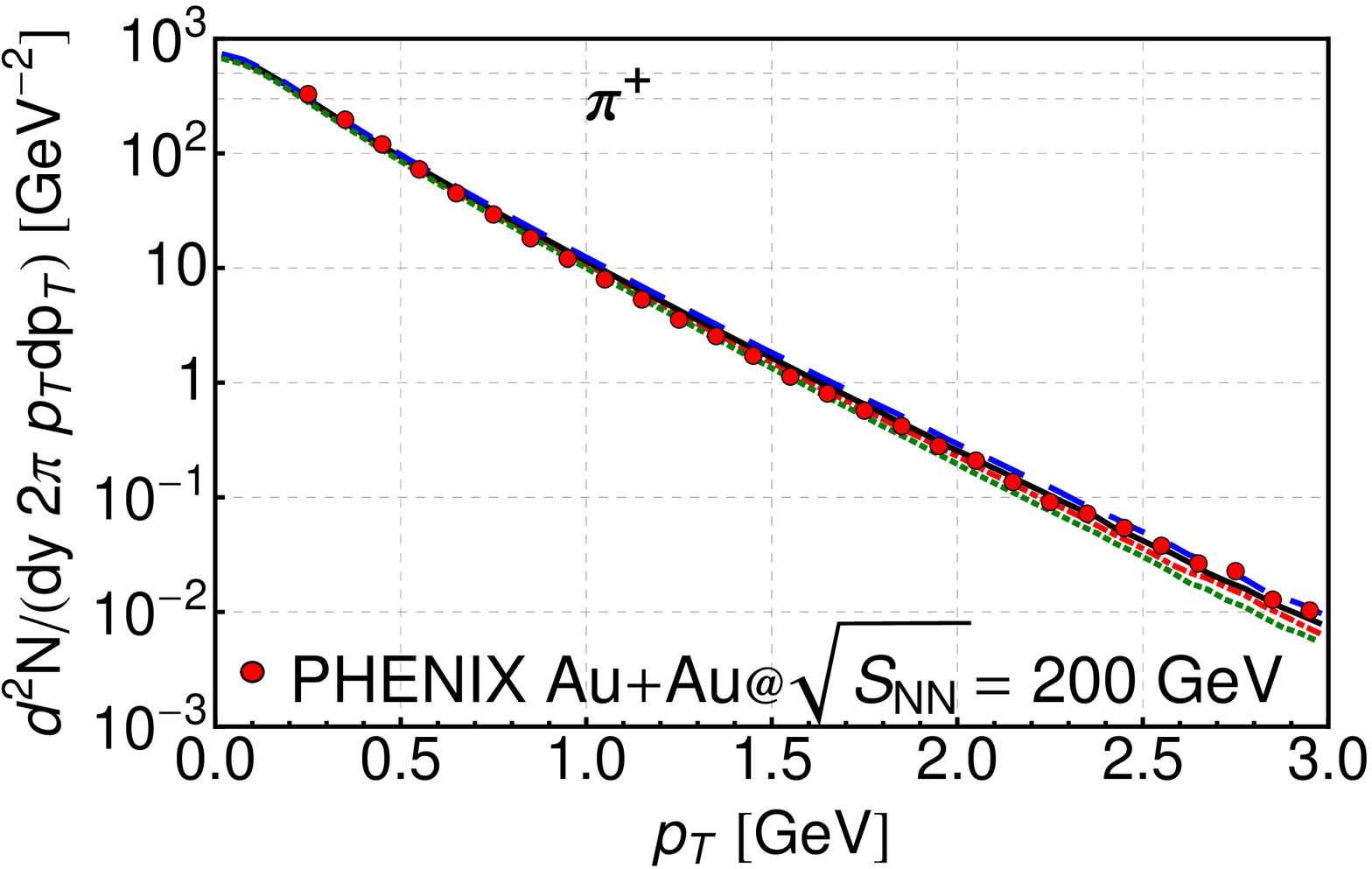}}\\
\subfigure{\includegraphics[angle=0,width=\textwidth]{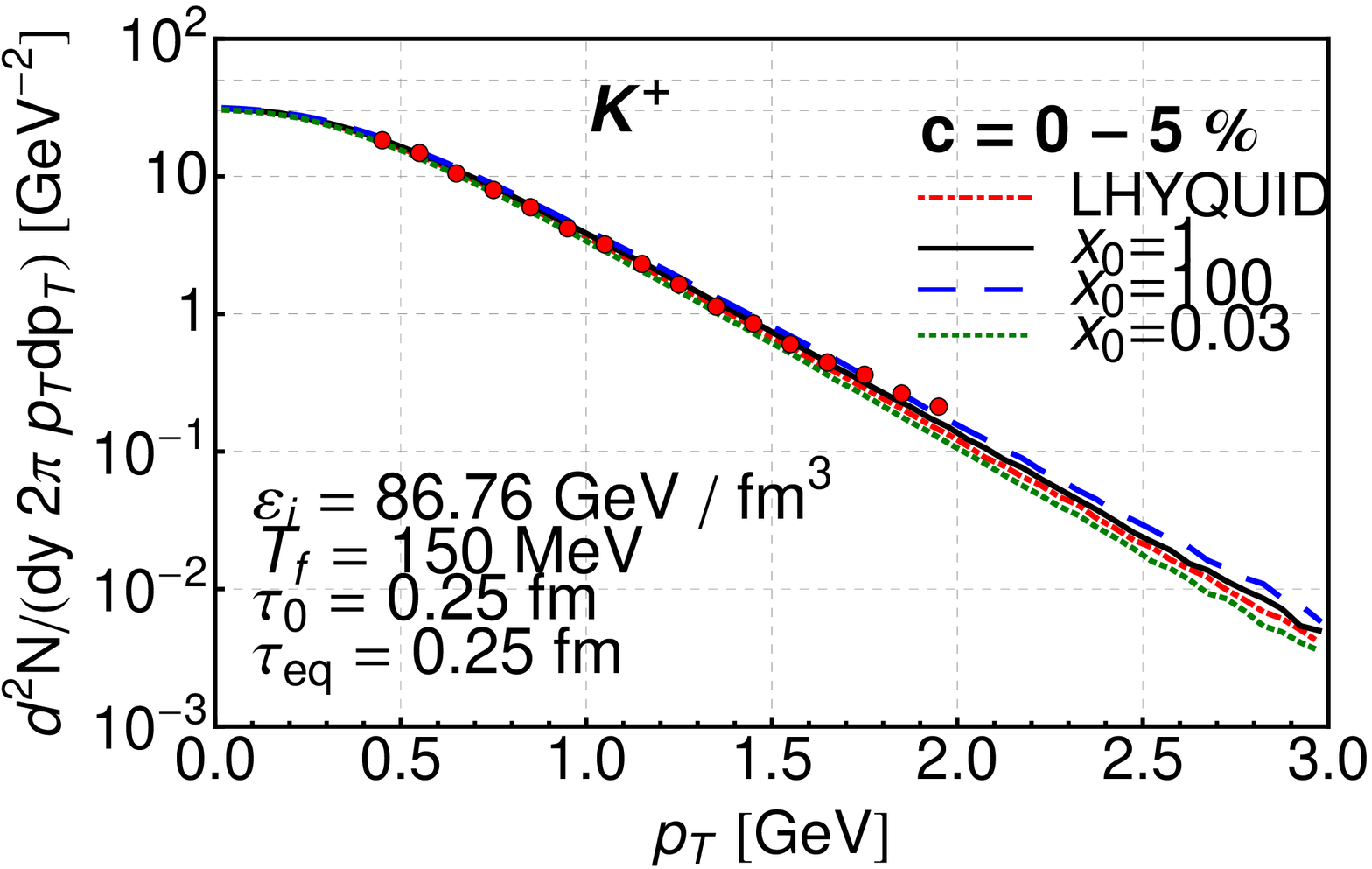}} \\
\subfigure{\includegraphics[angle=0,width=\textwidth]{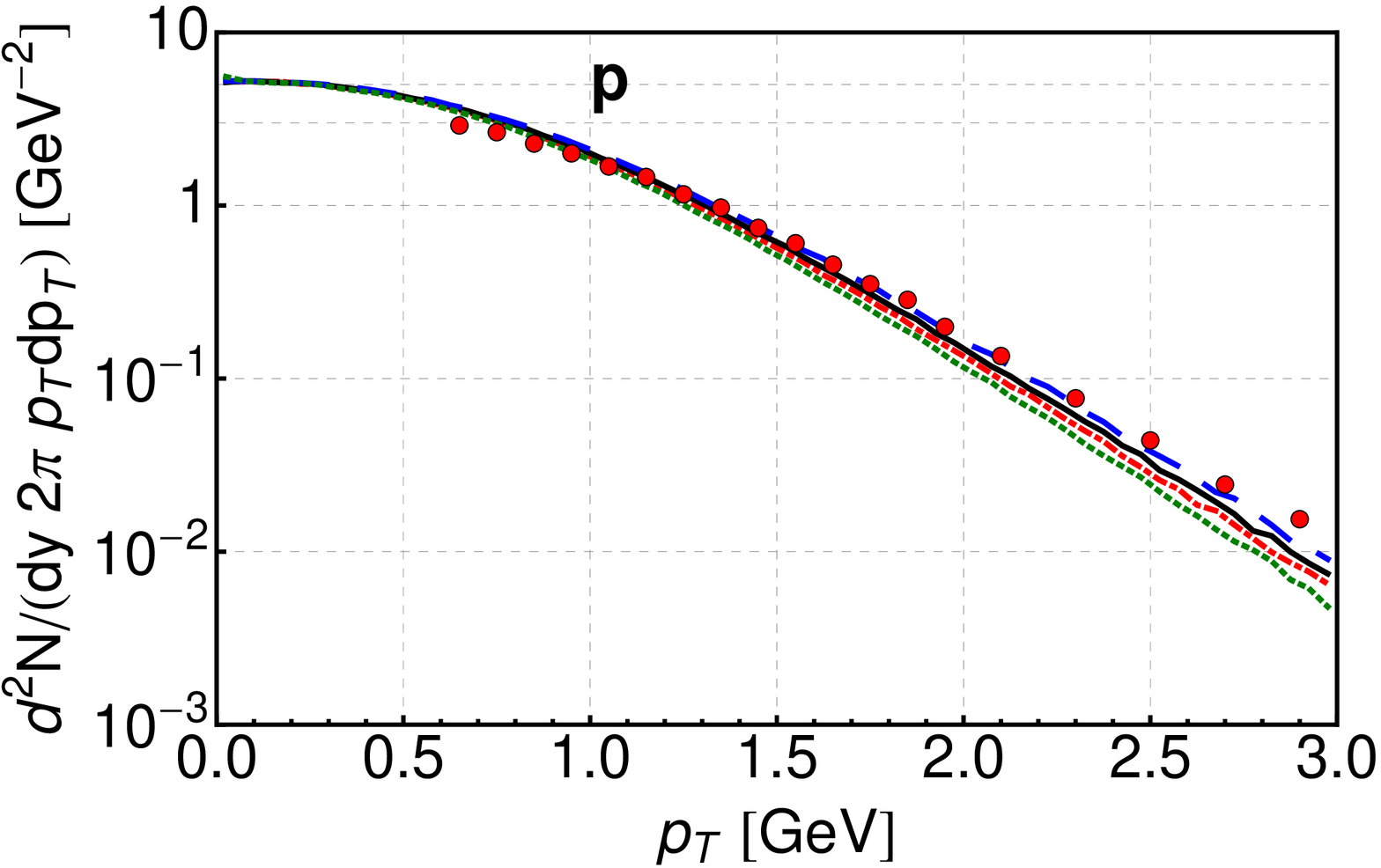}} 
\end{minipage}
\qquad
\begin{minipage}{5.75cm}
\subfigure{\includegraphics[angle=0,width=\textwidth]{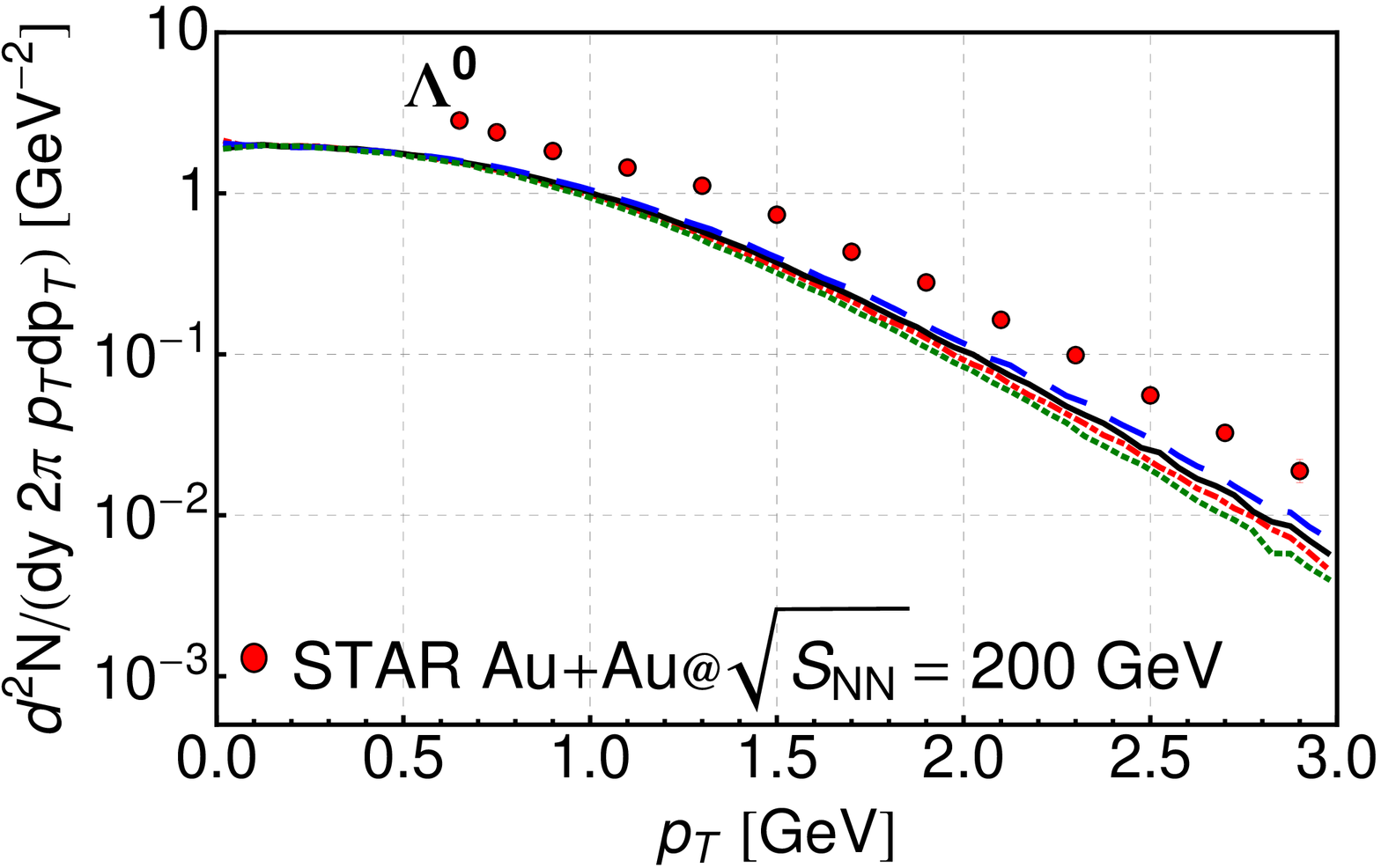}}\\
\subfigure{\includegraphics[angle=0,width=\textwidth]{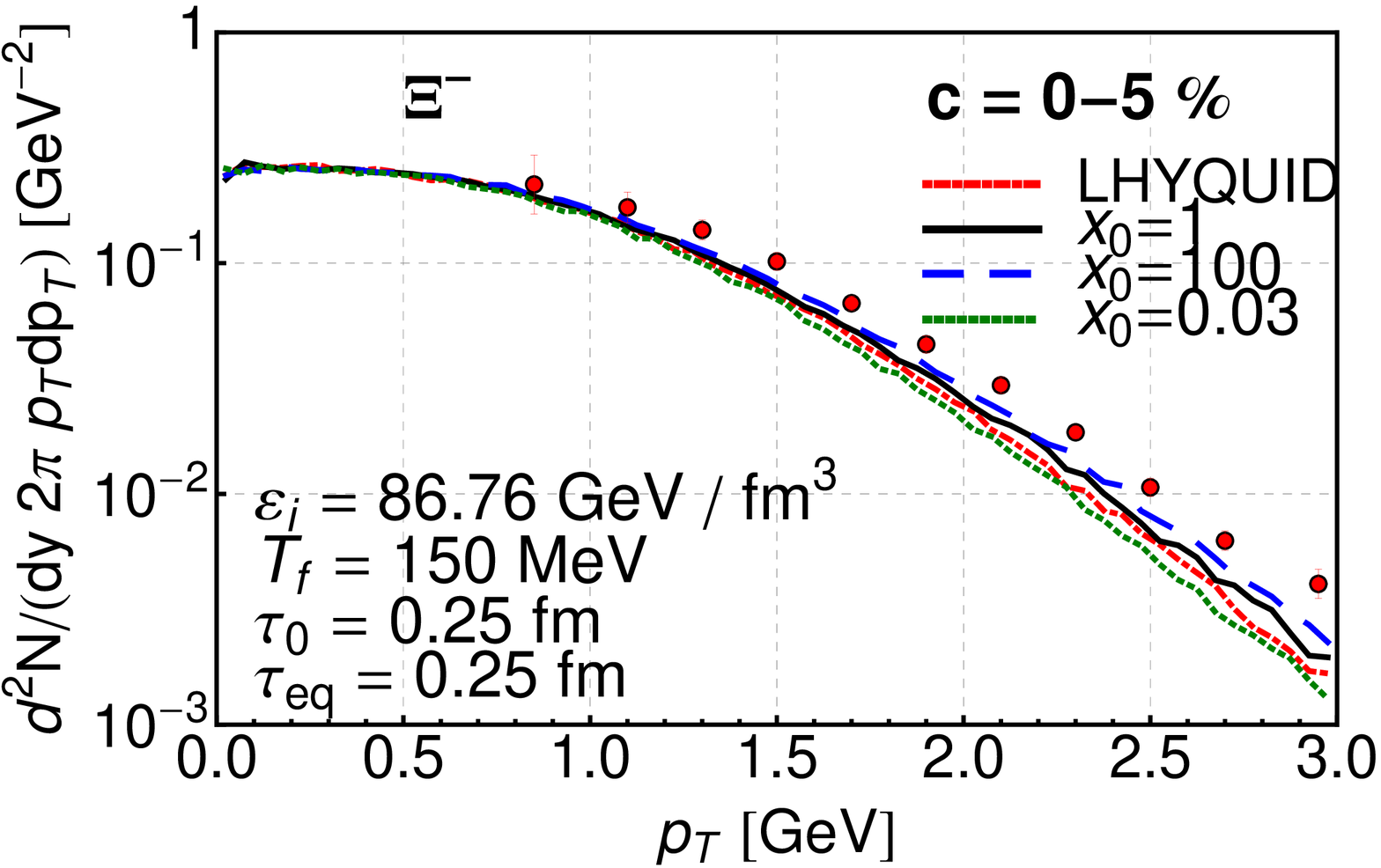}} \\
\subfigure{\includegraphics[angle=0,width=\textwidth]{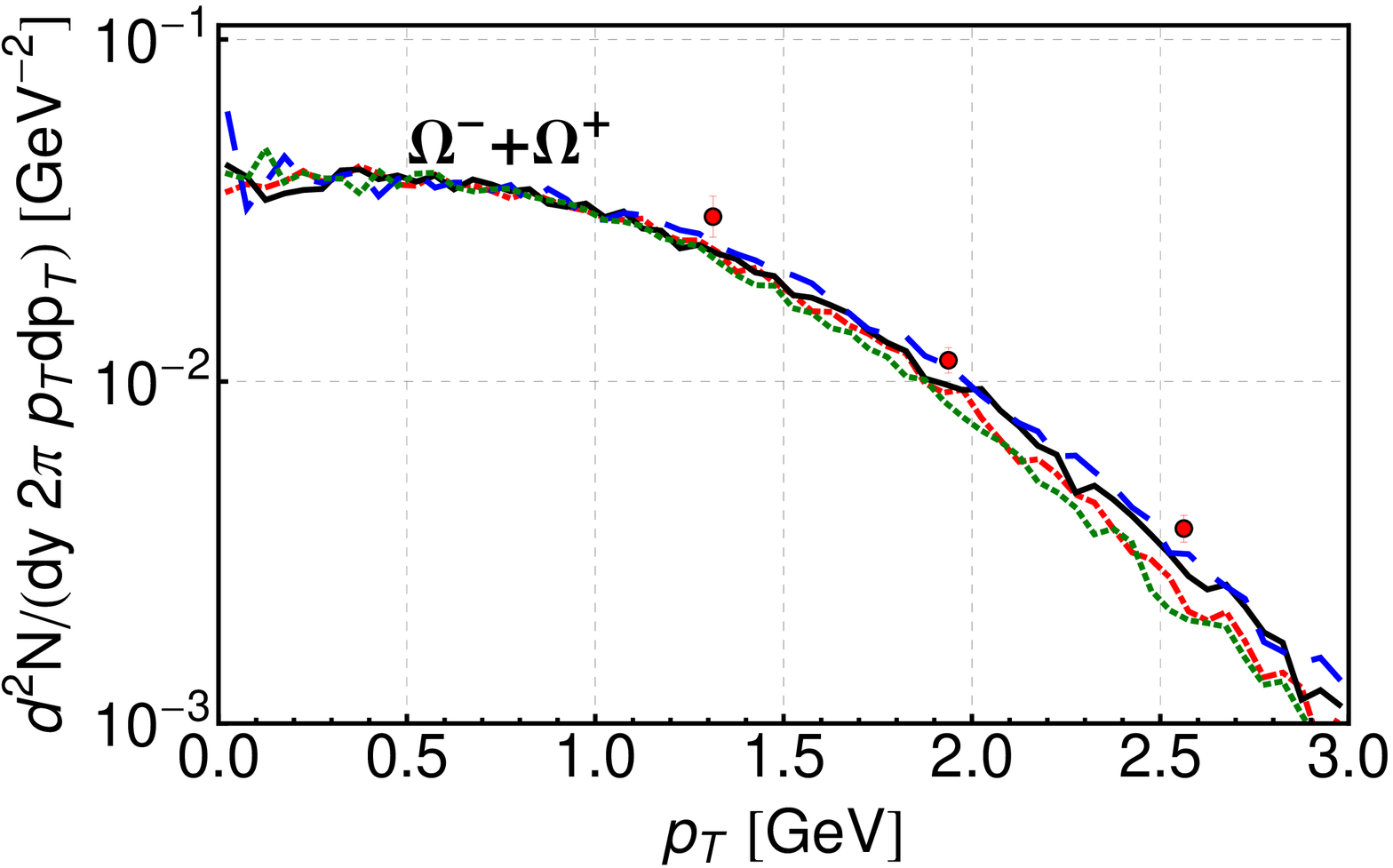}} 
\end{minipage}
\caption{\small Left panels: The calculated transverse-momentum spectra of $\pi^{+}$ (upper part), $K^{+}$ (middle part), and protons (lower part) for Au+Au collisions at $\sqrt{s_{\rm NN}}=200$ GeV and the centrality class $c=0-5$\% compared to the PHENIX data \cite{Adler:2003cb}. Protons are feed-down corrected for $\Lambda^{0}$. Right panels: The calculated transverse-momentum spectra of hyperons $\Lambda^{0}$ (upper part), $\Xi^{-}$ (middle part), and $\Omega^{+}+\Omega^{-}$ (lower part) for Au+Au collisions at $\sqrt{s_{\rm NN}}=200$ GeV and the centrality class $c=0-5$\% compared to the STAR data \cite{Adams:2006ke}. $\Lambda^{0}$ corrected for feed-down from $\Xi$ and $\Omega$.}
\label{fig:SpectraRHICc}
\end{figure}

\par In Fig.~\ref{fig:hiper0005} we present in- and out-of-plane freeze-out hypersurfaces determined from the numerical solutions of the \texttt{ADHYDRO} equations. The freeze-out condition is specified only for the locally equilibrated matter, thus, we exclude the space-time regions where $\tau < 1$ fm. In those regions the matter is anisotropic and consists mainly of gluons --- it is not likely that it contributes significantly to the hadron production. 

We use the freeze-out temperature value $T_{\rm f}=150$ MeV which has turned out very much satisfactory to reproduce the experimental data from RHIC using perfect-fluid hydrodynamics. After extracting the freeze-out hypersurface from the hydrodynamic code, we use {\tt THERMINATOR 2} \cite{Chojnacki:2011hb} in order to generate physical events. In Fig.~\ref{fig:hiper0005} we can see that for $x_0 = 100$ the system cools down a little bit slower than in the standard perfect-fluid hydrodynamics, so the freeze-out hypersurface is larger. At the same time we find that the values of the hydrodynamic flow are slightly larger. This, as will be shown later, leads to a slightly larger multiplicity and harder spectra. In the case $x_0 = 0.032$, we deal with the opposite situation. Despite these small differences, in all the cases the freeze-out hypersurfaces are similar. This suggests that the initial anisotropic stage of evolution does not modify significantly the freeze-out conditions, provided the initial entropy density is appropriately rescaled in each case.

\par In the left panels of Fig.~\ref{fig:SpectraRHICc} we show the transverse-momentum spectra of pions (upper part), kaons (middle part), and protons (lower part), while in the right panels of Fig.~\ref{fig:SpectraRHICc} we present the transverse-momentum spectra of $\Lambda^{0}$ (upper part), $\Xi^{-}$ (middle part), and $\Omega^{+}+\Omega^{-}$ (lower part). One can observe that for all values of the initial anisotropy the spectra are very similar and lie close to the (2+1)D perfect-fluid hydrodynamics results. The small differences can be noticed in the high $p_T$ region where the results for $x_0 = 100$ are the highest and for $x_0=0.032$ are the lowest. This is mainly due to larger values of transverse velocity of the fluid at the freeze-out (see discussion of Fig.~\ref{fig:hiper0005}) and slightly larger multiplicity (see left part of Fig.~\ref{fig:s}) in the case \textbf{(i)}. 

\begin{figure}[t]
\centering
\subfigure{\includegraphics[angle=0,width=0.4\textwidth]{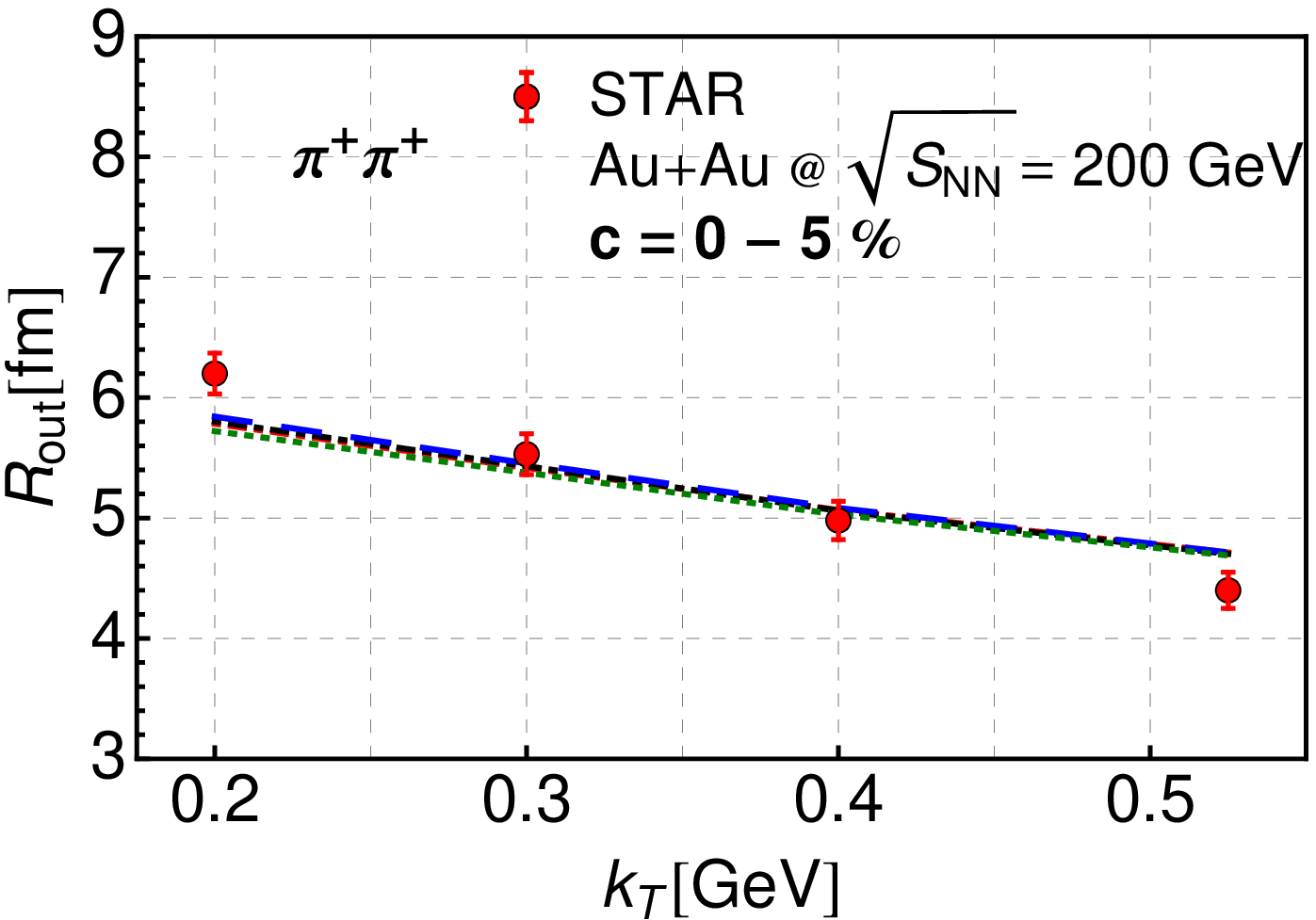}}
\subfigure{\includegraphics[angle=0,width=0.4\textwidth]{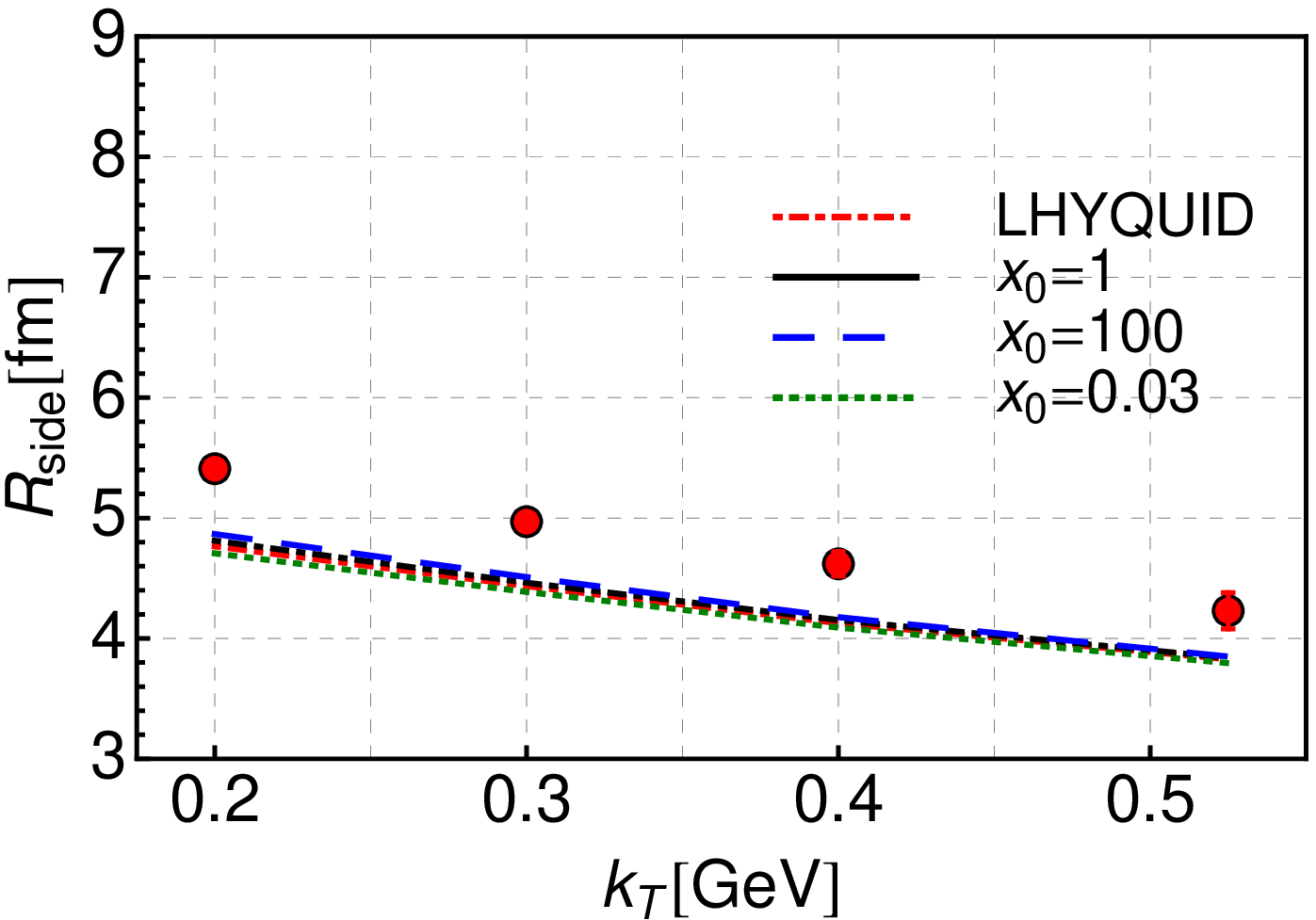}}
\subfigure{\includegraphics[angle=0,width=0.4\textwidth]{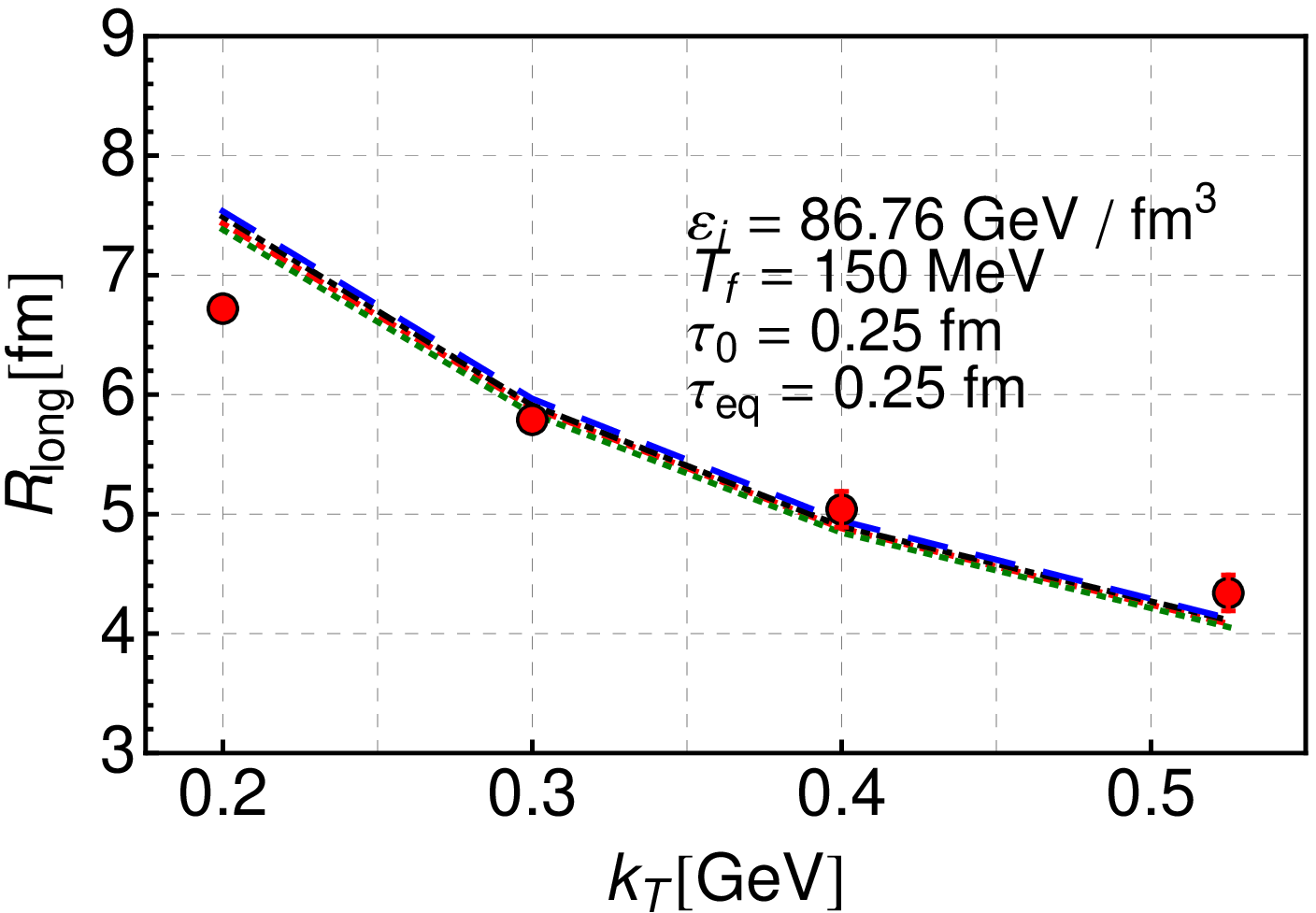}}
\caption{\small HBT radii: $R_{\rm out}$, $R_{\rm side}$, $R_{\rm long}$ as a function of total transverse momentum $k_{T}$ of the pion pair, calculated for positive pions with \texttt{THERMINATOR 2} \cite{Chojnacki:2011hb} for freeze-out hypersurfaces coming from \texttt{ADHYDRO} and \texttt{LHYQUID}. The results are presented together with the STAR data \cite{Adams:2004yc} (dots).}
\label{fig:Hbtc}
\end{figure}

\par We close the study of the central collisions with Fig.~\ref{fig:Hbtc} which presents the results for $k_{T}$-dependent HBT radii. We observe the overall agreement with the experimental data for the perfect-fluid calculation performed with \texttt{LHYQUID}. We also see that the correlation radii show almost no sensitivity to the initial anisotropic stage. All calculations yield similar results. This is understandable, since we obtain similar shapes of the freeze-out hypersurfaces for all considered initial values of $x_0$.

\par The analysis of the most central collisions in the framework of \texttt{ADHYDRO} and \texttt{LHYQUID} suggests that the considered observables are almost insensitive to the initial anisotropic stage as long as the initial entropy density is properly rescaled.

\section{Non-central collisions}
\label{sect:resbi_nc}
%
In this Section we turn to the analysis of non-central collisions. We take into consideration two centrality bins: \mbox{$c=20-30$\%} ($b=7.16$ fm) and \mbox{$c=20-40$\%} ($b=7.84$ fm). The same value of the initial central energy density, \mbox{$\varepsilon_{\rm i} = 86.76$ GeV/fm$^3$}, is used now, as before for central collisions. The initial entropy density at the center of the system is rescaled according to Eq.~(\ref{sig1}). In this way we find: \mbox{$\sigma_{\rm i} = 183.1$ fm$^{-3}$} for the case $x_0=1$, and  \mbox{$\sigma_{\rm i} = 120.3$ fm$^{-3}$} for the cases $x_0=100$ and $x_0=0.032$ if $c=20-30$\% [similarly: \mbox{$\sigma_{\rm i} = 171.3$ fm$^{-3}$} for the case $x_0=1$, and  \mbox{$\sigma_{\rm i} = 113.8$ fm$^{-3}$} for the cases $x_0=100$ and $x_0=0.032$ if $c=20-40$\%].

\begin{figure}[t]
\begin{center}
\subfigure{\includegraphics[angle=0,width=0.4\textwidth]{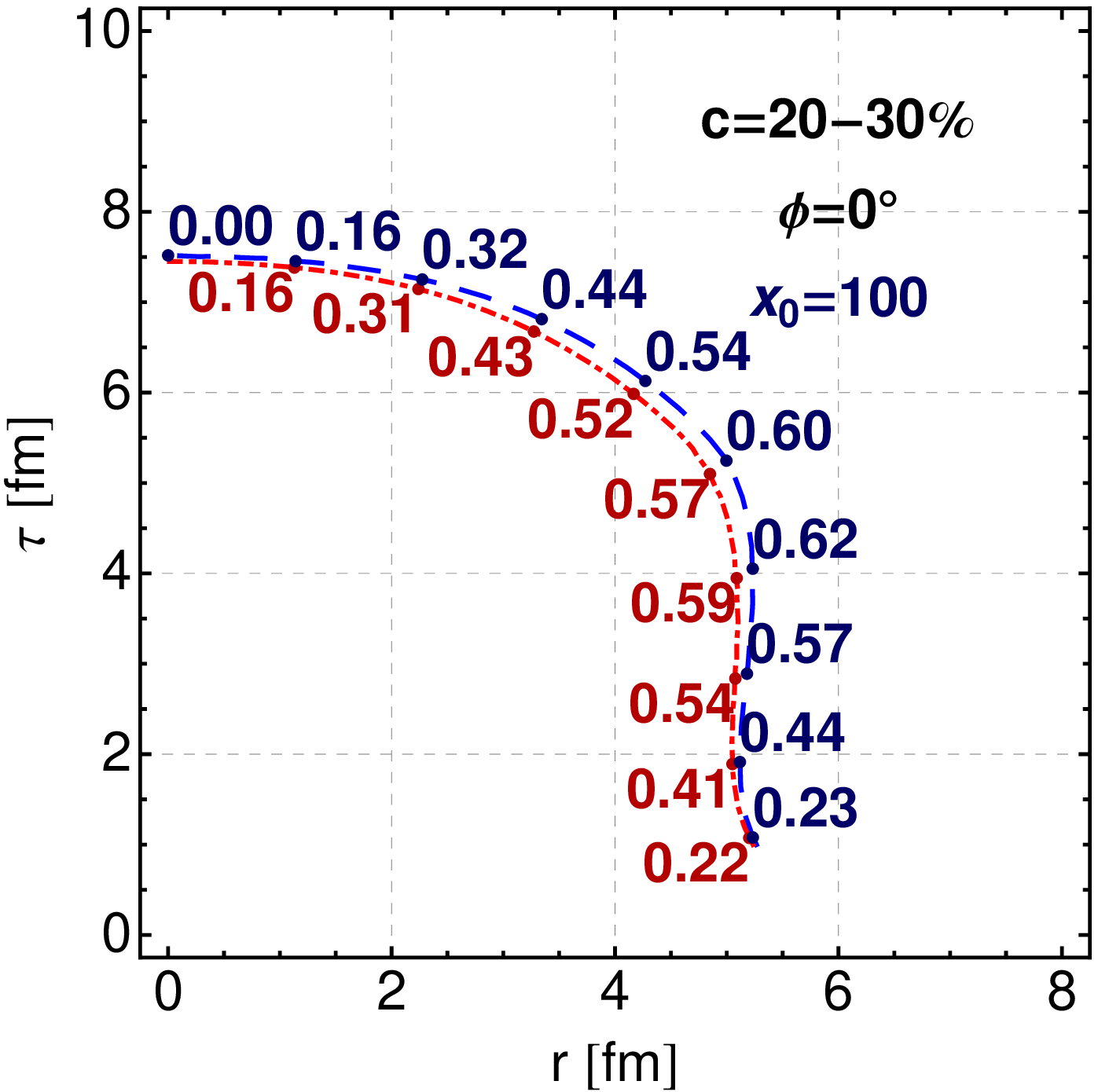}} 
\subfigure{\includegraphics[angle=0,width=0.4\textwidth]{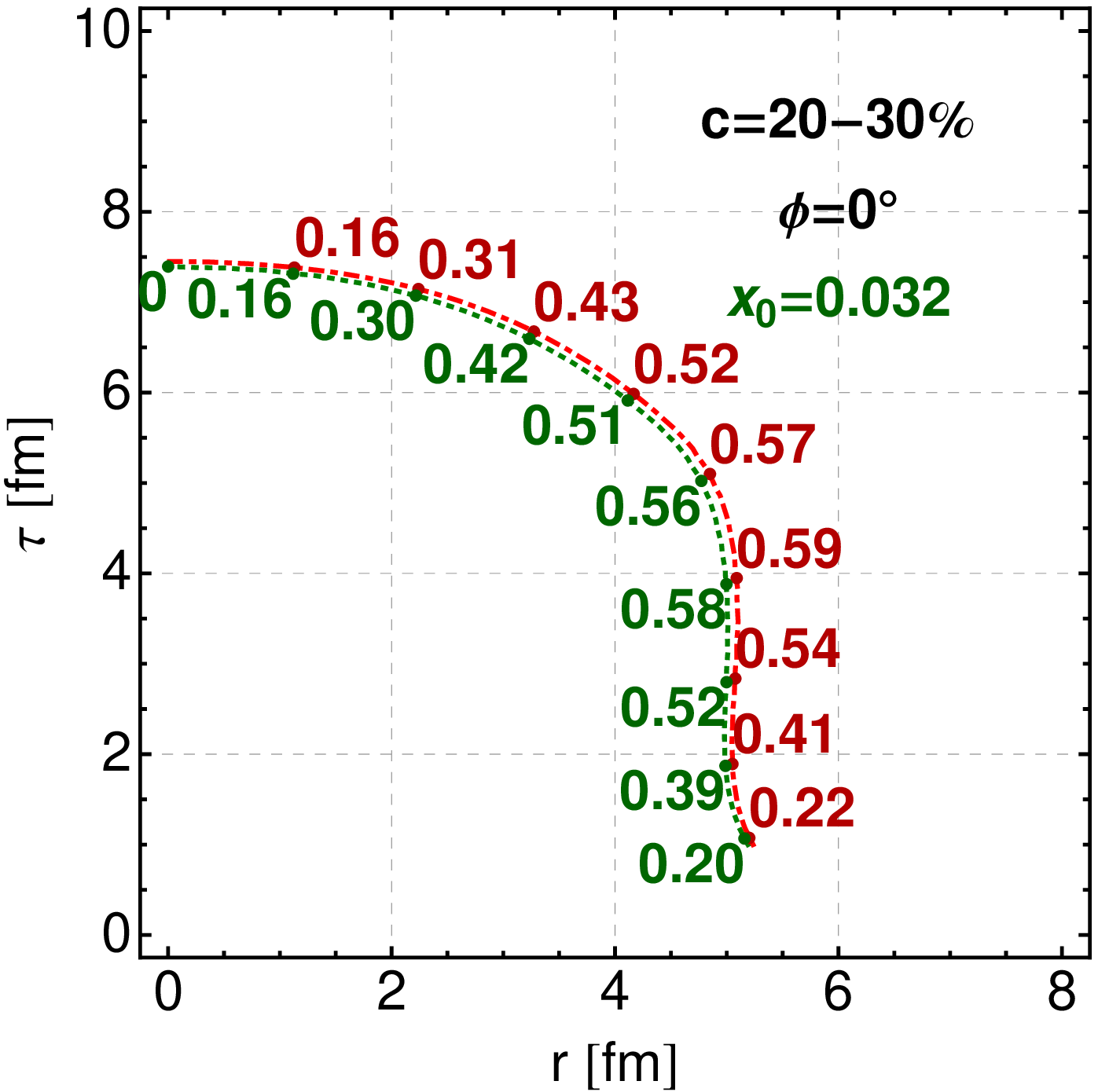}} \\
\subfigure{\includegraphics[angle=0,width=0.4\textwidth]{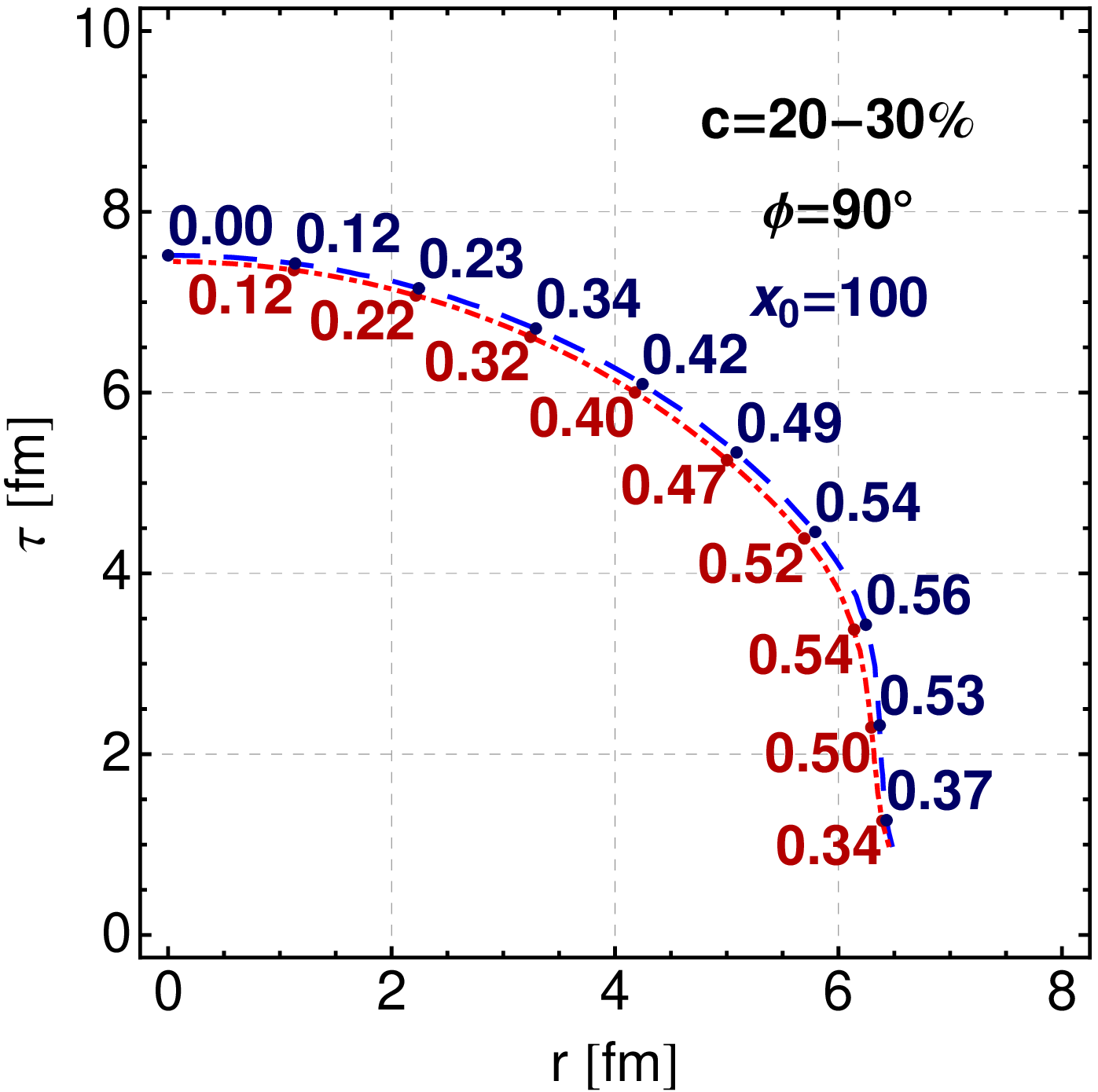}} 
\subfigure{\includegraphics[angle=0,width=0.4\textwidth]{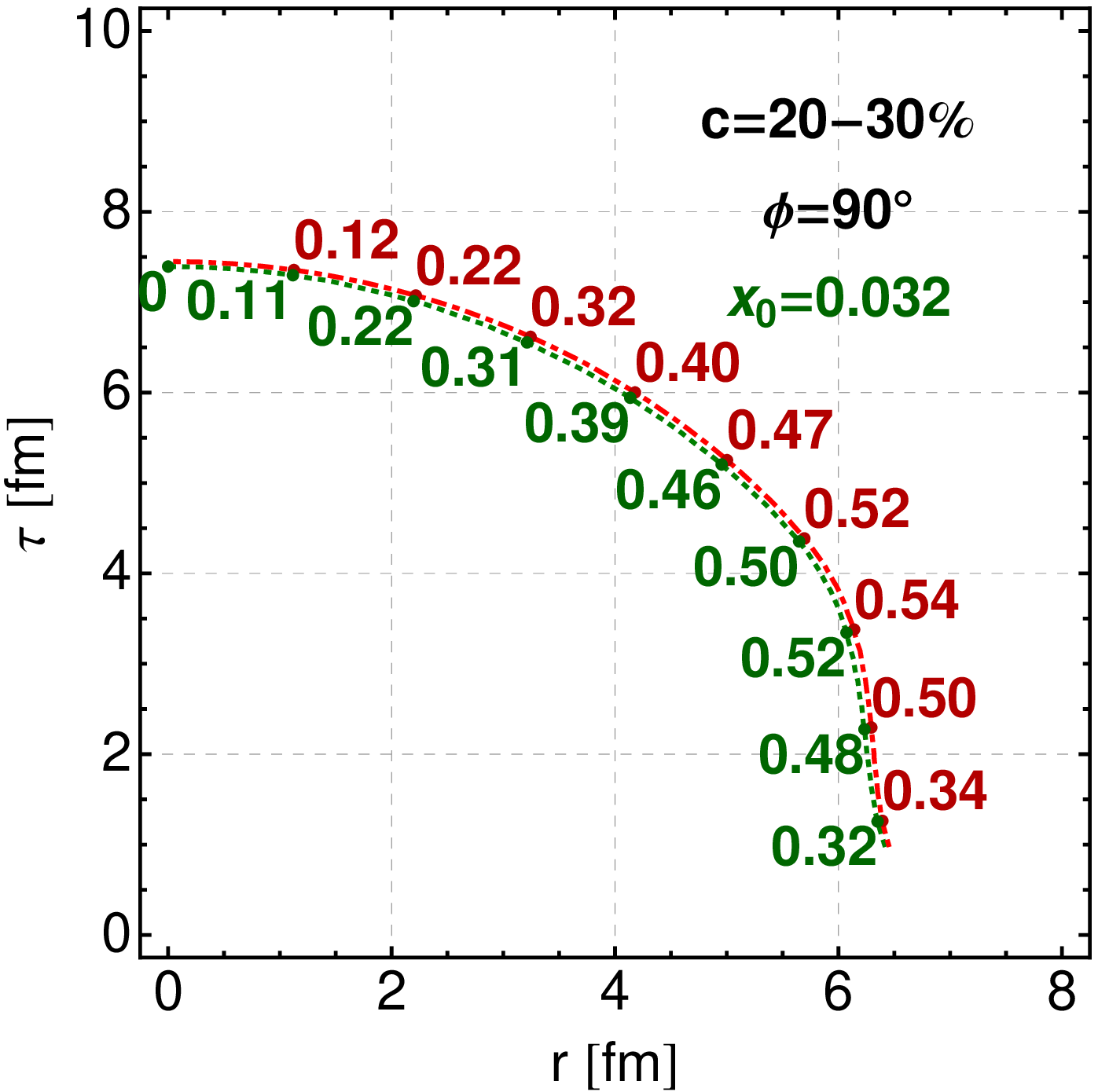}}
\end{center}
\caption{\small Same as Fig.~\ref{fig:hiper0005} but for the centrality class  \mbox{$c=20-30$\%} ($b=7.16$ fm).}
\label{fig:hiper2030}
\end{figure}

\begin{figure}[t]
\begin{center}
\subfigure{\includegraphics[angle=0,width=0.45\textwidth]{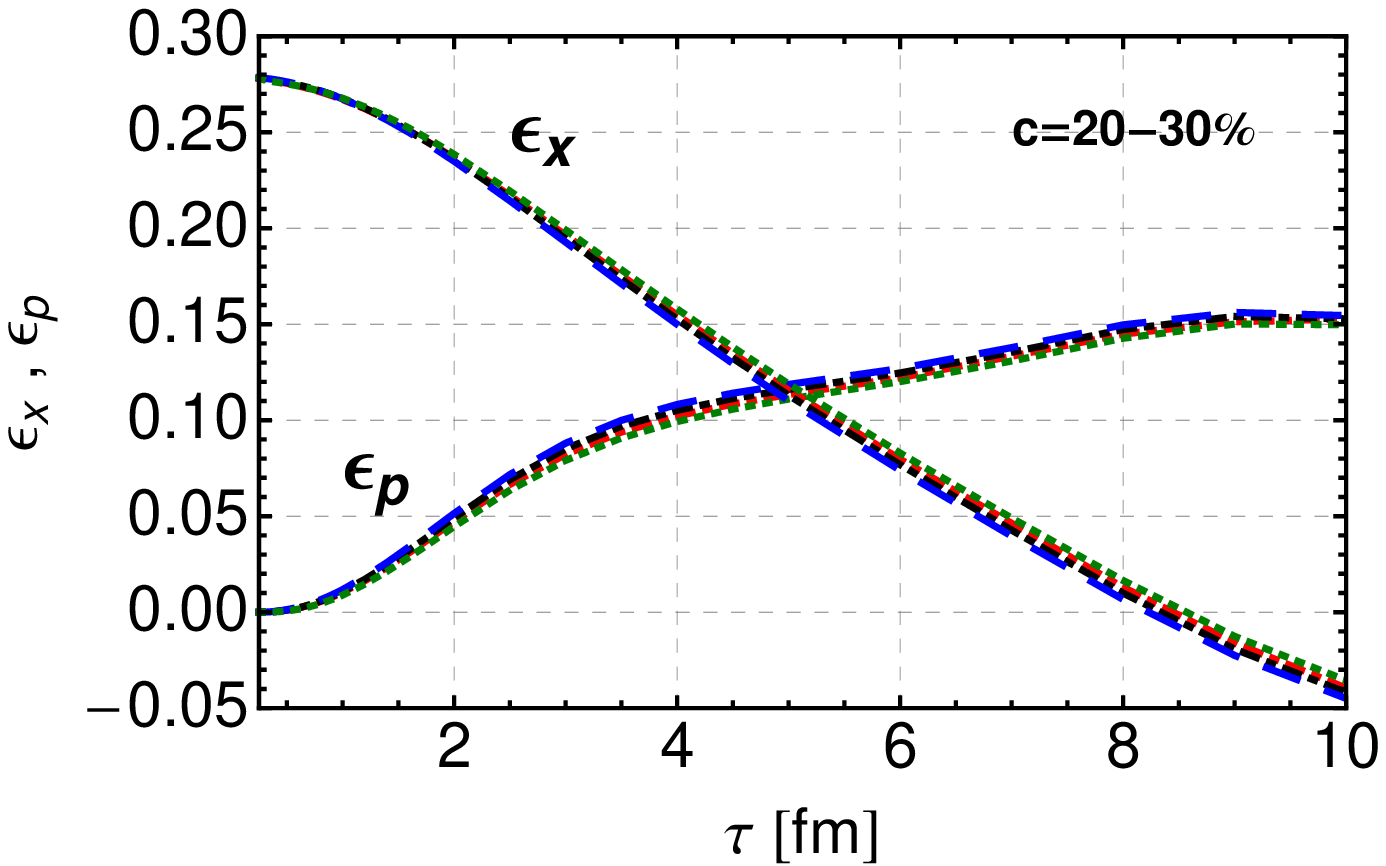}} 
\subfigure{\includegraphics[angle=0,width=0.4\textwidth]{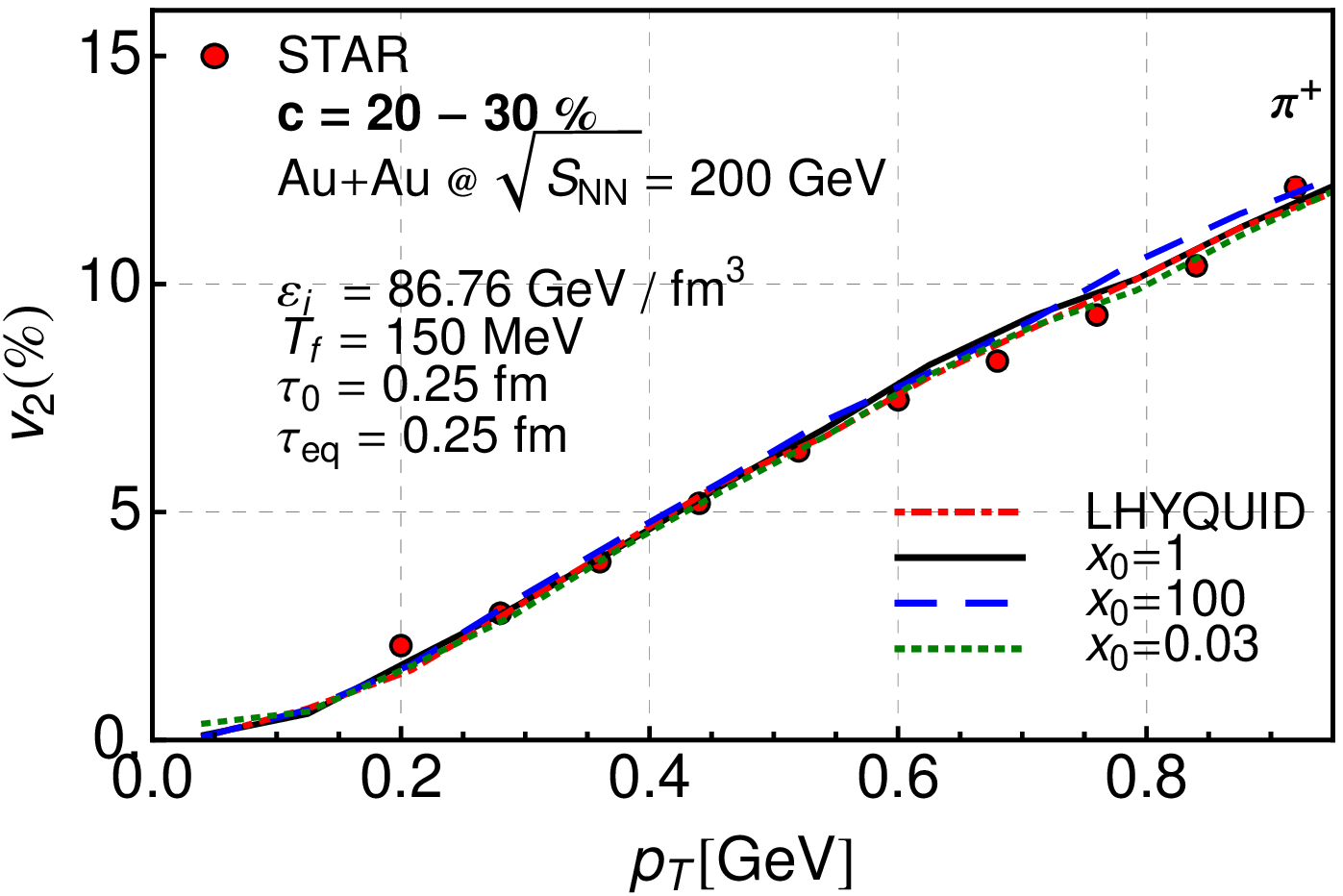}} 
\end{center}
\caption{\small Left panel: The spatial eccentricity  $\varepsilon_{x}(\tau) = (\langle y^2 -x^2 \rangle)/(\langle y^2 + x^2 \rangle)$  where the average is done with the energy density as a weight function, and the transverse-momentum anisotropy $\varepsilon_{p}(\tau) = (\int dx dy (T^{xx}-T^{yy}))/(\int dx dy (T^{xx}+T^{yy}))$ shown as functions of time. Calculation are done for centrality class $c=20-30$\%. Right panel: The elliptic flow of pions in Au+Au collisions at $\sqrt{s_{\rm NN}}=200$ GeV and the centrality class $c=20-30$\%. The model results are compared to the RHIC experimental data \cite{Adams:2004bi}.}
\label{fig:ecc}
\end{figure}
\par In the left part of Fig.~\ref{fig:hiper2030} we present the in-plane ($\phi = 0$) and out-of plane ($\phi = \pi/2$) freeze-out hypersurfaces for the centrality class \mbox{$c=20-30$\%}. We come to similar conclusions as in the case of central collisions. Moreover, we observe that the pressure acts in the same way in the in-plane and out-of-plane directions. This feature is expected since we introduced the anisotropy only between the longitudinal and transverse pressures. We also observe that the transverse spatial eccentricity of the fireball, $\varepsilon_{x}$, decreases faster with time in the $x_0=100$ case than in the $x_0=0.032$ case, see Fig.~\ref{fig:ecc}. This implies that the transverse momentum anisotropy $\varepsilon_{p}$ gets built faster. However, these effects are small if the equilibration time is short. 

\par In the right panel of Fig.~\ref{fig:ecc} we show the differential elliptic flow coefficient $v_2(p_T)$ of pions as a function of the transverse momentum, for the centrality class \mbox{$c=20-30$\%}. The model results describe quite well the data for \mbox{$p_T \leq 1$ GeV}. For different initial anisotropy parameters we obtain almost the same results. The slightly larger values of $v_2$ are observed in the case $x_0=100$, which is consistent with the conclusions coming from the analysis of the parameter $\varepsilon_{p}$. However, for small $\tau_{\rm eq}$ that is used by us, this effect is negligible. 
\begin{figure}[h!]
\centering
\begin{minipage}{5.75cm}
\subfigure{\includegraphics[angle=0,width=\textwidth]{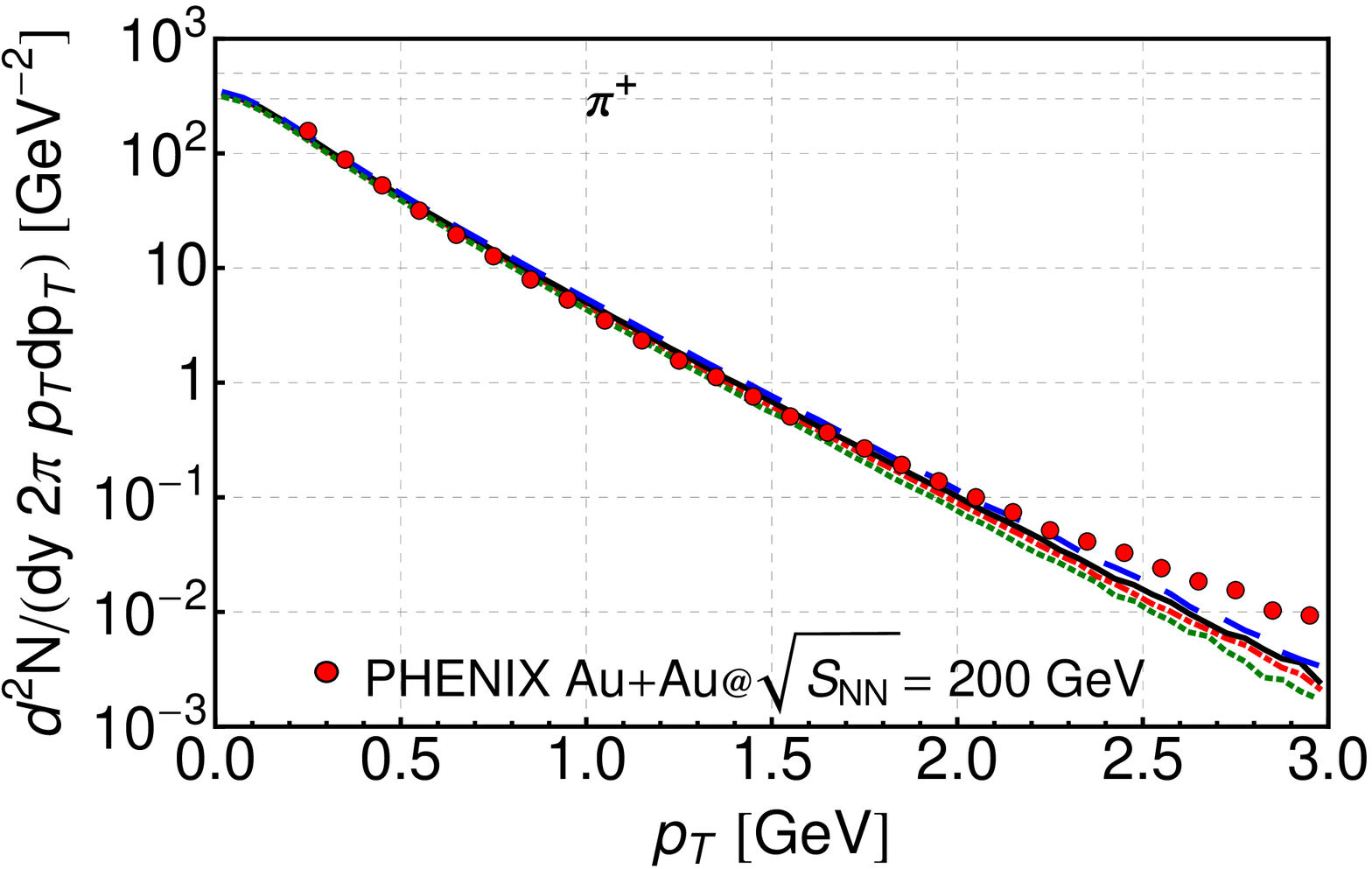}}\\
\subfigure{\includegraphics[angle=0,width=\textwidth]{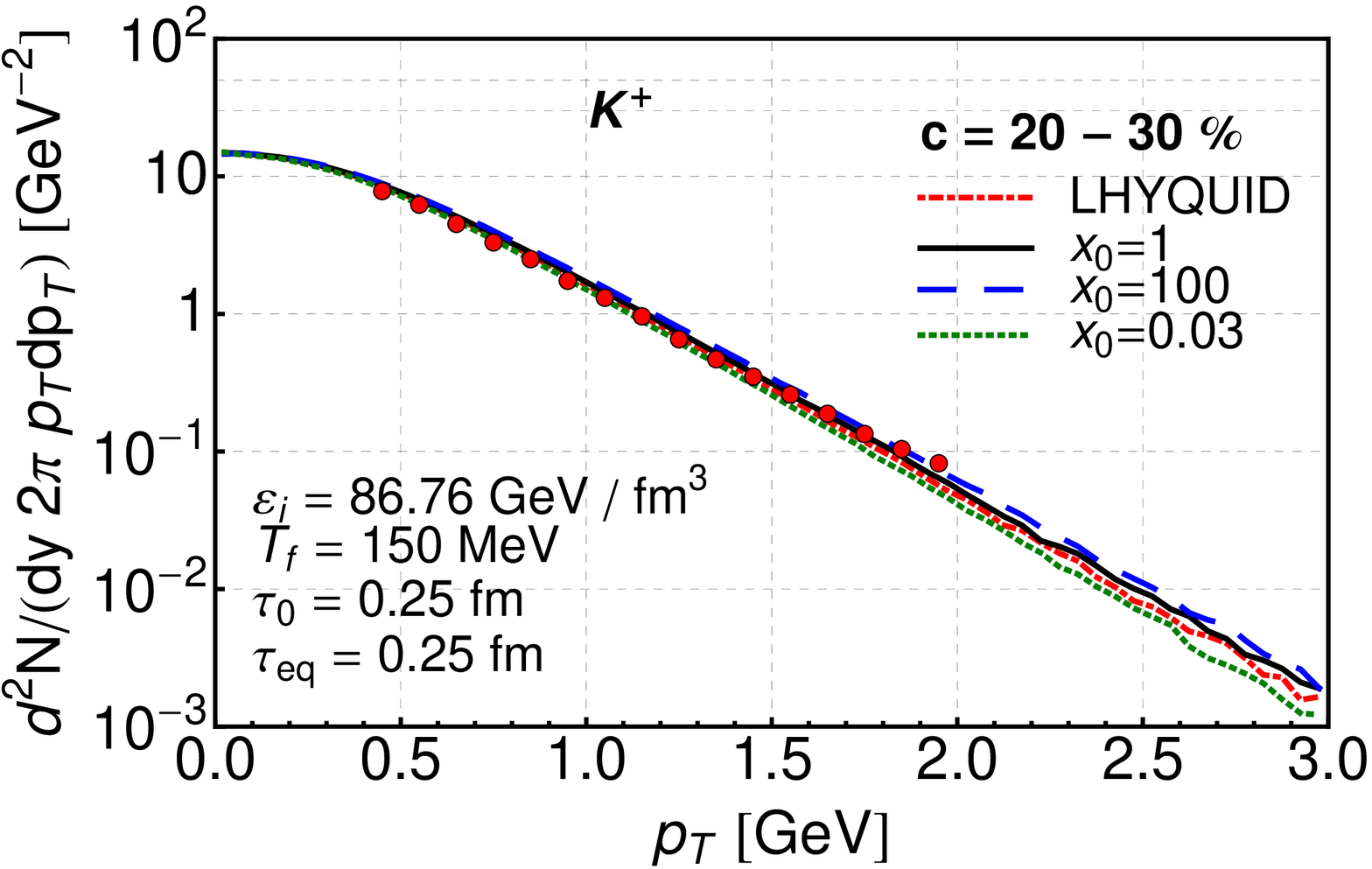}} \\
\subfigure{\includegraphics[angle=0,width=\textwidth]{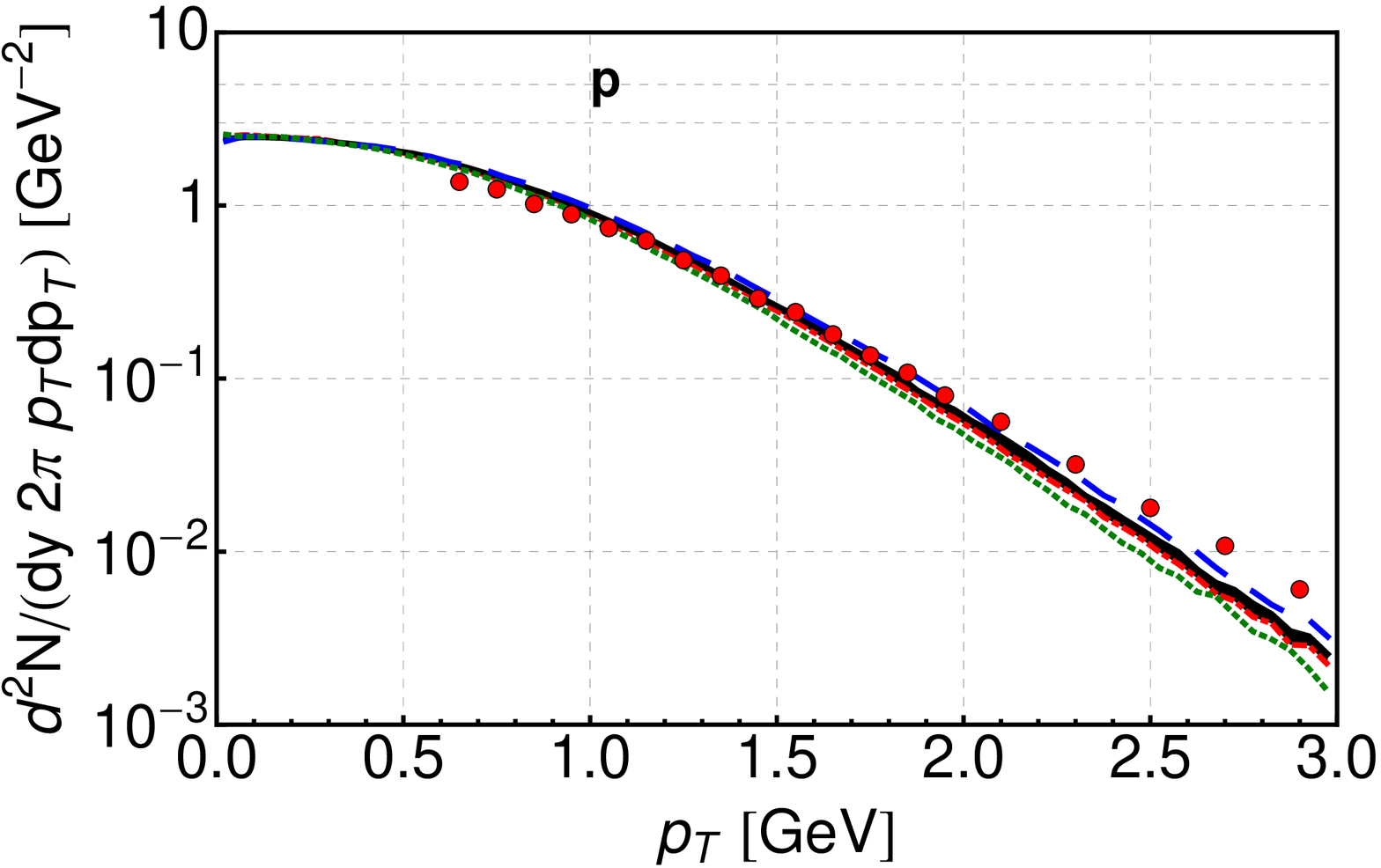}} 
\end{minipage}
\qquad
\begin{minipage}{5.75cm}
\subfigure{\includegraphics[angle=0,width=\textwidth]{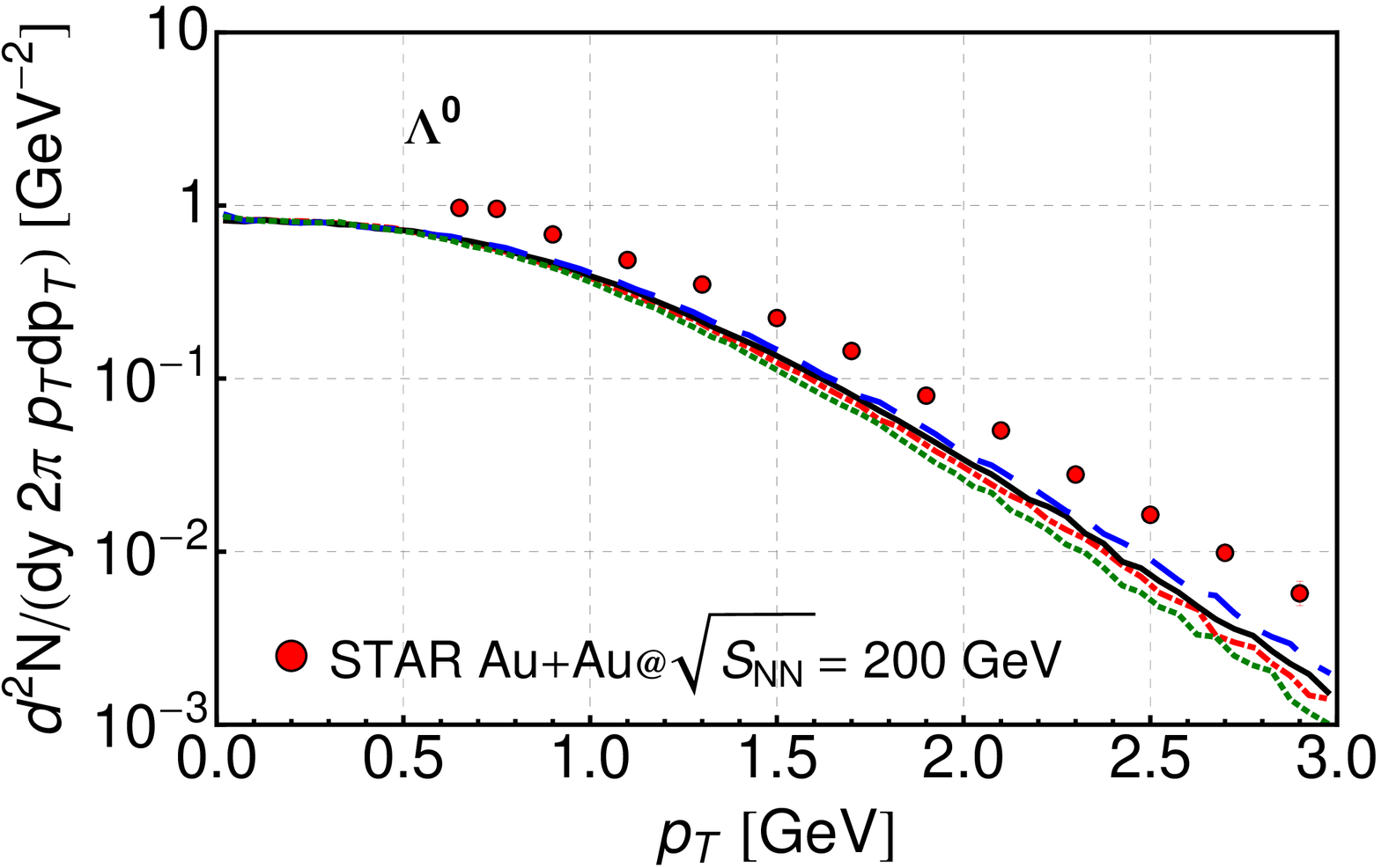}}\\
\subfigure{\includegraphics[angle=0,width=\textwidth]{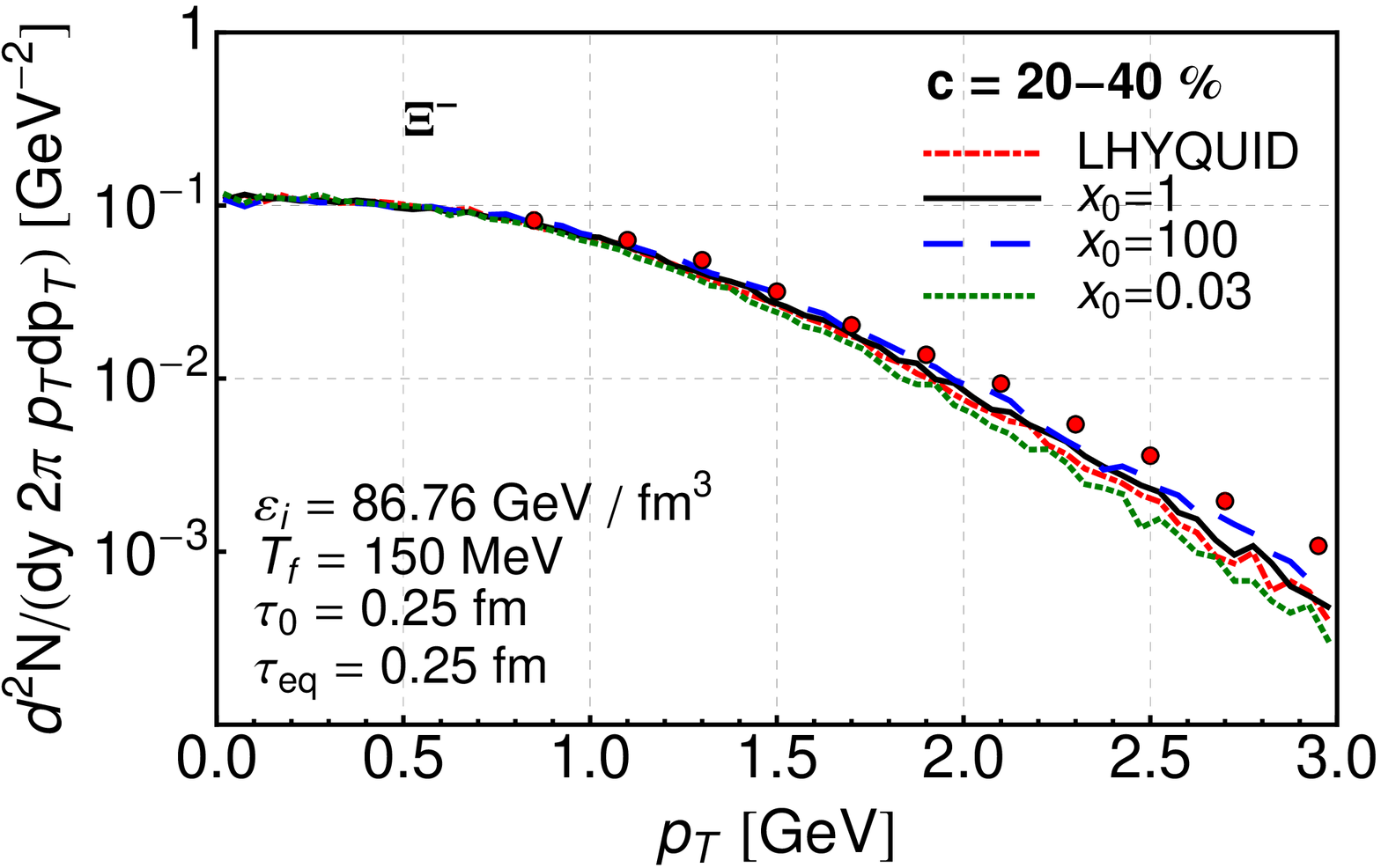}} \\
\subfigure{\includegraphics[angle=0,width=\textwidth]{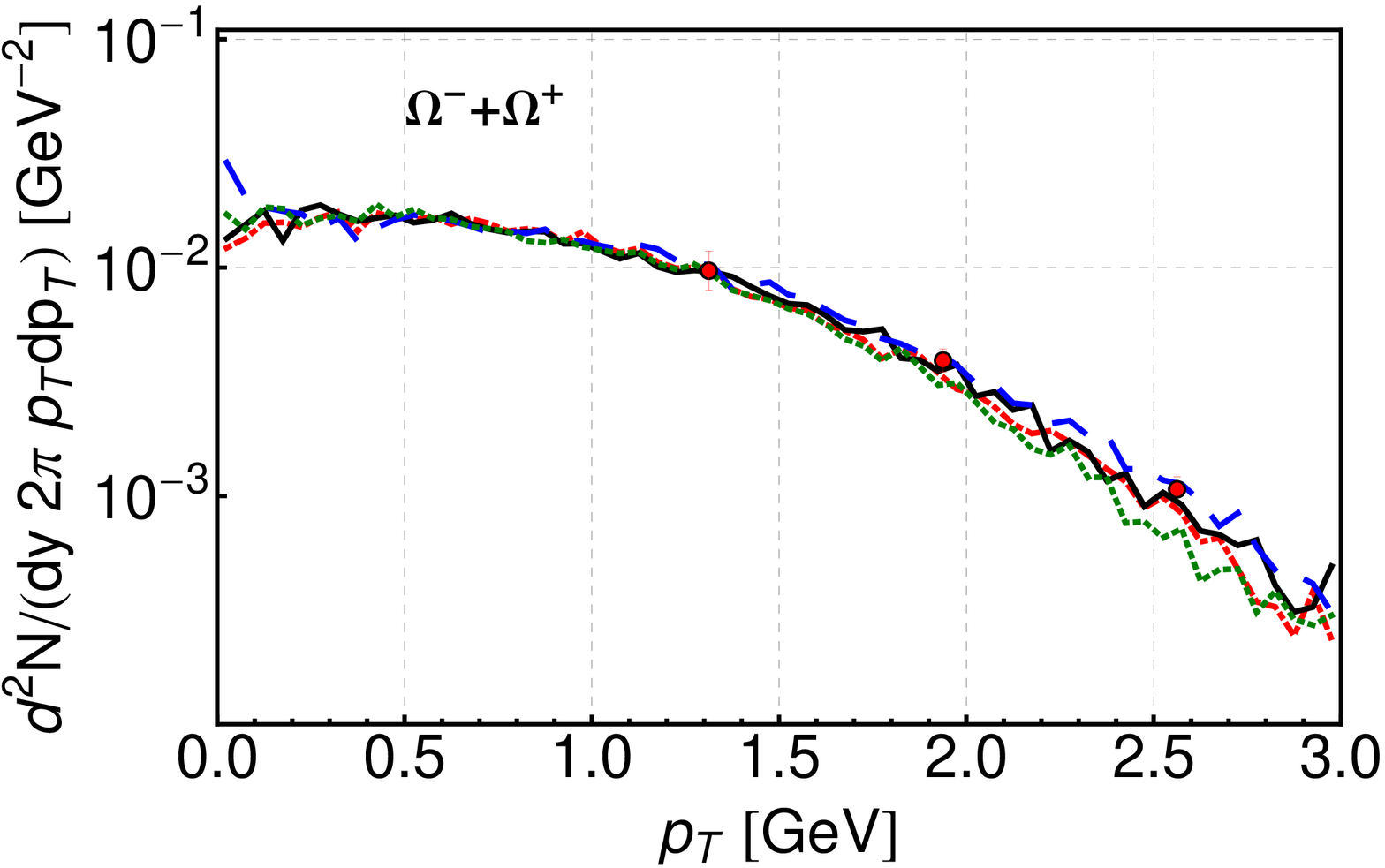}} 
\end{minipage}
\caption{\small Left panels: The calculated transverse-momentum spectra of $\pi^{+}$ (upper part), $K^{+}$ (middle part), and protons (lower part) for Au+Au collisions at $\sqrt{s_{\rm NN}}=200$ GeV and the centrality class $c=20-30$\% compared to the PHENIX data (dots) \cite{Adler:2003cb}. Protons are feed-down corrected for $\Lambda^{0}$. Right panels: The calculated transverse-momentum spectra of hyperons $\Lambda^{0}$ (upper part), $\Xi^{-}$ (middle part), and $\Omega^{+}+\Omega^{-}$ (lower part) for Au+Au collisions at $\sqrt{s_{\rm NN}}=200$ GeV and the centrality class $c=20-40$\% compared to the STAR data (dots) \cite{Adams:2006ke}. $\Lambda^{0}$ corrected for feed-down from $\Xi$ and $\Omega$.}
\label{fig:SpectraRHICnonc}
\end{figure}
\par In Fig.~\ref{fig:SpectraRHICnonc} we show our model results for the transverse-momentum spectra of pions (upper left part), kaons (middle left part), protons (lower left part), $\Lambda^{0}$ (upper right part), $\Xi^{-}$ (middle right part), and $\Omega^{+}+\Omega^{-}$ (lower right part) obtained for the centrality classes $c=20-30$\% and $c=20-40$\% and compared with the RHIC data from PHENIX \cite{Adler:2003cb} and STAR \cite{Adams:2006ke}. One can notice the very good agreement of the model results with the experimental data, except for $\Lambda^{0}$ whose normalization is too large. The model describes the experimental data up to \mbox{$p_T = 2$ GeV}, thus it works well in the region of applicability of hydrodynamics. The discrepancies for large $p_T$ are more significant than in the central collisions. Similarly to central collisions we observe improvement of the high $p_T$ tails for the case of $x_0=100$. 
\begin{figure}[h!]
\centering
\subfigure{\includegraphics[angle=0,width=0.4\textwidth]{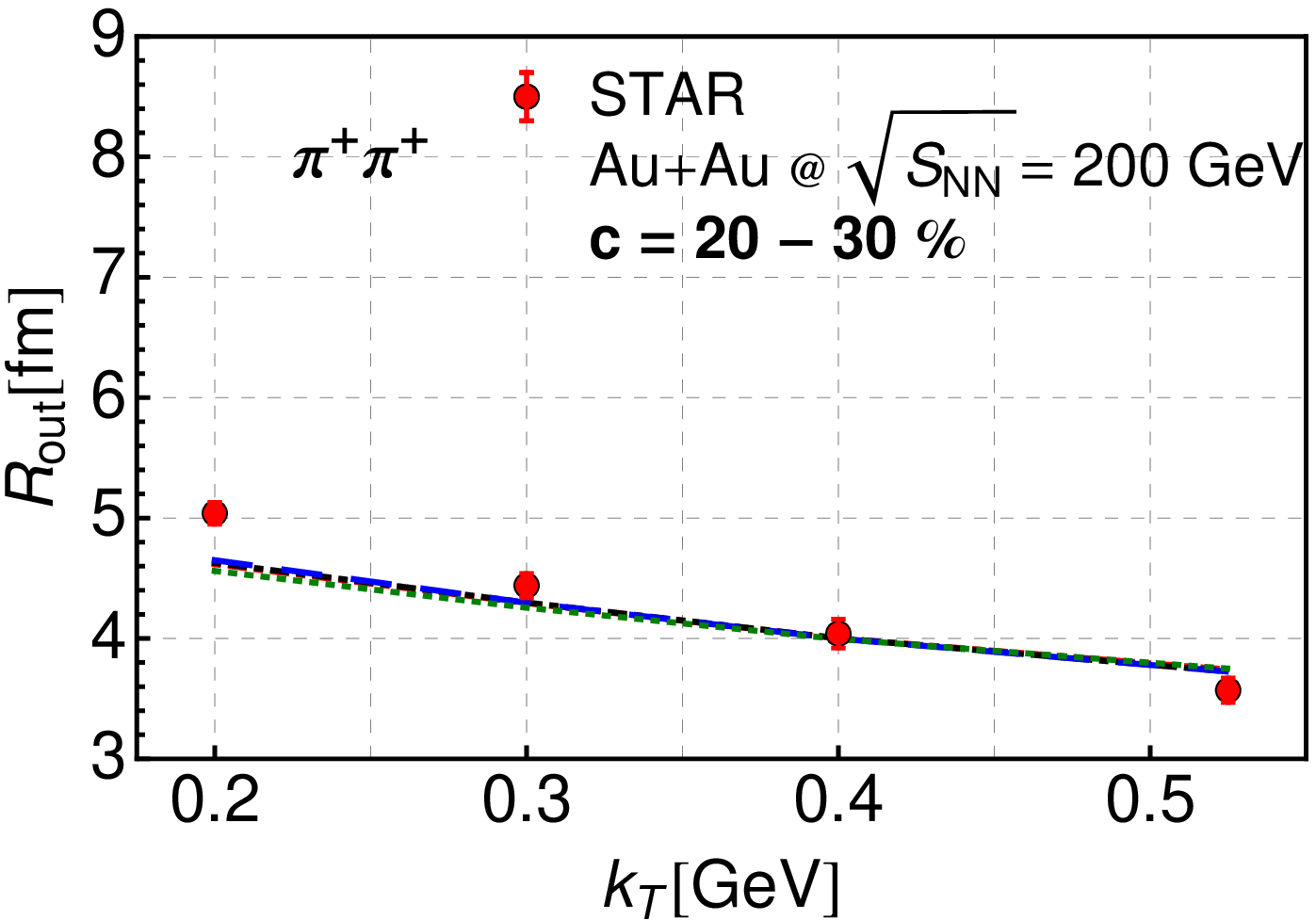}}
\subfigure{\includegraphics[angle=0,width=0.4\textwidth]{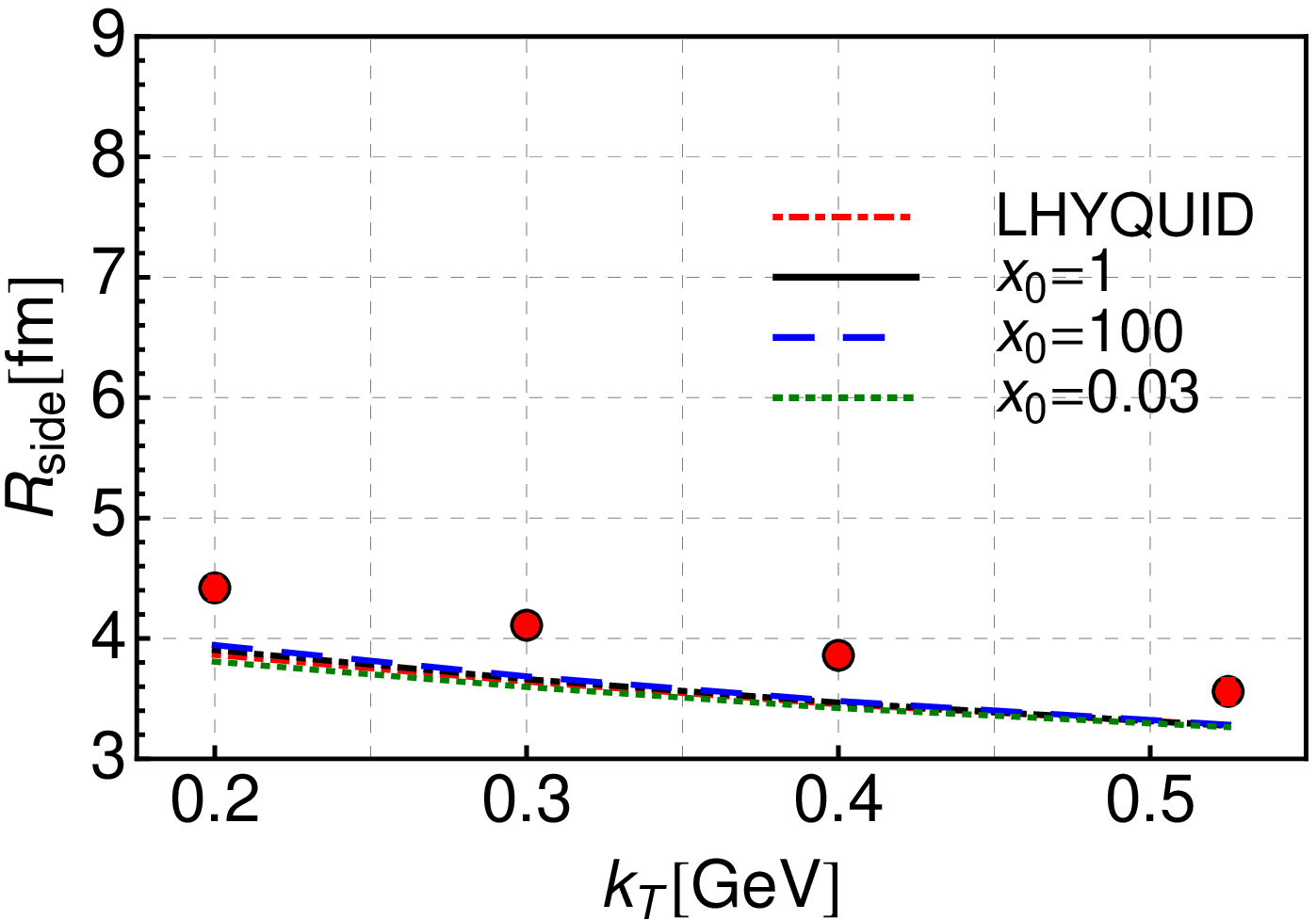}}
\subfigure{\includegraphics[angle=0,width=0.4\textwidth]{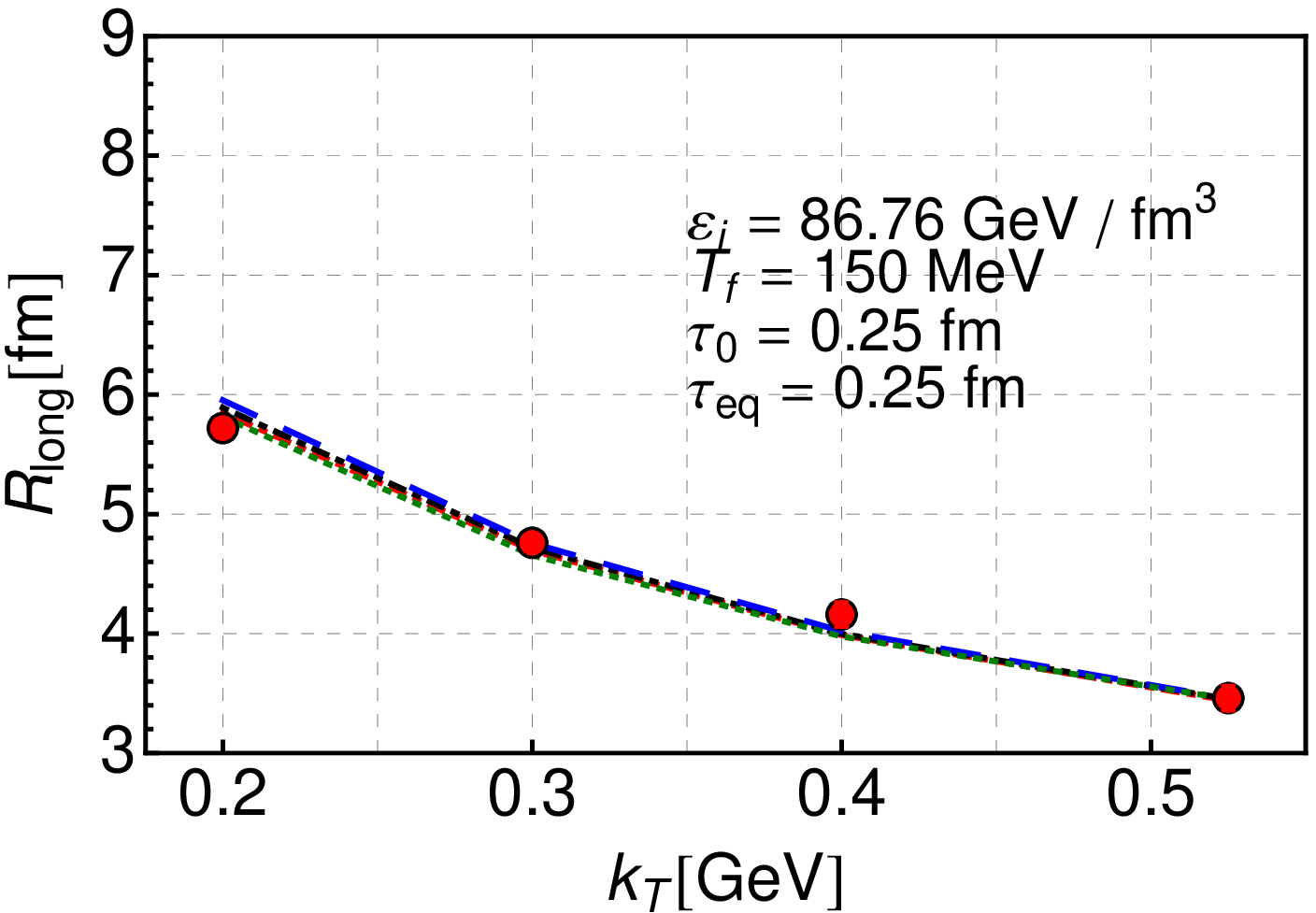}}
\caption{\small HBT radii: $R_{\rm out}$, $R_{\rm side}$, $R_{\rm long}$, calculated with \texttt{ADHYDRO} and \texttt{LHYQUID}. The model results for Au+Au collisions at $\sqrt{s_{\rm NN}}=200$ GeV and the centrality class \mbox{$c=20-30$\%} are compared to the STAR experimental data (dots) taken from \cite{Adams:2004yc}.}
\label{fig:Hbtnonc}
\end{figure}
\par Figure \ref{fig:Hbtnonc} presents the $k_T$ dependence of the HBT radii for the centrality class $c=20-30$\%. Similarly to central collisions the agreement with the data is quite satisfactory. There is almost no effects of the early anisotropic stages. 

\par Summarizing, the systematic study of the observables in the (2+1)D boost-invariant version of \texttt{ADHYDRO} model  brings us to the conclusion that the observables are almost insensitive to the initial highly-anisotropic stage, provided the initial entropy density is properly rescaled. Such results are consistent with the universality of the flow predicted in Ref.~\cite{Vredevoogd:2008id}.

\chapter[Non-boost-invariant description]{Non-boost-invariant description \\ of RHIC data}
\label{chapter:nonboostinv}
The RHIC results obtained outside the midrapidity region \cite{Back:2002wb,Bearden:2001qq,Bearden:2004yx,Back:2004mh,Abelev:2008jga,Back:2005pc}, in particular, the measured rapidity distributions of charged particles, suggest that any realistic description of the data should relax the assumption of Bjorken boost-invariance. Following this observation, in this Chapter we use the \texttt{ADHYDRO} model in 3+1 dimensions. We do not impose any symmetries to analyze the effects of initial anisotropic stages on the subsequent behaviour of matter. The analysis is done for Au+Au collisions studied at RHIC at the energy \mbox{$\sqrt{s_{\rm NN}}=200$ GeV}. In order to make a comparison with the experiment, we use the \texttt{ADHYDRO} model coupled to the Monte-Carlo model of statistical hadronization \texttt{THERMINATOR} \cite{Kisiel:2005hn,Chojnacki:2011hb} as described in Chapter \ref{chapter:freeze}.
\section{Constant initial momentum anisotropy}
\label{sect:cima}
First we analyze the simplest case where the initial anisotropy $x_{\rm 0}$ is constant in space. Once again, we perform calculations for three different cases: \textbf{i)} $x_{\rm 0}=100$ (dashed blue lines), \textbf{ii)} $x_{\rm 0}=1$ (solid black lines), and \textbf{iii)} $x_{\rm 0}=0.032$ (dotted green lines). We keep this notation all the way in this Chapter. We also use three values of the time-scale parameter: $\tau_{\rm eq}=0.25$ fm, $\tau_{\rm eq}=0.5$ fm, and $\tau_{\rm eq}=1.0$ fm to check how the results differ with respect to the length of the anisotropic stage. We show the results for the fixed freeze-out entropy density $\sigma_{\rm f} = 1.79\mathrm{/fm^3}$ which corresponds to the freeze-out temperature \mbox{$T_{\rm f} = 150$ $\mathrm{MeV}$}. This value of $T_{\rm f}$ correctly describes the slope of $p_{T}$ spectra of different hadron species, thus satisfactory separating the flow component from the thermal component in transverse-momentum spectra. The half-width of the central plateau in the initial profile, $\Delta\eta = 1$, and the half-width of gaussian tails,  $\sigma_\eta = 1.3$, are fixed, unless specified otherwise. They are fitted to approximately reproduce the shape of charged particle distributions in pseudorapidity in the case $x_{\rm 0}=1$. The initial energy density in the center of the fireball, $\varepsilon_{\rm i}$, is chosen in such a way as to reproduce the total experimental charged particle multiplicity  $N_{\rm ch}^{\rm exp}= 5060 \pm 250$ \cite{Back:2002wb} for the centrality class $c=0-6$\%. For the case $x_{\rm 0}=1$ we use the energy density $\varepsilon_{\rm i}=107.5$ $\mathrm{GeV/fm^3}$ which yields $N_{\rm ch}^{\rm theor}= 5020$. 

Since entropy grows during the hydrodynamic evolution due to non-zero entropy production source $\Sigma(\tau_{\rm eq}, \sigma, x)$, see Eq.~(\ref{en1}), the initial central energy density must be appropriately renormalized for each pair of parameters $x$ and $\tau_{\rm eq}$ to keep $N_{\rm ch}^{\rm theor}$ unchanged. In the case $x_{\rm 0}=100$ the total entropy produced during the hydrodynamic evolution grows with increasing $\tau_{\rm eq}$, while in the case $x_{\rm 0}=0.032$ the total produced entropy decreases with increasing $\tau_{\rm eq}$ (similarly to the one-dimensional behavior shown in Figs. \ref{fig:transvdom} and \ref{fig:longitdom} in Chap.  \ref{sect:hydro-r}).

The values of $\varepsilon_{\rm i}$ are summarized in Table~\ref{table:multiplicity}. The centrality dependence is reproduced by applying the scaling (\ref{dsourcest}). The initial conditions are given by the tilted source described in Chapter \ref{section:tini}.

\begin{table}[t]
  \begin{center}
    \begin{small}
      \begin{tabular}{lc@{\hskip 1cm}ccc@{\hskip 1cm}ccc}
      \hline \\ [-1ex]
$x_{\rm 0}$ & 1 & \multicolumn{3}{c@{\hskip 1cm}}{100} & \multicolumn{3}{c}{0.032} \\  [1ex] \hline \\ 
$\tau_{\rm eq} \quad \mathrm{[fm]}$ & 0.25 & 0.25 & 0.5 & 1.0 & 0.25 & 0.5 & 1.0 \\  [2ex]
$\varepsilon_{\rm i} \quad \mathrm{[GeV/fm^3]}$ & 107.5 & 58.3 & 53.4 & 48.8 & 72.5 & 75.9 & 80.1 \\ \\ \hline
      \end{tabular}
    \end{small}
  \end{center}
  \caption{\small Values of the initial central energy density, $\varepsilon_{\rm i}$, used in all the figures of this Chapter except for Fig.~\ref{fig:etadistr_RHIC_m}.
}

  \label{table:multiplicity}
\end{table}
%

\subsection{Particle spectra}
\label{sect:partspec}
%
\begin{figure}[!t]
\begin{center}
\subfigure{\includegraphics[angle=0,width=0.49\textwidth]{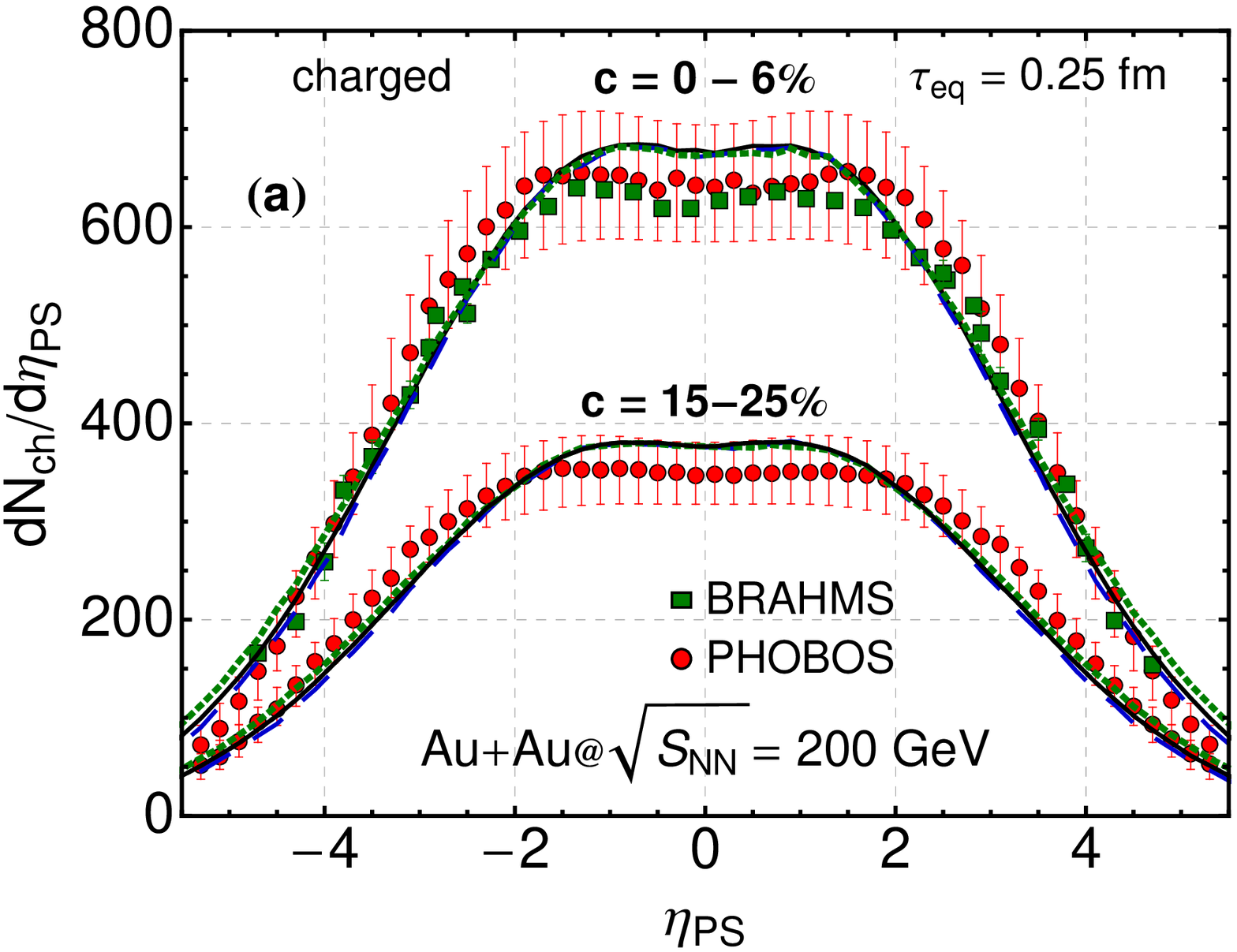}} 
\subfigure{\includegraphics[angle=0,width=0.49\textwidth]{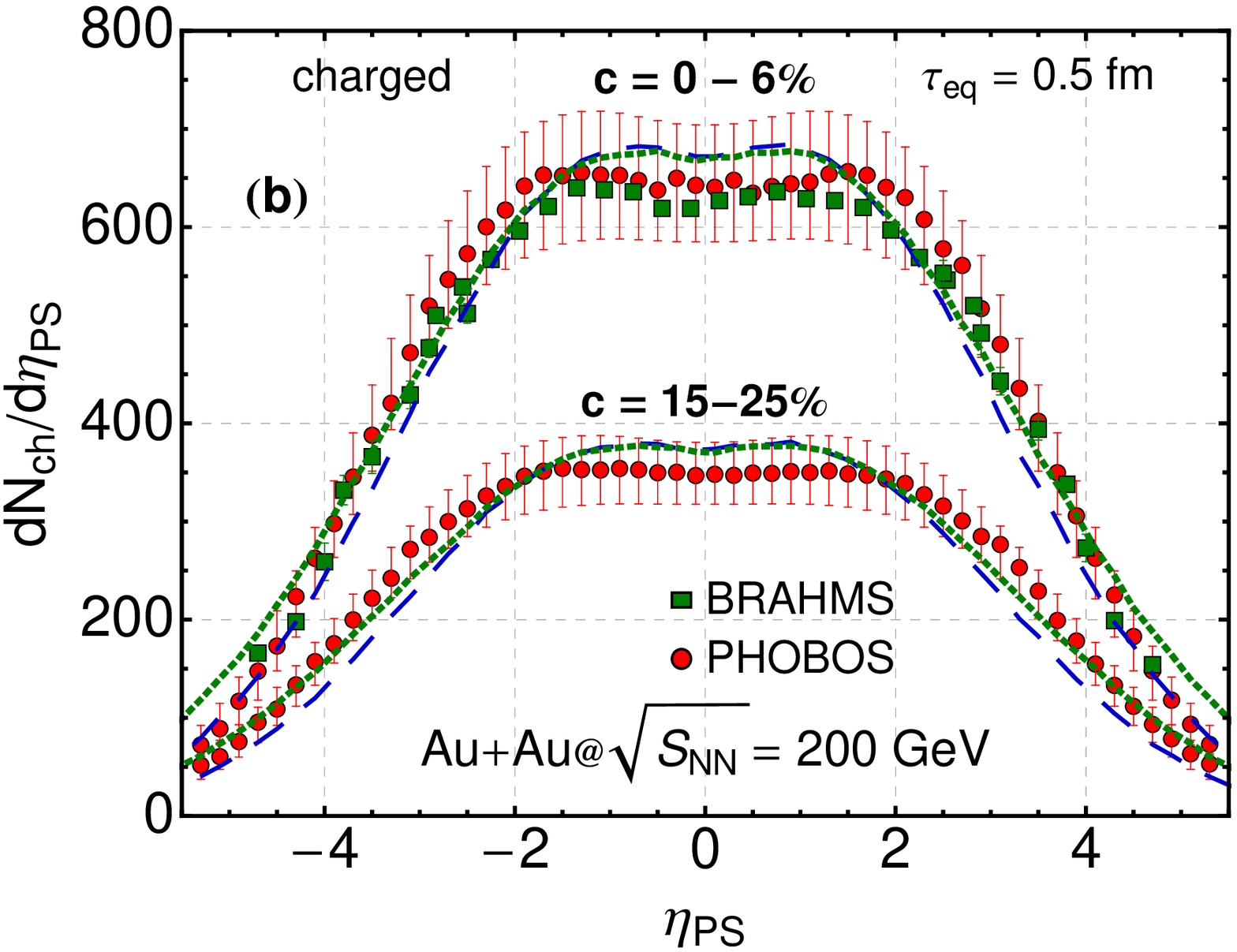}}  
\subfigure{\includegraphics[angle=0,width=0.49\textwidth]{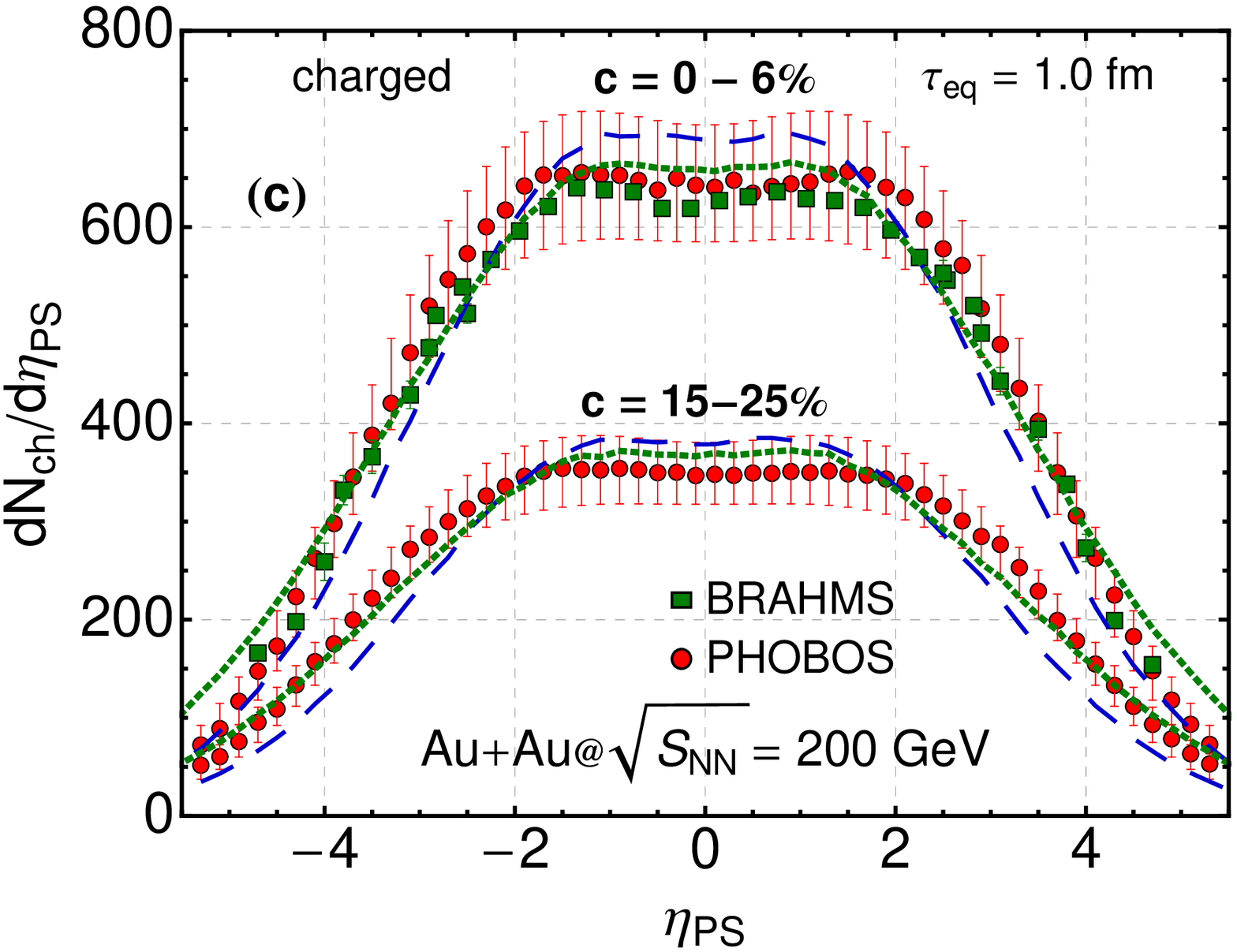}} 
\end{center}
\caption{\small The pseudorapidity distribution of charged particles for three values of the time-scale parameter: $\tau_{\rm eq}=0.25$ fm \textbf{(a)}, $\tau_{\rm eq}=0.5$ fm \textbf{(b)}, $\tau_{\rm eq}=1.0$ fm \textbf{(c)}. The results are obtained for three values of the initial momentum anisotropy: $x_{\rm 0}=100$ (dashed blue lines), $x_{\rm 0}=1$ (solid black lines) and $x_{\rm 0}=0.032$ (dotted green lines) and two centrality classes: $c=0-6$\% ($b=2.48$ fm) and $c=15-25$\% ($b=6.4$ fm). The results are compared to experimental data from PHOBOS \cite{Back:2002wb}(red dots) and BRAHMS \cite{Bearden:2001qq}(green squares). Vertical bars for PHOBOS data denote the systematic errors.}
\label{fig:etadistr_RHIC}
\end{figure}
\par We start our discussion of the results obtained with the \texttt{ADHYDRO} model with the analysis of the momentum distributions of charged and identified particles. These distributions include the feeding from unstable resonances whose  decays are implemented in \texttt{THERMINATOR} \cite{Kisiel:2005hn,Chojnacki:2011hb}. 

\subsubsection{Pseudorapidity distribution}
\label{sect:etadistr}

At first, we consider the pseudorapidity distributions of charged particles. The pseudorapidity is defined by the formula
\begin{eqnarray}
\eta_{PS} &=& \frac{1}{2} \ln \frac{p + p_{\parallel}}{p - p_{\parallel}}, \label{pseudoeta}
\end{eqnarray}
where $p$ is the momentum of a particle. In Fig.~\ref{fig:etadistr_RHIC} we show $dN_{\rm ch}/d\eta_{PS}$ for three different values of the initial momentum anisotropy: $x_{\rm 0}=100$ (dashed blue lines), $x_{\rm 0}=1$ (solid black lines), and $x_{\rm 0}=0.032$ (dotted green lines), and three values of the time-scale parameter: $\tau_{\rm eq}=0.25$ fm, $\tau_{\rm eq}=0.5$ fm, and $\tau_{\rm eq}=1.0$ fm. We observe that as long as the initial anisotropic stage is short its influence on pseudorapidity distributions of particles is small in both $x_{\rm 0}=100$ and $x_{\rm 0}=0.032$ cases, as compared to the case $x_{\rm 0}=1$. As the length of  the off-equilibrium stage increases, the difference of pressures lasts longer, and the difference between the longitudinal and transverse expansion rates becomes larger. For instance, in the case $x_{\rm 0}=100$ the pseudorapidity distribution becomes a bit steeper in the region of forward and backward rapidities due to reduced longitudinal pressure (see our discussion in Chapter \ref{sect:hydro-is}), but the width of the central plateau is nearly unchanged.

\begin{figure}[t]
\begin{center}
\subfigure{\includegraphics[angle=0,width=0.5\textwidth]{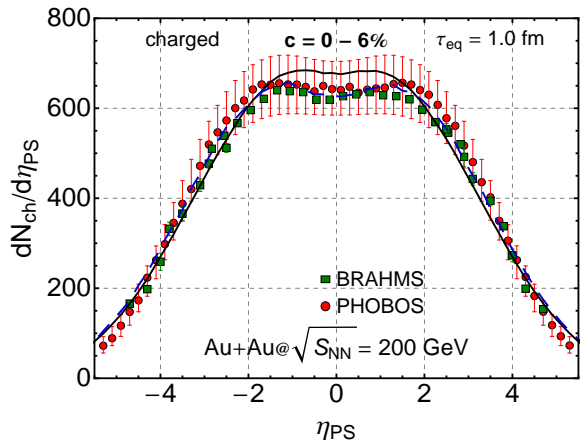}} 
\end{center}
\caption{\small The pseudorapidity distribution of charged particles for $x_{\rm 0}=1$ (solid black line) and $x_{\rm 0}=100$ (dashed blue line). In the latter case the initial conditions were modified to $\varepsilon_{\rm i} = 41.8$  $\mathrm{GeV/fm^3}$ (compare Table~\ref{table:multiplicity}) and $\Delta\eta = 1.5$ to obtain the same final total multiplicity. The results are compared to the experimental PHOBOS  \cite{Back:2002wb}(red dots) and BRAHMS \cite{Bearden:2001qq}(green squares) data.}
\label{fig:etadistr_RHIC_m}
\end{figure}

In Fig.~\ref{fig:etadistr_RHIC_m} we present the case $x_{\rm 0}=1$ with $\tau_{\rm eq}=0.25$ fm and the standard parameters of $f(\eta)$ together with the case $x_{\rm 0}=100$ with $\tau_{\rm eq}=1.0$ fm. In the latter case the initial energy density was {\it reduced} to $\varepsilon_{\rm i} = 41.8$  $\mathrm{GeV/fm^3}$ (from \mbox{$48.8$  $\mathrm{GeV/fm^3}$}, see Table~\ref{table:multiplicity}) and the width of the plateau of the initial profile $f(\eta)$ was {\it refitted} to $\Delta\eta = 1.5$. In this way we obtain the same total multiplicity. We can clearly see the improvement in the description of data as compared  to the case $x_{\rm 0}=100$ in Fig.~\ref{fig:etadistr_RHIC} \textbf{(c)}.
Thus, the shape of pseudorapidity distributions might be better described if the initial fireball is transversally thermalized and the profile parameters are properly readjusted. However, it should be stressed that the widths and forms of the {\it initial} longitudinal profiles (\ref{longitprof}) are not well known. Therefore, no definite conclusions about the length of the pre-equilibrium phase may be drawn from the correct description of the shape of the rapidity distributions alone, as it has been already noticed in  \cite{Bozek:2007qt}.
\begin{figure}[h]
\begin{center}
\subfigure{\includegraphics[angle=0,width=0.49\textwidth]{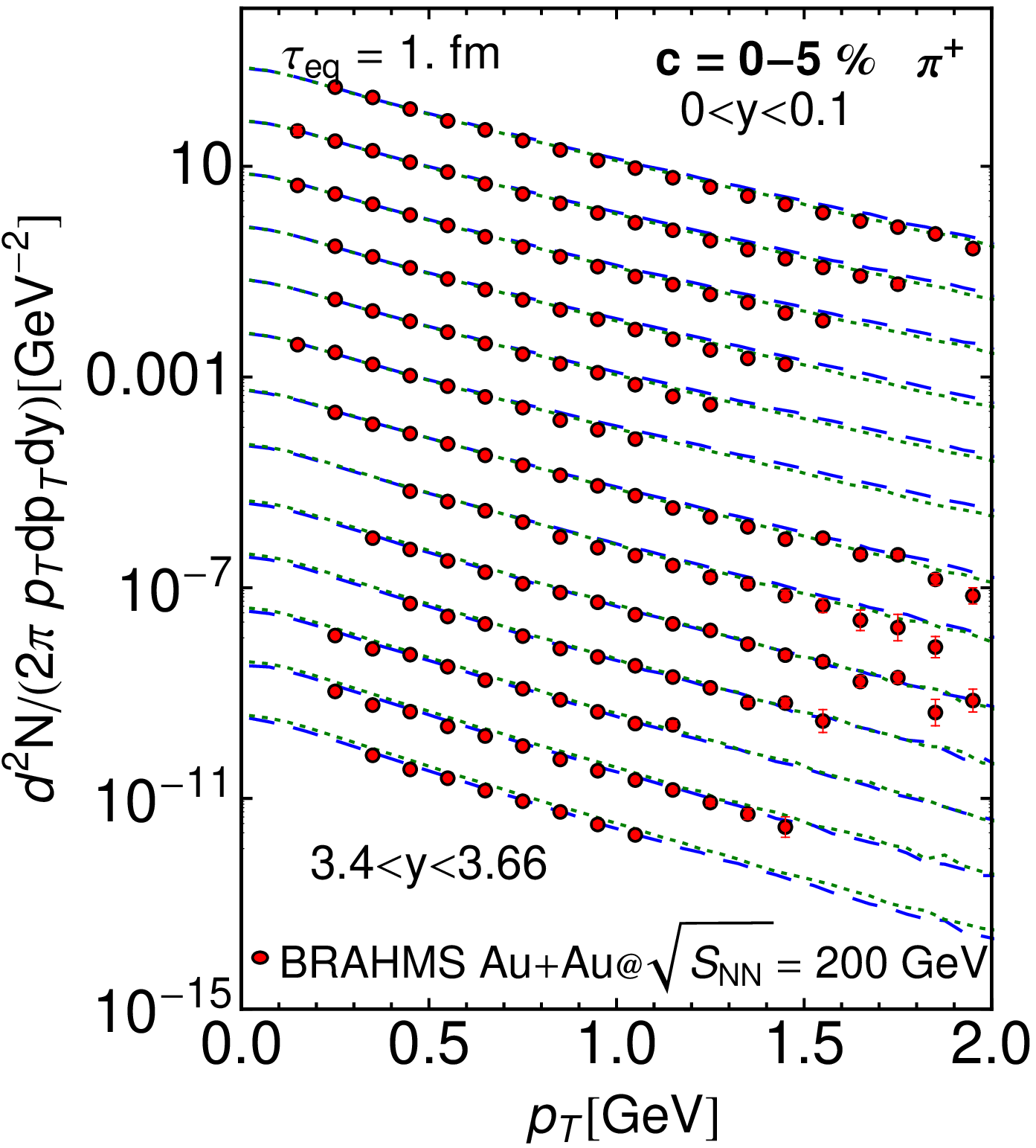}}
\subfigure{\includegraphics[angle=0,width=0.49\textwidth]{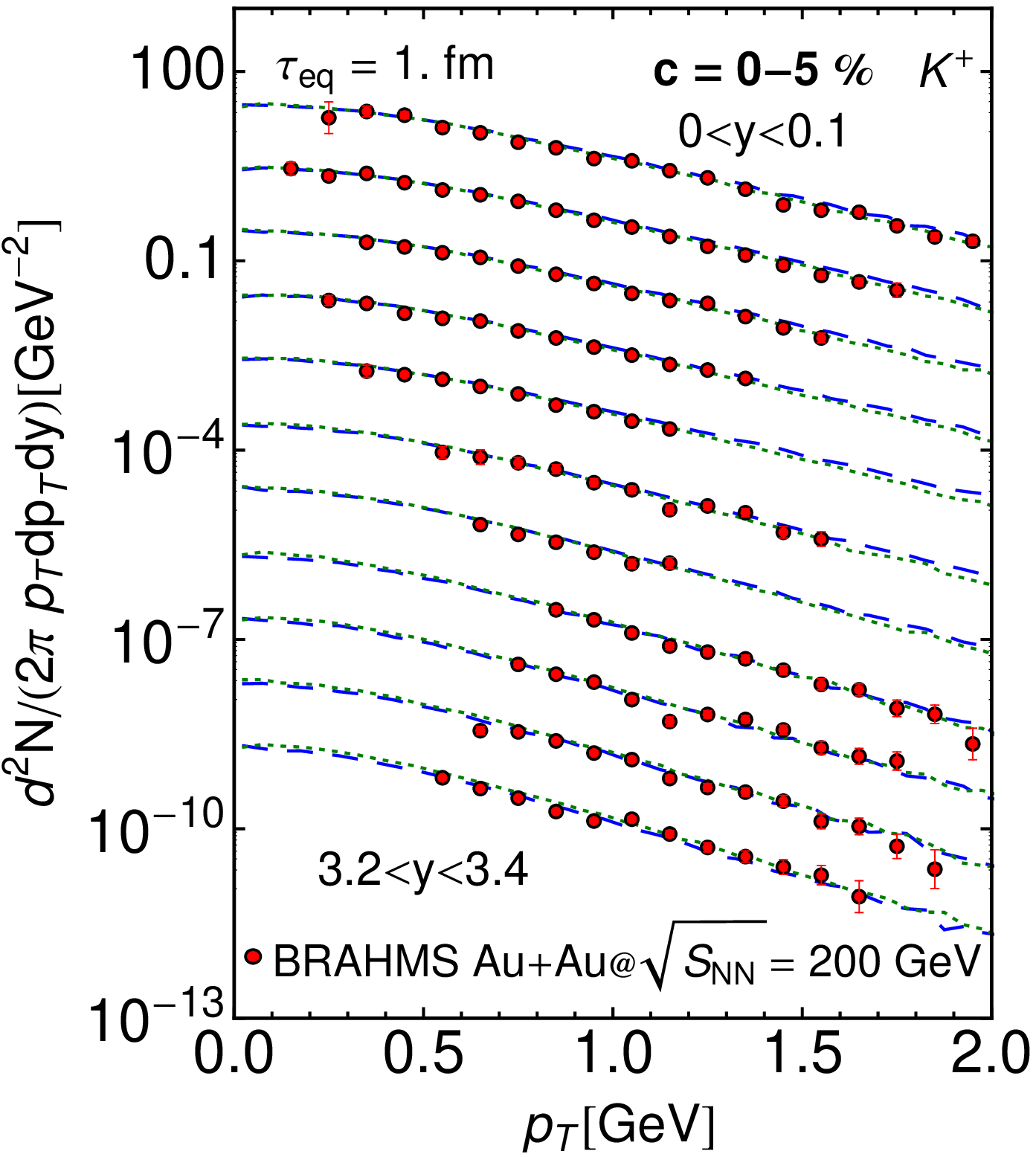}}
\end{center}
\caption{\small Transverse momentum spectra of $\pi^{+}$ (left panel) and $K^{+}$ (right panel) for the centrality class $c=0-5$\% ($b=2.23$ fm) and for different rapidity windows. The model results are obtained with the time-scale parameter $\tau_{\rm eq}= 1.0$ fm. The results are obtained for two values of the initial anisotropy parameter: $x_{\rm 0}=100$ (dashed blue lines) and $x_{\rm 0}=0.032$ (dotted green lines). The model results are compared to the experimental data from BRAHMS \cite{Bearden:2004yx}. The spectra for different rapidity windows are successively rescaled down by factor $0.1$. 
}
\label{fig:rapptdistr_RHIC3}
\end{figure}
%

\subsubsection{Transverse-momentum spectra}
\label{sect:transmomsp}

In Fig.~\ref{fig:rapptdistr_RHIC3} we present the transverse-momentum spectra of positive pions (left part) and kaons (right part) for the centrality class $c=0-5$\% ($b=2.23$ fm) and for different rapidity windows. The model calculations are done with $\tau_{\rm eq}= 1.0$ fm. The model results are compared to the BRAHMS data \cite{Bearden:2004yx}. We reproduce well the data in two different cases $x_{\rm 0}=100$ and $x_{\rm 0}=0.032$. The correct description of the  $p_{T}$ spectra in a broad range of rapidities is usually understood as an indication for a high level of thermalization of the entire fireball \cite{Bozek:2009ty}. After appropriate rescaling of the initial central energy density, the spectra are almost insensitive to initial values of anisotropy and to the length of the anisotropic stage (note a relatively large value of $\tau_{\rm eq}$ in Fig.~\ref{fig:rapptdistr_RHIC3}). The small discrepancies in normalizations between different rapidity windows are mainly due to discrepancies between theoretical and experimental $dN_{\rm ch}/d\eta_{PS}$ profiles (compare with the part \textbf{(c)} of Fig.~\ref{fig:etadistr_RHIC}).  
\begin{figure}[h]
\begin{center}
\subfigure{\includegraphics[angle=0,width=0.49\textwidth]{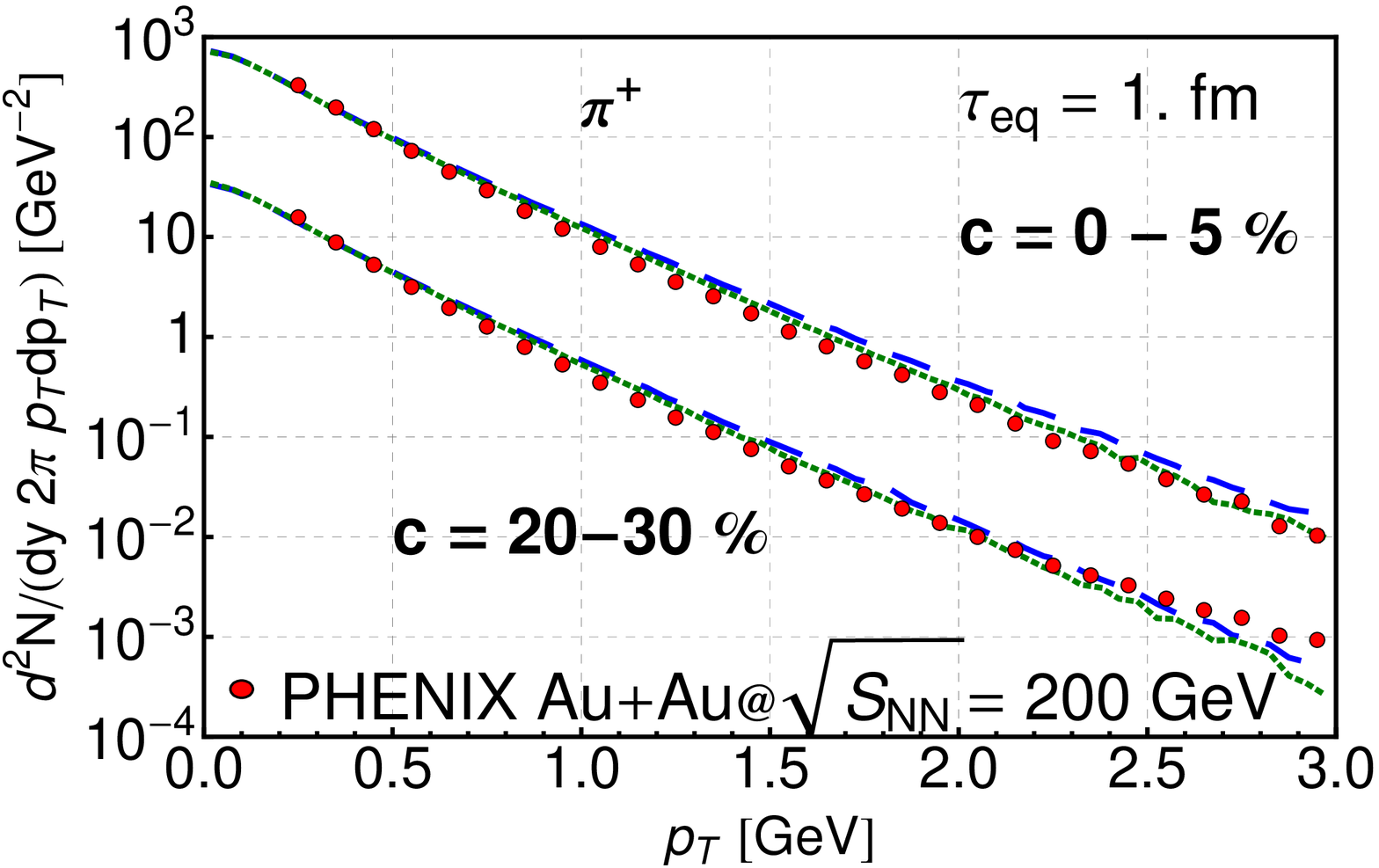}}
\subfigure{\includegraphics[angle=0,width=0.49\textwidth]{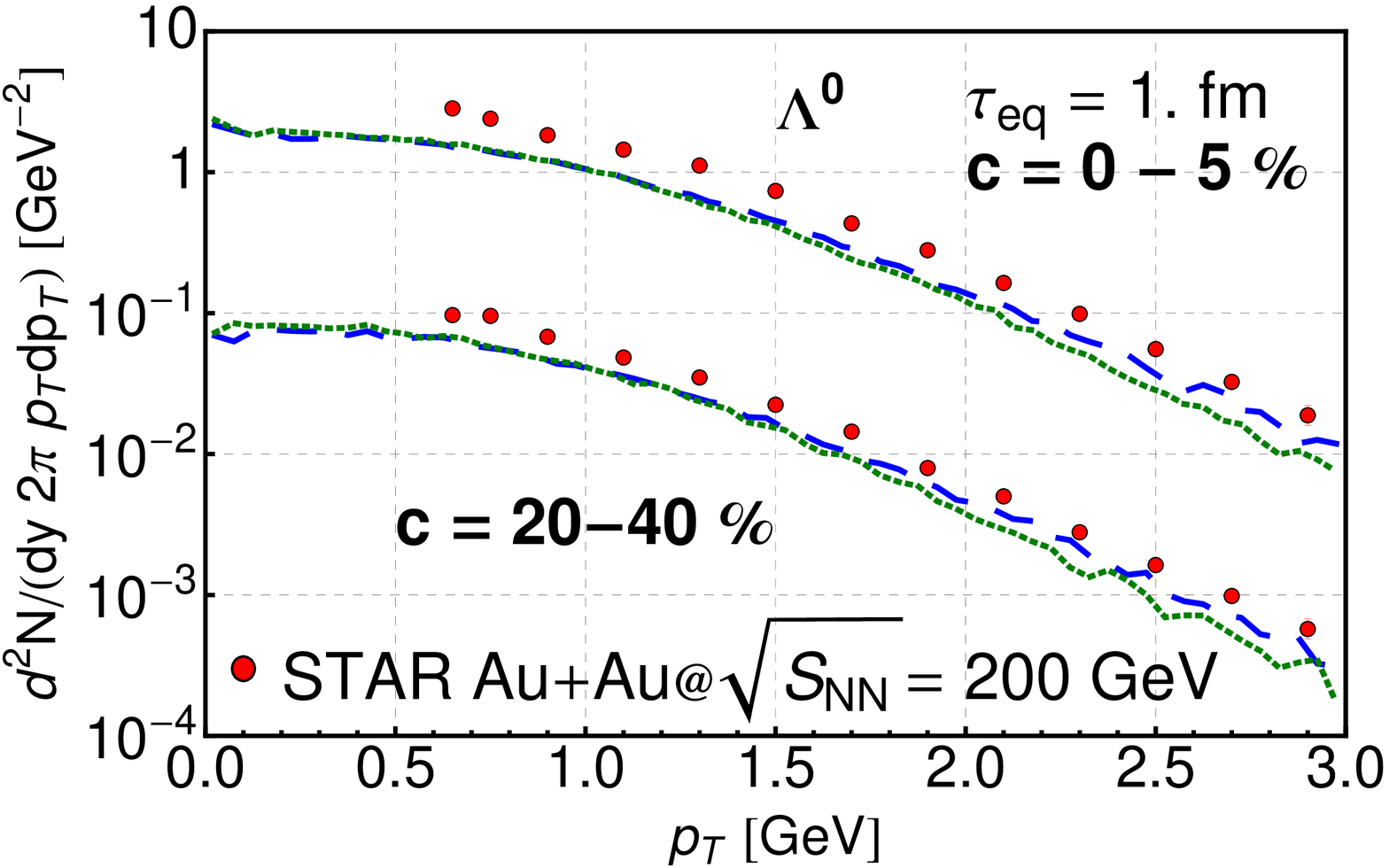}}
\subfigure{\includegraphics[angle=0,width=0.49\textwidth]{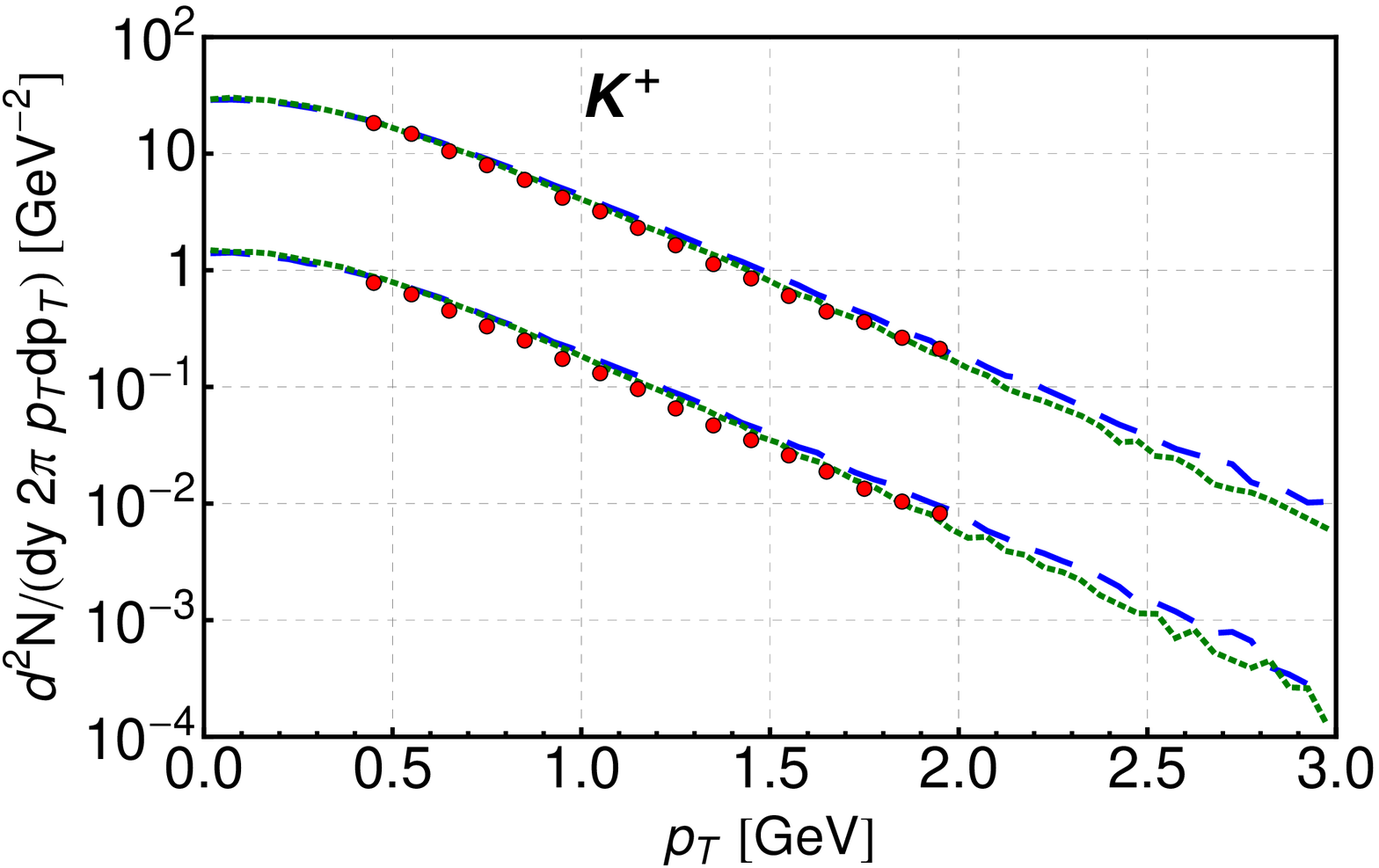}}
\subfigure{\includegraphics[angle=0,width=0.49\textwidth]{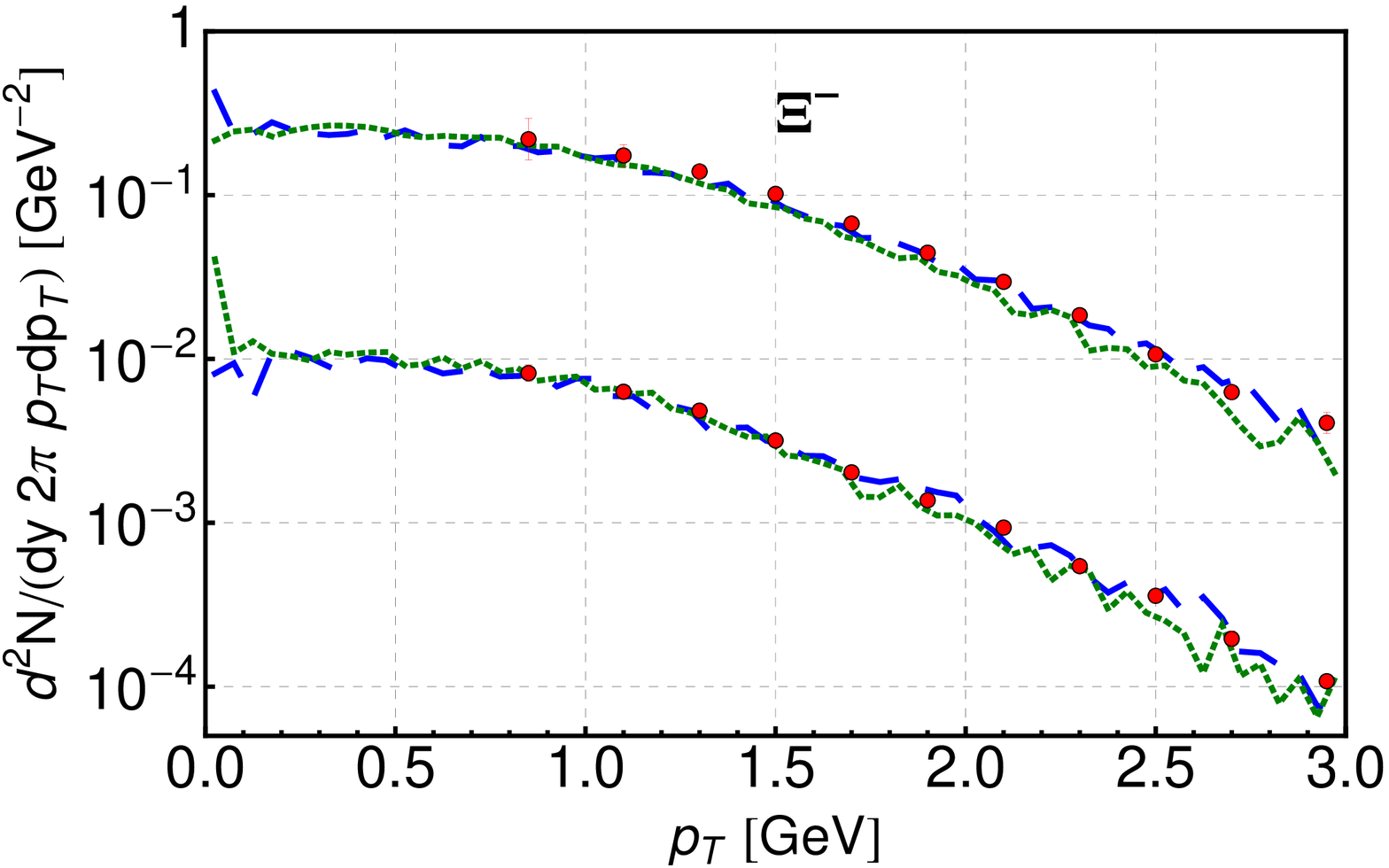}}
\subfigure{\includegraphics[angle=0,width=0.49\textwidth]{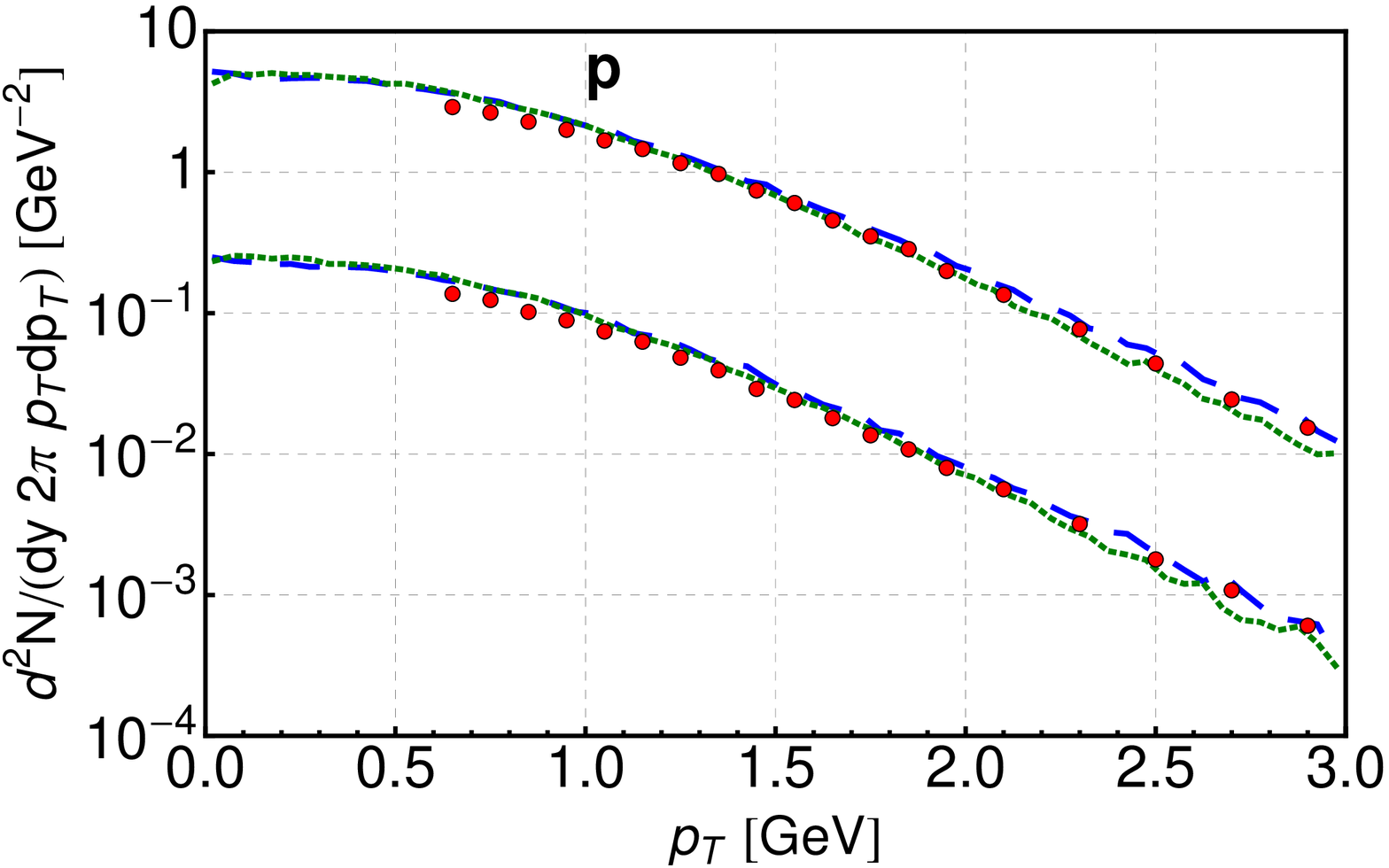}}
\subfigure{\includegraphics[angle=0,width=0.49\textwidth]{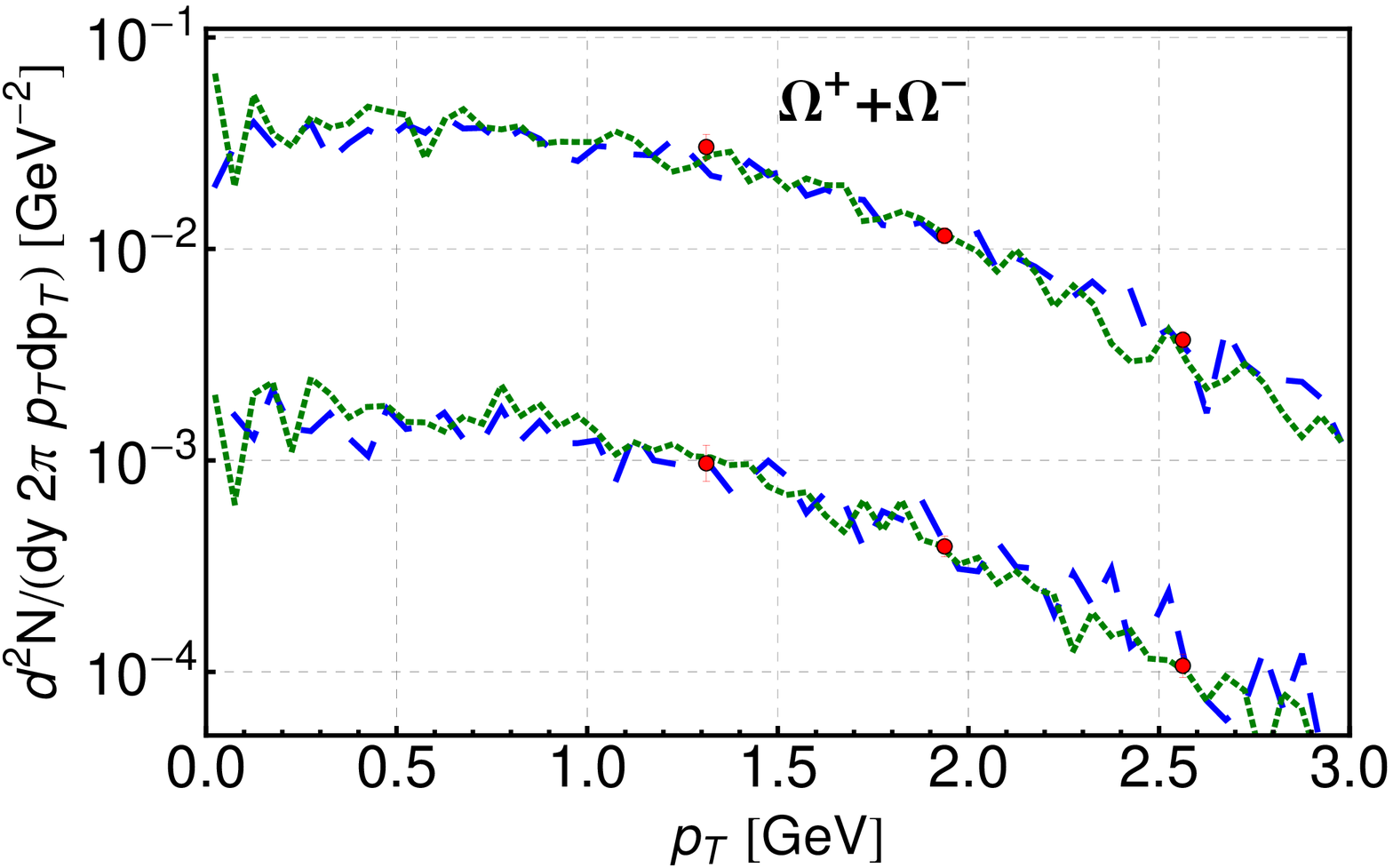}}
\end{center}
\caption{\small Left panels: Transverse momentum spectra of $\pi^{+}$ (top), $K^{+}$ (middle) and $p$ (bottom) for centralities $c=0-5$\% ($b=2.26$ fm) and $c=20-30$\% ($b=7.16$ fm) compared to the experimental data from PHENIX \cite{Adler:2003cb}. Right panels: Transverse-momentum spectra of $\Lambda^{0}$  (top), $\Xi^{-}$  (middle) and $\Omega^{-}+\Omega^{+}$ (bottom) for centralities \mbox{$c=0-5$\%} and $c=20-40$\% ($b=7.84$ fm) compared to the experimental data from STAR \cite{Adams:2006ke}. The spectra for $c=20-30$\% and $c=20-40$\% are scaled down by factor $0.1$. The results are obtained with time-scale parameter $\tau_{\rm eq}= 1.0$ fm. Errors are statistical. The $\Lambda$ feed-down corrections for proton spectra have be applied. The  $\Lambda$ spectra were feed-down corrected for $\Xi$ and $\Omega$.
}
\label{fig:centrptdistr_RHIC3}
\end{figure}

\FloatBarrier

\par In the left part of Fig.~\ref{fig:centrptdistr_RHIC3} we present the transverse-momentum spectra of positive pions (top), kaons (middle), and protons (bottom) at midrapidity ($|\mathrm{y}| < 1$) for the time-scale parameter $\tau_{\rm eq}= 1.0$ fm and for centralities $c=0-5$\% \mbox{($b=2.26$ fm)} and $c=20-30$\% ($b=7.16$ fm). The model results are compared to the data from PHENIX \cite{Adler:2003cb}. We can see that the model reproduces very well the slope of the spectra of $K^{+}$'s and protons up to $p_{T}=3$ GeV. The model pion spectra for \mbox{$c=20-30$\%} are underestimated for $p_{T}>2.5$ GeV. This is an expected feature, since hard particles require more rescatterings to thermalize and this region is out of applicability of hydrodynamics. 

The correct normalization of spectra for different hadron species is assured by introducing thermodynamic parameters at freeze-out, which were obtained in the framework of a thermal model using a complete treatment of resonances \cite{Baran:2003nm}. The reproduction of the convex curvature of the pion spectra at low $p_{T}$ is a consequence of feeding from resonance decays which are implemented in \texttt{THEMINATOR} \cite{Kisiel:2005hn,Chojnacki:2011hb}. 

On the right-hand-side of Fig.~\ref{fig:centrptdistr_RHIC3} we present the transverse-momentum spectra of hyperons for centralities $c=0-5$\% \mbox{($b=2.26$ fm)} and $c=20-40$\% \mbox{($b=7.84$ fm)} compared to the data from STAR \cite{Adams:2006ke}. Although our freeze-out temperature is a bit lower than that used in thermal models, the slopes of the hyperon spectra are reproduced very well. The normalization of $\Lambda$'s is too small, the effect that may indicate their higher freeze-out temperature. Fig.~\ref{fig:centrptdistr_RHIC3} shows again a small sensitivity of the model predictions to initial anisotropy, which is consistent with conclusions based on Fig.~\ref{fig:rapptdistr_RHIC3}. 

One would naively expect that the increase (decrease) of the pressure ratio $P_{\perp}/P_{\parallel}$ in the case $x_{\rm 0}=100$ ($x_{\rm 0}=0.032$),  compared  to the perfect-fluid case,  should lead to slower (faster) cooling of the system in the transverse direction. However the renormalization of the initial energy density (according to Table~\ref{table:multiplicity}) in order to obtain the same final multiplicity leads to a reduction (amplification) of the transverse flow produced during expansion in the anisotropic phase \cite{Bozek:2010aj}. This explicitly confirms the phenomenon of universality of the flow,  predicting that the overall growth of the flow is the same regardless of the pressure anisotropy of the energy-momentum tensor \cite{Vredevoogd:2008id}.

\subsection{Anisotropic flows}
\label{sect:anisoflow}
%

\begin{figure}[t]
\begin{center}
\subfigure{\includegraphics[angle=0,width=0.49\textwidth]{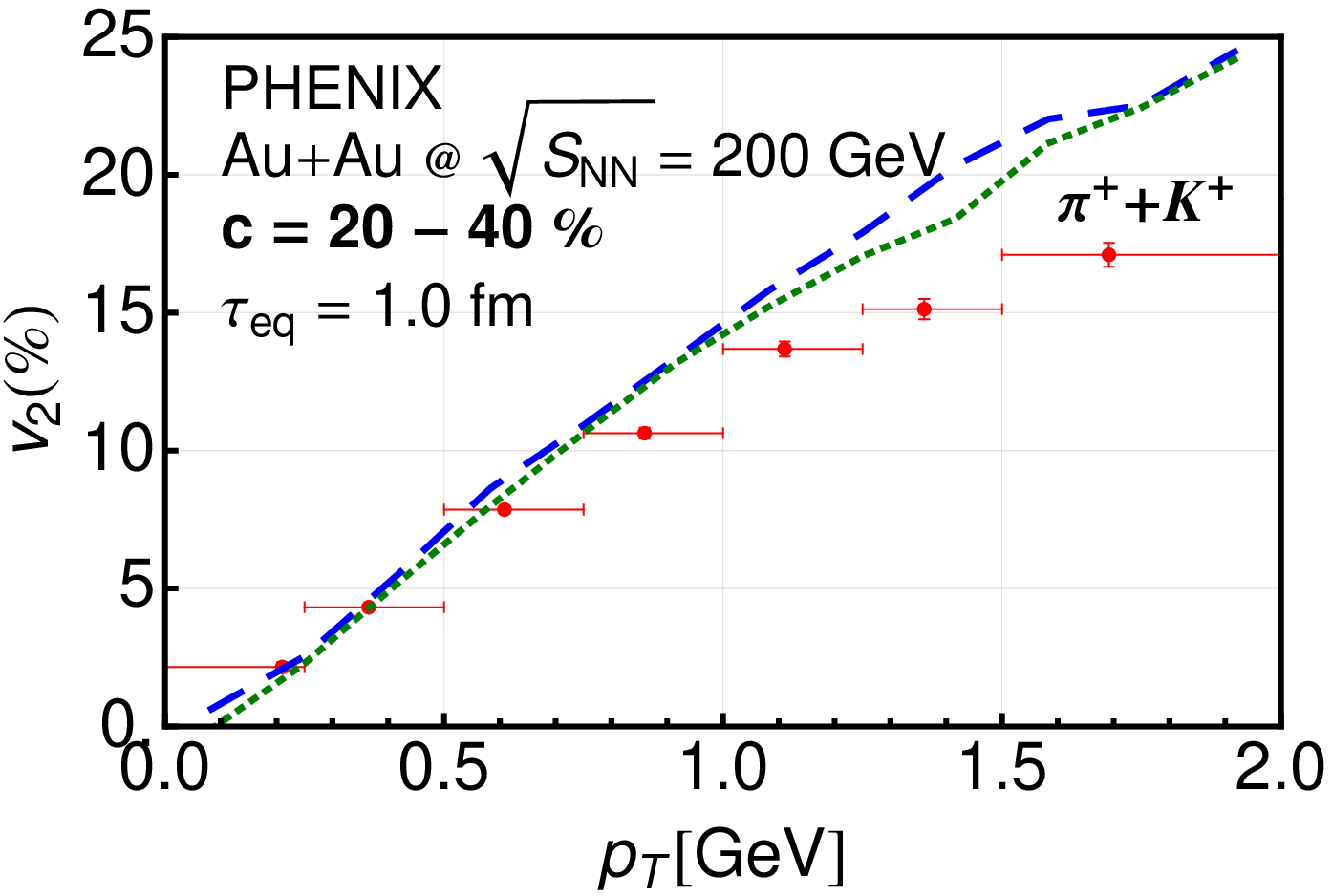}}
\subfigure{\includegraphics[angle=0,width=0.49\textwidth]{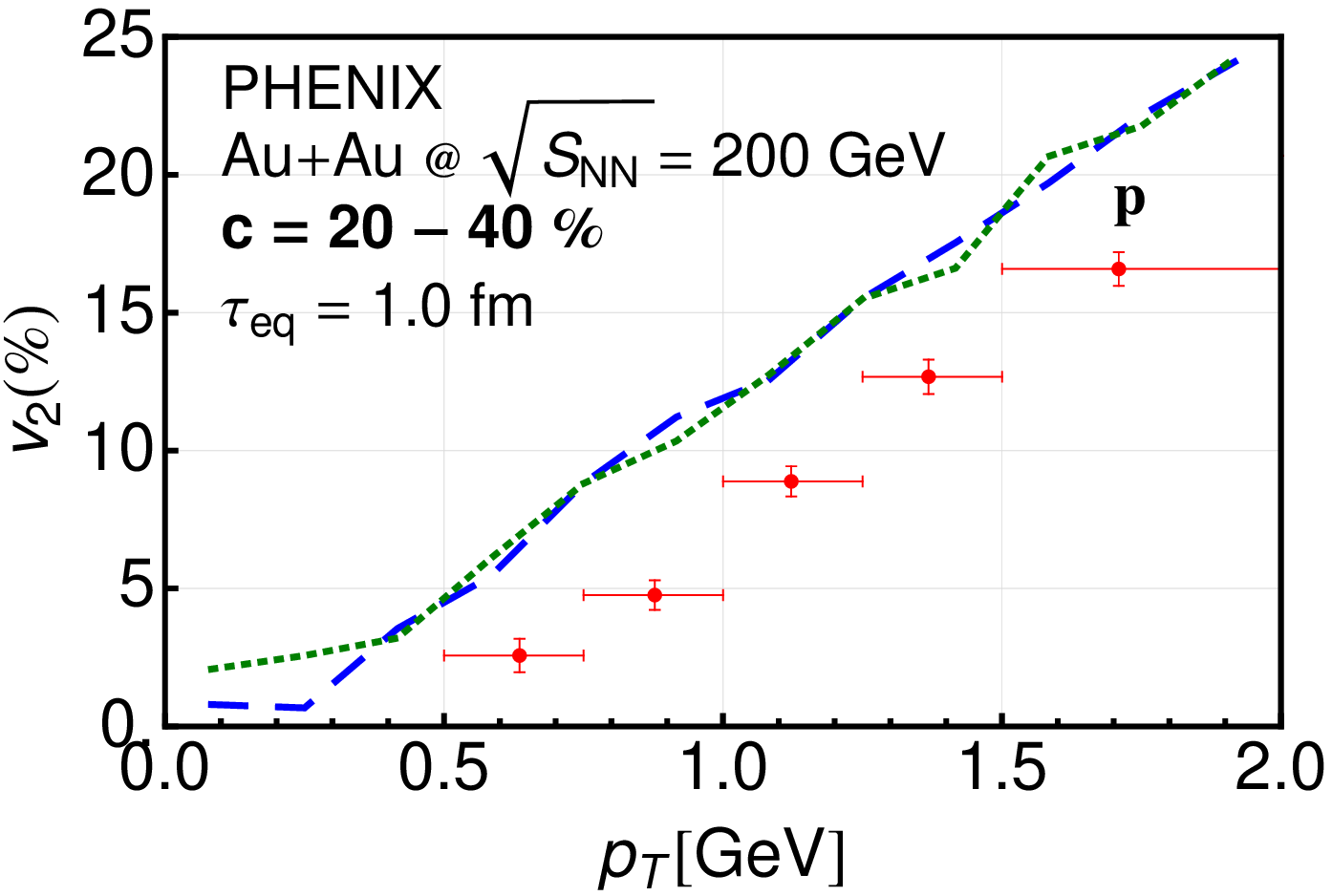}}
\end{center}
\caption{\small Transverse-momentum dependence of the elliptic flow coefficient $v_2$ of $\pi^{+}+K^{+}$ (left part) and protons (right part) calculated for the centrality $c=20-40$\% ($b=7.84$ fm) at midrapidity for the time-scale parameter $\tau_{\rm eq}= 1.0$ fm, and for two values of the initial anisotropy parameter. The results are compared to the PHENIX Collaboration data (red dots) \cite{Adler:2003kt}. The presented errors are statistical. Horizontal bars denote the $p_T$ bins.
}
\label{fig:ellflowpT}
\end{figure}
\begin{figure}[h]
\begin{center}
\subfigure{\includegraphics[angle=0,width=0.49\textwidth]{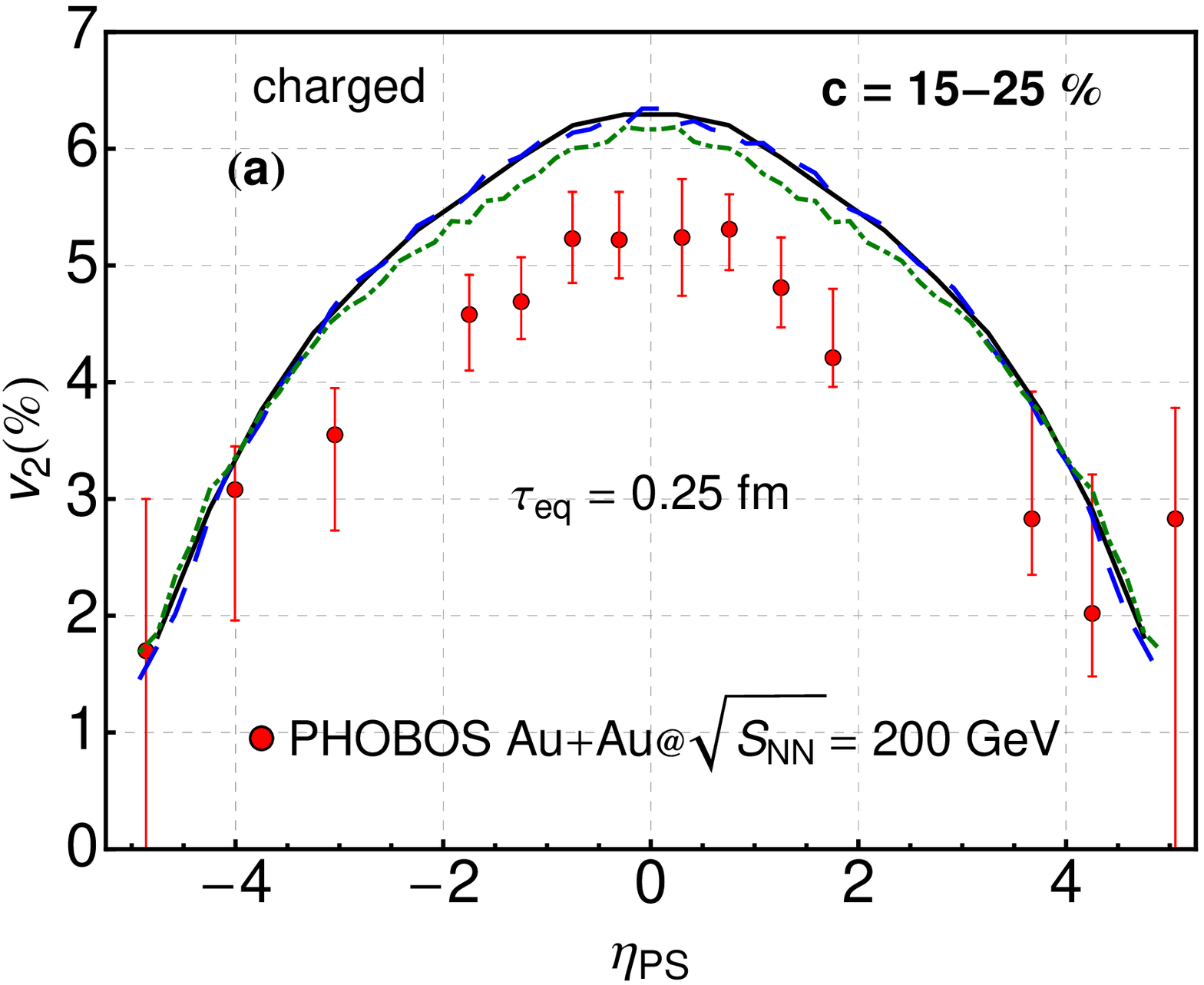}}
\subfigure{\includegraphics[angle=0,width=0.49\textwidth]{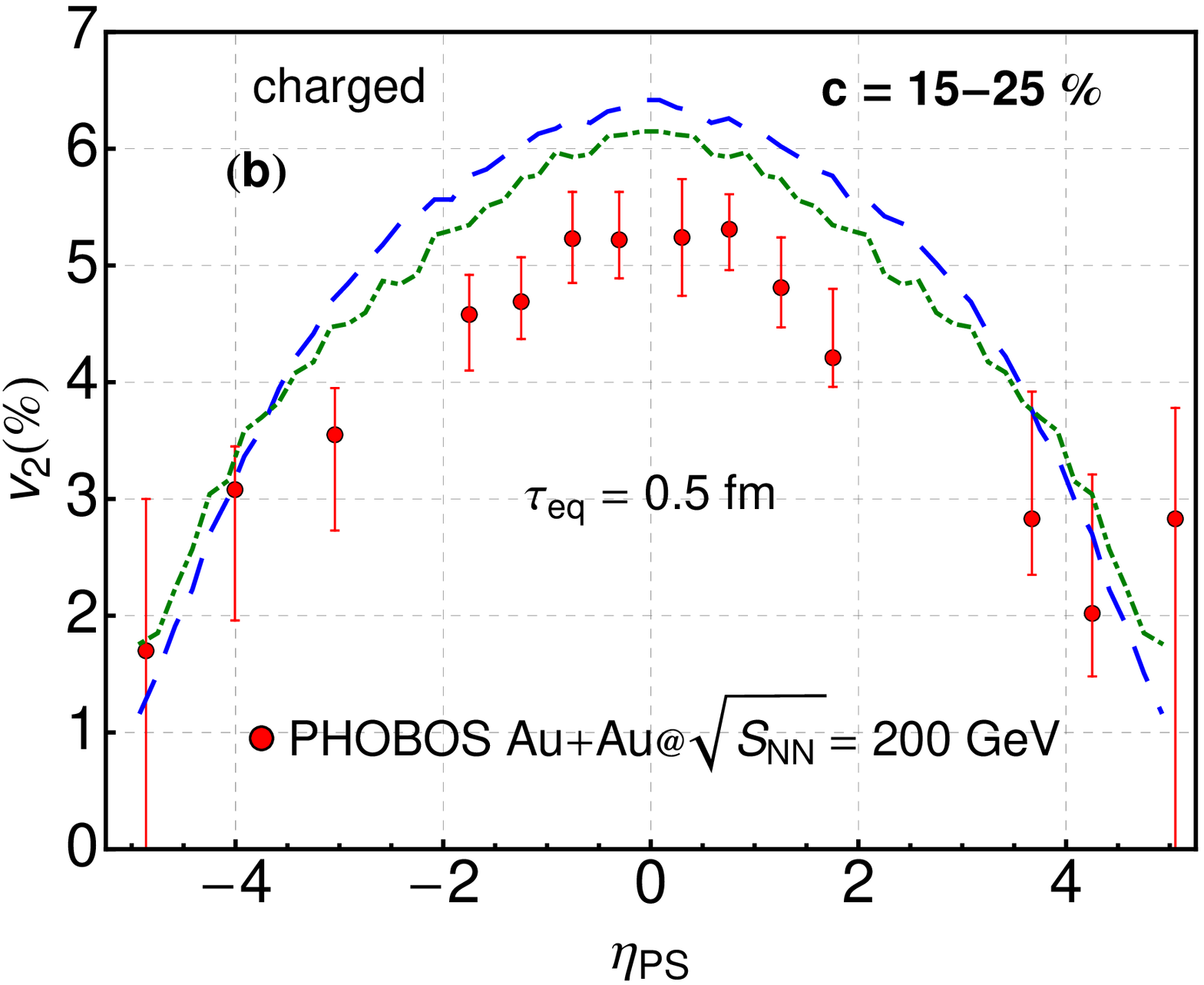}}
\subfigure{\includegraphics[angle=0,width=0.49\textwidth]{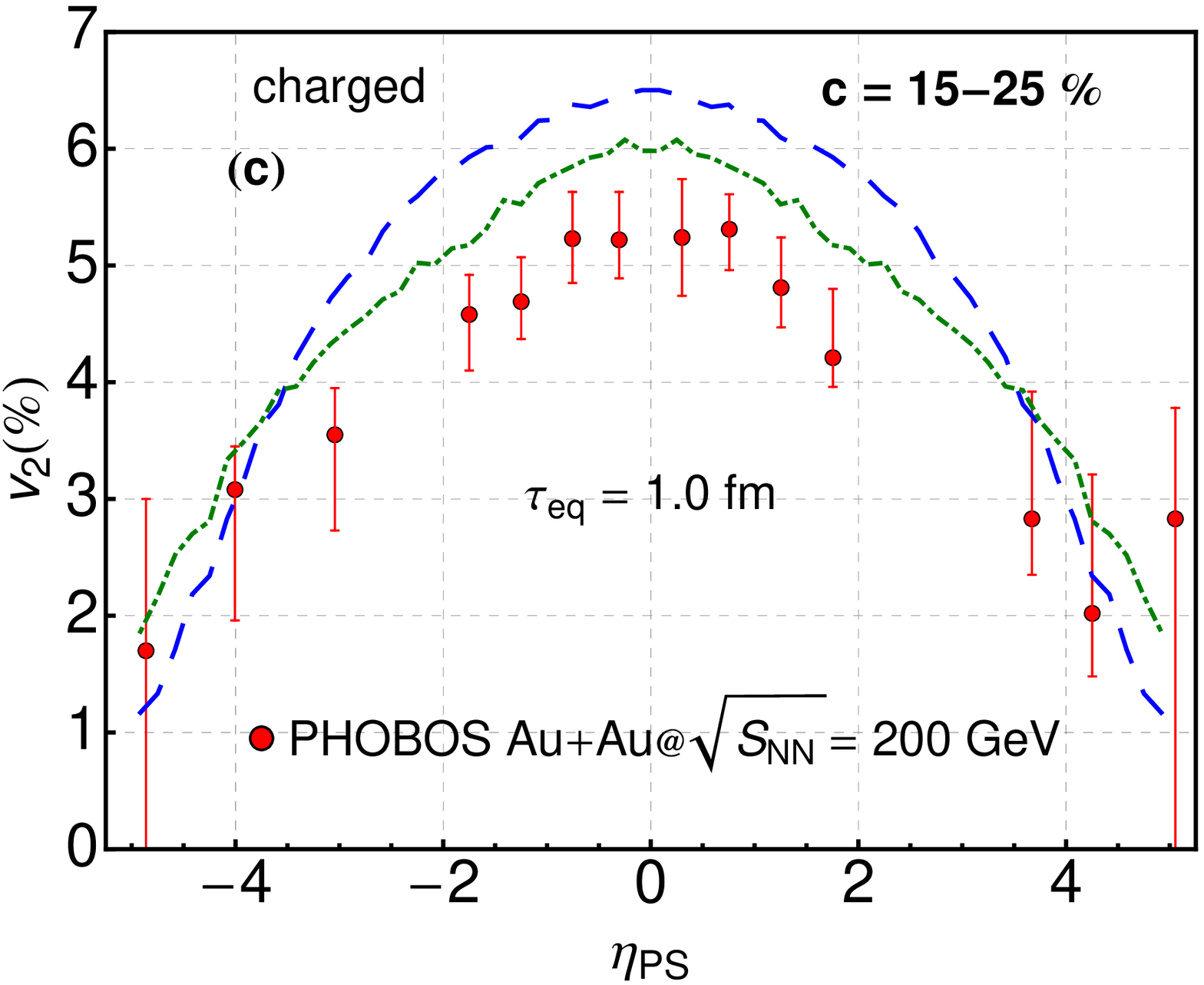}}
\end{center}
\caption{\small Pseudorapidity dependence of the $p_{T}$-integrated elliptic flow of charged particles calculated for centrality $c=15-25$\% for three time-scale parameters: $\tau_{\rm eq}=0.25$ fm \textbf{(a)}, $\tau_{\rm eq}=0.5$ fm \textbf{(b)}, $\tau_{\rm eq}=1.0$ fm \textbf{(c)} and three values of initial anisotropy parameter. The results are compared to the PHOBOS Collaboration data (red dots) \cite{Back:2004mh}. The error bars show statistical errors. }
\label{fig:ellfloweta}
\end{figure}

The significant amount of anisotropic flows is built at the very early stages of the evolution of matter, when the gradients of pressures are the largest due to the large {\it spacial} anisotropy  of the source. The inclusion of early non-equilibrium stages of the collisions may significantly change the momentum anisotropy observed in the final spectra, making $v_{\rm n}$ coefficients sensitive probes for the possible existence of non-equilibrium stages. 
\begin{figure}[h]
\begin{center}
\subfigure{\includegraphics[angle=0,width=0.49\textwidth]{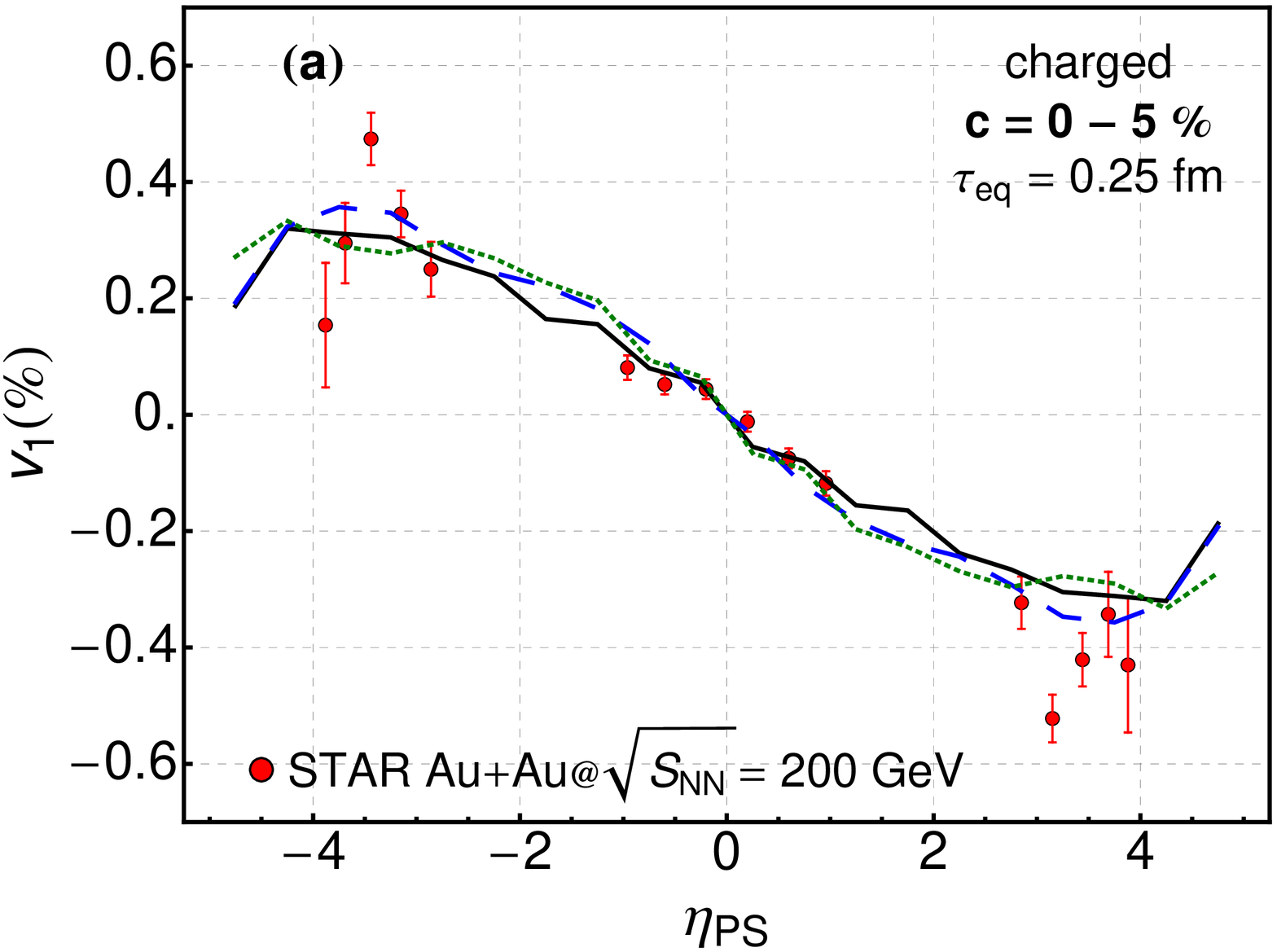}}
\subfigure{\includegraphics[angle=0,width=0.49\textwidth]{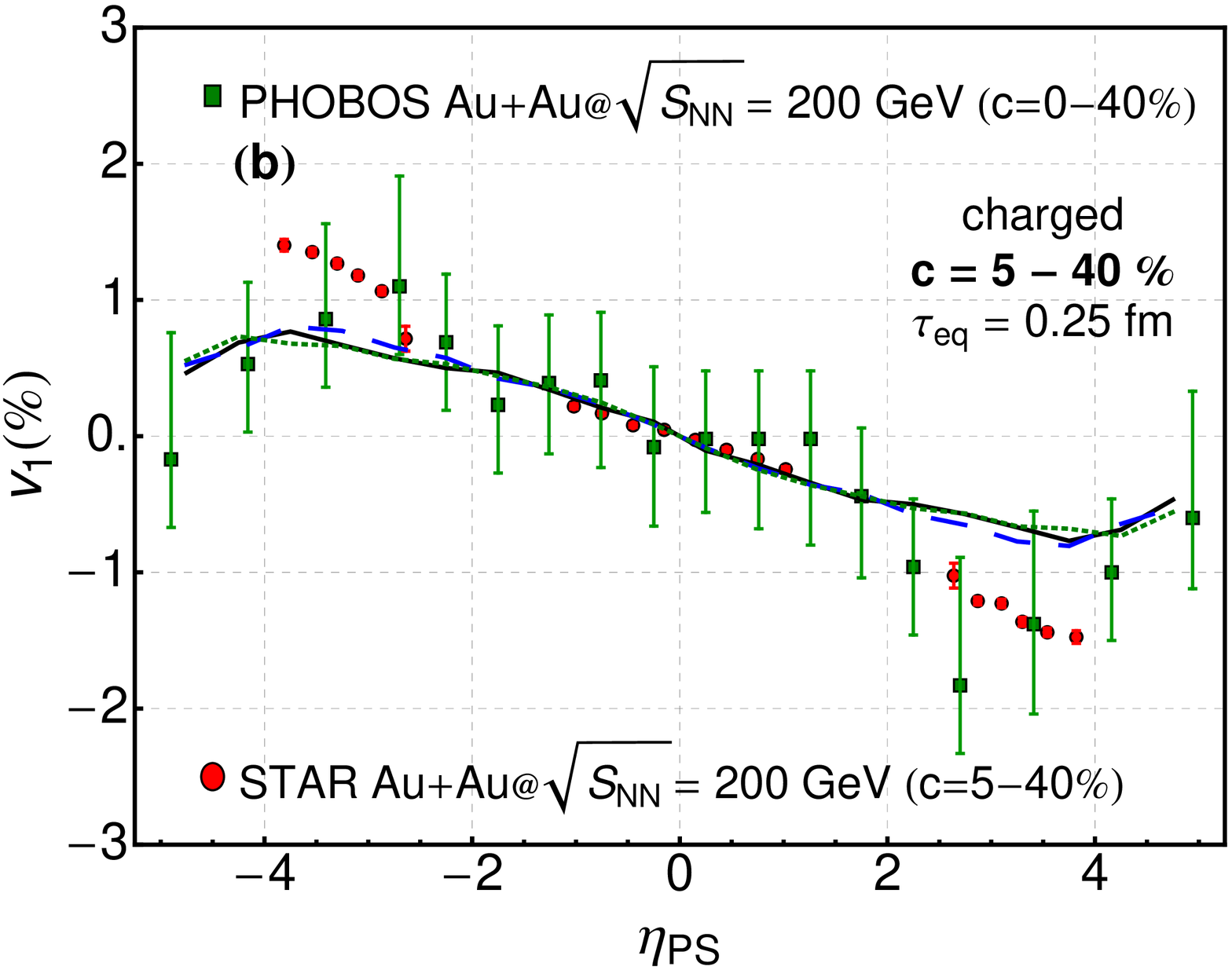}}
\subfigure{\includegraphics[angle=0,width=0.49\textwidth]{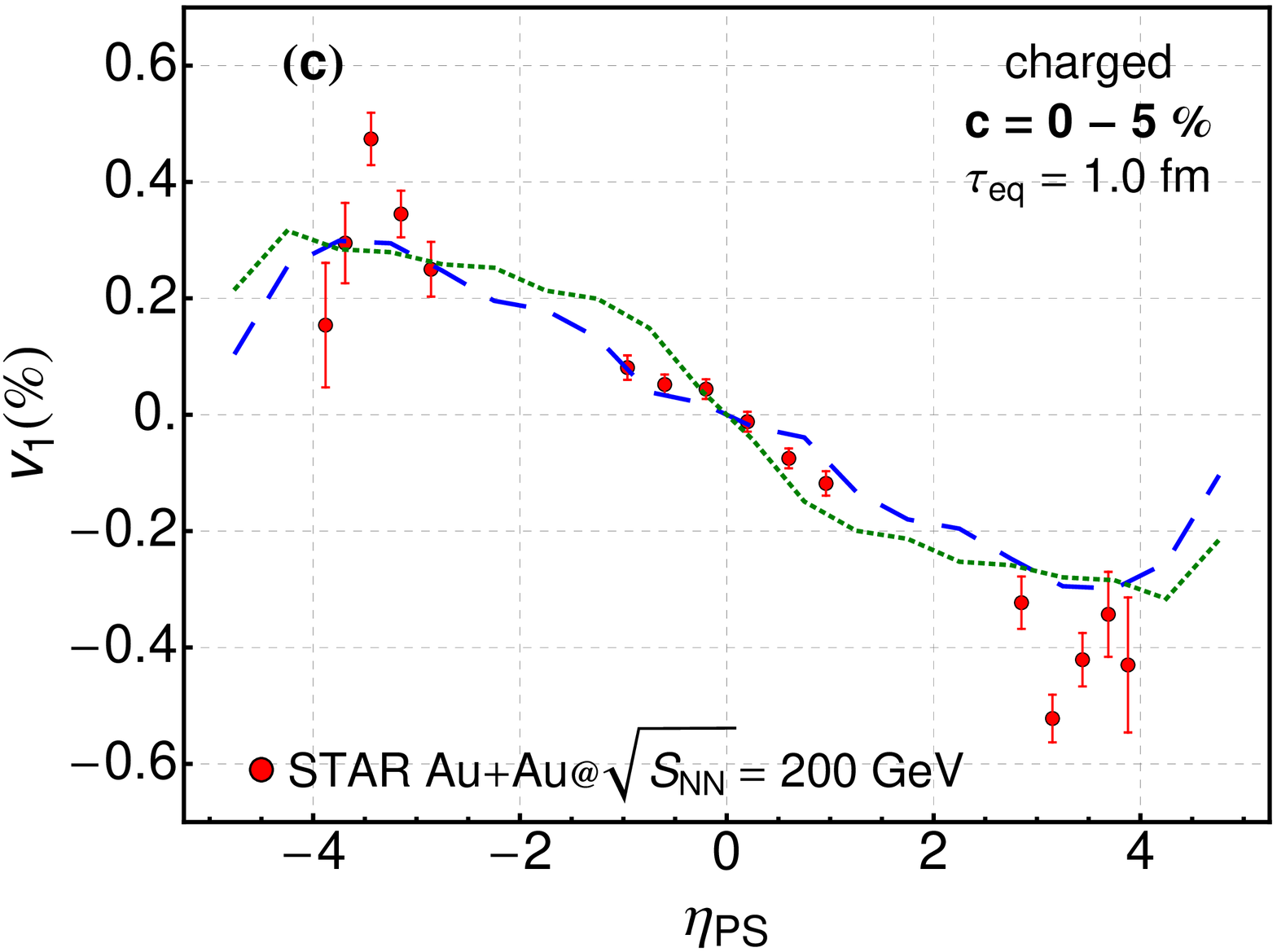}}
\subfigure{\includegraphics[angle=0,width=0.49\textwidth]{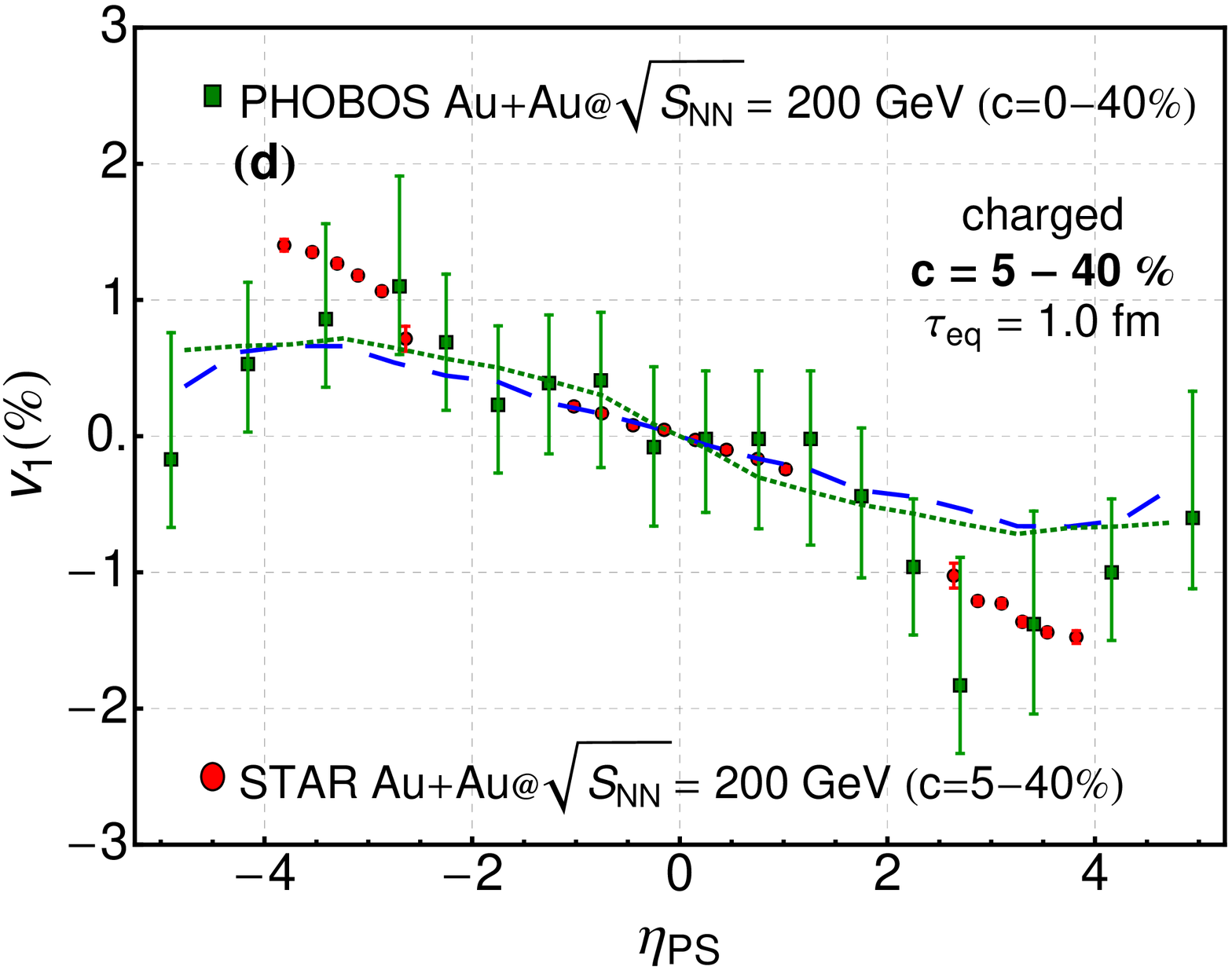}}
\end{center}
\caption{\small The pseudorapidity dependence of the directed flow of charged particles for two centrality bins  $c=0-5$\% (left part)  and $c=5-40$\% ($b=6.79$ fm)(right part) for time-scale parameters $\tau_{\rm eq}=0.25$ fm (\textbf{(a)} and \textbf{(b)}) and $\tau_{\rm eq}=1.0$ fm (\textbf{(c)} and \textbf{(d)}) and three values of initial anisotropy parameter. The results are compared to experimental data from STAR (red dots) \cite{Abelev:2008jga} and PHOBOS (green squares) \cite{Back:2005pc}. Errors are statistical only. 
}
\label{fig:dirfloweta1}
\end{figure}

\subsubsection{Elliptic flow}
\label{sect:3Dv2}

\par In the midrapidity region of non-central Au+Au collisions, the main information about the momentum anisotropy  of the produced particles is contained in the $v_2$ coefficient. In the left part of Fig.~\ref{fig:ellflowpT} we show the elliptic flow of $\pi^{+}+K^{+}$ as a function of the transverse momentum $p_T$. The model results obtained for the centrality $c=20-40$ \% ($b=7.84$ fm) and with the time-scale parameter $\tau_{\rm eq}= 1.0$ are compared to the data from PHENIX \cite{Adler:2003kt}. We observe that the model reproduces well the data up to $p_{T}=1$ GeV. The saturation for higher $p_{T}$ is not reproduced, since it requires inclusion of shear viscosity effects \cite{Bozek:2011ua}. Again, the model results weakly depend on the initial value of anisotropy, since the increase (decrease) of transverse pressure in the initial anisotropic stage is compensated by the decrease (increase) of the initial energy density. For completeness, in the right part of Fig.~\ref{fig:ellflowpT} we show the $p_{T}$-dependence of the elliptic flow of protons. The mass-splitting of $v_2$ is not reproduced. However, the elliptic flow of protons may be better described by introducing bulk viscosity \cite{Bozek:2009dw}. 

\par In Fig.~\ref{fig:ellfloweta} we present the $p_{T}$-integrated elliptic flow as a function of pseudorapidity. The model calculations are done for the centrality class $c=15-25$\%, three time-scale parameters, and three values of the initial anisotropy parameter. The model results are compared to the PHOBOS data \cite{Back:2004mh}. We observe that the model overshoots the data by about 20\% in the case of $x_{\rm 0}=1$ (solid black line). The peaked shape visible in the data is not well reproduced. 

We observe that the elliptic flow increases when the transverse pressure dominates. In this case, the elliptic flow is also steeper in pseudorapidity. It is easy to understand this behavior, since the reduction of longitudinal pressure leads to slower expansion of the fireball in the longitudinal direction. In the case of dominating longitudinal pressure the behavior is opposite. The strength of this effect depends on the length of the anisotropic stage. Thus, we conclude that the $p_{T}$-integrated $v_2$ is sensitive to the early anisotropic stages.

\subsubsection{Directed flow}
\label{sect:3Dv1}

\par When one considers non-central collisions of heavy-nuclei, one may observe non-zero directed flow in the region where $\eta_{PS} \neq 0$. This observable is sensitive to both transverse and longitudinal pressures, which makes it a perfect probe for measuring the early local anisotropies in momentum \cite{Bozek:2010aj}. In Fig.~\ref{fig:dirfloweta1} we show the directed flow of charged particles for the two centrality bins $c=0-5$\% (left part)  and $c=5-40$\% ($b=6.79$ fm, right part), and for the two time-scale parameters $\tau_{\rm eq}=0.25$ fm \textbf{(a)} and $\tau_{\rm eq}=1.0$ fm \textbf{(b)}. The results are plotted together with the experimental points from STAR \cite{Abelev:2008jga} and PHOBOS \cite{Back:2005pc}.

For different choices of our model parameters, the results describing $v_1$ are rather stable and consistent with the data. This leads to the conclusion that all the anisotropic stages described by \texttt{ADHYDRO} in this work are consistent with the $v_1$ data. Moreover, the agreement with the data suggests the validity of the idea of a tilted initial source.

\subsection{HBT correlations}
\label{sect:hbt1}
\par
\begin{figure}[t]
\begin{center}
\subfigure{\includegraphics[angle=0,width=0.49\textwidth]{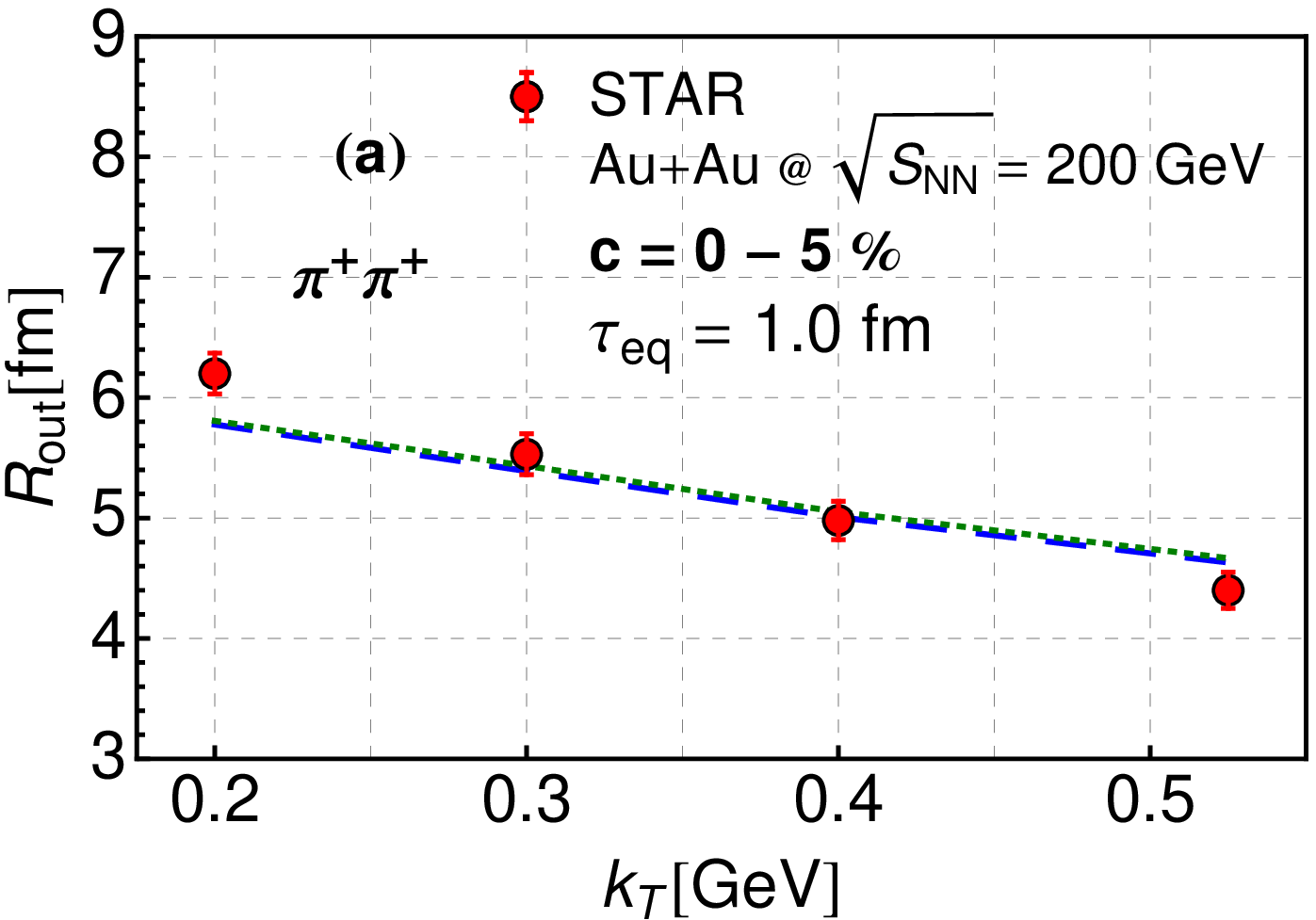}}
\subfigure{\includegraphics[angle=0,width=0.49\textwidth]{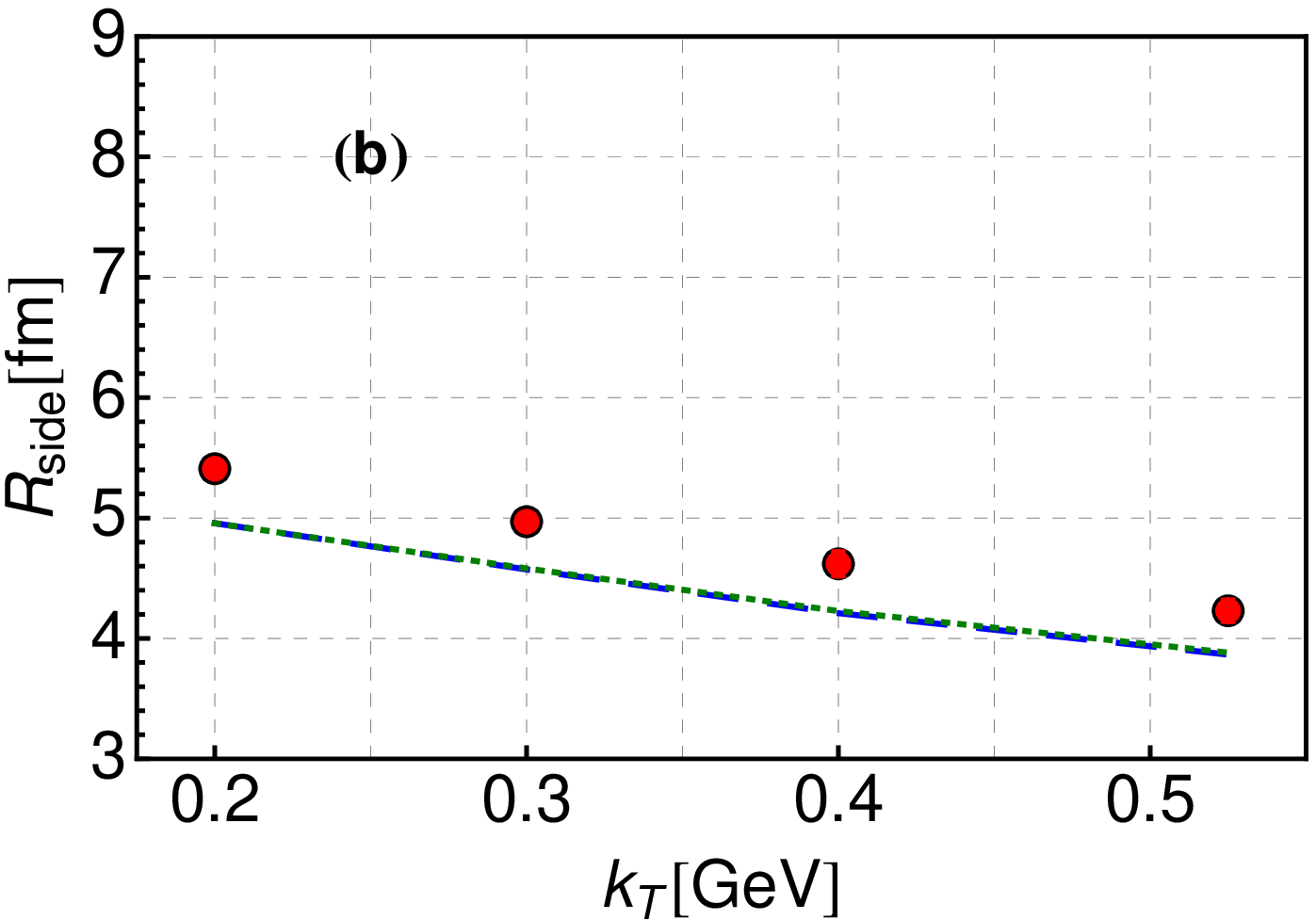}}
\subfigure{\includegraphics[angle=0,width=0.49\textwidth]{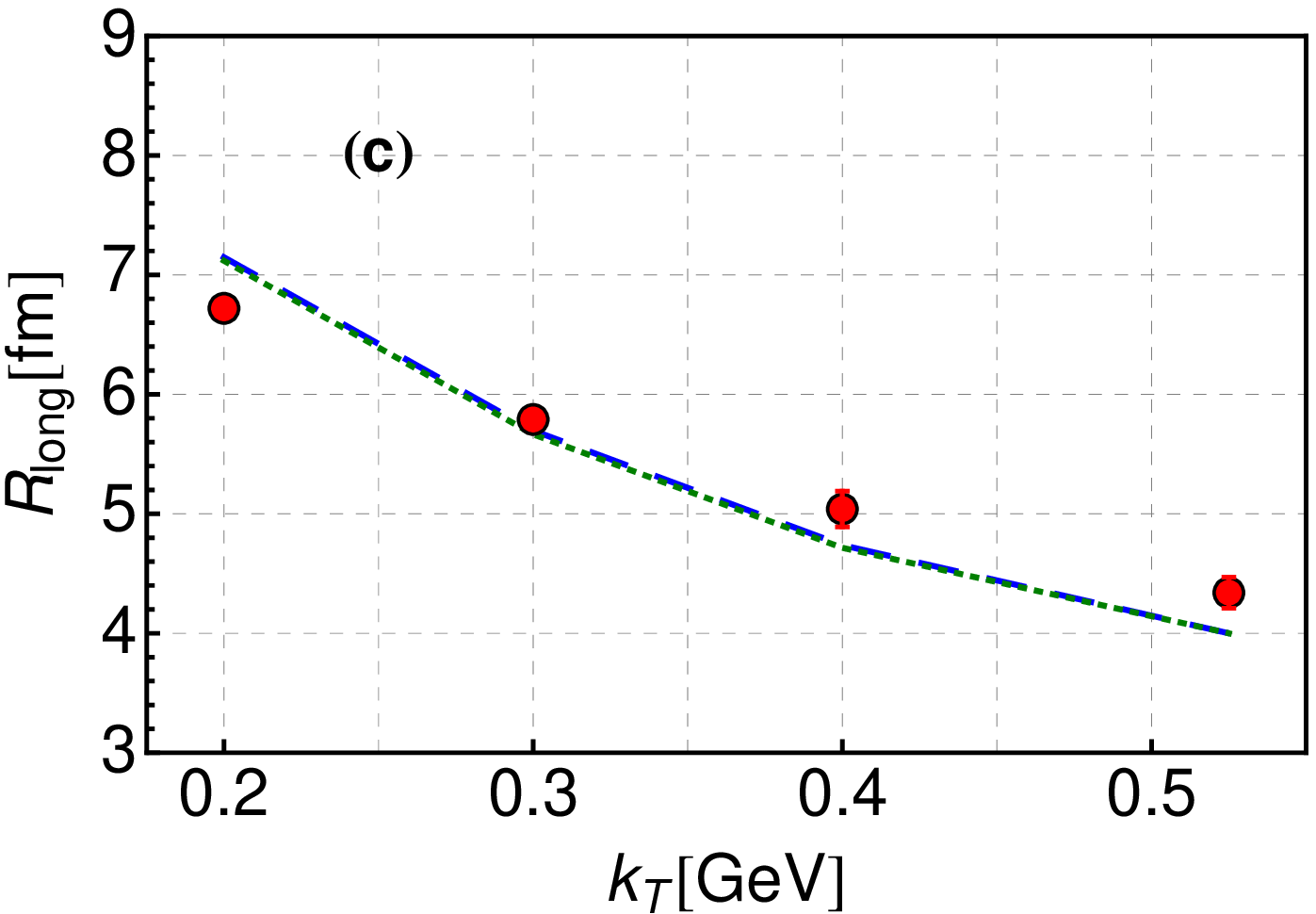}}
\subfigure{\includegraphics[angle=0,width=0.50\textwidth]{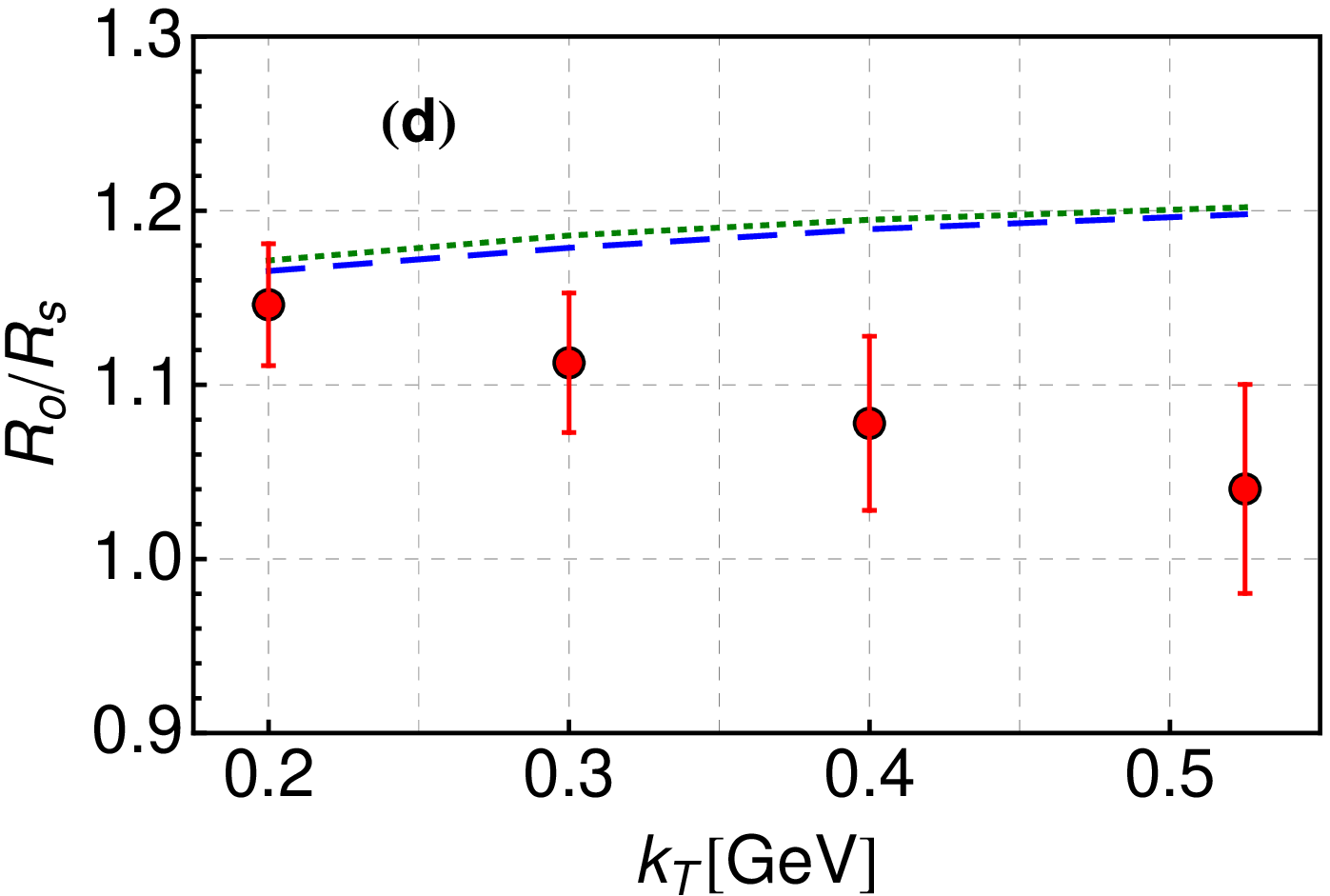}}
\end{center}
\caption{\small The HBT correlation radii: $R_{\rm out}$ \textbf{(a)}, $R_{\rm side}$ \textbf{(b)}, $R_{\rm long}$ \textbf{(c)} and ratio $R_{\rm out}/R_{\rm side}$ \textbf{(d)} of positive pions as a function of total transverse momentum of the pair for centrality  $c=0-5$\%  for time-scale parameter $\tau_{\rm eq}=1.0$ fm and three values of initial anisotropy parameter. The results are compared to experimental data from STAR Collaboration (red dots) \cite{Adams:2004yc}. Errors bars reflect statistical errors.
}
\label{fig:hbt1}
\end{figure}

 The identical particle interferometry is a useful tool used in the analysis of the system sizes at freeze-out. In Fig.~\ref{fig:hbt1} we present the model results together with the STAR data \cite{Adams:2004yc} for the HBT radii: $R_{\rm out}$ \textbf{(a)}, $R_{\rm side}$ \textbf{(b)}, and $R_{\rm long}$ \textbf{(c)}, all shown as functions of the pair total transverse momentum $k_T$. Differences between the data and the model results are smaller than 10\%. We observe that the slopes of the radii are quite well reproduced by our model. At large $k_T$, the side radius,  $R_{\rm side}$, is slightly smaller than that observed in the experiment, and the out radius, $R_{\rm out}$, is slightly larger. This is reflected in the overprediction of the $R_{\rm out}/R_{\rm side}$ ratio by about 20\%. The $k_T$ slope of $R_{\rm out}/R_{\rm side}$ is not reproduced.

For the all considered values of $x_{0}$, we obtain almost identical results. This is expected since the correct renormalization of the initial energy density results in the similar observed multiplicity and the flow on the freeze-out hypersurface (see our discussion in Sections \ref{sect:anisoflow} and \ref{sect:partspec}). The similarities in the description of the correlation radii for all $x_{0}$'s mean that the shapes of the freeze-out hypersurfaces are also very much similar. The last point was already explicitly highlighted in Fig.~\ref{fig:hiper0005} in Chapter \ref{chapter:boostinv}. 

\section{Space dependent initial anisotropy}
\label{sect:partspec}

\par In Ref.~\cite{Hirano:2001eu} it has been suggested that the system may be less equilibrated in forward and backward rapidities (i.e., in the fragmentation regions) than in the midrapidity region. On the other hand, in Ref.~\cite{Bozek:2005eu} it has been argued that the thermalization may depend on the density of participants, since the degree of  thermalization depends on the number of collisions between nucleons. Thus, one may expect that, in general, $x_{\rm 0}=x_{\rm 0}(\rho(\eta, \bf{x_{\perp}}))$, where $x_{\rm 0} \gg 1$ at the edges of the system. 

Here we skip the possible dependence on centrality and check two scenarios: \textbf{i)} a constant initial anisotropy profile $x_{0}=100$ with $\tau_{\rm eq}=1.0$ fm,  $\varepsilon_{\rm i} = 41.8$  $\mathrm{GeV/fm^3}$, $\Delta\eta = 1.5$, and $\sigma_\eta = 1.3$, see Fig.~\ref{fig:etadistr_RHIC_m} in Chapter \ref{sect:partspec}, and \textbf{ii)} a space dependent initial anisotropy profile $x_{\rm 0}=x_{\rm 0}(\eta, \bf{x_\perp})$ where
\begin{equation}
x_{\rm 0}(\eta, \bf{x_\perp})=\frac{1}{\tilde{\rho}(\eta, \bf{x_\perp})^2},
  \label{x0prof1}
\end{equation}
with $\tau_{\rm eq}=1.0$ fm,  $\varepsilon_{\rm i} = 73.8$  $\mathrm{GeV/fm^3}$, $\Delta\eta = 1$, and $\sigma_\eta = 1.3$.

\par
\begin{figure}[h]
\begin{center}
\subfigure{\includegraphics[angle=0,width=0.5\textwidth]{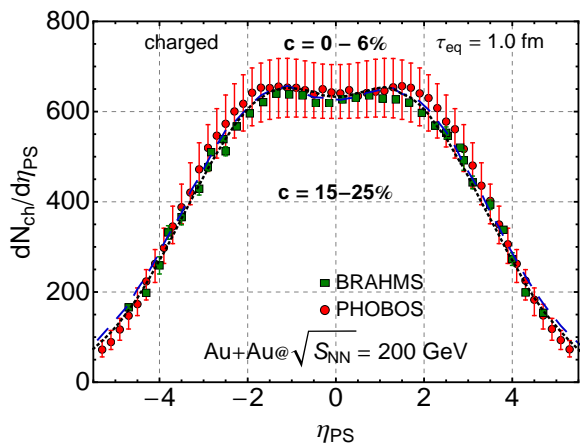}} 
\end{center}
\caption{\small The pseudorapidity distributions of charged particles for the case \textbf{i)} (blue dashed line)  and \textbf{ii)} (black dotted line). }
\label{fig:etadistr_RHIC_mp}
\end{figure}

 Fig.~\ref{fig:etadistr_RHIC_mp} shows the pseudorapidity dependence of charged particle multiplicity for the case \textbf{i)} (blue dashed line) and \textbf{ii)} (black dotted line). We observe that the distributions are similar despite different initial anisotropy profiles (note that in the case \textbf{ii)} $\Delta\eta$ is much smaller, since in the central rapidity region the matter evolves faster in the longitudinal direction than in the case \textbf{i)}). Moreover, moving to forward rapidities the anisotropy grows due to the ansatz (\ref{x0prof1}). Thus, the tails of the pseudorapidity distribution are described more precisely. The results concerning the transverse momentum spectra are very similar in both \textbf{i)}  and \textbf{ii)} cases. On the basis of conclusions from Section \ref{sect:transmomsp} this is an expected result.
\par
\begin{figure}[h]
\begin{center}
\subfigure{\includegraphics[angle=0,width=0.51\textwidth]{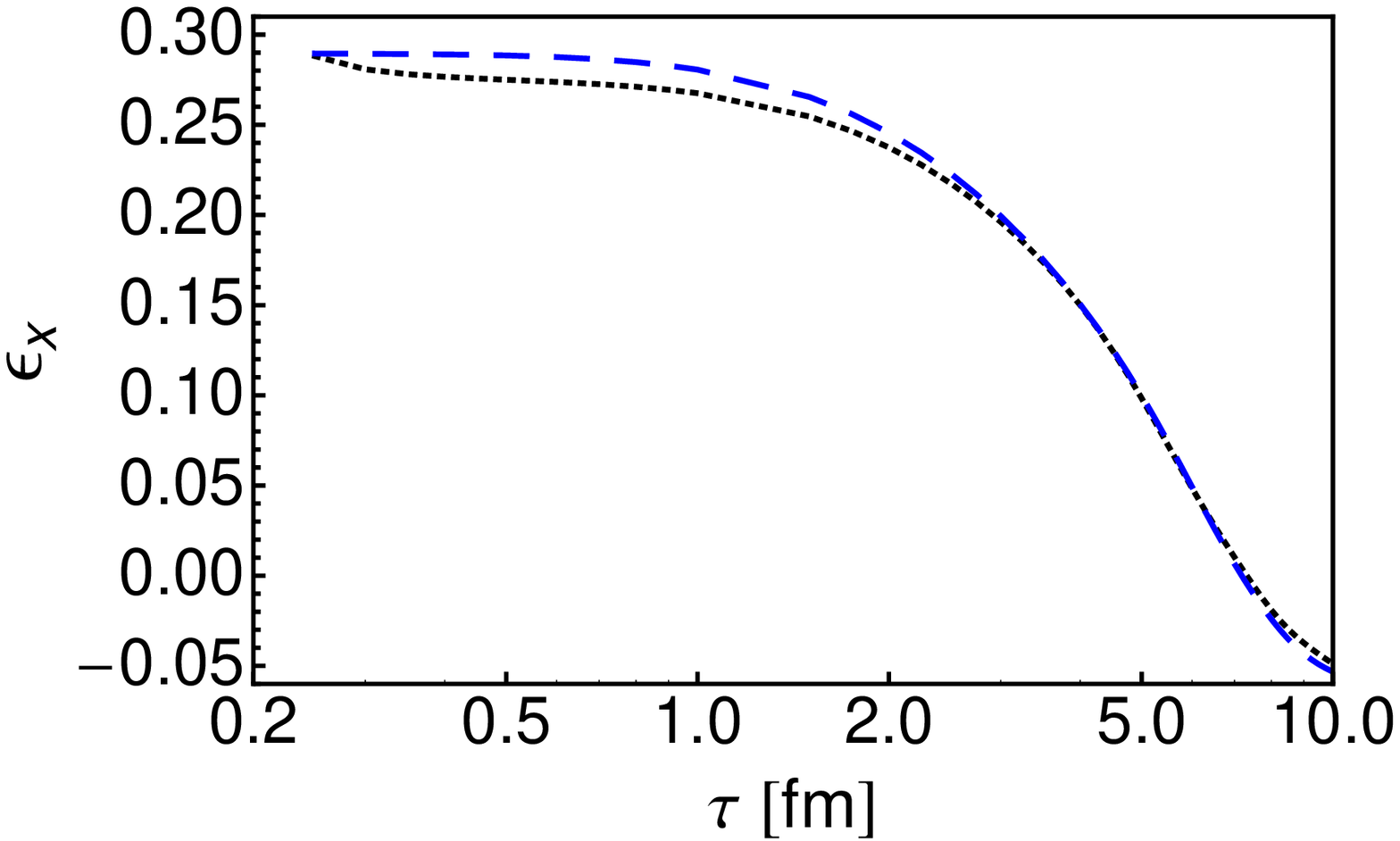}} 
\subfigure{\includegraphics[angle=0,width=0.46\textwidth]{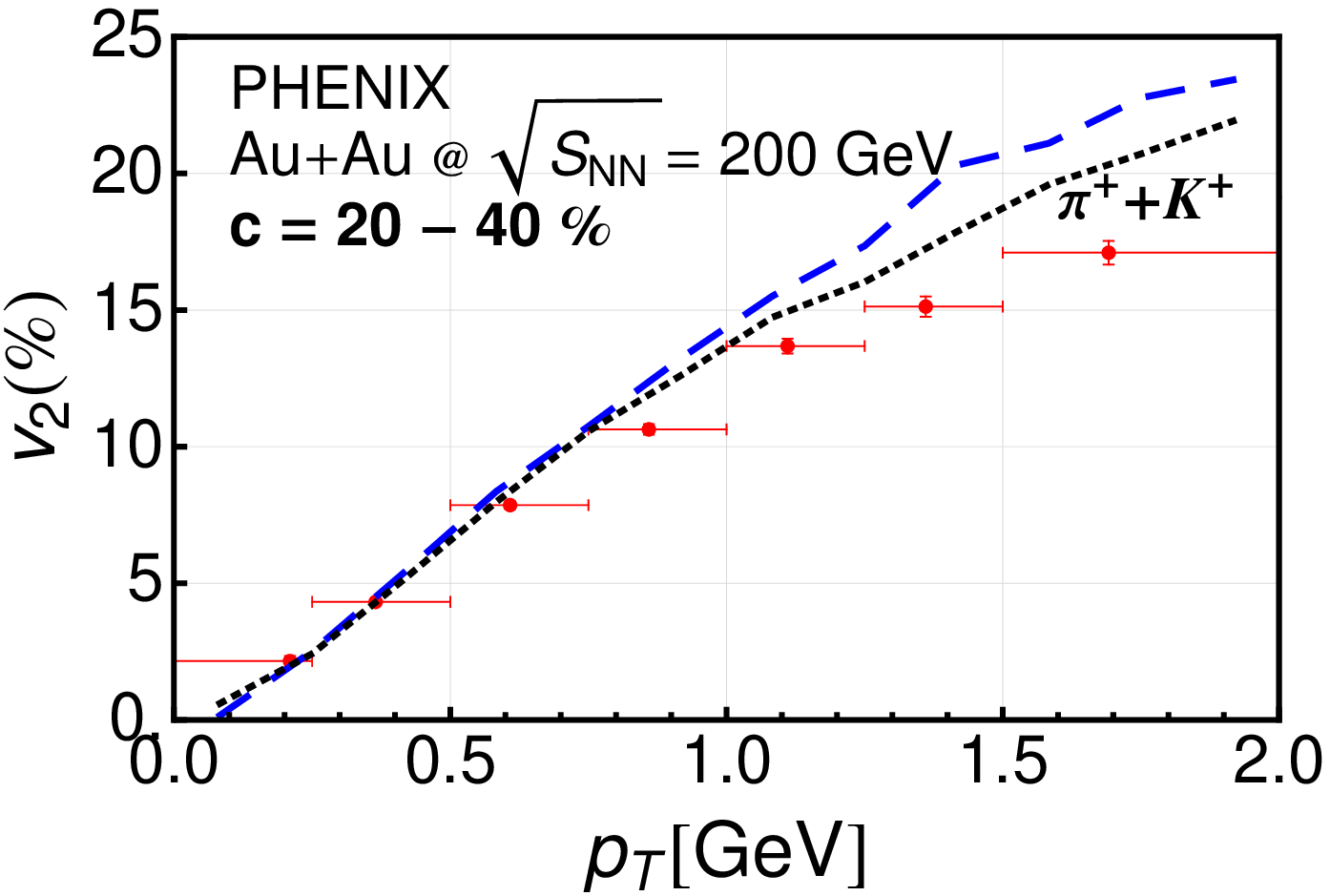}} 
\end{center}
\caption{\small Left panel: Time dependence of the spatial anisotropy of the fireball for $\eta=0$ for the case \textbf{i)} (blue dashed line)  and \textbf{ii)} (black dotted line) with the entropy density as a weight function. Right panel: Transverse-momentum dependence of the elliptic flow coefficient $v_2$ of pions and kaons for the case \textbf{i)} (blue dashed line)  and \textbf{ii)} (black dotted line). }
\label{fig:v2_mp}
\end{figure}

 Interesting observations may be done for the $v_2$ coefficient at midrapidity. In the case \textbf{i)} the imposed initial anisotropy is large and constant in entire fireball. The rescaling of the initial energy density cancels the effect of larger initial pressure in the transverse direction. Moreover the spatial anisotropy of the initially produced entropy in the case \textbf{i)}, $\Sigma(x_0,\sigma(\tau_0, \bf{x_\perp}))$, is the same as the spatial anisotropy of the initial entropy density profile. Hence the spatial anisotropy of the fireball transforms into the same momentum anisotropy as in the perfect-fluid case and we obtain similar results for $v_2$  (see also Section \ref{sect:3Dv2}). In the case \textbf{ii)} the initial anisotropy is space dependent. In the center of the system it is almost isotropic due to high density of wounded nucleons. The anisotropy is much larger in the regions of low density of wounded nucleons, thus the spatial anisotropy of the initially produced entropy in the case \textbf{ii)} is lower than that in the case \textbf{i)}. It results in a slightly lower spatial anisotropy of the whole fireball (see left part of Fig.~\ref{fig:v2_mp}) and in consequence it suppresses the generation of $v_2$ (see right part of Fig.~\ref{fig:v2_mp}).
\begin{figure}[h]
\begin{center}
\subfigure{\includegraphics[angle=0,width=0.5\textwidth]{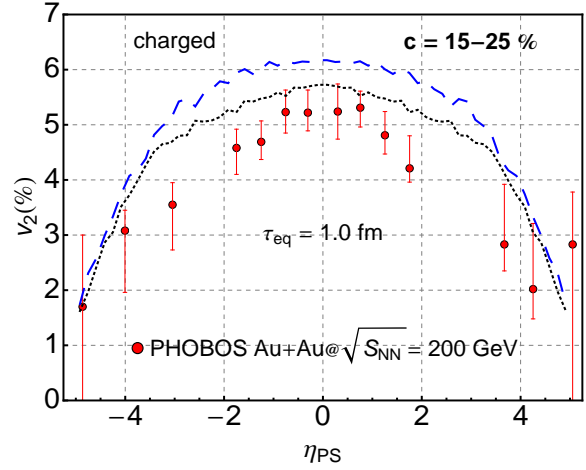}} 
\end{center}
\caption{\small Pseudorapidity dependence of the $p_{T}$-integrated elliptic flow of charged particles calculated for the case \textbf{i)} (blue dashed line)  and \textbf{ii)} (black dotted line). }
\label{fig:v2eta_mp}
\end{figure}
The effect of lowering of the $v_2$ is clearly seen in the Fig.~\ref{fig:v2eta_mp} where we present the pseudorapidity dependence of the $p_{T}$-integrated $v_2$. We observe better agreement of the calculations with the experimental data in the central rapidities in the case \textbf{ii)}. However the model still does not reproduce the peaked shape seen in the data.
\begin{figure}[t]
\begin{center}
\subfigure{\includegraphics[angle=0,width=0.49\textwidth]{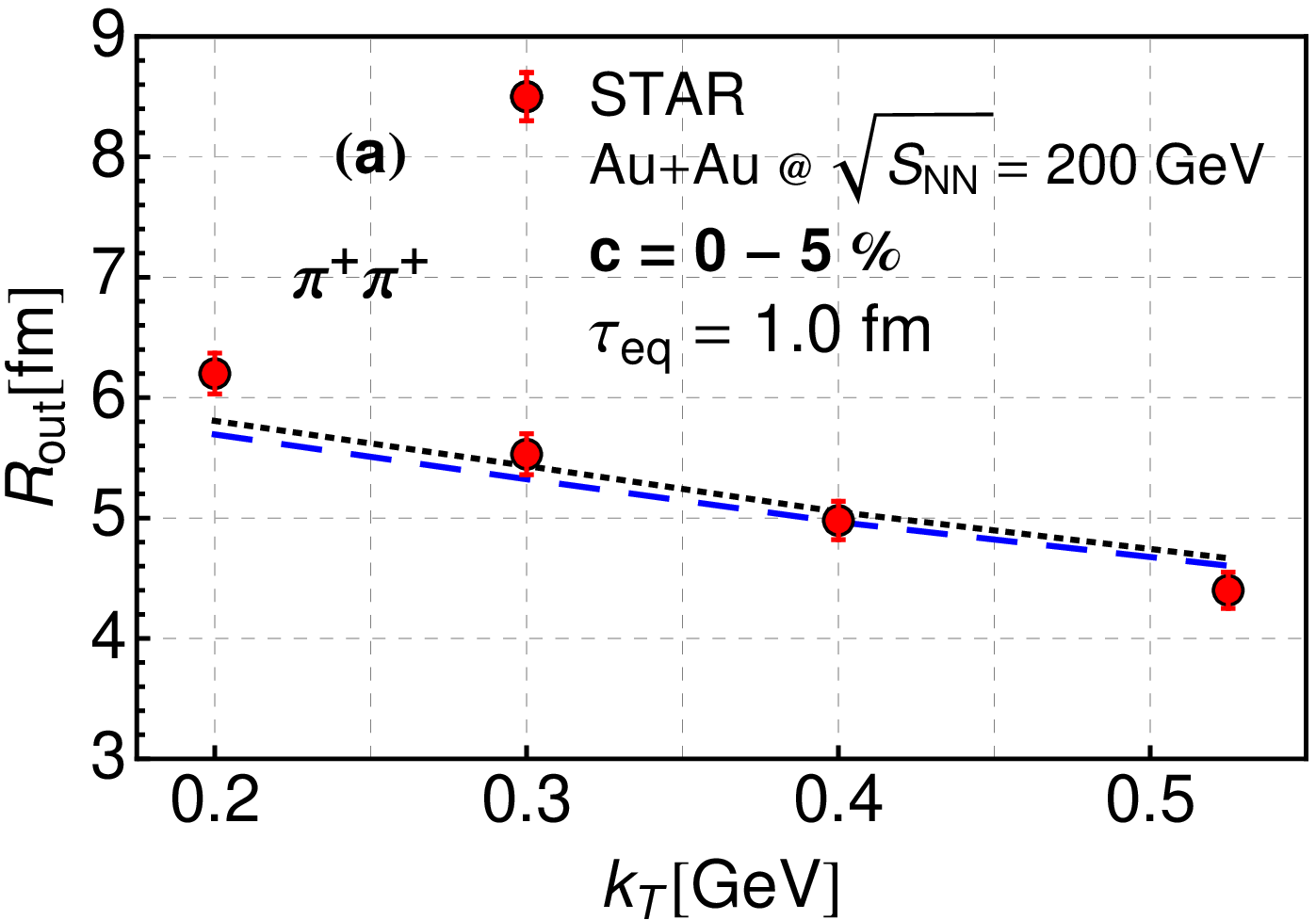}}
\subfigure{\includegraphics[angle=0,width=0.49\textwidth]{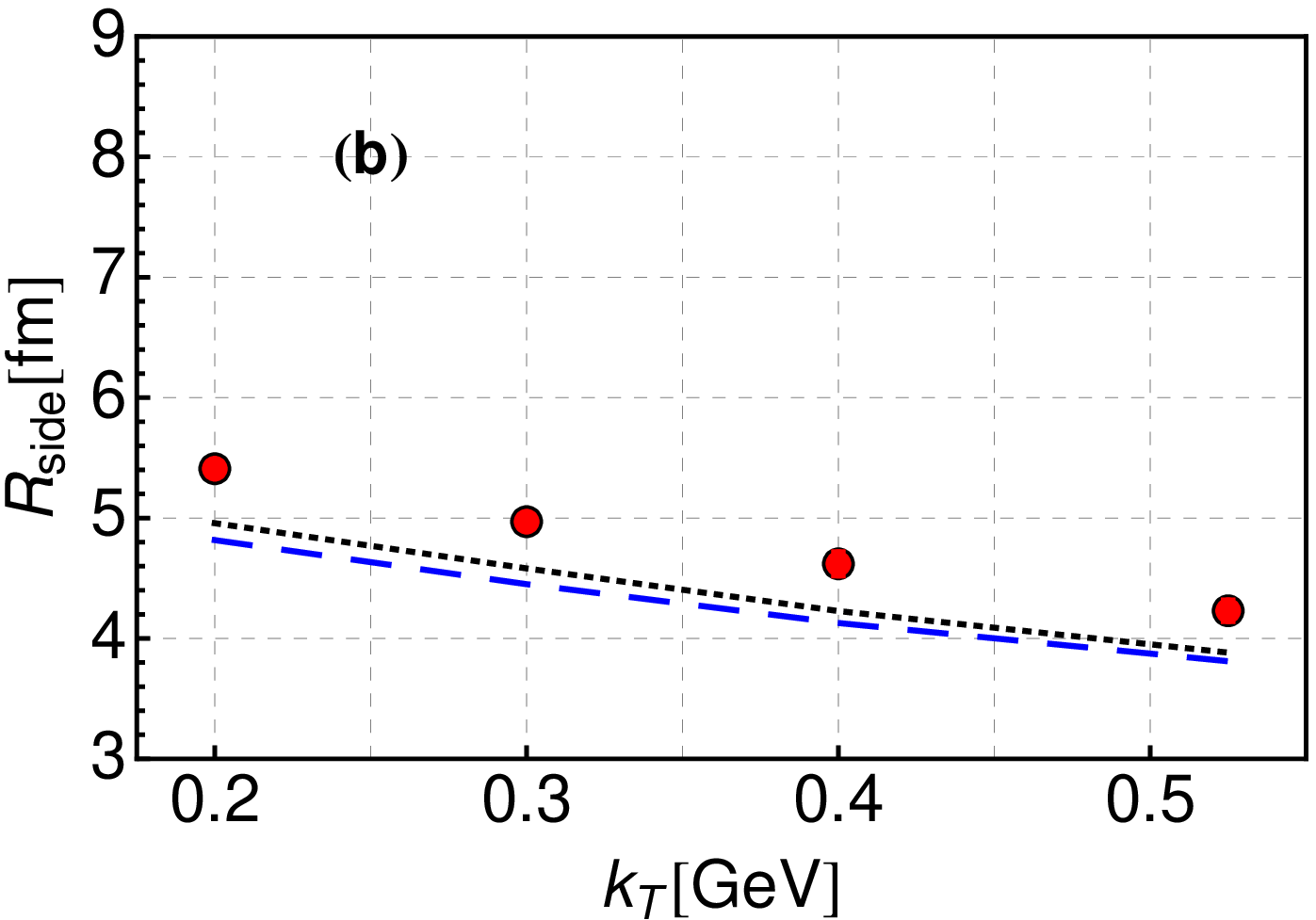}}
\subfigure{\includegraphics[angle=0,width=0.49\textwidth]{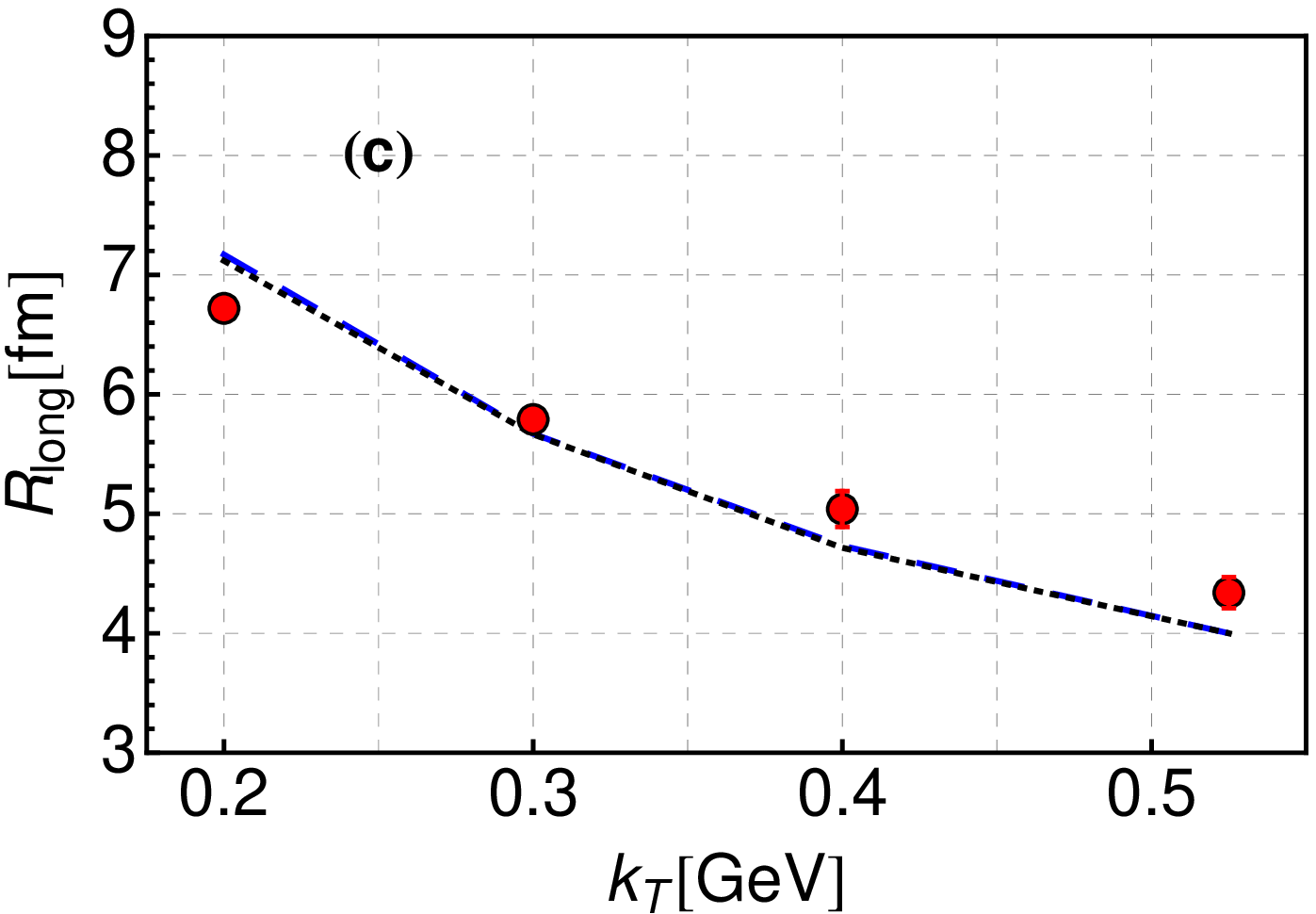}}
\subfigure{\includegraphics[angle=0,width=0.50\textwidth]{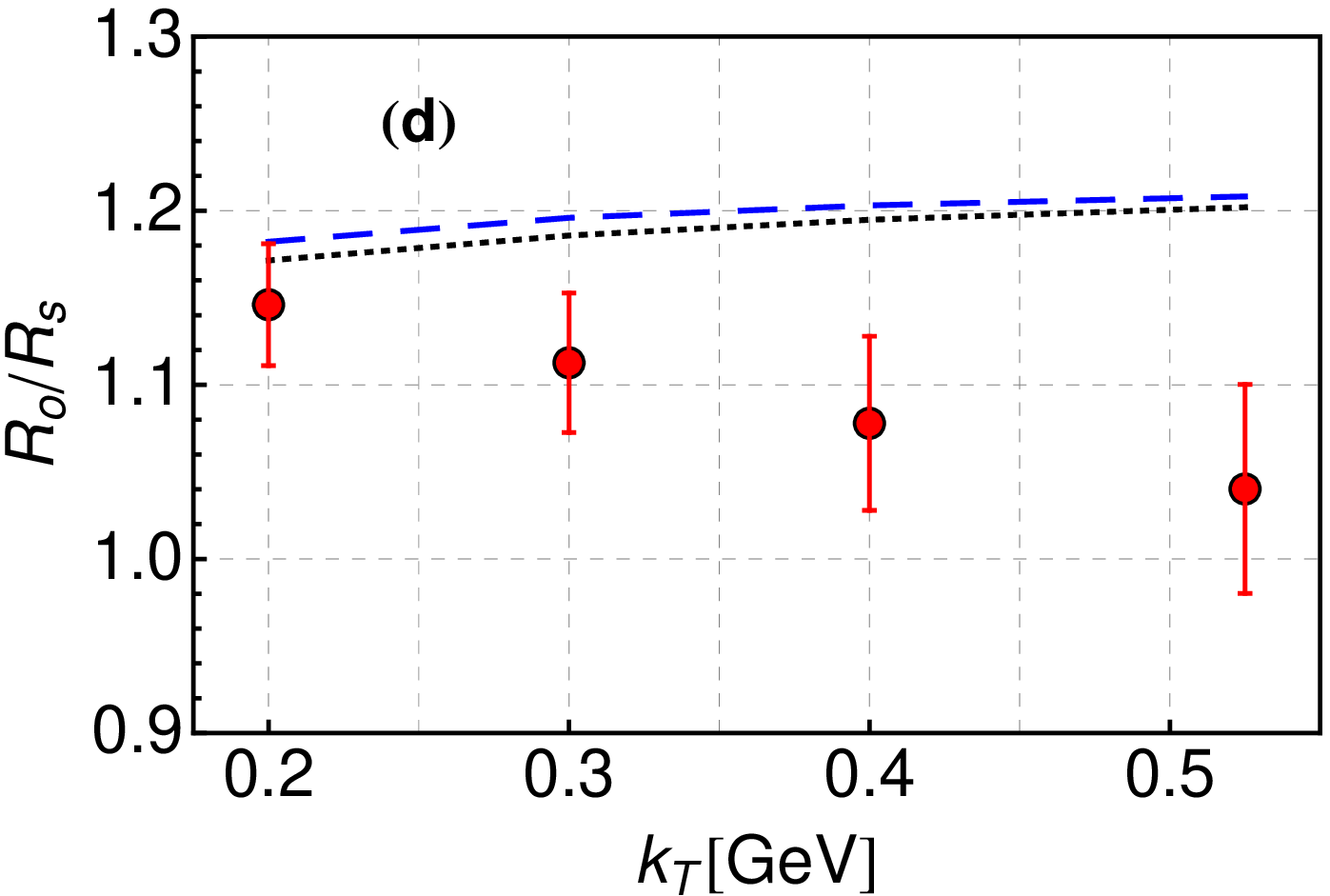}}
\end{center}
\caption{\small The HBT correlation radii: $R_{\rm out}$ \textbf{(a)}, $R_{\rm side}$ \textbf{(b)}, $R_{\rm long}$ \textbf{(c)} and ratio $R_{\rm out}/R_{\rm side}$ \textbf{(d)} of positive pions as a function of total transverse momentum of the pair for centrality  $c=0-5$\%  for the case \textbf{i)} (blue dashed line)  and \textbf{ii)} (black dotted line).
}
\label{fig:hbt_mod}
\end{figure}
The spatial dependence of the initial anisotropy has a small effect on the HBT radii. In Fig.~\ref{fig:hbt_mod} we present the HBT radii calculated for the case \textbf{i)} (blue dashed line)  and \textbf{ii)} (black dotted line). We observe a small improvement of $R_{\rm side}$ which results in better agreement of $R_{\rm out}/R_{\rm side}$ in the case \textbf{ii)} as compared to the case \textbf{i)}.
\chapter{Summary and conclusions}
\label{chapter:sum}
\par In this Thesis, a recently developed framework of highly-anisotropic and strongly-dissipative hydrodynamics -- \texttt{ADHYDRO} -- has been introduced and used to analyze the space-time evolution of matter produced in ultra-relativistic heavy-ion collisions. The main goal of this analysis was to study the effect of  initial highly-anisotropic stages on the final soft hadronic observables typically measured in the experiment. The study was done in the context of the heavy-ion measurements performed at RHIC (Relativistic Heavy Ion Collider) w Brookhaven National Laboratory. 
\par Starting from the general assumption about the form of the phase-space distribution function,  thermodynamic properties of locally anisotropic systems of particles have been studied and the form of the generalized equation of state has been formulated. The dynamic equations determining the evolution of a highly-anisotropic fluid have been introduced. The form of the entropy source related to the mechanisms leading to thermalization of the system has been defined. In the simplest case of purely-longitudinal and boost-invariant expansion, different features of the model have been analyzed. In particular, for small deviations from equilibrium,  the consistency of the \texttt{ADHYDRO} model with the Israel-Stewart theory has been demonstrated. 
\par Using the \texttt{ADHYDRO} model in the general (3+1)D and boost-invariant (2+1)D versions, the three possible scenarios of early stages of heavy-ion collisions have been analyzed: i) the system initially oblate in momentum space, ii)  the system initially prolate in momentum space, iii) the initially isotropic system. The results of the \texttt{ADHYDRO} model have been compared to the results obtained from the reference, perfect-fluid hydrodynamic model -- \texttt{LHYQUID}. The results of the hydrodynamic models were coupled to the statistical Monte-Carlo model \texttt{THERMINATOR} which has allowed us to perform systematic study of soft-hadronic observables: transverse-momentum spectra, directed and elliptic flows, and the HBT radii. \medskip

The following final conclusions may be drawn:
\begin{itemize}
  \item[a)] All studied observables are almost insensitive to the initial anisotropic stage provided the initial conditions of the evolution are properly readjusted.
  \vfill
  \item[b)] Complete thermalization of matter may take place only at the times of about 1 fm/c. In this way the early thermalization puzzle may be circumvented. 
\end{itemize}

\medskip
Finally, we have studied the systems where the initial anisotropy is spatially dependent. Our results suggest that the system may be highly anisotropic at the edges and simultaneously well thermalized in the center. It should be stressed, however, that the correct description of the data has been achieved only in the situations where the complete thermalization has been reached at about 1 fm/c.

\appendix
\chapter{Notation}
\label{chapter:Ysymbol}
%
Here we gather the symbols used in the Thesis.
\begin{table}[!ht]
{\footnotesize
  \begin{tabular}{cl}
    \hline
    \multicolumn{2}{c}{}\\
    \multicolumn{2}{c}{\bf Locally anisotropic systems of particles}\\
    \multicolumn{2}{c}{}\\
    \hline\\
    $\lambda_{\perp}, \lambda_{\parallel}$		& transverse and longitudinal temperature,\\
    $P_{\perp}, P_{\parallel}$		& transverse and longitudinal pressure,\\
    $m$			& mass,\\
    $m_T, m_\perp$			& transverse mass,\\
    $p_T, p_\perp$			& transverse momentum,\\
    $p_L, p_\parallel$			& longitudinal momentum,\\  
    $\displaystyle x=\left(\lambda_{\perp}/\lambda_{\parallel}\right)^2=\xi +1$			& pressure anisotropy parameter,\\  
    $\displaystyle U^\mu = \gamma\, (1,v_x,v_y,v_z)$	& four-velocity,\\
    $\displaystyle V^\mu = \gamma_z\, (v_z,0,0,1)$	& four-vector defining longitudinal direction,\\
    $\displaystyle \gamma = (1-v^2)^{-\frac{1}{2}}$	& Lorentz factor,\\
    $\displaystyle \gamma_z = (1-v_z^2)^{-\frac{1}{2}}$	& Lorentz factor,\\
    $p^{\mu}$ & particle four-momentum,\\
    $T^{\mu\nu}$	& energy-momentum tensor,\\
    $S^{\mu}$ & entropy current,\\
    $N^{\mu}$ & particle number current,\\ 
    $\varepsilon, \sigma, n$	& energy density, entropy density, and particle density\\ & in the anisotropic system,\\
    $\varepsilon_{\rm eq}, \sigma_{\rm eq}, n_{\rm eq}$	& energy density, entropy density, and particle density\\ & in equilibrium,\\
    $T$		& temperature in the isotropic system,\\
    $\epsilon$		& statistics identifier (+1 for FD, -1 for BE),\\
    $g_0$			& degeneracy factor related to internal quantum \\ & numbers of particles,\\
    $R$			& pressure relaxation function,\\
    $r(x) = \pi^2 R(x)/ (3 g_0)$			& rescaled pressure relaxation function,\\    
    $c_{\rm s}$		& sound velocity,\\
    $T_{\rm c}$		& critical temperature,\\
  \end{tabular}
}
\end{table}
\begin{table}[!ht]
{\footnotesize
  \begin{tabular}{cl}
    \hline
    \multicolumn{2}{c}{}\\
    \multicolumn{2}{c}{\bf Hydrodynamics}\\
    \multicolumn{2}{c}{}\\
    \hline\\
    $\Sigma$ & entropy source,\\
    $u_x, u_y$ & $x$ and $y$ components of the transverse vector ${\bf u}_\perp = (u_x,u_y)$,\\
    $\displaystyle \vartheta = \mathrm{arctanh} \,v_z$ & longitudinal fluid rapidity,\\
    $\displaystyle \tau = \sqrt{t^2-z^2}$		& longitudinal proper time,\\
    $\displaystyle \mathrm{y}=\ln \frac{E_p+p_{\parallel}}{E_p-p_{\parallel}}$ & rapidity,\\
    $\displaystyle \eta = \frac{1}{2} \ln \frac{t+z}{t-z}$ & space-time rapidity,\\
    $P^{\mu\nu}$ & pressure tensor,\\ 
    $\pi^{\mu\nu}$ & viscous shear tensor,\\ 
    $\Delta^{\mu\nu}=g^{\mu\nu}-U^{\mu}U^{\nu}$ & projection tensor,\\ 
    $\Pi$ & viscous bulk pressure,\\
    $\overline{\pi}=\pi^{33}$ & 33 component of the shear tensor,\\
    $\tau_{\pi}$ & shear relaxation time,\\
    $\eta_{\pi}$ & shear viscosity,\\
    $\displaystyle \theta=\partial_\mu U^\mu$ & volume expansion rate,\\
    $\tau_{\rm eq}$ & time-scale parameter,\\
    $\Gamma$ & inverse of the relaxation time,\\
    $\tau_0$ & initial proper time for starting hydrodynamics,\\
    
    \multicolumn{2}{c}{}\\
    \hline\\
    \multicolumn{2}{c}{\bf Initial conditions}\\
    \multicolumn{2}{c}{}\\
    \hline\\
    $\rho$		& density of sources,\\
    $\tilde{\rho}$		& normalized density of sources,\\    
    $\varepsilon_{\rm i}$	& initial energy density at the center of the system\\ & in most central collisions,\\
    $\rho_{\rm W}$	& wounded-nucleon density,\\
    $\rho_{\rm B}$	& binary-collisions density,\\
    $\rho_{\rm WS}$	& Woods-Saxon nuclear density profile,\\
    $\kappa$		& mixing factor between wounded-nucleon,\\ & and binary-collisions densities,\\    
    $\mathcal{T}$		& thickness function,\\
    $\sigma_{\rm in}$	& total inelastic $N N$ cross-section,\\
    $\rho_0, R_A, a$	& parameters for Woods-Saxon nuclear profile,\\
    ${\bf b}$		& impact vector,\\
    $\displaystyle b=|{\bf b}|$		& impact parameter,\\
    $f(\eta)$ & initial longitudinal profile,\\
    $\Delta\eta, \sigma_\eta$ & parameters of the initial longitudinal profile,\\
    $\theta$ & step function,\\
    $m_{\rm N}$ & nucleon mass,\\
    $\mathrm{y}_b$ & rapidity of the participant nucleon,\\
    $\eta_m$ & range of rapidity correlations,\\
  \end{tabular}
}
\end{table}
\begin{table}[!ht]
{\footnotesize
  \begin{tabular}{cl} 
    \multicolumn{2}{c}{}\\
    \hline\\
    \multicolumn{2}{c}{\bf Freeze-out}\\
    \multicolumn{2}{c}{}\\
    \hline\\
    $T_{\rm f}$ & freeze-out temperature,\\
    $d\Sigma^{\mu}$		& element of the three-dimensional freeze-out hypersurface,\\
    $\overline{g}$ & spin degeneracy factor,\\
    $\mu$		& chemical potential,\\
    $\mu_B, \mu_{I_3}, \mu_S, \mu_C$ & baryon, isospin, strange, and charm chemical potentials,\\
    $B, I_3, S, C$		& baryon number, third component of isospin, strangeness,\\ & and charm,\\
    $\displaystyle \phi_p = \arctan (p_y/p_x)$ &  azimuthal angle of the transverse momentum,\\
    $\zeta, \theta, \phi$		& angles parametrizing freeze-out hypersurface,\\
    $\Lambda$ & scale,\\
    $d(\zeta, \phi, \theta)$	& distance from the point $(\tau=\tau_{\rm 0},x=0,y=0,\eta=0)$\\ & to the hypersurface
                        point with coordinates $(\zeta, \phi, \theta)$.\\
  \end{tabular}
}
\end{table}

\bibliography{bibphd}
\end{document}